\documentclass[11pt,a4paper,twoside]{book}
\usepackage[english,spanish]{babel} 
\usepackage[latin1]{inputenc}
\usepackage{latexsym} % Símbolos
\usepackage{graphicx,type1cm} % Inclusión de gráficos. Soporte para \figura (más abajo)
\usepackage{epstopdf}
\usepackage{indentfirst}
\usepackage{amssymb,amsmath,amsthm}
\usepackage{enumitem}
\usepackage{cite}
\usepackage{verbatim}
\usepackage{epsfig}
\usepackage{xcolor}
\usepackage{color}
\usepackage[a4paper, inner=3.6cm, outer=2.5cm, top=3.5cm, bottom=3.5cm]{geometry}
\usepackage{subfigure}
\usepackage{xspace}
\usepackage{upgreek}
\usepackage{url}
\usepackage{slashbox,pict2e}
\usepackage{times}
\usepackage{latexsym}
\usepackage{graphicx}
\usepackage{ulem} 
\usepackage{booktabs}
\usepackage{multirow}
\usepackage{multicol}
\usepackage{array}
\usepackage{colortbl}
\usepackage{rotating}
\usepackage{changepage}
\usepackage{fancyvrb}
\usepackage{textcomp}  
\usepackage{fancyhdr}
\usepackage{marvosym}
\usepackage{setspace}
\usepackage{stfloats}
\usepackage[titletoc]{appendix}
\usepackage{hhline}
\usepackage{afterpage}
\usepackage{lscape}
\usepackage{listings}
\usepackage{caption}
\usepackage{datetime}
\usepackage[printonlyused]{acronym}
\usepackage{algorithm}
\usepackage{algorithmicx}
\usepackage{algpseudocode}
\newcommand\Algphase[1]{%
\vspace*{-.7\baselineskip}\Statex\hspace*{\dimexpr-\algorithmicindent-2pt\relax}\rule{\columnwidth}{0.4pt}%
\Statex\hspace*{-\algorithmicindent}\textbf{#1}%
\vspace*{-.7\baselineskip}\Statex\hspace*{\dimexpr-\algorithmicindent-2pt\relax}\rule{\columnwidth}{0.4pt}%
}

\usepackage{float}
\usepackage{blkarray} %added for matrix
\usepackage{tikz}%added to color  matrix element
\usetikzlibrary{calc}%added to color  matrix element
\newdateformat{mifecha}{Bilbao, a \THEDAY~de \monthname[\THEMONTH] de \THEYEAR}
\makeatletter
\renewcommand{\monthnamespanish}[1][\month]{%
  \@orgargctr=#1\relax
  \ifcase\@orgargctr
    \PackageError{datetime}{Invalid Month number \the\@orgargctr}{%
      Month numbers should go from 1 to 12}%
    \or Enero%
    \or Febrero%
    \or Marzo%
    \or Abril%
    \or Mayo%
    \or Junio%
    \or Julio%
    \or Agosto%
    \or Septiembre%
    \or Octubre%
    \or Noviembre%
    \or Diciembre%
    \else \PackageError{datetime}{Invalid Month number \the\@orgargctr}{%
      Month numbers should go from 1 to 12}%
  \fi}
\makeatother
\newcommand{\titFirstPart}{{D2D-Assisted Clustering}\xspace}
\newcommand{\titSecondPart}{{D2D for Network Optimization}\xspace}
\newcommand{\titFirstCh}{{Introduction}\xspace}
\newcommand{\titSecondCh}{{Background}\xspace}
\newcommand{\titThirdCh}{{Theoretical Analysis}\xspace}
\newcommand{\titForthCh}{{A ProSe-Compliant Opportunistic D2D Protocol}\xspace}
\newcommand{\titFifthCh}{{Experimental Evaluation}\xspace}
\newcommand{\titSixthCh}{{D2D Mode Selection}\xspace}
\newcommand{\titSeventhCh}{{D2D-Assisted Tie Breaking}\xspace}
\newcommand{\titEighthCh}{{Conclusions}\xspace}

\newtheorem{corollary}{Corollary}

\newcommand{\remove}[1]{}

\renewenvironment{proof}{\indent\textit{Proof}:}{\hfill$\blacksquare$}
\addto\captionsenglish{}

\newcommand{\thesisTitle}{Opportunistic Device-to-Device Communication in Cellular networks: From Theory to Practice\xspace}
\newcommand{\autor}{Arash Asadi\xspace}
\newcommand{\stdFilia}{IMDEA Networks Institute, University Carlos III of Madrid\xspace}
\newcommand{\stdEmail}{arash.asadi@imdea.org\xspace}
\newcommand{\supervisor}{Vincenzo Mancuso\xspace}
\newcommand{\supFilia}{IMDEA Networks Institute\xspace}
\newcommand{\supervisorS}{Ruben Cuevas Rumin\xspace}
\newcommand{\supFiliaS}{University Carlos III of Madrid\xspace}
\newcommand\clearemptydoublepage{\clearpage{\pagestyle{empty}\cleardoublepage}}

\newcommand{\tabincell}[2]{\begin{tabular}[t]{@{}#1@{}}#2\end{tabular}}

\newtheorem{proposition}{Proposition}
\newtheorem*{proposition*}{Proposition}
\newtheorem{defn}{Definition}
\newcolumntype{L}[1]{>{\raggedright\let\newline\\\arraybackslash\hspace{0pt}}m{#1}}
\newcolumntype{C}[1]{>{\centering\let\newline\\\arraybackslash\hspace{0pt}}m{#1}}
\newcolumntype{R}[1]{>{\raggedleft\let\newline\\\arraybackslash\hspace{0pt}}m{#1}}

\let\mathsetfont\mathcal

\newcommand\setD{\mathsetfont D}

\newcommand\setH{\mathsetfont H}

\newcommand\setL{\mathsetfont L}

\newcommand\setN{\mathsetfont N}

\newcommand\veczero{\boldsymbol 0}

\newcommand\vece{\boldsymbol e}

\newcommand\vecg{\boldsymbol g}
\newcommand\vech{\boldsymbol h}

\setitemize[0]{itemindent=25pt} % indentation of an itemize item

    \makeatletter
    \def\thebibliography#1{\chapter*{References\@mkboth
      {REFERENCES}{REFERENCES}}\list
      {[\arabic{enumi}]}{\settowidth\labelwidth{[#1]}\leftmargin\labelwidth
	\advance\leftmargin\labelsep
	\usecounter{enumi}}
	\def\newblock{\hskip .11em plus .33em minus .07em}
	\sloppy\clubpenalty4000\widowpenalty4000
	\sfcode`\.=1000\relax}
    \makeatother
 
\renewcommand{\thepart}{Part \Roman{part} :}

\usepackage{titlesec}
\titleformat{\part}{\huge\bfseries}{}{0pt}{\thepart\quad}
\begin{document}

\setcounter{secnumdepth}{4}
\setcounter{tocdepth}{2}
\sloppy
\onehalfspacing

% -------------------------------------------------
% title pages, Cover,...
% -------------------------------------------------

\begin{titlepage}
%\pagenumbering{alph}
\selectlanguage{spanish}
\vspace*{-2cm}
\begin{center}
        \centerline{\includegraphics[width=10em]{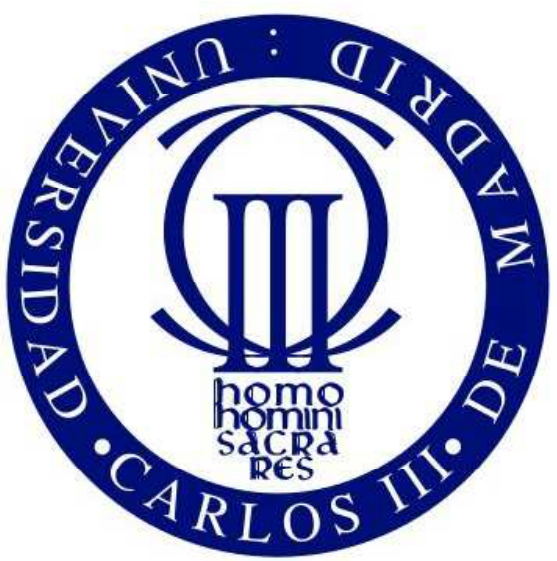}}
   %\vspace*{0.5cm}
   {\large UNIVERSIDAD CARLOS III DE MADRID} \\ \vspace*{0.25cm}
  	
   \vfill
    \vfill
   {\LARGE TESIS DOCTORAL}\\
    \vspace*{1cm}
			
    {\Large \sc \thesisTitle} 
			\vspace*{0.7cm}
			
			\vfill
			\begin{table}[!h]
			\hspace*{1.5cm}
				\begin{tabular}{rl}
					{\large Autor:} 		& {\large\autor, \stdFilia}\\
					{\large Director:} 	& {\large\supervisor, \supFilia} \\
					{\large Tutor:} 		&{\large\supervisorS, \supFiliaS} \\
%					& {\large{\supervisorT, \supFiliaT}} \\			
				\end{tabular}
			\end{table}	
			\vfill
   
\end{center}

  \vfill

\begin{center}
  
{\large DEPARTAMENTO DE INGENIER\'IA TELEM\'ATICA} \\ \vspace*{1cm}

  {\large Legan\'es (Madrid), \monthname~de \the\year} \\
  \vspace{0.2cm}

\end{center}

%\vspace*{-1cm}

\end{titlepage}

\clearemptydoublepage

\begin{titlepage}
\selectlanguage{english}
%\pagenumbering{alph}
\pagenumbering{roman}
\setcounter{page}{3}

\vspace*{-2cm}

\begin{center}
        \centerline{\includegraphics[width=10em]{./img/uc3m}}
   %\vspace*{0.5cm}
   {\large UNIVERSIDAD CARLOS III DE MADRID} \\ \vspace*{0.25cm}
  	
   \vfill
    \vfill
   {\LARGE PH.D. THESIS}\\
    \vspace*{1cm}
			
    {\Large \sc \thesisTitle} 
			\vspace*{0.7cm}
			
			\vfill
			
			\begin{table}[!h]
			\hspace*{1.5cm}
				\begin{tabular}{rl}
					{\large Author:} 		& {\large\autor, \stdFilia}\\
					{\large Director:} 	& {\large\supervisor, \supFilia} \\
					{\large Tutor:} 		&{\large\supervisorS, \supFiliaS} \\
%					& {\large{\supervisorT, \supFiliaT}} \\			
				\end{tabular}
			\end{table}	
			\vfill
   
\end{center}

  \vfill

\begin{center}
  
{\large DEPARTMENT OF TELEMATIC ENGINEERING} \\ \vspace*{1cm}

  {\large Legan\'es (Madrid), \monthname~\the\year} \\
  \vspace{0.2cm}
\end{center}

%\vspace*{-1cm}

\end{titlepage}

\thispagestyle{empty}
\cleardoublepage

\thispagestyle{empty}

\begin{flushleft}
\textit{\thesisTitle}\\[0.5in]

A dissertation submitted in partial fulfillment of the requirements for the degree of Doctor of Philosophy\\[0.5in]
Prepared by

\autor, \stdFilia\\[0.5in]
								
Under the advice of

\supervisor, \supFilia \\
\supervisorS, \supFiliaS \\
%\supervisorT, \supFiliaT \\	[0.5in]		

Departamento de Ingenier\'ia Telem\'atica,
Universidad Carlos III de Madrid

\end{flushleft}
\noindent\makebox[\linewidth]{\rule{\paperwidth}{0.4pt}}\\[0.5in]
Date: 		\monthname, \the\year\\[0.25in]
Web/contact:	\stdEmail\\[0.25in]
This work has been supported by IMDEA Networks Institute.

\begin{figure}[ht!]
%\centering
\includegraphics[width=50mm]{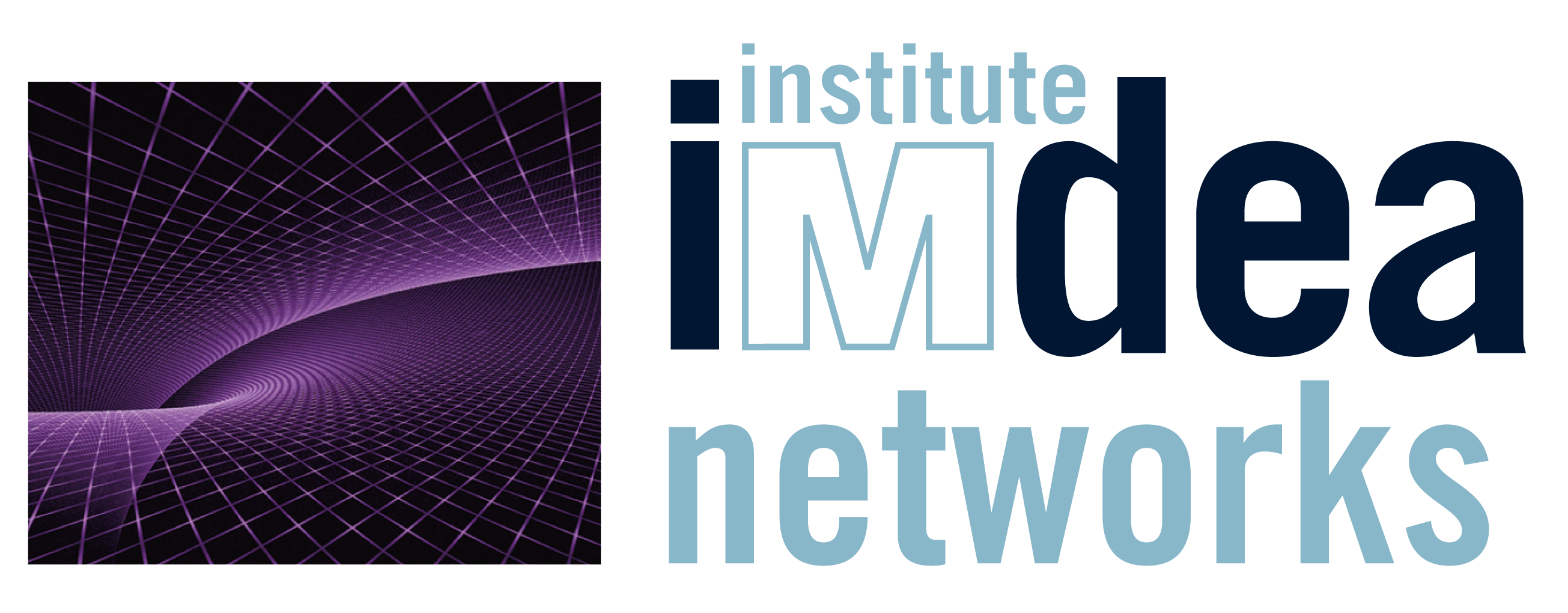}
%\caption{A simple caption \label{overflow}}
\end{figure}

\cleardoublepage

\thispagestyle{empty}

\begin{center}
{\large TESIS DOCTORAL}\\
\vspace*{1cm}
{\large \sc \thesisTitle\\} 
			\vspace*{1cm}
			\begin{table}[!h]
			\hspace*{2cm}
				\begin{tabular}{rl}
					{\large Autor:} 		& {\large\autor, \stdFilia}\\
					{\large Director:} 	& {\large\supervisor, \supFilia} \\
					{\large Tutor:} 		&{\supervisorS, \supFiliaS} \\
%					& {\large{\supervisorT, \supFiliaT}} \\			
				\end{tabular}
			\end{table}	

\end{center}   

\vspace{2 cm}

Firma del tribunal calificador:
\vspace{2cm}

\hspace{1.5 cm}        Presidente: Prof. Douglas Leith

        \vspace{2cm}
\hspace{1.5 cm}        Vocal:  Dr. Carla Fabiana Chiasserini
                                                                                     
        \vspace{2cm}
\hspace{1.5 cm}	Secretario: Prof. Albert Banchs

        \vspace{2cm}
        Calificaci\'on:
           \vspace{1 cm}
            \begin{flushright}
               Legan\'es, {\hspace{.5cm}} de {\hspace{3cm}} de
                 {\hspace{1.0cm}}
    \end{flushright}
\clearemptydoublepage
 
\thispagestyle{empty}
\selectlanguage{english}

% Cover:
%\maketitle

%abstract & acknowledgements
\chapter*{Acknowledgements}
\addcontentsline{toc}{chapter}{Acknowledgements}

First and foremost I want to thank my advisor Vincenzo Mancuso. It has been an honor to be his first Ph.D. student.  I appreciate all his contributions, the late nights spent on reviewing my papers, funding to make my Ph.D. experience productive and stimulating, and insightful discussions about my research.
I also have to thank the members of my PhD committee.

My special thanks goes to my other Ph.D. fellows, Ignacio De Castro Arribas, Thomas Nitsche, and Qing Wang with whom I had long meetings in our favorite meeting room (a.k.a, kitchen). I am grateful to Christian Vitale and Vincenzo Sciancalepore for their collaborations. 

I will forever be thankful to my life partner, Allyson Sim, whose support and enthusiasm was vital to my progress within the past 6 years. I hope I can return her support and kindness for years to come. 

Last, but by no means least, thanks go to mum, dad and my sister for almost unbelievable support. They are the most important people in my world and I dedicate this thesis to them.
\clearemptydoublepage

%!TEX root =  ../main.tex
\chapter*{Abstract}
\addcontentsline{toc}{chapter}{Abstract}

Cellular service providers have been struggling with users' demand since the emergence of mobile Internet. As a result, each generation of cellular network prevailed over its predecessors mainly in terms of connection speed. However, the fifth generation (5G) of cellular network promises to go beyond this trend by revolutionizing the network architecture. \ac{D2D} communication is one of the revolutionary changes that enables mobile users to communicate directly without traversing a base station. This feature is being actively studied in 3GPP with special focus on public safety as it allows mobiles to operate in adhoc mode. Although under the (partial) control of the network, \ac{D2D} communications open the door to many other use-cases. 

This dissertation studies different aspects of \ac{D2D} communications and its impact on the key performance indicators of the network. We design an architecture for the collaboration of cellular users by means of timely exploited \ac{D2D} opportunities. We begin by presenting the analytical study on opportunistic outband \ac{D2D} communications. The study reveals the great potential of opportunistic outband \ac{D2D} communications for enhancing energy efficiency, fairness, and capacity of cellular networks when groups of \ac{D2D} users can be form and managed in the cellular network. Then we introduce a protocol that is compatible with the latest release of IEEE and 3GPP standards and allows for implementation of our proposal in a today's cellular network. To validate our analytical findings, we use our experimental \ac{SDR}-based testbed to further study our proposal in a real world scenario. The experimental results confirm the outstanding potential of opportunistic outband \ac{D2D} communications. Finally, we investigate the performance merits and disadvantages of different \ac{D2D} ``modes''. Our investigation reveals, despite the common belief, that all \ac{D2D} modes are complementary and their merits are scenario based.

\clearemptydoublepage

\phantomsection
\addcontentsline{toc}{chapter}{Table of Contents}
% Index, Prologue, ...
\tableofcontents

%\addcontentsline{toc}{chapter}{Contents}
%\clearemptydoublepage

\listoftables
\addcontentsline{toc}{chapter}{List of Tables}
\clearemptydoublepage

\listoffigures
\addcontentsline{toc}{chapter}{List of Figures}
\clearemptydoublepage

\renewcommand*\listfigurename{List of Acronyms}
\listoffigures
\addcontentsline{toc}{chapter}{List of Acronyms}
\let\standardclearpage\clearpage
\let\clearpage\relax
%\chapter*{List of Acronyms}
\begin{acronym}
\acro{AADTR}{Average Absolute Deviation of Transmission Rate}
\acro{ADC}{Analog to Digital Conversion}
\acro{AMBR}{Aggregate Maximum Bit Rate}
\acro{ARP}{Allocation and Retention Priority}
\acro{BCH}{}
\acro{BE}{Best Effort}
\acro{BITS}{BS-drIven Traffic Spreading}
\acro{BS}{Base Station}
\acro{BSR}{Buffer Status Report}
\acro{C-RNTI}{Cell Radio Network Temporary Identifier}
\acro{CAC}{Call Admission Control}
\acro{CCH}{Control CHannel}
\acro{CDF}{Cumulative Distribution Function }
\acro{CEB}{Cooperative Eigen Beamforming}
\acro{CL(MR)}{Cluster Max Rate}
\acro{CL(WRR)}{Cluster Weighted Round Robin}
\acro{CQI}{Channel Quality Indicator}
\acro{CSI}{Channel State Information}
\acro{D2D}{Device-to-Device}
\acro{DAC}{Digital to Analog Conversion}
\acro{DCI}{Downlink Control Information}
\acro{DHCP}{Dynamic Host Configuration Protocol}
\acro{DORE}{D2D Opportunistic Relay with QoS-Enforcement}
\acro{DRONEE}{Dual Radio Opportunistic Networking for Energy Efficiency}
\acro{DRX}{Discontinuous Reception}
\acro{DSP}{Digital Signal Processing}
\acro{E-RAB}{E-UTRAN Radio Access Bearer}
\acro{EPC}{Evolved Packet Core}
\acro{EPS}{Evolved Packet System}
\acro{ES}{Equal Share}
\acro{FAM}{FlexRIO Adaptor Module}
\acro{FFT}{Fast Fourier Transform}
\acro{FPGA}{Field Programmable Gate Array}
\acro{FSMC}{Finite State Markov Chain}
\acro{GBR}{Guaranteed Bit Rate}
\acro{GO}{Group Owner}
\acro{GUTI}{Globally Unique Temporary UE Identity}
\acro{HARQ}{Hybrid Automatic Repeat Request}
\acro{ICI}{Inter Cell Interference}
\acro{IE}{Information Element}
\acro{iFFT}{inverse Fast Fourier Transform}
\acro{IM}{Implementation Margin}
\acro{IMEI}{International Mobile Equipment Identity}
\acro{IMSI}{International Mobile Subscriber Identity}
\acro{ISM}{Industrial, Scientific and Medical}
\acro{LTE}{Long Term Evolution}
\acro{M-LWDF}{Modified Largest Weighted Delay First}
\acro{MBR}{Maximum Bit Rate}
\acro{MCS}{Modulation and Coding Scheme}
\acro{MME}{Mobility Management Entity}
\acro{MR}{Max Rate}
\acro{NAS}{Non Access Stratum}
\acro{OFDM}{Orthogonal Frequency-Division Multiplexing}
\acro{OFDMA}{Orthogonal Frequency-Division Multiple Access}
\acro{OTA}{Over The Air}
\acro{P-GW}{PDN Gateway}
\acro{PCEF}{Policy and Charging Enforcement Function}
\acro{PCRF}{Policy and Charging Rules Function}
\acro{PDCCH}{Physical Downlink Control CHannel}
\acro{PDCP}{Packet Data Convergence Protocol}
\acro{PDN}{Packet Data Network}
\acro{PDSCH}{Physical Downlink Shared CHannel}
\acro{PDU}{Packet Data Unit}
\acro{PDU}{Packet Data Unit}
\acro{PF}{Proportional Fair}
\acro{ProSe}{Proximity-based Services}
\acro{PSNR}{Peak Signal-to-Noise Ratio} 
\acro{PUCCH}{Physical Uplink Control CHannel}
\acro{PUSCH}{Physical Uplink Shared CHannel}
\acro{QCI}{QoS Class Identifier}
\acro{QoE}{Quality of Experience}
\acro{QoS}{Quality of Service}
\acro{RF}{Radio Frequency}
\acro{RLC}{Radio Link Control}
\acro{RR}{Round Robin}
\acro{RRC}{Radio Resource Control}
\acro{S-GW}{Serving Gateway}
\acro{S-TMSI}{SAE-Temporary Mobile Subscriber Identity}
\acro{SAE}{System Architecture Evolution}
\acro{SC-FDMA}{Single Carrier-FDMA}
\acro{SCH}{Shared CHannel}
\acro{SDR}{Software Defined Radio}
\acro{SDU}{Service Data Unit}
\acro{SINR}{Signal-to-Interference-plus-Noise Ratio}
\acro{SNR}{Signal-to-Noise Ratio}
\acro{SR}{Scheduling Request}
\acro{SRB}{Signaling Radio Bearer}
\acro{SS}{Shapely Share}
\acro{SSIM}{Structural Similarity} 
\acro{TB}{Transport Block}
\acro{TDMA}{Time Division Multiple Access}
\acro{TEID}{Tunneling Endpoint ID}
\acro{TFT}{Traffic Flow Template}
\acro{TMSI}{Temporary Mobile Subscriber Identity}
\acro{UE}{User Equipment}
\acro{UIT}{Uplink Information Transfer}
\acro{WPS}{Wireless Protected Setup}
\acro{WRR}{Weighted Round Robin}
\acro{WS}{Weighted Share}
\acro{WSL}{Workload-base Scheduling with Learning}
\acro{}{}
\end{acronym}
\clearemptydoublepage
\let\clearpage\standardclearpage

\pagenumbering{arabic}
%\setcounter{page}{1}

%Chapters

\chapter{\titFirstCh}
\label{ch:intro}

The emergence of smartphones and their ever expanding role in our daily life transformed the cellular networks from an option to a necessity. This transformation created a flow of cash to the cellular operators' accounts and traffic to their infra-structure. Initially, the operators leveraged the former to support the latter.  Their first actions were increasing the number of base stations and buying more bandwidth to improve the cellular network capacity. However, these measures did not help the operators to catch up with the traffic demand of mobile users'. Next step was to exploit the cellular resources in a more efficient manner.  As a result, the 3rd and 4th generation (3G and 4G) of cellular technologies (e.g., UMTS and LTE) demonstrated more efficient resource utilization in comparison to their predecessors. In particular, use of opportunistic scheduling techniques~\cite{asadi2013survey} to leverage wireless channel diversities among cellular users become customary. Nevertheless, operators were still lagging behind mobile users' booming traffic demand. Finally, some realized that while the application of cellular networks has evolved tremendously within the past 20 years, its network architecture changed very little. This triggered an architectural revolution towards 5G cellular standards and introduced a new paradigm called \ac{D2D} communication~\cite{asadi2014survey}. 

\ac{D2D} communication in cellular networks is defined as direct communication between mobile users without traversing the \ac{BS} or core network. In a traditional cellular network, all communications must go through the \ac{BS} even if both communicating parties are in range for direct communication. This architecture suits the conventional low data rate mobile services such as voice call and text message in which end users are not usually close enough to have direct communication. However, mobile users in today's cellular networks use high data rate for services (e.g., video sharing, gaming, proximity-aware social networking) that involve users potentially in range of direct communications (i.e., \ac{D2D}). Hence, \ac{D2D} communications in such scenarios can highly increase the spectral efficiency of the network not only because of avoiding unnecessary transmissions to and from the \ac{BS}, but also because of the higher data rates achievable at lower power levels when users are in proximity. Nevertheless, the advantages of \ac{D2D} communication is not only limited to enhanced spectral efficiency.

The advent of \ac{D2D} communication has set off numerous proposals in industry and academia to improve the performance of cellular networks. 
As of today there are not only several proposals for cellular relaying, multicasting, cellular offloading, and content distribution leveraging \ac{D2D}~\cite{lin_multihop_2000, asadi2013survey}, but also entire system architectures based on \ac{D2D} to {\it complement}  cellular-based services in a scalable way with new types of applications~\cite{wu2013TON}.
In addition to academia and telecommunication companies, 3GPP is also investigating \ac{D2D} communications as Proximity Services (ProSe)~\cite{3GPPTR22.803}.
Indeed, 3GPP is actively studying the feasibility and the architecture of ProSe to finalize the standardization process for both inband and outband \ac{D2D} modes, in which inband \ac{D2D} uses the cellular spectrum, while outband \ac{D2D} uses unlicensed spectrum. In particular, the feasibility of ProSe and its use-cases in LTE are studied in~\cite{3GPPTR22.803} and the required architectural enhancements to accommodate such use-cases are investigated in~\cite{3GPPTR23.703}. Release 12 of 3GPP already specifies system overview and discovery procedure of  \ac{D2D} communications. Moreover, there are still ongoing studies on architecture enhancements and radio management aspects of \ac{D2D}~\cite{3GPP23.303,3GPP36.843}. 

So far, only the operator's infra-structure was involved in data delivery. By allowing mobile users to relay the cellular traffic for other users, \ac{D2D} communications unleash the true potentials of user cooperations. Hence, mobile users have a chance to join forces with the operators and leverage cooperative communication techniques to enhance the resource utilization efficiency using more aggressive opportunistic scheduling schemes. 

In this dissertation, we cover various aspects and use-cases of \ac{D2D} communication. Indeed, we pioneer to propose outband \ac{D2D} communications and to design adaptive multi-band multi-mode \ac{D2D} communication. The main contributions of this dissertation are:

\begin{itemize}
\item We provide a model for throughput and energy consumption for opportunistic outband D2D-clustering under LTE-A and WiFi Direct technologies; 
\item We use coalitional game theory to study the revenue distribution and fairness in presence of opportunistic outband D2D-clustering;
\item We propose the first protocol that accommodates network-assisted D2D clustering within the current LTE-A and WiFi Direct framework with minor changes to the existing network architecture and protocols;
\item We use the state-of-the-art \ac{SDR} platforms to implement the first outband D2D prototype in order to verify our analytical findings with real world experiments;
\item We propose {\it floating band D2D} technique in which D2D users leverage a multi-mode multi-band D2D environment;
\item We are the first to use the properties of D2D to improve fairness of MaxRate scheduler without any throughput penalty by exploiting smart tie-breaking schemes. 
\end{itemize}

\section{Roadmap}
This dissertation starts with a short background on opportunistic scheduling and \ac{D2D} communications in cellular networks to familiarize the reader with the terminology and the state-of-the-art. Next, we study our proposed {\it network-assisted \ac{D2D} clustering} techniques in which we show the merits of simple opportunistic scheduling and cooperative communication techniques in \ac{D2D} architecture. Then, we address the problem of {\it mode selection} in \ac{D2D}-enabled networks. Finally, we investigate our proposed {\it \ac{D2D} tie-breaking} mechanism to improve user fairness without impacting the aggregate system throughput. In the following, we provide more details about each of these parts composing this dissertation.

\subsection{Network-Assisted D2D Clustering}
From Chapter~\ref{ch:clus_theo}~to~\ref{ch:clus_exp}, we present a channel-opportunistic architecture that leverages outband \ac{D2D} communications and opportunistic clustering techniques. Specifically, we build on top of the forthcoming \ac{D2D} features of LTE-A networks, and on WiFi Direct. In our proposal, mobile users form clusters opportunistically, in which only the user with the best channel condition communicates with the base station on behalf of the entire cluster. Within the cluster, WiFi Direct is used to relay traffic. Our architecture benefits \ac{D2D} users in terms of throughput and energy efficiency, for which we provide an analytical model. In addition, we use coalitional game theory to find a suitable payoff distribution among \ac{D2D} users. Focusing on the implementation feasibility of \ac{D2D} communications in a network controlled by an operator, we introduce a \ac{D2D} protocol based on the features of LTE-A and WiFi Direct. We use simulations and real experiments to validate the superiority of \ac{D2D}-based cluster communication schemes over conventional cellular communications in terms of throughput, delay, fairness and energy efficiency. In particular, we develop \ac{DORE} algorithm in our experimental evaluation that is based on algorithms used in our simulation with an additional delay-awareness feature.
The results of the work presented in these chapters appeared in three peer-reviewed journal articles~\cite{asadi2014survey, asadi2013survey, asadi2014ComCom}, one magazine article~\cite{asadi2015Commag},  six conference papers~\cite{asadi2013WD,asadi2013MSWIM, asadi20155garch, asadi2015WInnComm, vitale2015Greencom,asadi2016Infocom}, three posters~\cite{asadi2013energy, asadi2013IMDEA, asadi2015IMDEA}, and one demo~\cite{asadi2015EUCNC}.

\subsection{D2D for Network Optimization}
In Chapter~\ref{ch:mode}, we propose Floating Band \ac{D2D}, an adaptive framework to exploit the full potential of \ac{D2D} transmission modes. We show that inband and outband \ac{D2D} modes exhibit different pros and cons in terms of complexity, interference, and spectral efficiency. Moreover, none of these modes are suitable as a one-size-fits-all solution for today's cellular networks, due to diverse network requirements and variable users' behavior. Therefore, we unveil the need for going beyond traditional single-band mode-selection schemes. Specifically, we model and formulate a general and adaptive multi-band mode selection problem, namely Floating Band \ac{D2D}. The problem is NP-hard, so we propose simple yet effective heuristics. Our results show the superiority of the Floating Band \ac{D2D} framework, which dramatically increases network utility and achieves near complete fairness.
The results of the work presented in this chapter appeared into two conferences~\cite{asadi2015wowmom,asadi2014Mama} and one peer-reviewed journal~\cite{asadi2014Per}. 

%\subsection{\ac{D2D} Tie-Breaking}
Opportunistic schedulers such as MaxRate and Proportional Fair are known for trading off throughput and fairness of users in cellular networks. In Chapter~\ref{ch:tie}, we show how to achieve maximum fairness without sacrificing throughput. We propose a novel solution that integrates opportunistic scheduling design principles and cooperative \ac{D2D} communications capabilities in order to improve both fairness and capacity in cellular networks. We develop a mathematical approach and design smart tie-breaking mechanisms which enhance the fairness achieved by the MaxRate scheduler. We show that users that cooperatively form clusters benefit from both higher throughput and fairness. Our scheduling mechanism is simple to implement and scales linearly with the number of clusters, and is able to achieve equal or better fairness than Proportional Fair schedulers. 
The results of the work presented in this chapter is still under review process in a journal and a conference. However, part of the results is already presented in one demo~\cite{asadi2014IMDEA}.

\section{Publications}
The research performed in this dissertation resulted in eight conference/workshop papers~\cite{asadi2013WD, asadi2013MSWIM, asadi2015wowmom, asadi20155garch, asadi2015WInnComm, asadi2014Mama,vitale2015Greencom,asadi2016Infocom}, four journal articles~\cite{asadi2014survey, asadi2013survey, asadi2014ComCom, asadi2014Per}, one magazine article~\cite{asadi2015Commag},  four posters~\cite{asadi2013energy, asadi2013IMDEA, asadi2014IMDEA, asadi2015IMDEA} and one demo~\cite{asadi2015EUCNC}. In this section, we elaborate on the goal of each paper and the author's contribution.

The following surveys are the outcome of our literature review and investigation into open problems in opportunistic scheduling and \ac{D2D} communications in cellular networks. 
\begin{itemize}
\item {\textit{\textbf{Arash Asadi } and Vincenzo Mancuso }, {``A Survey on Opportunistic Scheduling in Wireless Communications''},  IEEE Surveys and Tutorials on Communications, 2013}.

\item {\textit{\textbf{Arash Asadi }, Qing Wang, and Vincenzo Mancuso }, {``A Survey on Device-to-Device Communication in Cellular Networks''},  IEEE Surveys and Tutorials on Communications, 2014}.
\end{itemize}

We published our proposed architecture for opportunistic outband D2D-clustering and the analytical models for the throughput and the energy consumption of dual-radio devices using LTE-A and WiFi direct in the the following papers. 
\begin{itemize}
\item{\textit{\textbf{Arash Asadi } and Vincenzo Mancuso}, {``Energy Efficient Opportunistic Uplink Packet Forwarding in Hybrid Wireless Networks''}, in Proceeding of the ACM International Conference on Future Energy Systems (e-Energy), Berkeley, CA, USA, 2013}.

\item {\textit{\textbf{Arash Asadi } and Vincenzo Mancuso}, {``On the Compound Impact of Opportunistic Scheduling and \ac{D2D} Communications in Cellular Networks''}, in Proceeding of the ACM International Conference on Modeling, Analysis and Simulation of Wireless and Mobile Systems (MSWIM), Barcelona, Spain, 2013}.

\item {\textit{\textbf{Arash Asadi } and Vincenzo Mancuso }, {``DRONEE: Dual-radio Opportunistic Networking for Energy Efficiency''},  Elsevier Computer Communications, 2014}.
\end{itemize}

Our Wireless Days paper is dedicated to design of a protocol for D2D-clustering that can be easily adapted to the current LTE-A and WiFi Direct architecture. 
In this paper, we also evaluate the performance of our proposal using packet-level simulation. 
\begin{itemize}
\item {\textit{\textbf{Arash Asadi } and Vincenzo Mancuso}, {``WiFi Direct and LTE D2D in Action''}, in Proceeding of the IEEE Wireless Days, Valencia, Spain, 2013 }.
\end{itemize}

Our INFOCOM paper focuses on a 3GPP compliant structure for outband D2D communications. We also present the first SDR-based outband D2D communication platform and its performance metrics in this paper. 
\begin{itemize}
\item {\textit{\textbf{Arash Asadi }, Vincenzo Mancuso and Rohit Gupta}, {``An SDR-based Experimental Study of Outband D2D Communications'}, in Proceeding of the IEEE INFOCOM, San Francisco, USA, 2016}.
\end{itemize}

The floating band idea is presented in the following papers. These papers highlight the importance of adaptability and use of all \ac{D2D} modes in \ac{D2D} communications. Our contributions in these papers are the idea, evaluation and assistance in the development of the mathematical analysis. 
\begin{itemize}
\item {\textit{\textbf{Arash Asadi }, Peter Jacko, and Vincenzo Mancuso }, {``Modeling D2D Communications with LTE and WiFi''},  ACM SIGMETRICS Performance Evaluation Review, 2014}.

\item {\textit{\textbf{Arash Asadi}, Vincenzo Mancuso, and Peter Jacko}, {``Floating Band D2D: Exploring and Exploiting the Potentials of Adaptive D2D-enabled Networks''}, in Proceeding of the IEEE International Symposium on a World of Wireless, Mobile and Multimedia Networks (WoWMoM), Boston, MA, USA, 2015}.
\end{itemize}

We collaborated with many researchers within the course of this dissertation, especially in the framework of EU FP7 CROWD\footnote{\url{www.ict-crowd.eu}} project. The following papers are the results of these collaborations. Our contributions are mainly the analysis and evaluation of \ac{D2D} communications and opportunistic scheduling. 
\begin{itemize}
\item {\textit{\textbf{Arash Asadi}, Vincenzo Sciancalepore, and Vincenzo Mancuso}, {``On the Efficient Utilization of Radio Resources in Extremely Dense Wireless Networks''}, IEEE Communication Magazine, 2015}.

\item {\textit{Christian Vitale, Vincenzo Sciancalepore, \textbf{Arash Asadi}}, {``Two-level Opportunistic Spectrum Management for Green 5G Radio Access Networks''}, Accepted for publication in IEEE Online International Conference on Green Computing and Communications, 2015}.

\item {\textit{Rohit Gupta, Bjoern Bachmann, Andreas Kruppe, Russell Ford, Sundeep Rangan, Nikhil Kundargi, Amal Ekbal, Karamavir Rathi, \textbf{Arash Asadi}, Vincenzo Mancuso, and Arianna Morelli}, {``Labview Based Software-Defined Physical/Mac Layer Architecture For Prototyping Dense LTE Networks''}, in Proceeding of the Wireless Innovation Forum Conference on Wireless Communications Technologies and Software Defined Radio, San Diego, CA, USA, 2015}.

\item {\textit{Maria Isabel Sanchez, \textbf{Arash Asadi}, Martin Draexler, Rohit Gupta, Vincenzo Mancuso, Arianna Morelli, Antonio De la Oliva, Vincenzo Sciancalepore}, {``Tackling the Increased Density of 5G Networks; the CROWD Approach''}, in Proceeding of the 1st International Workshop on 5G Architecture co-located with VTC-spring, 2015}.
\end{itemize}

   % Introduction

\chapter{\titSecondCh}
\label{ch:background}

In this chapter, we provide the reader with a thorough review on the-state-of-the art in opportunistic scheduling and \ac{D2D} communications in cellular networks based on our propose taxonomy. This chapter include various tables which comparison a handful of related work which may have not been elaborated in details for brevity. Nevertheless, interested reader can find further details in~\cite{asadi2013survey, asadi2014survey} for such articles. In addition to literature review, we conclude each group of works with a discussion on short-comings of the current works, open issues and the future trends. 

\section{Opportunistic Scheduling}

Opportunistic schedulers take into account information such as the channel quality in terms of \ac{QoS} metrics (i.e., throughput, delay, jitter) that allows the scheduler to find the proper transmission resources for each user. The notion of opportunistic scheduling was first introduced by Knopp and Humblet in~\cite{knopp95ICC}. They showed that using the multiuser diversity in scheduling process can significantly improve the capacity. In a pure opportunistic approach, the scheduler always chooses the user in the best channel condition to use the resources. This approach is referred to as MaxRate scheduling in the literature. The gain in opportunistic scheduling depends on the multiuser diversity due to random wireless channel impairments such as fading and multipath. After~\cite{knopp95ICC}, researchers aimed to take advantage of diversity caused by the channel impairments instead of eliminating it. Some authors even propose techniques such as opportunistic beamforming to increase the multiuser diversity~\cite{27,48,91}. With this technique, the same signal is transmitted over multiple antennas with different transmission powers. This increases the channel diversity of users, which leads to improved opportunistic gain. 
MaxWeight~\cite{andrews2004IS} is another opportunistic scheduler that selects the user with the highest product of queue length and transmission rate. MaxWeight was considered throughput-optimal before the authors in~\cite{23} prove otherwise under flow level dynamics. However, MaxWeight is throughput-optimal with fully backlogged queues. Exp rule schedulers~\cite{jang2004INFOCOM} are throughput-optimal schedulers that prioritize users based on an exponential formula using queue size and transmission rate of every user. 
%Table~\ref{tb:optimal} shows the scheduling policy of the aforementioned throughput-optimal schedulers. 
Opportunistic scheduling has been proposed not only to improve capacity or \ac{QoS}. For instance, Wong {\it et al.} proposed an opportunistic scheduling strategy to leverage multiuser diversity in an \ac{OFDM} systems and do attempt to minimize the overall transmission power~\cite{89}.

\begin{figure*} [!t]
\begin{center}
\includegraphics[width=0.8\columnwidth] {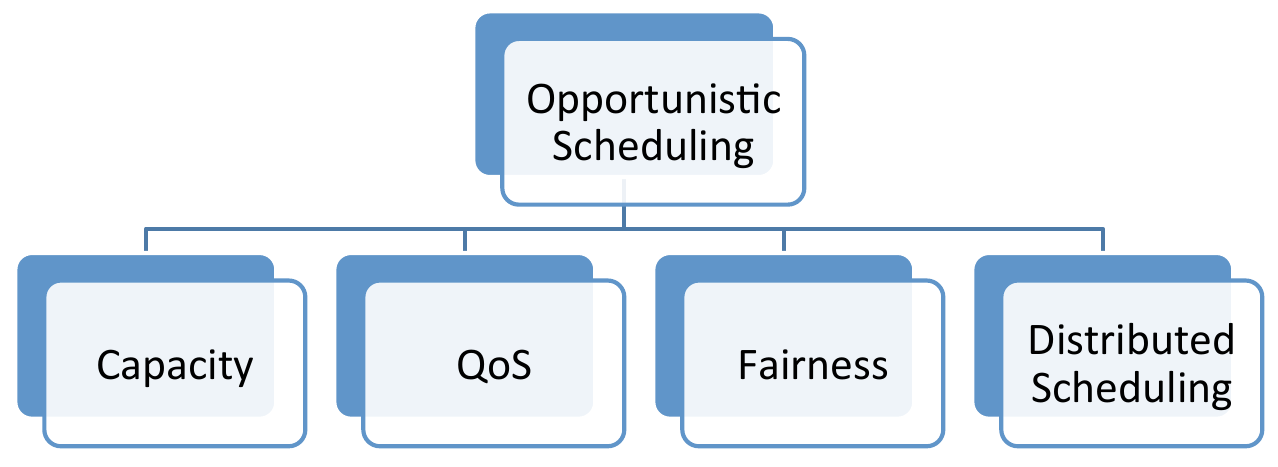}
\caption{Opportunistic scheduler classification.}
\label{fig:1}
\end{center}
\end{figure*}

The available literature on opportunistic scheduling tackles the issue of scheduling from different aspects. Most of these proposals are subclasses of four major categories: capacity, \ac{QoS}, fairness, and distributed scheduling. Proposals that purely improve network capacity regardless of \ac{QoS} or fairness implications are listed under the first category.
% We further categorize these proposals into schedulers with full/non-full channel state information and schedulers for cognitive radio networks. The second category covers proposals that aim to improve specific quality indices such as delay, jitter, average throughput, etc. The works under this category are also divided into subcategories of single \ac{QoS} objective and multiple \ac{QoS} objective. Works under the third category tackle the fairness issue in opportunistic scheduling. In fact, opportunistic scheduling of users can lead to highly unfair treatment toward different users. Therefore, fairness is always a concern in opportunistic scheduling because users with low channel quality can be sacrificed due to the greedy nature of pure opportunistic approaches. Eventually, most of the proposals focus on mechanisms that can be implemented at the base station of a cellular network. However, distributed mechanisms have been proposed as well. We classify these distributed opportunistic algorithms in a separate class because they usually aim at a different network configuration. 
Here, we review the existing literature based on the proposed taxonomy, which is schematically depicted in Figure~\ref{fig:1}.

\subsection{Capacity-oriented Opportunistic Schedulers}
\label{s:capacity}

In many proposals, opportunistic scheduling is employed as a solution to enhance the total capacity of wireless networks. 
%As shown in Figure~\ref{fig:1}, we further classify these proposals based on the assumption of {\it full \ac{CSI}} availability (i.e., base station has instantaneous knowledge of all users' \ac{CSI}). Hence, the scheduler always knows which user has the best channel state at every time instant~\cite{7,10,13,14,17,28,29,30}. However, note that \ac{CSI} of mobile users is acquired via feedback in cellular networks, so that the perfect and instantaneous ({\it full}) knowledge of user's \ac{CSI} is hardly practical in real deployments. To cope with this issue, many proposals leverage the knowledge of user's channel statistical behavior, and \ac{CSI} samples, rather than the knowledge of the exact and instantaneous \ac{CSI}. Significantly different is the case of opportunistic schedulers improving capacity by using cognitive radio techniques. Indeed, with cognitive radio, full or partial \ac{CSI} can be available, but opportunism mainly focuses on which resources can be used and when, given that some other primary user has strictly priority access to the wireless resources. Therefore, in the following we describe first those proposals relaying on full \ac{CSI} availability, then proposals not using full \ac{CSI}, and eventually proposals based on cognitive radio.

%\subsubsection{Resource allocation with full channel state information}
%\label{ss:withCSI}
%Here, we survey the work attempting to enhance wireless network throughput under the assumption of full \ac{CSI} availability at the base station.

In many wireless technologies, users can transmit over more than one carrier. This ability extends the opportunistic scheduling decision process to carrier allocation among users.
Andrews and Zhang~\cite{7} tackle the problem of scheduling in a multi-carrier wireless system. Their work is dedicated to adapt the MaxWeight algorithm for multi-carrier scenarios 
for which they define three objective functions that emulate the MaxWeight behavior. The first objective function simply maximizes the product of queue size and feasible rate for each user over all subcarriers. The second and third objective functions are NP-hard problems that account for the ignorance of MaxWeight algorithm towards users with small queue and bad channel quality, as discussed in~\cite{23}. To serve this purpose, the second objective function prioritizes the users with small queues and the third objective function maximizes the negative drift of a Lyapunov function~\cite{78} (i.e., maximizes the queue length variation in every slot).  
The authors propose five algorithms based on the objective functions defined for MaxWeight. The algorithms which are derived from second and third objectives inherit the NP-hardness. The authors solve the NP-hard algorithms via approximations and prove their stability. The simulation results showed that the algorithms based on the second and third objective provide better performance. They also show that the algorithms which optimize the scheduling decision over all carriers instead of local carrier optimization outperform the rest.

%\subsubsection{Resource allocation without full channel state information}
%\label{ss:noCSI}
%Here, we describe the proposals considering that base stations do not have direct and instantaneous access to user's \ac{CSI}, but they can periodically acquire \ac{CSI} from mobile user's reports, see~Figure~\ref{fig:1}. These {\it non-full-\ac{CSI}} proposals are inherently different for the ones surveyed in the previous \red{subsection}, and therefore lead to substantially different mechanisms.

In~\cite{4}, Liu {\it et al.} propose a throughput-optimal scheduler that does not require any prior knowledge of channel state and user demands. This can be achieved using the so called \ac{WSL}. 
The authors define the flows that continuously inject traffic as {\it long-lived} and those with finite number of bits as {\it short-lived}.  In order to find the maximum possible data rate of short-lived flows, their data rate is monitored for a learning period. The authors of~\cite{4} also provide the necessary conditions for stability of a scheduler which is: $(i)$ the service allocated to each user should not be less than what was requested if the service were supportable at all. $(ii)$ the total airtime allocated to short-lived and long-lived flows should be less than or equal than the total available time . The authors prove that \ac{WSL} is throughput-optimal.

In the same work, Liu {\it et al.} discuss the basic problem of MaxWeight, e.g., a flow with small backlog may never be served. A solution for this problem is to use the product of the head of line delay (delay-based scheduling). However, in~\cite{4} it is shown that the delay-based scheduler is also not stable in presence of short-lived flows. The authors conclude with a set of simulations to evaluate the performance of \ac{WSL}, MaxWeight, and delay-based scheduler. They show that \ac{WSL} can sustain a zero blocking probability while admitting almost 20\% more traffic. \ac{WSL} also shows better delay performance.

Table~\ref{tb:sum-cap} shows each proposal mainly focusing on capacity with details regarding the assumptions taken by the authors, analytical tools used for the proposals, the scenario in which the proposal in applicable, and other considerations taken into account besides capacity improvement.

\begin{table*}[ht!]
\center
\caption{Summary of proposals with main focus on {\it capacity}}
\label{tb:sum-cap}
\resizebox{\columnwidth}{!}	{$
\begin{tabular}{|l|l|l|l|l|}
\hline
\textbf{Proposal}	& \textbf{Assumptions} 	& \textbf{Analytical tools}	& \textbf{Topology}	& \textbf{Other focus} \\
\hline
Scheduler for multicarrier			&Full \ac{CSI}				& Greedy algorithms						&Single cell (OFDMA)	& Fairness		\\
wireless system~\cite{7}			&Not fully backlogged	& Dynamic programming					&Multi-carrier			& Stability			\\
							&Traffic: Generic		& Linear programming					&Downlink 			& 				\\
							&					& Lyapunov drift						&					& 				\\
\hline
Scheduler for multichannel		&Full \ac{CSI}				& Markov chain							&Single cell (OFDM)		& Throughput-optimal\\
wireless systems with 			&Not fully backlogged	& Lagrangian multipliers					&Multi-carrier			& Stability			\\
flow-level dynamics~\cite{10}		&Traffic: Generic		& Lyapunov drift						&Downlink 			& 				\\
\hline
Optimal scheduler				&Full \ac{CSI}				& Finite state Markov chain				& Single cell (HSDPA)	& Throughput		\\
for HSDPA networks~\cite{17}		&Not fully backlogged	& Dynamic programming					& Downlink			& Delay			\\
							&Error free transmission	& Markov decision process 				&		 			& Fairness		\\
							&Traffic: Bernoulli		& 									&					& 				\\
\hline
Throughput-optimal scheduler		&Non-full \ac{CSI}			& Markov chain							& Single cell			& Throughput-optimal\\
that accounts for				&Not fully backlogged	& Lagrangian multipliers					& Downlink			& Stability			\\
flow-level dynamics~\cite{4}		&Traffic: Flow level with 	& Lyapunov drift						&					& 				\\
							& two class of flows		&									&					& 				\\
\hline
Joint channel estimation			&Non-full \ac{CSI}			& Markov chain							& Single cell			& 				\\
and scheduler for				&					& Restless multi-armed bandit process		& Downlink			& 				\\
wireless networks~\cite{6}		&					& Partially observable Markov 		 		&					& 				\\
							&					& decision process						&					& 				\\
\hline 
Throughput-optimal 				&Non-full \ac{CSI}			& Markov decision process				& Single cell			& Throughput-optimal\\
scheduling with limited			&Not fully backlogged	& Lyapunov drift						& Downlink			& Stability			\\
channel information~\cite{11}		&Traffic: Generic		& Optimal stopping theory					&					& 				\\
\hline
Optimal feedback allocation		&Non-full \ac{CSI}			& Markov chain							& Single cell (FDD)		& Near throughput-optimal		\\
in multichannel wireless			&Not fully backlogged	& 									& Downlink			& Stability			\\
networks~\cite{12}				&Traffic: Poisson		& 									& Multi channel			& 				\\
\hline
Flow-level scheduler			&Non-full \ac{CSI}			& Markov decision process				& Single cell			& Stability			\\
for wireless networks~\cite{63}		&Not fully backlogged	& Gilbert-Elliot model					& Downlink			& 				\\
							&Traffic: Bernoulli		& Lagrangian multipliers					&					& 				\\
							&					& Dynamic programming					&					& 				\\
\hline
Opportunistic scheduler			&Non-full \ac{CSI}			& Markov chain							& Multi-cell			& Throughput-optimal\\
for cognitive radio				&Not fully backlogged	& Lyapunov drift						& 					& 				\\
networks~\cite{Luo2011}				&Traffic: Bernoulli		& Lyapunov optimization					&					& 				\\
							&					& Maximum weight match					&					& 				\\
\hline
Optimal scheduler				&Full \ac{CSI}				& Lyapunov drift						& Single cell			& Stability			\\
for cooperative cognitive			&Fully backlogged		& Lyapunov optimization					& 					& 				\\
radio networks~\cite{14}			&					&									&					& 				\\
\hline
\end{tabular}
$}
\end{table*}

\subsection{Quality of Service-oriented Opportunistic Schedulers}
\label{s:QOS}

With the recent advent of applications such as VoIP and video conferencing, \ac{QoS} gained popularity in both research and industry. There are several \ac{QoS} objectives defined such as throughput, delay, jitter, packet loss, error rate, latency and so on. Among the \ac{QoS} metrics, the opportunistic scheduling proposals pay more attention to delay and throughput as depicted in Figure~\ref{fig:1} . %Some of the proposed schedulers may adopt a single \ac{QoS} objective~\cite {1,3,5} while others use multiple objectives~\cite{16,19}. In the following, we review the proposals with single \ac{QoS} objective. Then we continue with proposals taking multiple \ac{QoS} objectives into account. 

%\subsubsection{Single \ac{QoS} objective}
%In this subsection, we focus on proposals with one single \ac{QoS} objective. Among available \ac{QoS} objectives, delay and average throughput are the most common used.

In~\cite{1}, Kim and de Veciana investigate the performance of opportunistic scheduling with heterogenous traffic (i.e., \ac{QoS} and \ac{BE} flows).  
They show that traffic integration---i.e., the coexistence of \ac{QoS} and \ac{BE} traffic in the same network---deteriorates the performance of the system in terms of capacity, stability, and delay. This {\it performance anomaly} was previously dealt with at packet-level in~\cite{20,21,22}.

Kim and de Veciana~\cite{1} studied the interaction of \ac{QoS} and \ac{BE} traffic at flow-level for the first time. They find necessary and sufficient stability conditions for the traffic integration models that was previously provided using a 2-dimentional Markov chain in~\cite{38,39,40}. The proposed opportunistic scheduler is designed in a way that \ac{QoS} flows receive a fixed average throughput per slot. Other \ac{QoS} objectives such as delay or jitter are not considered. The authors argue that allocating an average throughput to \ac{QoS} flows in every slot reduces the chance of starvation in the long period. 
\ac{BE} traffic is modeled as finite file transfers using HTTP or FTP
and its performance is evaluated through the average time needed to complete a file transfer.

Additionally, the authors of~\cite{1} propose an opportunistic scheduling scheme that monitors the number of \ac{QoS} and \ac{BE} flows. In order to be able to guarantee the fixed average throughput in every slot, the maximum number of \ac{QoS} flows is limited such that the total promised bandwidth remains less than the total available bandwidth. It should be noted that average channel quality of users affects the total capacity of the network and the maximum number of \ac{QoS} flows. Kim and de Veciana also propose a bandwidth borrowing/lending scheme that allows \ac{QoS} services to borrow bandwidth from \ac{BE} services when required. Therefore, each \ac{QoS} flow borrows bandwidth from \ac{BE} flows to maintain the promised average throughput. Similarly, \ac{QoS} flows can borrow their extra bandwidth to \ac{BE} flows. They show that integration of \ac{QoS} and \ac{BE} flows reduces the system capacity and leads to the so called {\it loss in opportunism} phenomenon (33\% capacity reduction in the example provided in~\cite{1}). This loss is due to \ac{QoS} requirements of flows, which forces the opportunistic scheduler to transmit packets of \ac{QoS} flows, although that was not the {\it opportunistic} choice at that moment. \ac{QoS} flows also affect the delay experienced by \ac{BE} flows. This effect magnifies with lower \ac{SNR}, higher guaranteed bandwidth, and larger number of \ac{QoS} flows. If \ac{QoS} flows remain in the system for a long time, \ac{BE} flows are under-served until they have a chance to recover, i.e., \ac{QoS} sessions leave the system. This phenomenon is called {\it local instability} which is caused by the coexistence of \ac{QoS} and \ac{BE} flows. Kim and de Veciana propose a \ac{CAC} for \ac{BE} flows which solves the local instability issue. Using numerical evaluation, they show that using CAC reduces the local instability and improves the delay for \ac{BE} flows.

In~\cite{5}, Neely proposes an opportunistic scheduling algorithm with delay guarantees. He develops a novel virtual queue technique (i.e., the {\it $\epsilon$-persistent service queue}) which guarantees a worst case delay for each users.  He further uses Lyapunov drift and optimization techniques to obtain a throughput-optimal scheduling algorithm that guarantees bounded worst-case delay. The proposed scheduler is compatible with both ergodic and possibly non-ergodic channel and arrival settings. Moreover, it can be used for both single-hop and multi-hop scenarios. Finally, the author proves that  the performance of the proposed algorithm is comparable to schedulers that have advance knowledge of channel variations (i.e., full \ac{CSI}).

%\subsubsection{Multiple \ac{QoS} objectives}
%In this subsection, we list the proposals that intend to meet more than one \ac{QoS} metric.

In opportunistic scheduling, it is common to observe that users with low channel quality frequently experience transmission rate fluctuations. These fluctuations result in larger queues and longer delays. Choi {\it et al.} proposed the \ac{AADTR}  metric to be able to measure and control these fluctuations and their resulting delays~\cite{16}.
The algorithm proposed by Choi {\it et al.} in~\cite{16}  targets \ac{OFDMA} wireless networks in which users can transmit over different subcarriers at the same time. Their proposal maximizes system throughput while meeting the {\it required average transmission rate} and the \ac{AADTR}. 
The latter is a metric to control the transmission rate fluctuations. \ac{QoS} flows have both average transmission rate and AADTR objectives. Average transmission rate is the only objective for \ac{BE} flows. The proposal addresses both real time (i.e., video conferencing) and \ac{BE} traffic. 
The authors formulate the problem of scheduling in the \ac{OFDMA} wireless communications which can be solved using the dual optimization technique~\cite{43}. The proposed algorithm calculates the optimal solution for every frame which guarantees average throughput with bounded fluctuations over time. The proposal performance is illustrated using computer simulations in both stationary and non-stationary channel conditions. In the simulations, it is assumed that the queues are fully backlogged and users move with the speed of 50 km/h invariably. Results show that the throughput of the proposed algorithm is on average 30\% higher than that of  \ac{M-LWDF}~\cite{35}. \ac{M-LWDF} is a heuristic that was originally designed for \ac{TDMA} systems. It selects users based on a simple metric, taking into account both the current channel state and the head-of-line packet delay. Unlike \ac{M-LWDF}, packet drop rate of the \ac{AADTR}-based algorithm remains the same with increasing number of users.

Table~\ref{tb:sum-QoS} shows each proposal mainly focusing on \ac{QoS} with details regarding the assumptions taken by the authors, analytical tools used for the proposal, the scenario in which the proposal in applicable, and other considerations taken into account besides \ac{QoS}.

\begin{table*}[ht!]
\center
\caption{Summary of proposals with main focus on {\it QoS} }
\label{tb:sum-QoS}
\resizebox{\columnwidth}{!}	{$
\begin{tabular}{|l|l|l|l|l|}
\hline
\textbf{Proposal}	& \textbf{Assumptions} 	& \textbf{Analytical tools}	& \textbf{Topology}	& \textbf{Other focus} \\
\hline
Flow-level scheduler			&Non-full \ac{CSI}			& Markov decision process& Single cell			& Stability			\\
for wireless networks~\cite{63}		&Not fully backlogged	& Gilbert-Elliot model	& Downlink			& 				\\
							&Traffic: Bernoulli		& Lagrangian multipliers	&					& 				\\
							&					& Dynamic programming	&					& 				\\
\hline
Scheduler for wireless			&Not fully backlogged	& Markov chain			&Single cell (TDMA)		&Stability			\\
systems with integrated			&Traffic: Poisson		& Lyapunov drift		&					& 				\\
traffic~\cite{1}					&					& Foster theorem		&		 			& 				\\
\hline
Delay-optimal Log rule based		&Not fully backlogged	& Markov decision process &Single cell (HDR)		&Throughput-optimal\\
scheduler for wireless			&Traffic: Poisson		& Dynamic programming	&Downlink			&Delay-optimal 	\\
networks~\cite{3}				&					& Lyapunov drift		&		 			& 				\\
							&					& Foster theorem		&					& 				\\
\hline
Opportunistic scheduler for		&No \ac{CSI}				& Lyapunov drift		&Single/multi-hop		&				\\
wireless networks with			&Not fully backlogged	& Lyapunov optimization	&					&	 			\\
worst-case guarantee~\cite{5}		&					& 					&		 			& 				\\
\hline
Scheduler for OFDMA			&Fully backlogged		& Duality theory		&Single cell (OFDMA)	&				\\
systems with multimedia			&					& Lagrangian multipliers	&Downlink			&	 			\\
support~\cite{16}				&					& Convex optimization	&		 			& 				\\
\hline
Adaptive QoS scheduler			&Not fully backlogged 	&Unity cube mapping	&Single cell			&				\\
for wireless networks~\cite{19}		&Traffic: Poisson \& Pareto& 					&Downlink			& 				\\
\hline
\end{tabular}
$}
\end{table*}

\subsection{Fairness-oriented Opportunistic Schedulers}
\label{s:fairness}

Due to the greedy behavior of opportunistic schedulers, their fairness performance is always a concern. Scheduling users opportunistically can result in under-serving some users due to their poor channel quality, while the rest are over-served because they are in a better channel conditions. As a result, it is essential to monitor the way a scheduler allocates the resources to avoid unfairness among users in the long term.

There are different metrics defined for fairness (e.g., Jain's index, temporal fairness, utilitarian fairness). Jain's index is one of the popular fairness metrics for studying fairness performance of the schedulers. For  a given set $X=\{x_1, x_2, \dots, x_n\}$ Jain's index is computed as follows~\cite{95}:

\begin{equation*}
\label{eq:jain}
\text {Jain's index} = \frac{ \left( \sum\limits_i^n x_i\right)^2 } {n~\sum\limits_i^n x_i^2}.
\end{equation*} 

In~\cite{86}, authors introduce optimal policies for opportunistic scheduling in \ac{OFDM} systems with three different fairness criteria, namely {\it temporal fairness}, {\it utilitarian fairness}, and {\it minimum-performance guarantees}. Under temporal fairness criteria, all users are given at least a certain share of airtime, whereas under utilitarian fairness criterion users are given a certain share of throughput~\cite{90}. The policies with minimum-performance guarantees, as the name implies, aim to maximize the network performance while satisfying minimum user requirements. 
Temporal and utilitarian fairness methods oblige the scheduler to allocate a predefined share of resources (i.e., time, throughput) to every user. In contrast, with minimum-performance guarantees the scheduler is restrained to satisfy the minimum service requirement of the users. The authors of~\cite{86} interpret the optimal policies as bipartite matching problem and solve it using the Hungarian algorithm~\cite{94}. Simulation results shows that temporal, utilitarian, and minimum-performance guarantee policies provide 46\%, 32\%, and 31\% gain over Round Robin, respectively.

One of the most diffused opportunistic approaches with fairness constraints is the proportional fair scheduler~\cite{liu2008WCNC,yang2004GLOB}. This scheduler assigns priorities to users based on the ratio of two functions: the first function accounts for the rate potentially achievable in the current slot, while the second function accounts for historical average of the user's throughput.

The authors of~\cite{84} adapt the analytical model proposed for PF by Liu {\it et al.}~\cite{liu2008WCNC} to an \ac{OFDMA} networks with more realistic assumptions. Their model accounts for multiple subcarriers, but also for less realistic Poisson traffic arrivals. The adapted PF scheduler computes a matrix containing user rankings over all subcarriers. For every subcarrier, the user with the highest rank and {\it non-empty buffer} is scheduled. Non-empty buffer condition accounts for the fact that in real world a users can be eligible to be scheduled when it has no packets to transmit. In such cases, the scheduler selects the next best user with non-empty buffer to avoid wasting airtime. The proposed analytical model is validated in terms of average throughput and Jain's fairness index by simulation.

In~\cite{88}, an adaptive resource allocation for \ac{OFDM} system is proposed that accounts for each user's required data rate as a fairness measure. The authors formulate the optimization problem for subchannel and power allocation with a proportional fairness constraint. Since their proposed optimization problem requires linearization of nonlinear constraints, the authors propose a suboptimal solution with lower complexity. The suboptimal solution carries out the subchannel allocation and power allocation separately. Via simulations, it is shown that the suboptimal solution can achieve 95\% of the optimal capacity with much lower complexity.

In~\cite{18}, Kwon {\it et al.} tackle the fairness issue in \ac{TDMA} networks. The proposed opportunistic fair scheduler prior to~\cite{18} are either based on average rate-based utility functions or instantaneous rate-based utility functions. Average rate-based utility functions are suitable for elastic services (e.g., HTTP, Email, and FTP) for which instantaneous data rate does not affect the \ac{QoS}. On the contrary, satisfaction of services such as video streaming depends on the instantaneous data rate. For such services, the utility function should be based on instantaneous rate. Kwon {\it et al.} propose a framework that can accommodate both elastic and non-elastic services. Instead of deterministic scheduling, they use a probabilistic scheduling policy that randomly schedules a user per time-slot with a certain probability. Kwon {\it et al.} model the channel using a \ac{FSMC}, formulate the scheduling problem based on convex optimization~\cite{96} and solve it using a Lagrangian function, the duality theorem~\cite{83}. An iterative algorithm is also proposed that can find the optimum solution in every time slot. It is shown via numerical simulations that their proposed scheduler meets the required fairness objective for users with elastic and non-elastic services.

\subsection{Distributed Opportunistic Scheduling Algorithms}
\label{s:distributed}

In a centralized scheduling approach, the scheduler is aware of all user's channel condition or it will acquire an estimate of that information to make the scheduling decision. On the other hand, in a distributed scheduling approach, users make scheduling decisions independent of the central entity and possibly without an overall knowledge of the network.

Tang {\it et al.}~\cite{54} propose two joint multi-cell scheduling and beam coordination schemes, namely {\it \ac{SINR} feedback} and {\it ABC}. 
In the {\it \ac{SINR} feedback} scheme, each base station selects $m$ beams randomly and sends beam-pilots within the cell. The users will send their feedback with respect to their \ac{SINR} which are also affected by the beam pilots received from the neighboring cells. Each base station makes scheduling decisions based on the received \ac{SINR} feedbacks. Intra-cell scheduling decisions are made based on the PF algorithm proposed in~\cite{56}. 

In the {\it ABC} scheme, the cellular network is partitioned dividing base stations into A/B/C subsets. Although the network is partitioned, it operates as a reuse factor $1$. In the first step, base stations tagged as $A$ select $m$ beams, send beam pilots and make their scheduling decision. Note that users of subset B and C also listen to beam pilots to be able to estimate their \ac{SINR}. After the scheduling decision has been made, the identity of the scheduled users and the beams assigned to them is sent to the neighboring base stations tagged as subset B. This helps the base stations in subset B to avoid choosing beams that interfere with subset A.  Base station in subset B will perform the same operation and inform to those in subset C. Due to priority (in term of frequency selection) given to the base stations in subset A, they will provide better service to their users in comparison to the base stations in subset B and C. To avoid such unfairness, the authors propose to assign the A/B/C tags in a Round Robin fashion. 

In their work, Tang {\it et al.} compare the proposed schemes with two other schemes. In the first scheme, network operates using a  frequency reuse partition with a factor $3$. In the second scheme, each base station operates fully independently and without considering \ac{ICI}. Using simulations, it is shown that {\it \ac{SINR} feedback} and {\it ABC} outperform the two other schemes. {\it ABC} scheme provides more than 100\% throughput gain in comparison to schemes without \ac{ICI} control mechanism. In addition, users on the edge of the cell receive higher throughput with {\it ABC} scheme.

In~\cite{55}, Bendlin {\it et al.} propose a distributed multi-cell scheduling, namely \ac{CEB}, that is tolerant to delay and capacity limitations of backhaul links. In many proposals, authors assume that backhaul links have zero delay and unlimited capacity, which is not a realistic assumption. In \ac{CEB}, authors assume that each base station schedules one user per slot and it has full \ac{CSI} knowledge of its users. Scheduling is performed in two steps in \ac{CEB}.  In the first step, each base station chooses the proper beamforming which minimizes \ac{ICI} and in the second step, a user will be scheduled. The main advantage of \ac{CEB} is the low amount of data exchanged among base stations. Bendlin {\it et al.} show that \ac{CEB} can perform very close to schemes that disseminate the full \ac{CSI} information in the network. \ac{CEB} also exhibits robustness towards delay in backhaul links.

\subsection{Future Trends and Thesis Contribution}

After two decades of research, we can say that opportunistic scheduling became very mature. This maturity calls for new opportunistic schedulers whose designs are backed up with analysis. Indeed, many authors not only show the performance advantage of their proposal, but also prove the stability of the schedulers. Currently, researchers are active towards two major directions. First, researchers evaluate the performance of existing proposals under more realistic scenarios such as flow-level dynamics, multi-user multi-carrier scheduling, and mobility. This helps us to have a better overview of the system performance in a real world scenario. Second, researchers seek for novel applications and new challenges for opportunistic scheduling. Use of opportunistic scheduling in cooperative communications is one of the newly explored areas which has attracted the interest of many researchers. Indeed, in this dissertation, we integrate opportunistic scheduling with \ac{D2D} communications. We propose to form \ac{D2D} clusters in which the member experiencing the best channel handles the cluster traffic resulting in higher throughput for all cluster members.

\section{\ac{D2D} Communications}

As telecom operators are struggling to accommodate the existing demand of mobile users, new data intensive applications are emerging in the daily routines of mobile users (e.g., proximity-aware services). Moreover, 4G cellular technologies (WiMAX~\mbox{\cite{wimax}} and LTE-A~\mbox{\cite{LTEadv36213}}), which have extremely efficient physical and MAC layer performance, are still lagging behind mobile users' booming data demand. Therefore, researchers are seeking for new paradigms to revolutionize the traditional communication methods of cellular networks. \ac{D2D} communication is one of such paradigms that appears to be a promising component in next generation cellular technologies. 
\ac{D2D} communication in cellular networks is defined as direct communication between two mobile users without traversing the \ac{BS} or core network. \ac{D2D} communications are generally non-transparent to the cellular network and it can occur on the cellular spectrum (i.e., {\it inband}) or unlicensed spectrum (i.e., {\it outband}). 
%In a traditional cellular network, all communications must go through the BS even if both communicating parties are in range for \ac{D2D} communication. This architecture suits the conventional low data rate mobile services such as voice call and text message in which users are not usually close enough to have direct communication. However, mobile users in today's cellular networks use high data rate services (e.g., video sharing, gaming, proximity-aware social networking) in which they could potentially be in range for direct communications (i.e., \ac{D2D}). Hence, \ac{D2D} communications in such scenarios can highly increase the spectral efficiency of the network. Nevertheless, the advantages of \ac{D2D} communications are not only limited to enhanced spectral efficiency. In addition to improving spectral efficiency, \ac{D2D} communications can potentially improve throughput, energy efficiency, delay, and fairness.

\begin{figure} [!h]
\centering
\includegraphics[scale=0.45]{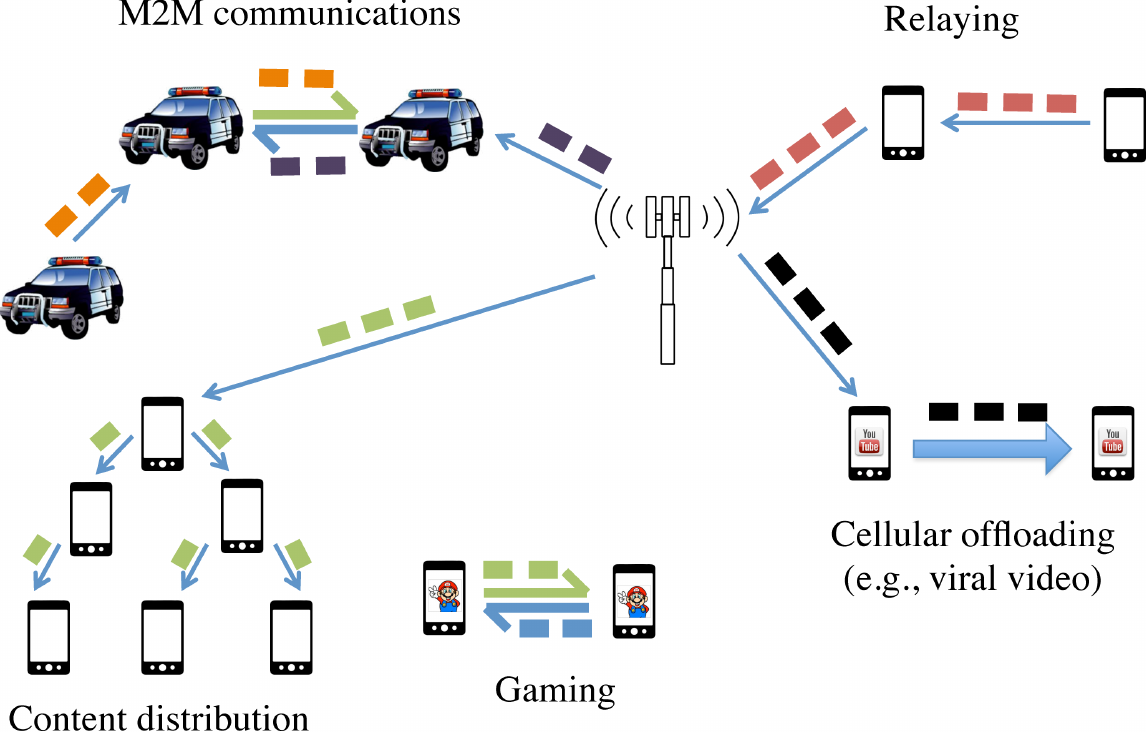}
\caption{Representative use-cases of \ac{D2D} communications in cellular networks.}
\vspace{-2mm}
\label{fig:use-case}
%\vspace{-4mm}
\end{figure}

In academia, \ac{D2D} communication was first proposed in~\cite{lin_multihop_2000} to enable multihop relays in cellular networks. Later the works in~\cite{kaufman_cellular_2008,doppler2009COMMAG,doppler_device_2009A,osseiran_advances_2009A,peng_interference_2009} investigated the potential of \ac{D2D} communications for improving spectral efficiency of cellular networks. Soon after, other potential \ac{D2D} use-cases were introduced in the literature such as multicasting~\cite{du_compressed_2012,zhou2013TVT}, peer-to-peer communication~\cite{lei_operator_2012}, video dissemination~\cite{golrezaei2012Icc,golrezaei2012Globecom,li_device_2012,doppler2009COMMAG}, machine-to-machine communication~\cite{pratas_low_2013}, cellular offloading~\cite{bao_dataspotting_2010}, and so on. The most popular use-cases of \ac{D2D} communications are shown in Figure~\ref{fig:use-case}.
The first attempt to implement \ac{D2D} communications in a cellular network was made by Qualcomm's FlashLinQ~\cite{wu2013TON} which is a PHY/MAC network architecture for \ac{D2D} communications underlaying cellular networks. FlashLinQ takes advantage of OFDM/\ac{OFDMA} technologies and distributed scheduling to create an efficient method for timing synchronization, peer discovery, and link management in \ac{D2D}-enabled cellular networks.
In addition to academia and telecommunication companies, 3GPP is also investigating \ac{D2D} communications as \ac{ProSe}. In particular, the feasibility of \ac{ProSe} and its use-cases in LTE are studied in~\cite{3GPPTR22.803} and the required architectural enhancements to accommodate such use-cases are investigated in~\cite{3GPPTR23.703}. 
%Currently, \ac{ProSe} is supposed to be included in 3GPP Release $12$ as a public safety network feature with focus on one to many communications~\cite{3GPPTR23.703}. 
A brief overview of standardization activities and the fundamentals of 3GPP \ac{ProSe} can be found in~\cite{lin2014ComMag}.

\begin{figure} [!ht]
\centering
\includegraphics[scale=0.42]{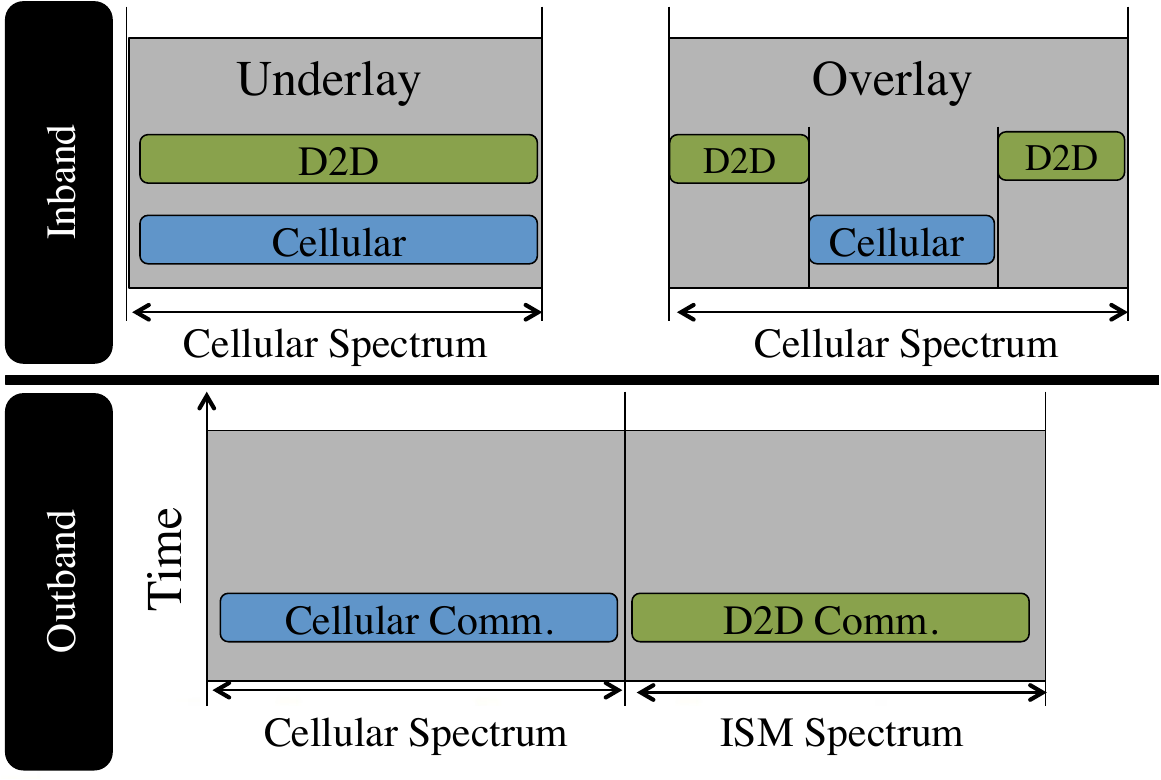}
\caption{Schematic representation of overlay inband, underlay inband, and outband \ac{D2D}.}
\vspace{-2mm}
\label{fig:in-outband}
\vspace{-4mm}
\end{figure}

The majority of the literature on \ac{D2D} communications proposes to use the cellular spectrum for both \ac{D2D} and cellular communications (i.e., {\it underlay inband} \ac{D2D}). These works usually study the problem of interference mitigation between \ac{D2D} and cellular communication~\cite{yu_performance_2009,xu_interference_2012,xu_effective_2010,xu_performance_2012,peng_interference_2009,zhang_interference_2013,janis_interference_2009,min_capacity_2011,elkotby_exploiting_2012}.
In order to avoid the aforementioned interference issue, some propose to dedicate part of the cellular resources only to \ac{D2D} communications (i.e., {\it overlay inband} \ac{D2D}).
Here resource allocation gains utmost importance so that dedicated cellular resources be not wasted~\cite{pei_resource_2013}.
Other researchers propose to adopt outband rather than inband \ac{D2D} communications in cellular networks so that the precious cellular spectrum be not affected by \ac{D2D} communications. In outband communications, the coordination between radio interfaces is either controlled by the \ac{BS} (i.e, {\it controlled}) or the users themselves (i.e., {\it autonomous}).
Outband \ac{D2D} communications face a few challenges in coordinating the communication over two different bands because usually \ac{D2D} communications happen on a second radio interface (e.g., WiFi Direct~\cite{WifiDirect2013} and Bluetooth~\cite{bluetooth2001bluetooth}). The studies on outband \ac{D2D} investigate issues such as power consumption~\cite{asadi2013energy,asadi2013MSWIM,asadi2014ComCom, asadi2013WD, wang2013WOWMOM} and inter-technology architectural design. Figure~\ref{fig:in-outband} graphically depicts the difference among underlay inband, overlay inband, and outband communications.

\subsection{Taxonomy}
\label{s:taxo}
In this section, we categorize the available literature on \ac{D2D} communication in cellular networks based on the spectrum in which \ac{D2D} communications occur. In the following, we provide a formal definition for each category and sub-category. Next, we provide a quick overview of the advantages and disadvantages of each \ac{D2D} method.
%\add[Q]{Add figures to illustrate inband (underlay/overlay) and outband??}

{\bf Inband \ac{D2D}:} The literature under this category, which contains the majority of the available work,  proposes to use the cellular spectrum for both \ac{D2D} and cellular links. The motivation for choosing inband communication is usually the high control over cellular (i.e., licensed) spectrum. Some researchers (see, e.g.,~\cite{akkarajitsakul_mode_2012, doppler_device_2009A}) consider that the interference in the unlicensed spectrum is uncontrollable which imposes constraints for \ac{QoS} provisioning. Inband communication can be further divided into underlay and overlay categories. In underlay \ac{D2D} communication, cellular and \ac{D2D} communications share the same radio resources. In contrast, \ac{D2D} links in overlay communication are given dedicated cellular resources. 
Inband \ac{D2D} can improve the spectrum efficiency of cellular networks by reusing spectrum resources (i.e., underlay) or allocating dedicated cellular resources to \ac{D2D} users that accommodates direct connection between the transmitter and the receiver (i.e., overlay). The key disadvantage of inband \ac{D2D} is the interference caused by \ac{D2D} users to cellular communications and vice versa. This interference can be mitigated by introducing high complexity resource allocation methods, which increase the computational overhead of the \ac{BS} or \ac{D2D} users.
%\add[Q]{The main issues in inband \ac{D2D} are interference, resource allocation, mode selection, and so forth. (expand the main problems/issues in inband \ac{D2D}???)} 

{\bf Outband \ac{D2D}:} Here the \ac{D2D} links exploit unlicensed spectrum. The motivation behind using outband \ac{D2D} communication is to eliminate the interference issue between \ac{D2D} and cellular links. Using unlicensed spectrum requires an extra interface and usually adopts other wireless technologies such as WiFi Direct~\cite{WifiDirect2013}, ZigBee~\cite{alliance2006zigbee} or Bluetooth~\cite{bluetooth2001bluetooth}. Some of the work on outband \ac{D2D} (see, e.g.,~\cite{golrezaei2012Icc, golrezaei2012Globecom, asadi2013MSWIM, asadi2013energy}) suggest to give the control of the second interface/technology to the cellular network (i.e., controlled). In contrast, others (see, e.g.,~\cite{wang2013WOWMOM}) propose to keep cellular communications controlled and leave the \ac{D2D} communications to the users (i.e., autonomous). 
Outband \ac{D2D} uses unlicensed spectrum which makes the interference issue between \ac{D2D} and cellular users irrelevant. On the other hand, outband \ac{D2D} may suffer from the uncontrolled nature of unlicensed spectrum. It should be noted that only cellular devices with two wireless interfaces (e.g., LTE and WiFi) can use outband \ac{D2D}, and thus users can have simultaneous \ac{D2D} and cellular communications.

\begin{figure*} [!t]
\centering
\includegraphics[scale=0.65]{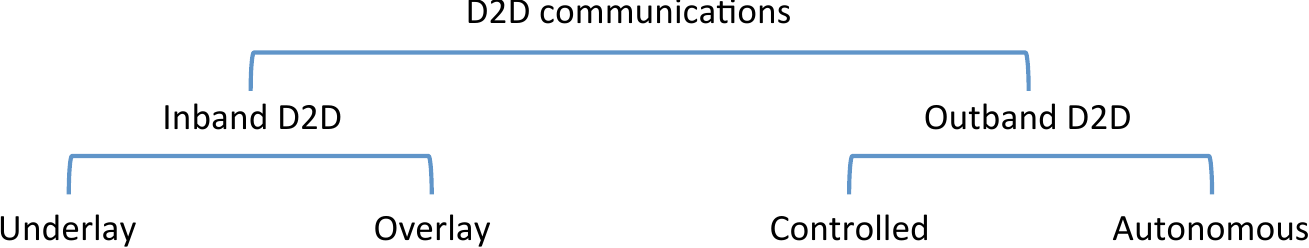}
\caption{Device-to-Device communication classification.}
\vspace{-2mm}
\label{fig:taxo}
\vspace{-4mm}
\end{figure*}

Figure~\ref{fig:taxo} illustrates the taxonomy introduced for \ac{D2D} communications in cellular networks. In the following sections, we review the related literature based on this taxonomy. 

\subsection{Underlay Inband \ac{D2D}}
\label{ss:underlay}
Early works on \ac{D2D} in cellular networks propose to reuse cellular spectrum for \ac{D2D} communications. To date, the majority of available literature is also dedicated to inband \ac{D2D}, especially \ac{D2D} communications underlaying cellular networks. In this section, we review the articles that employ underlaying \ac{D2D} to improve the performance of cellular networks, in terms of \emph{spectrum efficiency, energy efficiency, cellular coverage}, and other performance targets.

Zhang {\it et al.}~\cite{zhang_interference_2013} propose a graph-based resource allocation method for cellular networks with underlay \ac{D2D} communications. They mathematically formulate the optimal resource allocation as a nonlinear problem which is NP-Hard. The authors propose a suboptimal graph-based approach which accounts for interference and capacity of the network. In their proposed graph, each vertex represents a link (\ac{D2D} or cellular) and each edge connecting two vertices shows the potential interference between the two links. The simulation results show that the graph-based approach performs close to the throughput-optimal resource allocation.

Janis {\it et al.} address a similar solution in~\cite{janis_interference_2009}, where the \ac{D2D} users also measure the signal power of cellular users and inform the \ac{BS} of these values. The \ac{BS} then avoids allocating the same frequency-time slot to the cellular and \ac{D2D} users which have strong interference with each other, which is different from~\cite{xu_effective_2010}. The proposed scheme of~\cite{janis_interference_2009} minimizes the maximum received power at \ac{D2D} pairs from cellular users. The authors first show via numerical results that \ac{D2D} communications with random resource allocation can increase the mean cell capacity over a conventional cellular system by $230\%$. Next, they show that their proposed interference-aware resource allocation scheme achieves $30\%$ higher capacity gain than the {random resource allocation strategy}.

The work in~\cite{min_capacity_2011} proposes a new interference management in which the interference is not controlled by limiting \ac{D2D} transmission power as in the conventional \ac{D2D} interference management mechanisms. The proposed scheme defines an interference limited area in which no cellular users can occupy the same resources as the \ac{D2D} pair. Therefore, the interference between the \ac{D2D} pair and cellular users is avoided. The disadvantage of this approach is reducing multi-user diversity because the physical separation limits the scheduling alternatives for the \ac{BS}. However, numerical simulations prove that the capacity loss due to multi-user diversity reduction is negligible compared to the gain achieved by their proposal. In fact, this proposal provides a gain of $129\%$ over conventional interference management schemes.

The authors of~\cite{jung_joint_2012} propose an algorithm for power allocation and mode selection in \ac{D2D} communication underlaying cellular networks. The algorithm measures the power efficiency, which is a function of transmission rate and power consumption, of the users in different modes (cellular  and \ac{D2D}). After computing the power efficiency, each device uses the mode in which it achieves higher power efficiency. The drawback of this algorithm is that the controller should perform an exhaustive search for all possible combinations of modes for all devices. The authors benchmark their algorithm against the scheme of~\cite{hakola_device_2010} in which two users communicate over \ac{D2D} link only if their pathloss is lower than the pathlosses between each user and the \ac{BS}. The simulation results indicate that their algorithm achieves up to $100\%$ gain over the scheme proposed in~\cite{hakola_device_2010}. 

The authors of \cite{su_resource_2013} consider the mode selection and resource allocation in \ac{D2D} communications underlay cellular networks, where several pairs of \ac{D2D} links co-exist with several cellular users. They formulate the problem of maximizing the system throughput with minimum data rate requirements, and use the \emph{particle swarm optimization}~\cite{kennedy2010particle} method to obtain the solutions. 
The simulation results show that the proposed method has $15\%$ throughput gain over the orthogonal resource sharing scheme (i.e., overlay \ac{D2D} which will be explained later), where the achievable gain varies with the distance of \ac{D2D} users. Simulation results also show that this method can improve the system performance under the constraint of minimum data rate of users.

The authors of \cite{han_subchannel_2012} consider the scheduling and mode selection problem for \ac{D2D} in \ac{OFDMA} networks. They assume that the system time is slotted and each channel is divided into sub-channels. They formulate the problem of maximizing the mean sum-rate of the system with \ac{QoS} satisfaction as a stochastic optimization problem, and use the stochastic sub-gradient algorithm to solve it. From the solution, they design a sub-channel opportunistic scheduling algorithm that takes into account the \ac{CSI} of \ac{D2D} and cellular links as well as the \ac{QoS} requirement of each \ac{D2D} user. The numerical results show that the mean sum-rate can be improved by up to $500\%$. This gain increases when the average \ac{D2D} pair distance reduces. Moreover, with the \ac{D2D} communication, the fairness among users can be achieved with the \ac{QoS} requirement specified for each user.

Finally, a  summary of the works on underlay \ac{D2D} communications in cellular networks is provided in Table~\ref{tb:underlay_inband}, in terms of metrics, use-cases, analytical tools, evaluation method, scope, and achieved performances.

\begin{table}[h!]
\centering
\caption{Summary of the literature proposing underlay inband \ac{D2D}}
\label{tb:underlay_inband}
\resizebox{\columnwidth}{!}	{$
\begin{tabular}{| l | l | l | l | l | l | l |}
\hline
Proposal & Analytical tools	& Platform & Direction & Use-case & Evaluation & Achieved performance \\
\hline \hline
\tabincell{l}{Improving spectrum\\efficiency \cite{peng_interference_2009,xu_effective_2010}\\
	\cite{janis_interference_2009, kaufman_cellular_2008, min_capacity_2011, 
	chen_downlink_2012}\\ 
 	\cite{yu_device_2012, doppler2009COMMAG, Doppler2010WCNC, osseiran_advances_2009A}\\	
 	\cite{zulhasnine2010WiMob, pei_resource_2013, liu_optimal_2012, xu_resource_2012}}
	& \tabincell{l}{-Chen-Stein method\\ -Zipf distribution\\ -Integer/linear \\~programming\\
	-Mixed integer \\~nonlinear\\~programming\\ -Convex optimization\\ -Bipartite Matching\\ 
	-Kuhn-Munkres \\~algorithm\\ -Han-Kobayashi\\ -Newton's method\\ -Lagrangian\\~multipliers\\ 
	-Graph theory\\ -Auction algorithm\\ -Particle swarm \\~optimization}
	& \tabincell{l}{-WiMax\\ -CDMA\\ -LTE\\ -LTE-A} 
	& \tabincell{l}{-Uplink\\ -Downlink\\ -Uplink/\\~downlink}
	& \tabincell{l}{-Content \\~distribution\\ -File sharing\\ -Video/file\\~exchange}
	& \tabincell{l}{-Numerical \\simulation\\ -System-level \\simulation}
	& \tabincell{l}{-System throughput can be improved 
		\\~from $16\%$ to $374\%$ compared with 
		\\~conventional cellular networks under 
		\\~common scenarios 
		\\-Throughput can be improved up to
		\\~$650\%$ when \ac{D2D} users are far away
		\\~from the BS
		\\-Number of admitted \ac{D2D} users 
		\\~can be increased up to $30\%$} \\ \hline	
\tabincell{l}{Improving power\\efficiency \cite{xiao_qos_2011,jung_joint_2012}\\ 		
	\cite{belleschi_performance_2011,yu_power_2009}}
	& \tabincell{l}{-Heuristic algorithms\\ -Exhaustive search\\ -Linear programming}
	& \tabincell{l}{-LTE\\ -LTE-A\\ -OFDMA} & \tabincell{l}{-Uplink\\ -Downlink\\ 
	-Uplink/\\~downlink}
	&  & \tabincell{l}{-System-level \\simulation}
	& \tabincell{l}{-Power efficiency can be improved
		\\~from $20\%$ to $100\%$ compared with
		\\~conventional cellular networks} \\ \hline
\tabincell{l}{Improving performance\\ with QoS/power
	\\ constraints \cite{feng2013TOC,su_resource_2013}\\ 
	\cite{han_subchannel_2012,le_fair_2012, yu_power_2009, yu2011TWC}}
	& \tabincell{l}{-Heuristic algorithms\\ -Bipartite Matching\\ -Kuhn-Munkres \\~algorithm}
	& \tabincell{l}{-LTE\\ -LTE-A} & \tabincell{l}{-Uplink\\ -Downlink}	
	& \tabincell{l}{-VOIP/FTP}
	& \tabincell{l}{-Numerical \\simulation \\-System-level \\simulation}  
	& \tabincell{l}{-From $15\%$ to $70\%$ throughput gain
		\\~with QoS constraint 
		\\-From $45\%$ to $500\%$ sum-rate gain
		\\~with QoS/power constraint} \\ \hline
\tabincell{l}{Improving fairness~\cite{xu_interference_2012}}\!\!\!
	& -Auction algorithm &  & -Downlink &  & 
	\tabincell{l}{-System-level \\simulation}
	& \tabincell{l}{ -A fairness index around $0.8\!\!$} \\ \hline	
\tabincell{l}{Improving cellular\\ coverage \cite{vanganuru_system_2012}}
	&  & \tabincell{l}{-LTE\\-LTE-A} & \tabincell{l}{-Uplink/\\~downlink} &  & 
	\tabincell{l}{-Numerical \\simulation}
	& \tabincell{l}{ -Throughput of cell edge users can 
		\\ ~be improved up to $300\%$
		\\ -Cell coverage is also enlarged
		\\ ~up to $20\%$} \\ \hline	
\tabincell{l}{Supporting setup of\\ \ac{D2D} \cite{yang_solving_2013}}
	& -Protocol & -LTE-A & \tabincell{l}{-Uplink/\\~downlink} & 
	\tabincell{l}{-\ac{D2D} link\\~setup} &
	& \tabincell{l}{}  \\ \hline
\tabincell{l}{Improving reliability\\ \cite{min_reliability_2011,kaufman2013ton}}
	& &	 &  &  & \tabincell{l}{-Numerical \\simulation}
	& \tabincell{l}{ -Outage probability reduces by~$99\%$} \\ \hline
\tabincell{l}{Increasing the number\\ of concurrent \ac{D2D}\\ links\cite{han_uplink_2012}}
	& \tabincell{l}{-Mixed-integer \\~nonlinear\\~programming \\ -Hungarian algorithm 
	\\ -Heuristic algorithm} & -LTE & -Uplink &  
	& \tabincell{l}{-System-level \\simulation}
	& \tabincell{l}{ -Number of admitted \ac{D2D} links is 
		\\ ~increased up to $10\%$ compared to  
		\\~random \ac{D2D} link allocation} \\ \hline
\tabincell{l}{Offloading traffic \cite{bao_dataspotting_2010}}
	& &	 &  & \tabincell{l}{-Offloading\\~traffic} & \tabincell{l}{-System-level 
	\\simulation}
	& \tabincell{l}{} \\ \hline			
\tabincell{l}{Improving performance\\ of multicast~\cite{du_compressed_2012, 
	seppala2011WNCN}} 
	& & \tabincell{l}{-LTE\\ -LTE-A} & \tabincell{l}{-Uplink/\\~downlink} & -Multicast
	& \tabincell{l}{-Numerical \\simulation\\ -System-level \\simulation}
	& \tabincell{l}{ -Frame loss ratio of feedback is 
		\\ ~reduced by $80\%$} \\ \hline
\end{tabular}
  $}
\end{table}

\subsection{Overlay Inband \ac{D2D}}
\label{ss:overlay}
Different from the works reviewed in the previous subsection, the authors of~\cite{fodor_design_2012,li_device_2012,zhou2013TVT} propose to allocate dedicated resources for \ac{D2D} communications. This approach eliminates the concerns for interference from \ac{D2D} communications on cellular transmissions, but reduces the amount of achievable resources for cellular communications.

In~\cite{fodor_design_2012}, Fodor {\it et al.} elaborate on the challenges of \ac{D2D} communications in cellular networks and suggest to control \ac{D2D} communications from the cellular network. They claim that network assistance can solve the inefficiencies of \ac{D2D} communications in terms of service and peer discovery, mode selection, channel quality estimation, and power control. In a conventional peer and service discovery method, \ac{D2D} users should send beacons in short intervals and monitor multiple channels which is very energy consuming. However, this process can become more energy efficient if the \ac{BS} regulates the beaconing channel and assists \ac{D2D} users so that they do not have to follow the power consuming random sensing procedure. \ac{BS} assistance also improves the scheduling and power control which reduces the \ac{D2D} interference. The authors use simple Monte-Carlo simulation to evaluate the performance of \ac{D2D} communications. The results show that \ac{D2D} can increase the energy efficiency from $0.8$ bps/Hz/mW to $20$ bps/Hz/mW in the best case scenario where the distance between \ac{D2D} users is $10$m.

The authors of~\cite{li_device_2012} propose the incremental relay mode for \ac{D2D} communications in cellular networks. In the incremental relay scheme, \ac{D2D} transmitters multicast to both the \ac{D2D} receiver and \ac{BS}. In case the \ac{D2D} transmission fails, the \ac{BS} retransmits the multicast message to the \ac{D2D} receiver. The authors claim that the incremental relay scheme improves the system throughput because the \ac{BS} receives a copy of the \ac{D2D} message which is retransmitted in case of failure. Therefore, this scheme reduces the outage probability of \ac{D2D} transmissions. Although the incremental relay mode consumes part of the downlink resources for retransmission, the numerical simulation results show that this scheme still improves the cell throughput by $40\%$ in comparison to underlay mode.

In~\cite{zhou2013TVT}, \ac{D2D} communication is used to improve the performance of multicast transmission in cellular networks. Due to wireless channel diversity, some of the multicast group members (i.e., cluster) may not receive the data correctly. The authors propose to use \ac{D2D} communications inside the clusters to enhance the multicast performance. Specifically, after every multicast transmission, some of the members which manage to decode the message will retransmit it to those which could not decode the message. Unlike the prior work in~\cite{spinella2009integration}~and~\cite{fitzek2007cognitive} where there is only one predefined retransmitter, the number of retransmitters in~\cite{zhou2013TVT} changes dynamically to maximize the spectral efficiency. The authors show via numerical simulations that their proposed algorithm consumes $90\%$  less spectrum resources in comparison to the scenario with only one retransmitter. 

A summary of the works on overlay \ac{D2D} communication in cellular networks is provided in Table~\ref{tb:overlay_inband}.

\begin{table*}[h!]
\centering
\caption{Summary of the literature proposing overlay inband \ac{D2D}}
\label{tb:overlay_inband}
\resizebox{\columnwidth}{!}	{$
\begin{tabular}{| l | l | l | l | l | l | l |}
\hline
Proposal & Analytical tools	& Platform & Direction & Use-case & Evaluation & 
Achieved performance \\
\hline \hline
\tabincell{l}{Increasing energy \\efficiency \cite{fodor_design_2012}}
	&  & -LTE & -Uplink &  &\tabincell{l}{-Numerical\\~simulation }
	& \tabincell{l}{-Energy efficiency can be increased
		\\~from 0.8 bps/Hz/mW to 20 bps/Hz/mW} \\ \hline	
\tabincell{l}{Improving spectrum \\efficiency \cite{li_device_2012}}
	& \tabincell{l}{-Convex\\~Optimization} & & -Uplink &  &\tabincell{l}{-Numerical\\~simulation }
	& \tabincell{l}{-Cell throughput is improved by
		\\ ~$40\%$ over underlay mode} \\ \hline	
\tabincell{l}{Improving performance \\of multicast \cite{zhou2013TVT}}
	&  &  & -Downlink & \tabincell{l}{-Video \\~transmission} & \tabincell{l}{-Numerical\\~simulation }
	&  \tabincell{l}{-$90\%$ gain in bandwidth compared to
		\\ ~the method using only one retransmitter} \\ \hline
\end{tabular}
$}
\end{table*}

\subsection{Outband \ac{D2D}}
\label{s:outband}
In this section, we review the articles in which \ac{D2D} communications occur on a frequency band that is not overlapping with the cellular spectrum. Outband \ac{D2D} is advantageous because there is no interference issue between \ac{D2D} and cellular communications. Outband \ac{D2D} communications can be managed by the cellular network (i.e., controlled) or it can operate on its own (i.e., autonomous).

\subsubsection{Controlled}
In works that fall under this category, the authors propose to use the cellular network advanced management features to control \ac{D2D} communications to improve the efficiency and reliability of \ac{D2D} communications. They aim to improve system performance in terms of \emph{throughput, power efficiency, multicast}, and so on.

The authors of ~\cite{zhou2013Wcnc} propose to use \ac{ISM}  band for \ac{D2D} communications in LTE. They state that simultaneous channel contention from both \ac{D2D} and WLAN users can dramatically reduce the network performance. Therefore, they propose to group \ac{D2D} users based on their \ac{QoS} requirements and allow only one user per group to contend for the WiFi channel. The channel sensing between groups is also managed in a way that the groups do not sense the same channel at the same time. They show via simulation that their approach increases the \ac{D2D} throughput up to $25\%$ in comparison to the scenario in which users contend for the channel individually.   

Golrezaei {\it et al.}~\cite{golrezaei2012Icc,golrezaei2012Globecom} 
point out the similarities among video content requests of cellular users. They propose to cache the popular video files (i.e., viral videos) on smartphones and exploit \ac{D2D} communications for viral video transmissions in cellular networks. They partition each cell into clusters (smaller cells) and cache the non-overlapping contents within the same cluster. When a user sends a request to the \ac{BS} for a certain content, the \ac{BS} checks the availability of the file in the cluster. If the content is not cached in the cluster, the user receives the content directly from the \ac{BS}. If the content is locally available, the user receives the file from its neighbor in the cluster over the unlicensed band (e.g., via WiFi). The authors claim that their proposal improves the video throughput by one or two orders of magnitude.

The authors of~\cite{ji_wireless_2013} propose a method to improve video transmission in cellular networks using \ac{D2D} communications. This method exploits the property of asynchronous content reuse by combining \ac{D2D} communications and video caching on mobile devices. Their objective is to maximize per-user throughput constrained to the outage probability (i.e., the probability that a user's demand is unserved). They assume devices communicate with each other with a fixed data rate and there is no power control over the \ac{D2D} link. Through simulations, the authors show that their proposed method outperforms the schemes with conventional unicast video transmission as well as the coded broadcasting~\cite{maddah2012fundamental}. The results show that their proposed method can achieve at least $10000\%$ and $1000\%$ throughput gain over the conventional and coded broadcasting methods, respectively, when the outage probability is less than $0.1$.

Wang \textit{et al.} \cite{wang2014WOWMOM} propose a \ac{BITS} algorithm to exploit both the cellular and \ac{D2D} links. BITS leverages devices' instantaneous channel conditions and queue backlogs to maximize the \ac{BS}'s scheduling options and hence increases the opportunistic gain. The authors model the \ac{BITS} policy with the objective to maximize delay-sensitive utility under an energy constraint. They develop an online scheduling algorithm using stochastic Lyapunov optimization and study its properties. 
Through simulations, they show that under \ac{BITS} the utility can be improved greatly and the average packet transfer delay can be reduced by up to $70\%$. The authors also evaluate BITS using realistic video traces. The results show that \ac{BITS} can improve the average peak signal-to-noise ratio of the received video by up to $4$ dB and the frame loss ratio can be reduced by up to $90\%$.

\begin{figure}[!ht]
  \centering
  \includegraphics[width=88mm]{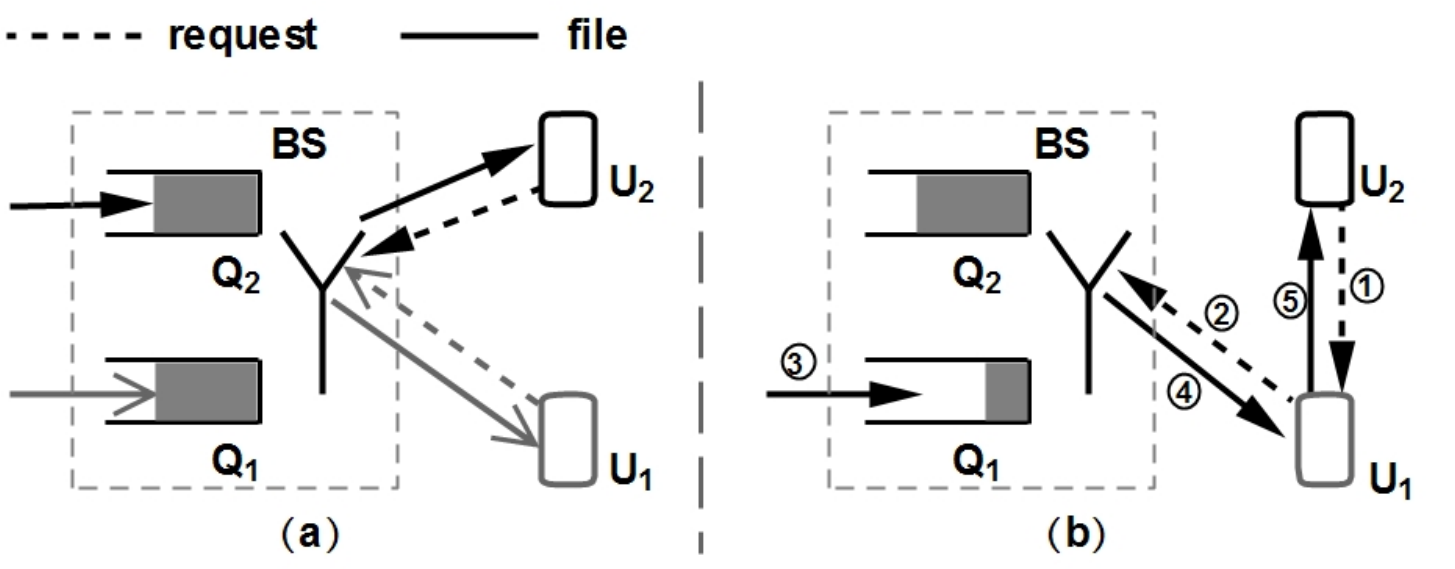}
  \caption{An example of the BS-transparent traffic spreading: (a) No traffic spreading; (b) Traffic spreading from $U_2$ to $U_1$.}
  \label{fig:wang2013wowmom_scheme}
\end{figure}

\subsubsection{Autonomous}
Autonomous \ac{D2D} communications are usually motivated by reducing the overhead of cellular networks. It does not require any changes at the \ac{BS} and can be deployed easily. Currently, there are very few works in this category. Wang \textit{et al.}~\cite{wang2013WOWMOM} propose a downlink \ac{BS}-transparent dispatching policy where users spread traffic requests among each other to balance their backlogs at the \ac{BS}, as shown in Figure~\ref{fig:wang2013wowmom_scheme}. They assume that users' traffic is dynamic, i.e., the \ac{BS} does not always have traffic to send to all the users at any time. They illustrate the dispatching policy by considering a scenario with two users, $U_1$ and $U_2$ being served by the \ac{BS}. The queues $Q_1$ and $Q_2$ depict the numbers of files at user's \ac{BS} queues. In Figure~\ref{fig:wang2013wowmom_scheme}(a), since the queues at the \ac{BS} are balanced, the dispatchers at each user would detect that traffic spreading is not beneficial. Thus, users send their new requests to the \ac{BS} directly. In Figure~\ref{fig:wang2013wowmom_scheme}(b), there are more files in $Q_2$ than $Q_1$. The dispatcher of $U_2$ would detect that traffic spreading is beneficial, because in the near future $Q_1$ would be empty and thus the opportunistic scheduling gain is lost. Therefore, $U_2$ asks $U_1$ to forward its new file requests to the \ac{BS}. After receiving the corresponding files from the \ac{BS}, $U_1$ forwards them to $U_2$. 
This dispatching policy is user-initiated (i.e., it does not require any changes at the \ac{BS}) and works on a per-file basis. This policy exploits both the time-varying wireless channel and users' queueing dynamics at the \ac{BS} in order to reduce average file transfer delays seen by the users.  The users perceive their channel conditions to the \ac{BS} (i.e., cellular channel conditions) and share them among each other. The authors formulate the problem of determining the optimal file dispatching policy under a specified tradeoff between delay performance and energy consumption as a Markov decision problem. Next, they study the properties of the corresponding optimal policy in a two-user scenario. A heuristic algorithm is proposed which reduces the complexity in large systems by aggregating the users. 
%\off{Based on realistic Rayleigh fading channels, the authors provide simulation results demonstrating that file transfer delays can be reduced by up to 50\% using the proposed methodology and that significant gains (up to 78\% of the gain) are typically achieved at only $20\%$ of the power expenditure of the performance-centric algorithm, which achieves the best performance at high power expenditure.}
The simulation results demonstrate that the file transfer delays can be reduced by up to $50\%$ using the proposed methodology. In addition, their proposal consumes $80\%$ less power than performance-centric algorithms while achieving significant gains (up to $78\%$).

\begin{table*}[t!]
\centering
\caption{Summary of the literature proposing outband \ac{D2D}}
\label{tb:outband}
\resizebox{\columnwidth}{!}	{$
\begin{tabular}{| l | l | l | l | l | l | l |}
\hline
Proposal & Analytical tools	& Platform & Direction & Use-case & Evaluation & Achieved performance\\
\hline \hline
\tabincell{l}{Improving throughput\\ of video distribution\\ \cite{golrezaei2012Icc, 
	golrezaei2012Globecom}} & \tabincell{l}{-Game theory\\ -Chen-Stein method}  & & -Downlink 
	& -Content distribution & \tabincell{l}{-Numerical \\~simulation} 
	& \tabincell{l}{-Video throughput is 
		\\~improved by up to two 
		\\~orders of magnitude} \\ \hline	
\tabincell{l}{Reducing channel\\ sensing overhead~\cite{zhou2013Wcnc}}
	&  & -LTE & & -Relaying & \tabincell{l}{-System-level\\~simulation}
	& \tabincell{l}{-Throughput is improved 
		\\~by up to $25\%$} \\ \hline	
\tabincell{l}{Improving  throughput,\\ energy efficiency, and \\fairness~\cite{asadi2013MSWIM, 
	asadi2013energy, Cai2013}} & -Game theory  & \tabincell{l}{-LTE\\ -CDMA} & -Downlink 
	& \tabincell{l}{-Relaying\\ -Video transmission} 
	& \tabincell{l}{-Numerical\\ ~simulation} 
	& \tabincell{l}{-Throughput and energy 
		\\~efficiency are improved 
		\\~by $50\%$ and $30\%$ over 
		\\~classical Round Robin 
		\\~scheduler, respectively} \\ \hline
\tabincell{l}{Designing a protocol\\ for outband \ac{D2D} \\communications~\cite{asadi2013WD}} 
	&   & -LTE & \tabincell{l}{ -Downlink \\-Uplink}
	&-Relaying  & \tabincell{l}{-System-level \\~simulation}
	& \tabincell{l}{-$50\%$ delay improvement
		\\~compared to Round Robin
		\\~scheduler } \\ \hline	
\tabincell{l}{Improving video\\ transmission~\cite{ji_wireless_2013}}
	&  & -LTE & -Downlink & -Video transmission 
	& \tabincell{l}{-System-level \\~simulation} 
	& \tabincell{l}{-Throughput is improved by
		\\~$10000\%$ and $1000\%$ over 
		\\~conventional and coded
		\\~broadcasting methods, 
		\\~respectively} \\ \hline
\tabincell{l}{Improving delay\\sensitive utility~\cite{wang2014WOWMOM}}
	& \tabincell{l}{-Lyapunov optimization} & -LTE & -Downlink
	& \tabincell{l}{-Online gaming\\ -Live video} 
	& \tabincell{l}{-System-level \\~simulation}
	& \tabincell{l}{-Average packet delay is
		\\~reduced by up to $70\%$
		\\-Utility can be improved 
		\\~greatly} \\ \hline
\tabincell{l}{Reducing average file\\ transfer delay~\cite{wang2013WOWMOM}}
	& \tabincell{l}{-Dynamic programming\\ -Heuristic algorithm\\ 
	-Distributed algorithm\\ -Queueing theory} & -LTE & -Downlink
	& \tabincell{l}{-Web browsing\\ -HTTP live streaming} 
	& \tabincell{l}{-System-level \\~simulation}
	& \tabincell{l}{-Average file transfer delay 
		\\~is reduced by up to $50\%$ 
		\\~compared to methods
		\\~without traffic spreading} \\ \hline
\end{tabular}
$}
\end{table*}

A summary of the works on outband \ac{D2D} communications in cellular networks is provided in Table~\ref{tb:outband}.

\subsection{Discussions}
\label{s:discussions}

So far we have reviewed the available literature on \ac{D2D} communications in cellular networks. In this section, we will shed light on some important factors such as common assumptions, scope of the works, and common techniques. 

\subsubsection{Common Assumptions} 

Most of the papers in the literature assume the \ac{BS} is aware of the instantaneous \ac{CSI} of cellular and/or \ac{D2D} links, e.g.,~\cite{min_capacity_2011,xiao_qos_2011,feng2013TOC, yu_device_2012,yu_power_2009,su_resource_2013,han_uplink_2012}. This assumption is essential because their proposed solutions need the \ac{BS}'s participation to make scheduling decisions for cellular and \ac{D2D} users. Alternatively, when the \ac{D2D} users decide on the their transmission slots, the common assumption is that \ac{D2D} users are aware of the cellular and \ac{D2D} links. On the other hand, there are also papers such as~\cite{yu_performance_2009} and \cite{wang2013WOWMOM} that assume the \ac{BS} or \ac{D2D} users are only aware of the statistical \ac{CSI} of the links. With this assumption, the large overhead for reporting instantaneous \ac{CSI} can be avoided.
To mitigate possible interference from \ac{D2D} transmissions to cellular transmissions, \cite{kaufman_cellular_2008} assumes that \ac{D2D} users are aware of minimum interference threshold of cellular users. With the latter assumptions, the \ac{D2D} users can opportunistically choose the transmission slots in which they do not interfere with the cellular users. 

The proposals which involve in clustering users commonly assume that the cluster are far enough so that there is no or negligible interference among different clusters, e.g.,~\cite{du_compressed_2012, cheng_resource_2012,asadi2013energy,asadi2013MSWIM}. This assumption may not hold in populated areas or dense deployments. A very interesting observation from the reviewed literature is that the majority of articles assume that the \ac{BS} or \ac{D2D} users always have traffic to send, therefore they use throughput as a common metric. However, the authors of~\cite{wang2013WOWMOM, wang2014WOWMOM} consider a scenario with dynamic traffic load and evaluate the average file transfer delay and delay-sensitive utility under their proposed traffic spreading mechanism, respectively. Since the latter assumption is more realistic, it would be interesting to see the performance of the aforementioned works under dynamic traffic flows. 

\subsubsection{Inband or Outband?}
Majority of the papers propose to reuse the cellular resources for \ac{D2D} communications (i.e., inband)~\cite{kaufman_cellular_2008,doppler2009COMMAG,doppler_device_2009A,Doppler2010WCNC,doppler_advances_2011}. However, outband communications is attracting more and more attention in the past few years~\cite{asadi2013MSWIM,asadi2013energy,wang2013WOWMOM,wang2014WOWMOM,wang2014arXiv}.  Before comparing the two approaches, we summarize the advantages and disadvantages of each approach. 

\begin{table*}[t!]
\centering
\caption{Advantages and disadvantages of  different types of \ac{D2D} communications}
\label{tb:adv-disadv}
\resizebox{\columnwidth}{!}	{$
\begin{tabular}{|l|c|c|c|c|}
\cline{2-5}
\multicolumn{1}{c|}{}							&\multicolumn{2}{|c|}{Inband} 			&\multicolumn{2}{|c|}{Outband}	 		\\
\cline{2-5}
\multicolumn{1}{c|}{}							&Underlay		&Overlay			&Controlled		&Autonomous		\\	
\hline
\hline
Interference between \ac{D2D} and cellular users 		&$\checkmark$		&$\times$			&$\times$			&$\times$			\\
\hline
Requires dedicated resources for \ac{D2D} users 		&$\times$			&$\checkmark$		&$\times$			&$\times$			\\
\hline
Controlled interference environment 				&$\checkmark$		&$\checkmark$		&$\times$			&$\times$			\\
\hline
Simultaneous \ac{D2D} and cellular transmission		&$\times$			&$\times$			&$\checkmark$		&$\checkmark$		\\
\hline
Requires inter-platform coordination				&$\times$			&$\times$			&$\checkmark$		&$\checkmark$		\\
\hline
Requires devices with more than one radio interface&$\times$			&$\times$			&$\checkmark$		&$\checkmark$		\\
\hline
Introduces extra complexity to scheduler			&$\checkmark$		&$\checkmark$		&$\checkmark$		&$\times$			\\
\hline
\end{tabular}
$}
\end{table*}		

{\bf Inband.} Inband \ac{D2D} is advantageous in the sense that: $(i)$ underlay \ac{D2D} increases the spectral efficiency of cellular spectrum by exploiting the spatial diversity; $(ii)$ any cellular device  is capable of using inband \ac{D2D} communications (the cellular interface usually does not support outband frequencies); and $(iii)$ \ac{QoS} management is easy because the cellular spectrum can be fully controlled by the \ac{BS}. The disadvantages of inband \ac{D2D} communications are: $(i)$ cellular resources might be wasted in overlay \ac{D2D}; $(ii)$ the interference management among \ac{D2D} and cellular transmissions in underlay is very challenging; $(iii)$ power control and interference management solutions usually resort to high complexity resource allocation methods; and $(iv)$ a user cannot have simultaneous cellular and \ac{D2D} transmissions. 
%\add[comments]{(Do you mean for a user, \ac{D2D} and cellular communications can not happen simultaneous?) } 
It appears that underlay \ac{D2D} communications is more popular than overlay. The authors who propose to use overlay \ac{D2D} usually try to avoid the interference issue of underlay~\cite{fodor_design_2012,li_device_2012,zhou2013TVT}. However, allocating dedicated spectrum resources to \ac{D2D} users is not as efficient as underlay in terms of spectral efficiency. We believe that the popularity of underlay \ac{D2D} is due to its higher spectral efficiency. 

{\bf Outband.} This type of \ac{D2D} communications has merits such as: $(i)$ there is no interference between cellular and \ac{D2D} users; $(ii)$ there is no need for dedicating cellular resources to \ac{D2D} spectrum like overlay inband \ac{D2D}; $(iii)$ the resource allocation becomes easier because the scheduler does not require to take the frequency, time, and location of the users into account; and $(iv)$ simultaneous \ac{D2D} and cellular communications is feasible. Nevertheless outband \ac{D2D} has some disadvantages which are: $(i)$ the interference in unlicensed spectrum is not under the control of the \ac{BS}; $(ii)$ only cellular devices with two radio interfaces (e.g., LTE and WiFi) can use outband \ac{D2D} communications; $(iii)$ the efficient power management between two wireless interfaces is crucial, otherwise the power consumption of the device can increase; and $(iv)$ packets (at least the headers) need to be decoded and encoded because the protocols employed by different radio interfaces are not the same. 

Although the literature on inband \ac{D2D} is wider than that of outband, it seems that researchers have started to explore the advantages of outband \ac{D2D} and they are considering it as a viable alternative to inband \ac{D2D}. We believe that with the evolutionary integration of smartphones in phone market, the majority of mobile devices will be equipped with more than one wireless interface which makes it possible to implement outband \ac{D2D} schemes. Moreover, the standards such as 802.21~\cite{802.21} are looking into handover to and from different platforms (e.g., WiMAX and LTE) which could significantly reduce the complexity of coordination between different wireless interfaces in outband \ac{D2D}.  Table~\ref{tb:adv-disadv} summarizes the above mentioned merits and disadvantages.

\subsubsection{Maturity of \ac{D2D} in Cellular Networks}
We believe \ac{D2D} communications in cellular networks is a relatively young topic and there is a lot to be done/explored in this field.
We support this belief by looking into the analytical techniques and evaluation methods which are used in the available literature. 

{\bf Analytical techniques.}
In comparison to other fields such as opportunistic scheduling~\cite{asadi2013survey}, the number of techniques used in the literature and their popularity is very low. The majority of the literature only proposes ideas, architectures, or simple heuristic algorithms. 
%\remove[Q]{In a few cases an optimization problem is introduces which is left unsolved due to NP-Hardness.}
Some of the papers formulate their objectives as optimization problems but leave them unsolved due to NP-hardness.
Therefore, we believe there is room for investigating optimal solutions for interference coordination, power management, and mode selection.  Table~\ref{tb:math} summarizes the mathematical techniques used in the \ac{D2D} related literature. 

\begin{table}[h!]
\centering
\caption{Analytical tools used in the literature}%{\red are all the papers here?} {\blue Qing: to be updated tmr}}
\label{tb:math}
\begin{tabular}{|l|c|}
\hline
Tools 							& { Related literature}\\
\hline
Discrete Time Markov chain			& \cite{akkarajitsakul_mode_2012} \\
Merge and split algorithm			& \cite{akkarajitsakul_mode_2012} \\
Distributed algorithms				& \cite{wang2013WOWMOM,akkarajitsakul_mode_2012, 
	han_uplink_2012}\\
Coalitional game theory				& \cite{akkarajitsakul_mode_2012}\\
Poisson point process				& \cite{lin_comprehensive_2013} \\
Queueing theory						& \cite{wang2013WOWMOM} \\
Alzer's inequality					& \cite{lin_comprehensive_2013} \\
Fubini's theorem					& \cite{lin_comprehensive_2013} \\
Laplace transform					& \cite{lin_comprehensive_2013,Erturk_distributions_2013} \\
Slivnyak's theorem					& \cite{lin_comprehensive_2013} \\
Heuristic algorithm					& \cite{wang2013WOWMOM,xiao_qos_2011, 
	han_uplink_2012} \\
Convex Optimization					& \cite{li_device_2012,liu_optimal_2012} \\
Chen-Stein Method					& \cite{golrezaei2012Globecom} \\
Maximum Weight Bipartite Matching	& \cite{feng2013TOC}\\
Kuhn-Munkres algorithm				& \cite{feng2013TOC}\\
Han-Kobayashi						& \cite{yu_device_2012} \\
Jensen's Inequality					& \cite{Erturk_distributions_2013} \\
Mixed integer nonlinear programming	& \cite{zulhasnine2010WiMob}\\
Integer programming					& \cite{le_fair_2012}\\
Linear programming					& \cite{belleschi_performance_2011, le_fair_2012}\\
Nonlinear programming				& \cite{han_uplink_2012} \\
Dynamic programming					& \cite{wang2013WOWMOM} \\
Newton's method						& \cite{le_fair_2012} \\
Lagrangian multipliers				& \cite{le_fair_2012} \\
Graph theory						& \cite{zhang_interference_2013} \\
Auction algorithms					& \cite{xu_interference_2012,xu_resource_2012} \\	
Exhaustive search 					& \cite{jung_joint_2012} \\
Geometrical probability				& \cite{kaufman2013ton}\\
Evolution theory					& \cite{cheng_resource_2012} \\
Particle swarm optimization			& \cite{su_resource_2013} \\
Sub-gradient algorithm				& \cite{han_subchannel_2012} \\
Hungarian algorithm					& \cite{han_uplink_2012} \\
Lyapunov optimization				& \cite{wang2014WOWMOM} \\
\hline
\end{tabular}
\end{table}

{\bf Evaluation method.}
Another metric for maturity of a field is the evaluation method. The more realistic the evaluation method, the more mature the study of that field. Table~\ref{tb:eval} shows different evaluation methods used in the literature. As we can see, majority of the papers use numerical evaluation and some use simple home-grown simulators. There is no paper using experimental evaluation. This is mainly due to the fact that experimental testbeds for cellular network are extremely costly and do not have support for \ac{D2D} yet. The literature rarely uses popular network simulators such as NS3~\cite{ns3}, OPNET~\cite{opnet}, Omnet++~\cite{varga2007omnet++}.  In turn, currently available network simulators do not support \ac{D2D} communications. 

\begin{table}[h!]
\centering
\caption{Evaluation methods in the literature}
\label{tb:eval}
\begin{tabular}{|l|c|}
\hline
Evaluation method		& {Related literature}\\
\hline
Numerical simulation	& \cite{min_capacity_2011, lin_comprehensive_2013, du_compressed_2012, 
	golrezaei2012Icc, kaufman_cellular_2008, doppler2009COMMAG, feng2013TOC, 
	yu_device_2012} \\
	& \cite{Erturk_distributions_2013, fodor_design_2012, chen_downlink_2012, le_fair_2012, 
	janis_interference_2009, zhou2013TVT, pratas_low_2013, Doppler2010WCNC} \\
	& \cite{cheng_resource_2012,pei_resource_2013,yu2011TWC,han_subchannel_2012, 
	golrezaei2012Globecom, zhang_interference_2013, yu_performance_2009, yu_power_2009} \\
	& \cite{yu_performance_2009,min_reliability_2011,xu_resource_2012,su_resource_2013, 
	vanganuru_system_2012,xu_resource_2012,su_resource_2013, vanganuru_system_2012} \\ \hline
%{\red Not sure} & {\red \cite{osseiran_advances_2009A,li_device_2012,
%	xing_investigation_2010}} \\ 
%	\hline
System-level simulation & \cite{zulhasnine2010WiMob, belleschi_performance_2011, 
	zhou2013Wcnc, peng_interference_2009,xu_resource_2012,liu_optimal_2012, 	
	doppler_device_2009A,janis_interference_2009} \\
	& \cite{xiao_qos_2011,rasmussen2000matrix,akkarajitsakul_mode_2012, ji_wireless_2013, 
	jung_joint_2012,bao_dataspotting_2010, seppala2011WNCN, koskela2010WCNC} \\
	& \cite{zhu_qos_2011,wang2013WOWMOM,han_uplink_2012,wang2014WOWMOM} \\
\hline
Experiment	& No experimental study\\			
\hline
\end{tabular}
\end{table}

%\begin{itemize}
%\item why uplink resources are used more than downlink
%\item discussion on self organized \ac{D2D} or autonomous \ac{D2D}.
%\item a table or graph showing the focus of paper (e.g., power, interference, throughput.)
%\item a table for advantageous and disadvantageous of inband and outband. 
%\end{itemize}

\subsubsection{How Far is \ac{D2D} from a Real World Implementation?}
Although \ac{D2D} communication is not mature yet, it is already being studied in the 3GPP standardization body~\cite{3GPPTR22.803,3GPPTR23.703}.  3GPP recently decided that the focus of \ac{D2D} in LTE would be on public safety networks~\cite{lin2014ComMag}.  Moreover, Qualcomm has shown interest in this technology and they also built a prototype for \ac{D2D} communications in cellular network which can be used in different scenarios such as social networking, content sharing, and so on~\cite{corson_toward_2010}. This confirms that \ac{D2D} communications are not only a new research topic in academia, but also that there is interest in such a technology in the industry. There are various obstacles to implement \ac{D2D} in cellular networks. For example, the operators are used to having control of their spectrum and the way it is used. As a result, a successful \ac{D2D} implementation should allow \ac{D2D} communications in a manner that operators are not stripped off their power to control their network. Moreover, there are physical challenges such as suitable modulation format and \ac{CSI} acquisition which should be addressed efficiently.  Therefore, we believe that thanks to \ac{ProSe}, \ac{D2D} communications will become an essential part of cellular communications in the next few years. 

\subsection{Future Trends and Thesis Contribution}
%\subsubsection{\ac{D2D} Implementation Challenges in Real World}
Although \ac{D2D} communication triggered a lot of attention and interest in academia, industry, and standardization bodies, it is not going to be integrated into the current communication infrastructure until the implementation challenges are resolved. Here, we explain some of the major challenges faced by \ac{D2D} communications. 

{\bf Interference management.}  Under inband \ac{D2D} communications, UEs can reuse uplink/downlink resources in the same cell. Therefore, it is important to design the \ac{D2D} mechanism in a manner that \ac{D2D} users do not disrupt the cellular services.  Interference management is usually addressed by power and resource allocation schemes, although the characteristics of \ac{D2D} interference are not well understood yet.

{\bf Power allocation.} In inband \ac{D2D}, the transmission power should be properly regulated so that the \ac{D2D} transmitter does not interfere with cellular UE communications while maintaining the minimum \ac{SINR} requirement of the \ac{D2D} receiver. In outband \ac{D2D}, the interference between \ac{D2D} and cellular user is not of concern. Therefore, power allocation may seem irrelevant in outband \ac{D2D}. However, with increased occupancy of \ac{ISM}  bands, efficient power allocation becomes crucial for avoiding congestion,  collision issues, and inter-system interference.

{\bf Resource allocation.} This is another important aspect of \ac{D2D} communications specially for inband \ac{D2D}. Interference can be efficiently managed if the \ac{D2D} users communicate over resource blocks that are not used by the nearby interfering cellular UEs. Resource allocation for outband \ac{D2D} simply consists in avoiding \ac{ISM}  bands which are currently used by other \ac{D2D} users, WiFi hotspots, etc.  

{\bf Modulation format.} This is one of the challenges which is rarely addressed by researchers. The existing LTE UEs use an \ac{OFDMA} receiver in downlink and a \ac{SC-FDMA} for uplink transmission. Thus, for using downlink (resp. uplink) resources, the \ac{D2D} UE should be equipped with \ac{OFDMA} transmitter (resp. \ac{SC-FDMA} receiver)~\cite{lin2014ComMag}. 

{\bf Channel measurement.} Accurate channel information is indispensable to perform efficient interference management, power allocation, and resource allocation. Conventional cellular systems only need the downlink channel information from UEs and the uplink channel information is readily computed at the base station. Unfortunately, \ac{D2D} communications require information on the channel gain between \ac{D2D} pairs, the channel gain between \ac{D2D} transmitter and cellular UE, and the channel gain between cellular transmitter and \ac{D2D} receiver. The exchange of such extra channel information can become an intolerable overhead to the system if the system needs instantaneous \ac{CSI} feedback. The trade-off between accuracy of \ac{CSI} and its resulting overhead is to be further investigated. 

{\bf Energy consumption.} \ac{D2D} communications can potentially improve the energy efficiency of the UE. However, this highly depends on the protocol designed for device discovery and \ac{D2D} communications. For example, if the protocol forces the UE to wake up very often to listen for pairing requests or to transmit the discovery messages frequently, the battery life of the UE may significantly reduces. The trade-off between UE's power consumption and discovery speed of the UEs should be better studied.

{\bf \ac{HARQ}.} Considering the complexity of interference management in \ac{D2D} communications, \ac{HARQ}  appears to be a viable technique to increase the robustness.  
\ac{HARQ}  can be sent either directly (i.e., from the \ac{D2D} receiver to the transmitter) or indirectly (i.e., from the \ac{D2D} receiver to the eNB, and from the eNB to the \ac{D2D} transmitter)~\cite{lin2014ComMag}. The direct mode poses less overhead to the eNB in comparison to indirect mode. Moreover, benefits from the ACK/NACK messages arrive to the transmitter with shorter delay. 

%\subsubsection{Thesis Contribution}

In this thesis, we contributed to several open problems in \ac{D2D} communications. We investigated energy consumption of the mobile devices participating in opportunistic \ac{D2D} clustering techniques. Moreover, we explored the implementation feasibility of \ac{D2D} clustering from a protocol point of view in LTE-A and WiFi Direct technologies. Last but not least, we prototyped the first experimental testbed for outband \ac{D2D} communications.

\part{\titFirstPart}
%\label{part:struct}
%\acresetall
\chapter{\titThirdCh}
\label{ch:clus_theo}

%\vspace{-9mm}
\section{Introduction}

%The emergence of \acf{D2D} communications has set off numerous proposals in industry and academia to improve the performance of cellular networks.  As of today there are not only several proposals for cellular relaying, multicasting, cellular offloading, and content distribution leveraging \ac{D2D}~\cite{lin_multihop_2000, asadi2013survey}, but also entire system architectures based on \ac{D2D} to  {\it complement}  cellular-based services in a scalable way with new types of applications~\cite{wu2013TON}. Indeed, Release 12 of 3GPP already specifies system overview and discovery procedure of  \ac{D2D} communications. Moreover, There are still ongoing studies on architecture enhancements and radio management aspects of \ac{D2D}~\cite{3GPP23.303,3GPP36.843}. To date, LTE-specific  \ac{D2D} data communication protocols  are left for future standardization activities~\cite{Flynn3GPPProse}.

In 3GPP's definition~\cite{3GPP23.203}, \ac{D2D} is a flexible paradigm that is open to use cellular platforms (i.e., inband D2D) or 802.11-like platforms (i.e., outband D2D) for direct communication~\cite{asadi2014survey}. 
Flexibility also extends to a variety of use-cases such as public safety, content distribution, advertisement, and network offloading using 802.11 protocols. In particular, network offloading based on \ac{D2D} has been studied in terms of grouping/clustering techniques~\cite{doppler2013patent, seppala2011WNCN, zhou2013TVT, bao2013infocom, asadi2013MSWIM, asadi2013WD}. The analytical and simulation results indicate that clustering cellular users to relay traffic to each other leads to lower signaling overhead, higher spectral efficiency, and better energy efficiency.
However, while the literature on inband D2D communications is abundant~\cite{asadi2014survey}, outband D2D has been less studied. This is mostly due to the skepticism of the research community towards acceptance of 802.11 as a possible platform for {\it network-controlled} \ac{D2D} communications in 3GPP \ac{ProSe}.  Nevertheless, the slow standardization process of inband D2D is driving industry to look at outband D2D as a more tangible implementation option.

\begin{figure} [h!]
\begin{center}
\includegraphics[scale=0.6]{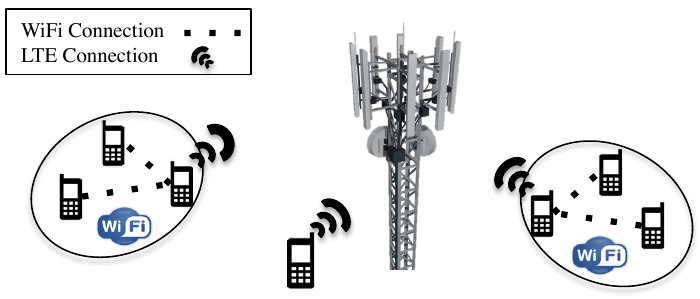}
%\vspace{-4mm}
\caption{Example scenario of D2D-clustering architecture with LTE and WiFi Direct coexistence.}
\label{fig:net_model}
\end{center}
%\vspace{-13mm}Figure
\end{figure}

In this chapter, we investigate an outband D2D clustering scheme that {\it opportunistically} leverages the flexibility of D2D communications in cellular networks for improving network performance under the control of the cellular operator. As shown in Figure~\ref{fig:net_model}, in our vision, cellular devices can form stable clusters using WiFi Direct~\cite{WifiDirect2013} and the cluster member with the highest channel quality acts as relay for other cluster members. The {\it opportunism} in our proposal is twofold: first, the relay node changes over time within the same cluster, to follow signal quality variations; second, the throughput gain obtained due to clustering is shared within the cluster according to the contribution of each member. In the proposal detailed in this article, we aim at maximizing the efficiency of cellular resource utilization, although we also account for the impact of per-user performance in general and for cluster formation policies in particular, which we analyze by using game theory.  
%We also prototype the first SDR-based experimental platform for a more realistic analysis of D2D proposals. Our evaluation results indicate that joining a cluster is beneficial for all members, not just for average system performance. Indeed, our experimental evaluation showed that our proposal enhances the system capacity by up to $71\%$ with clusters as small as five users. 
%The main contributions of our work consist in the design, analysis, simulation and SDR-based experimental validation of our network-controlled opportunistic outband D2D clustering scheme. 

\section{System Model}
\label{s:system_model}
%\vspace{-1mm}
This section details the architecture of our proposed D2D-clusters, the model assumptions, and the throughput and power model of the proposed scheme.
In our proposed scheme, mobiles can form clusters and receive downlink traffic through the {\it cluster head}, i.e., the node enabled to exchange data with the base station (see Figure~\ref{fig:net_model}).
Note that each node is a potential candidate to act as cluster head, and cluster heads are selected on a per-frame basis. A cluster consists of several mobiles that form a WiFi Direct group and all intra-cluster communications take place over the WiFi Direct network. Since the base station is aware of  and controls  clustering decisions,  whenever a packet is destined to a cluster member, the base station simply sends it to the cluster head. This maximizes the throughput at that  scheduling  epoch. Thereby, the base station schedules entire clusters as if they were regular users. From a modeling perspective, a cluster can be considered as a user whose  \ac{SNR} is the highest of the \ac{SNR} values of cluster members. As for intra-cluster resource sharing, unless otherwise specified, we assume that the extra throughput gained from clustering is equally distributed among users.

\subsection{Assumptions}
We model downlink transmissions in a single LTE-like in FDD mode. 
There are $N$ mobile users in the cell. %served either with legacy user-based schedulers or with our proposed cluster-based scheduling algorithms. 
Since we are interested in network capacity and fairness under heavy load conditions, we study the case of fully backlogged downlink flows in the analysis. Nevertheless, our analysis can be easily applied for uplink communications as well. The number of OFDM symbols in downlink is denoted by $S_{tot}$, and
%with the maximum achievable rate (i.e., \cyan{using} 64QAM-4/5)
%\footnote{We neglect LTE protocol overhead, and consider that all available downlink resources can be used for data traffic.} 
we assume that the D2D link (i.e., WiFi Direct) does not become a bottleneck in the data flow. 
%
%without carrier aggregation is $80.64$ Mbps
%
In fact, considering the short-range nature of D2D communications, the available WiFi capacity exceeds per-cluster achievable throughput over LTE.%
\footnote{There are at least 26 non-overlapping WiFi channels in 2.4GHz and 5GHz band with per-channel nominal capacity of $433$ Mbps, which is much more than the $80$ Mbps achievable in a SISO LTE-A system or the $298$ Mbps of a $4 \times 4$ MIMO system.}
It is also assumed that all mobiles belong to the same operator. The downlink channel of mobile node $i$ is characterized by stationary Rayleigh fading.
 %as in~\cite{rong2012TranIT,rong2009wiopt}. 
 Therefore, the \ac{SNR} can be described as a r.v. $C_i$ with average \ac{SNR} $\gamma_i$, so that the \ac{CDF} of the \ac{SNR} has the following expression:
\begin{eqnarray}
%	f_i(z) = \frac{1}{\gamma_i} e^{-\frac{z}{\gamma_i}} u(z), 
%	\quad 
	F_i(z) = 1 - e^{-\frac{z}{\gamma_i}}, z \geq 0, \forall i \in \{1\dots N\}.  
\end{eqnarray}
We assume that user channels are independently distributed but not identically, and the \ac{CSI} is available at the base station.
Transmissions occur at different rates according to $M$ available \acp{MCS}. We assume that the \ac{MCS} for user $i$ is a function of the instantaneous \ac{SNR}, i.e.: 
\begin{eqnarray}
MCS_i = k  \iff  C_i \in \left[th_k; th_{k+1} \right[, \; k=1 \dots M; th_1 = 0;   th_p < th_q \iff p < q;  th_{M+1} = \infty. \nonumber
\end{eqnarray} 
%\begin{eqnarray}
%th_1 = 0;  & th_p < th_q \iff p < q; & th_{M+1} = \infty. \nonumber
%\end{eqnarray} 
Hence, the probability that scheduled transmissions to 
%scheduled 
user $i$ 
%receives a frame 
are encoded 
%according to 
 with the $k$th \ac{MCS} is:
\begin{eqnarray}
\label{eq:pi}
\pi_k^{(i)} =  \int_{th_k}^{th_{k+1}} dF_i(z) =  e^{-\frac{th_k}{\gamma_i}} -  e^{-\frac{th_{k+1}}{\gamma_i}}.
\end{eqnarray}

The number of data bits transferred in one \ac{OFDMA} symbol using the $k$th \ac{MCS} is denoted by $b_k$. 

%\renewcommand{\arraystretch}{1.2}
%\newcolumntype{H}{>{\setbox0=\hbox\bgroup}c<{\egroup}@{}}%this is used to delete the SINR column from the table
%\begin{table}[t!]
%\vspace{-3mm}
%\caption{Modulation and coding schemes and their thresholds}
%\vspace{1mm}
%\scriptsize
%\centering
%\label{tb:MCS}
%\begin{tabular}{|@{}c@{}|@{ }c@{ }|@{ }c@{ }|@{ }c@{ }|@{ }c@{ }|H @{ }c@{ }|}
%%\begin{tabular}{|c|c|c|c|c|H c|}
%\hline
%\scriptsize \textbf{ Modulation } & \scriptsize \textbf{ Coding } & \scriptsize \textbf{ \ac{SNR} } & \scriptsize \textbf{ IM } & \scriptsize \textbf{ \ac{SNR}+IM } & \scriptsize \textbf{\ac{SNR}} &\scriptsize $\textit{\textbf{b}}_\textit{\textbf{k}}$(Bits per  \\
% & \scriptsize \textbf{Rate} &\scriptsize (dB)&\scriptsize (dB)&\scriptsize (dB) && \scriptsize symbol) \\
%\hline
%No transmission & - & $-\infty$ & - & $-\infty$ &  & 0\\
%\hline
%\multirow{8}{*}{QPSK}
%& 1/8 & -5.1 &\multirow{8}{*}{2.5} & -2.6 & 0.54 &0.25\\
%& 1/5 & -2.9 & & - 0.4 & 0.91 &0.4\\
%& 1/4 & -1.7 & & 0.8 & 1.2 &0.5\\
%& 1/3 & -1 & & 1.5 & 1.41 &0.67\\
%& 1/2 & 2 & & 4.5 & 2.81 &1 \\
%& 2/3 & 4.3 & & 6.8 & 4.78 &1.3\\
%& 3/4 & 5.5 & & 8.0 & 6.3 &1.5\\
%& 4/5 & 6.2 & & 8.7 & 7.41 &1.6\\
%\hline
%\multirow{4}{*}{16QAM}
%& 1/2 & 7.9 & \multirow{4}{*}{3}& 10.9& 12.30 & 2 \\
%& 2/3 & 11.3 & &14.3 & 26.91 & 2.66\\
%& 3/4 & 12.2 & &15.2 & 33.13 & 3\\
%& 4/5 & 12.8 & &15.8 & 38.01 & 3.2\\
%\hline
%\multirow{3}{*}{64QAM}
%& 2/3 & 15.3 &\multirow{3}{*}{4} &19.3 & 85.11 & 4\\
%& 3/4 & 17.5 & & 21.5 & 141.25 &4.5\\
%& 4/5 & 18.6 & & 22.6 & 181.97 &4.8\\
%\hline
%\end{tabular}
%\vspace{-7mm}
%\end{table}

\subsection{Throughput Model for D2D-Clusters}
\label{s:sched}
%\vspace{-1mm}
In the following, the throughput of D2D clusters is modeled under two simple opportunistic cluster head selection (cluster scheduling) schemes.
%, namely, \red{\ac{CL(WRR)} and \ac{CL(MR)}.} 
Opportunistic schemes commonly result in unfairness which is often resolved at the cost of increased complexity. However, the practicality of such schemes  
is often doubted due to high computation overhead imposed to the base station. We intentionally opt for opportunistic schemes with low complexity to pave the way towards a practical proposal. We choose to resolve the unfairness issue by leveraging cooperative nature of D2D clusters instead of increasing the complexity. We consider the case in which the base station schedules $N_c$ clusters instead of $N$ normal users. 
This means that the base station decides which cluster has to be served, and then transmissions are managed by the 
current cluster head. Defining $X_n$ as the \ac{SNR} of cluster $n$ (${\textit CL_n}$), we have:  
$%\begin{eqnarray}
\label{eq:Xi}
X_n \!\!=\!\!  \max\{C_j, j:u_j \in {\textit CL_n} \}, \; n \in \{1 \dots N_c\},
$%\end{eqnarray}
where $u_j$ is user $j$, $ j\in \{1\dots N\}$. 
The \ac{CDF} of $X_n$ can be readily computed considering that the random
variables  $C_j$ are all independent: 
\begin{eqnarray}
\label{e:RRCFx}
F_{X_n}(z) & \!\!\!\! = \!\!\!\!& \prod_{ j \in{\textit CL_n} } \!\! F_j(z) = \! \prod_{j \in {\textit CL_n} } \!\!\!\! \left (  1 - e ^ { - \frac { z } { \gamma _ j} } \right), z \geq 0.
\end{eqnarray}
%the corresponding pdf, $f_{X_n}(z)$, can be obtained by derivation
%from $F_{X_n}(z)$. 
The adopted \ac{MCS}, for each transmission, only depends on 
the instantaneous \ac{SNR} of the best channel in the scheduled cluster, i.e., it 
only depends on $X_n$ at the scheduling epoch: 
\begin{eqnarray}
\pi_k^{({\textit CL_n})} = \int_{th_k}^{th_{k+1}} f_{X_n}(z)dz. 
\end{eqnarray}

%Note that the values for $\pi_k^{(CL_i)}$ are not affected by $n_s$, namely the number of users scheduled in a frame. 

%
%Because of round robin, the probability to schedule each cluster is $1/2$.  More 
%in general, in case that $N_c$ clusters are present and scheduled 
%in round robin, the cluster scheduling probability is $1/N_c$.
%
%The throughput of each cluster is then:
%\begin{eqnarray}
%\label{e:TCLi}
%T_{CL_i} & = & \frac{1}{N_c} \sum_{k=1}^{M} \pi_k^{(CL_i)}b_k , \quad i \in \{1..N_c\}.
%\end{eqnarray}
%
%Finally, the average per-node throughput is obtained by dividing the throughput 
%of the $i$th cluster by the number of users in the cluster, namely $N_i$:
%
%\begin{eqnarray}
%\label{e:uTCLi}
%T_{j: U_j \in CL_i} & = & \frac{1}{N_c N_i} \sum_{k=1}^{M} \pi_k^{(CL_i)}b_k , \quad i \in \{1..N_c\}.
%\end{eqnarray}

{\bf   \ac{CL(WRR)}.}
This scheme chooses the cluster member with the best channel quality as the cluster head and it schedules the cluster heads in a \ac{WRR} fashion. Hence, each cluster $n$ receives a portion of airtime which corresponds to its weight $w_n$, $n \!\! \in \!\! \{1 \dots N_c\}$. 
%The weights can be computed using a variety of methods. 
In this dissertation, the weight of  ${\textit CL_n}$ is calculated using $w_n\!\! =\!\! N_n / N$, where $N_n$ denotes the number of cluster members of ${\textit CL_n}$. 
In other words, each cluster receives an amount of airtime which is proportional to its size. 

In such a system, the per-cluster scheduling probability is exactly $w_n$, while  
the average symbol rate only depends on the selected \ac{MCS}. Since available frame resources $S_{tot}$ are allocated in a \ac{WRR} manner, the average cluster and the per-user throughput are given by the Propositions 1 and 2, whose proofs are immediate so we omit them. 

\begin{proposition}
\label{p:Tc-W}
Under \ac{CL(WRR)}, the average throughput received by cluster $CL_n$ is 
\begin{equation}
E[T_{{\textit CL_n}}]  \!=\!\
w_n S_{tot}\,  \sum_{k=1}^{M} \pi_k^{({\textit CL_n})}b_k, \;\; n \in \{1 \dots N_c\}.
\end{equation}  
\end{proposition}

\begin{proposition}
\label{p:Tu-W}
Under \ac{CL(WRR)}, the average throughput of user $i \in CL_n$ can be expressed as 
\begin{equation}
\label{eq:Tu-W}
E[T_ i] \! =\! \frac{S_{tot}}{N} \sum_{k=1}^{M} \pi_k^{({\textit CL_n})}b_k , \; i \in {\textit CL_n}, \; n\! \in \! \{1 \dots N_c\}.
\end{equation}  
\end{proposition} 

The following proposition gives the probability that a user $i$ is scheduled.  
\begin{proposition}
\label{p:Ph-W}
Under \ac{CL(WRR)}, a user $i \in CL_n$ is scheduled with probability 
\begin{equation}
\label{eq:Ph-W}
P_{h}^{(i)} \!\!=\! w_n \!\! \sum_{k=1}^M \pi_k^{({\textit CL_n})} \!\!\!\! \int_0^{\infty} \!\!\!\!\!\left[ 1 - F_i(z|MCS_i \! = \! k)\right] dF_{Y_i}(z), 
\end{equation} 
where $Y_i  =  \max_{j \in {\textit CL_n}\setminus\{i\}}\{C_j\}, \; i \in CL_n$.
\end{proposition}

The proof of Proposition~\ref{p:Ph-W} is reported in Appendix~A. 
Note that, under Rayleigh fading assumptions, 
the conditional probability $F_i(z|MCS_i \! = \! k)$ is simply given by the following formula:
\begin{align}
\label{eq:Fik}
\!\!\!\!F_i (z | MCS_i \! = \! k )  \!\!=\!\!  
 	\frac{F_i \left ( \min \left( z,  th_{k+1} \right )  \right )\!\!  - \!\! F_i  \left ( th_k  \right ) }
	{ \pi_k^{(i)}},  
	\, z \geq th_k.
\end{align}

\textbf{\ac{CL(MR)}.}
Here, the cluster heads are scheduled in a pure \ac{MR} fashion~\cite{knopp95ICC}. In this scheme, each frame resources $S_{tot}$ are allotted to the cluster whose cluster head is experiencing the best \ac{SNR} in the system. 
Propositions~\ref{p:Tc-M} and \ref{p:Tu-M} express the cluster throughput and average per-user throughput achieved using \ac{CL(MR)}.

\begin{proposition}
\label{p:Tc-M}
Under \ac{CL(MR)}, the average throughput received by cluster $CL_n$ is 
\begin{align}
E[T_{{\textit CL_n}}] 
& =  S_{tot} \sum_{k=1}^{M} \left [   \pi_k^{({\textit CL_n})}b_k \right.  \!\!\! \left . \int_0^{\infty}\!\!\!\!\!\! \left[ 1 - F_{X_n} (z|MCS_{{\textit CL_n}} =k)\right] dF_{Y_n}(z) \right ], 
\end{align}  
where $n\!\! \in \!\! \{1 \dots N_c\}$, $X_n$ is defined in~\eqref{eq:Xi}, and $Y_n \!\! =\!\!  \max\limits_{j \not \in {\textit CL_n}}\{C_j\}$.
\end{proposition}
\noindent The proof of Proposition~\ref{p:Tc-M} is reported in Appendix~A.%\ref{s:appA}. 

\begin{proposition}
\label{p:Tu-M}
Under \ac{CL(MR)},  the average throughput received by user $i \in CL_n$ is 
\begin{align}
\label{eq:Tu-M}
E[T_i] 
& =  
	\frac{S_{tot}}{N_n} \sum_{k=1}^{M} \left [   \pi_k^{({\textit CL_n})}b_k  \right .    
	\left . \int_0^{\infty}\!\!\!\!\!\! \left[ 1 - F_{X_n} (z|MCS_{{\textit CL_n}} =k)\right] dF_{Y_n}(z) \right ], 
\end{align}  
where $X_n$ is defined in Eq.~\eqref{eq:Xi},  and $Y_n  =  \max_{j \not \in {\textit CL_n}}\{C_j\}$.
\end{proposition} 

\noindent The proof of Proposition~\ref{p:Tu-M} is like the proof of Proposition~\ref{p:Tc-M}. 
The probability that a user $i$ is scheduled as a cluster head is given in the following proposition, which is proven in Appendix~A.%\ref{s:appA}. 

\begin{proposition}
\label{p:Ph-M}
Under \ac{CL(MR)}, a user $i$ is scheduled with probability 
\begin{equation}
\label{eq:Ph-M}
P_{h}^{(i)} =  \sum_{k=1}^{M} \pi_k^{(i)}
\int_0^{\infty}\!\!\!\!\!\! \left[ 1 - F_i (z|MCS_i =k)\right] dF_{Y_i}(z),
\end{equation} 
where $\!Y_i \!\!=\!\!  \max\limits_{j \not = i }\{C_j\!\}\!$ and $\!F_i (z|MCS_i \!\! = \!\! k\!)$ is given by Eq.\eqref{eq:Fik}.
\end{proposition}

\subsection{Power Consumption Analysis}
\label{s:power_model}
We derive the power consumption of mobiles from the empirical power models proposed for LTE and WiFi in~\cite{huang2012Mobisys} and \cite{serrano2014TON}. Those studies show: $(i)$ how to include the baseline power required to keep the interface
up and running; and $(ii)$ how to account for the variability of power consumption with transmission rate, and differentiate transmission from reception. 
However, unlike the existing models, we account for practical details such as power consumption of the UE in active and idle periods, and the difference between transmission and reception power. 

Before further elaboration on power model of D2D-clusters, we want to differentiate between the average throughput $E[T]$ and the data rate $R$ of a user. 
$E[T]$ is the user-application local data received by a user directly via LTE or via WiFi relay, and it is computed via~\eqref{eq:Tu-W} and \eqref{eq:Tu-M}. 
$R$ is the amount of data received by a user and it includes non-local traffic to be relayed.

\subsubsection{Power Saving in LTE and WiFi}
LTE allows the UE to switch to idle mode in order to save energy. The mechanisms that handle idle periods are discontinuous reception
and discontinuous transmission~\cite{LTEdrxSpecification}.
In WiFi, users can turn off the wireless interface during idle periods and only switch it on to 
receive beacons~\cite{gupta2007Secon}. 
In both LTE and WiFi, interfaces in power saving mode periodically wake up to transmit/receive control information even if there is no data traffic to handle. However, it has been shown that the periodic wake-up of power saving mechanisms in LTE and WiFi impacts at most $ 5\%$ of the idle time~\cite{huang2012Mobisys}. Therefore, for simplicity, we ignore the periodic wake-up operation. We assume that wireless interfaces instantaneously switch to power saving mode in absence of packets to be tranceived. With the arrival of a new packet in the transmission queue, the interfaces switch back to active mode instantly. 
%Therefore, in our model, interfaces stay in power saving mode during the entire idle interval.Throughout this paper, we use the expressions {\it power saving mode} and {\it idle mode} interchangeably.

\subsubsection{LTE Consumption} 
\label{sss:lte}

Based on \cite{huang2012Mobisys}, the downlink power consumption of user $i$ in the cellular network consists of the sum of a baseline power and a term which is proportional to the transmission rate of the device. As mentioned earlier, we extend the existing model to account for active/idle periods. The probability that the LTE interface is in active mode is equivalent to the probability $P_h^{(i)}$ of being the cluster head, see Eqs.~\eqref{eq:Ph-W} and \eqref{eq:Ph-M}.
 Therefore, the LTE interface power consumption of a device can be expressed as follows: 
 \begin{equation}
\label{eq:pow-lte}
W_{lte}^{(i)} = P_h^{(i)} \, \beta_{lte} + \left(1-P_h^{(i)} \right) \beta_{lte}^{idle} + \alpha_{rx} \, R_{rx}^{(i,\, lte)},
\end{equation}
where, $\beta_{lte}$ and $\beta_{lte}^{idle}$ are the baseline powers in active and idle mode, respectively; $\alpha_{rx}$ is the power consumption per Mbps  in uplink, and $R_{rx}^{(i,\, lte)}$ is the average data rate transmitted by user $i$ over the LTE interface.
The value of $R_{rx}^{(i,\, lte)}$ is computed using the following two propositions. 

\begin{proposition}
\label{p:RLTE-W}
Using \ac{CL(WRR)}, the uplink LTE data rate of user $i \in CL_n$ is given by  
\begin{equation}
\label{eq:avg-rate1}
R_{rx}^{(i, \, lte)} \!= \!w_n  S_{tot} \!\! \sum_{k  = 1}^M \!\! \pi_{k}^{(\textit CL_n)}  b_k \!\!\! \int_0^{\infty} \!\!\!\!\!\!\left[ 1\!\! -\!\! F_i(z|\text{MCS}_i \! = \! k)\right] dF_{Y_i}(z),
\end{equation}
where $Y_i  =  \max_{j \in {\textit CL_n}\setminus\{i\}}\{C_j\}$, $ i \in CL_n$.
\end{proposition}
\noindent The proof of Proposition~\ref{p:RLTE-W} is omitted due to its similarity to the proof of Proposition~\ref{p:Ph-W}.
 
\begin{proposition}
\label{p:RLTE-M}
Using \ac{CL(MR)}, the uplink LTE data rate of user $i \in CL_n$ is given by  
\begin{eqnarray}
\label{eq:avg-rate2}
R_{rx}^{(i, \, lte)} = S_{tot} \sum_{k  = 1}^M \!\! \pi_{k}^{(i)}  b_k \!\!\! \int_0^{\infty} \!\!\!\!\!\!\left[ 1\!\! -\!\! F_i(z|\text{MCS}_i \! = \! k)\right] dF_{Y_i}(z),
\end{eqnarray}
where $Y_i  =  \max_{j \not = i }\{C_j\}$, $ i \in CL_n$.
\end{proposition}
\noindent The proof of Proposition~\ref{p:RLTE-M} is omitted due to its similarity to the proof of Proposition~\ref{p:Ph-M}.

\subsubsection{WiFi Consumption}
\label{sss:wifi}
We use the accurate model of~\cite{serrano2014TON}, which 
accounts for the power required for packet processing as well as for transmission. We extend the model to include the probability that user $i$ is in active mode $P_a^{(i)}$. The power consumption of WiFi interface is:
\begin{align}
\label{eq:wifi_pow}
W_{wifi}^{(i)} & =  P_a^{(i)} \, \beta_{wifi} + \left(1-P_a^{(i)} \right) \beta_{wifi}^{idle}  +  \zeta_{tx}  \tau_{tx} + \zeta_{rx}  \tau_{rx} + \kappa_{tx}  \lambda_{tx}+ \kappa_{rx}  \lambda_{rx},
\end{align}
where $\beta_{wifi}$ and $\beta_{wifi}^{idle}$ are the WiFi baseline powers in active and idle mode, respectively; $\zeta_{tx}$ and $\zeta_{rx}$ represent the power consumptions due to transmission and reception, respectively; $\tau_{tx}$ and $\tau_{rx}$ are the fractions of time spent in transmission and reception, respectively (i.e., $\tau_{tx}^{(i)} = R_{tx}^{(i, wifi)} / R_{wifi}$ and $\tau_{rx}^{(i)} = R_{rx}^{(i, wifi)} / R_{wifi}$); $\kappa_{tx}$ and $\kappa_{rx}$ are the power consumptions due to packet processing in transmission and reception, respectively; eventually, $\lambda_{tx}$  and $\lambda_{rx}$ are the packet rates, respectively in transmission and reception.   

%The power consumption over the WiFi network depends on the total traffic exchanged over the 
%WiFi interface. 

The WiFi power related parameters introduced in Eq.~\eqref{eq:wifi_pow} are computed as follows: 
%\red{ $\tau_{tx}$ is the ratio between the transmission rate over the WiFi interface and the achievable rate of the WiFi connection, i.e., for user $i$, we have $\tau_{tx}^{(i)} = R_{tx}^{(i, wifi)} / R_{wifi}$.   }
%Similarly, $\tau_{rx}^{(i)}$ is obtained from $R_{rx}^{(i, wifi)} / R_{wifi}$.
$\lambda_{tx}^{(i, wifi)}$ is computed as the ratio between the rate $R_{tx}^{(i, wifi)}$ and the average packet size $L_p$; and
similarly, user $i$ transmits  $\lambda_{rx}^{(i, wifi)} = R_{rx}^{(i, wifi)} / L_p$  packets per second.
It is assumed that the achievable WiFi rate is independent from the cellular network status and its average 
value $R_{wifi}$ is the same for all clusters (i.e., this is an input parameter for our problem). If the achievable WiFi rate is larger than the intra-cluster traffic (i.e., 
 $R_{wifi} > \sum_{i \in {\textit CL_n}} R_{rx}^{(i, wifi)} = \sum_{i \in {\textit CL_n}} R_{tx}^{(i, wifi)}$),
then to evaluate the WiFi power consumption, we need to compute the WiFi data rates $R_{rx}^{(i, wifi)}$ and $R_{tx}^{(i, wifi)}$, and 
the probability $P_a^{(i)}$ that the WiFi interface of user $i$ be active. 
$R_{rx}^{(i, wifi)}$ and $R_{tx}^{(i, wifi)}$ can be computed using Proposition~\ref{p:RWIFI}, whose proof is reported in the Appendix~A.%\ref{s:appA}.
\begin{proposition}
\label{p:RWIFI}
The WiFi data rate of user $i \in CL_n$ is given by the following expressions, which hold for the received and transmitted traffic, respectively:  
 \begin{align}
 \label{eq:RWIFI-rx}
R_{tx}^{(i, wifi)}  = (1  - \delta_i) \cdot R_{rx}^{(i, \, lte)}, \\
 \label{eq:RWIFI-tx}
R_{rx}^{(i, wifi)} = \delta_i \cdot \sum_{j \in {\textit CL_n} \!\setminus \!\{ i \}} R_{rx}^{(j, lte)},
\end{align}
where 
\begin{equation}
\label{eq:delta}
\delta_i = \frac{E[T_i]}{E[T_{{\textit CL_n}}]}. 
\end{equation}
\end{proposition}
%\noindent The proof of Proposition~\ref{p:RWIFI} is given in Appendix.%~\ref{s:appA}.

Finally, the probability $P_a^{(i)}$ that the WiFi interface of user $i$ is in active mode is given by the following proposition, whose proof is reported in the Appendix~A.%\ref{s:appA}.

\begin{proposition}
\label{p:Pa}
The WiFi interface of user $i$ is active with probability $P_a^{( i )}$ that is computed as:
\begin{equation}
\label{eq:pa2}
P_a ^ { ( i ) }
		= 	
		\frac { E [ T _ i ] + ( 1 - 2  \delta_i ) R _ { rx } ^ { ( i, \, lte ) } }  { R _ { w i f i } }, 
\end{equation}
with $\delta_i$ defined in~\eqref{eq:delta}.	
\end{proposition}
%\noindent The proof of Proposition~\ref{p:Pa} is reported in Appendix.%~\ref{s:appA}.

\textbf{Total Power Consumption.}
Combining the results for LTE and WiFi consumptions, the resulting total power consumption of a clustered user 
is as follows:
\begin{align}
\label{eq:Ptot}
W_{tot}^{(i)} 
& =  
\beta_{lte}^{idle} + 
\beta_{wifi} ^ {idle} +
\left ( \beta_{lte}  - \beta_{lte}^{idle} \right ) P_h^{(i)} % \nonumber \\
%& 
+ 
\left ( \beta_{wifi}  - \beta_{wifi} ^ {idle} \right )  \frac { E [ T _ i ] + ( 1 - 2  \delta_i ) R _ { rx } ^ { ( i, \, lte ) } }  { R _ { w i f i } }  \nonumber \\
& +  \alpha_{rx} \, R_{rx}^{(i,\, lte)}% \nonumber \\
%& 
+
 \left ( \zeta_{tx} + \frac{\kappa_{tx}} {L_p} \right ) 
 ( 1 - \delta _ i ) \frac { R _ { rx } ^ { ( i, \, lte ) } } { R _ {wifi} } %\nonumber \\
%& 
+
\! \left ( \zeta_{rx} \!+\! \frac{\kappa_{rx}} {L_p} \right )  
 \frac { E [ T _ i ] - \delta_i R _ { rx } ^ { ( i, \, lte ) } }  { R _ {wifi} }.
\end{align}

The first term in Eq.~\eqref{eq:Ptot} represents the baseline power consumption of WiFi and LTE interfaces in idle mode;
the second and third terms express the baseline power consumption of the interfaces in active mode; the fourth term accounts for LTE downlink transmissions, while 
the fifth term is due to the WiFi transmissions when the user is cluster head; finally, the last term in Eq.~\eqref{eq:Ptot} represents the power spent to receive WiFi traffic from the cluster head. 

\textbf{Energy efficiency.}
To evaluate the beneficial impact of clustering under fully backlogged traffic assumption, we use as metric the energy efficiency, i.e., the amount of data (bits) that can be transferred to the final user per energy unit (Joule), e.g., for user $i$, the energy efficiency is given by $\eta_i =E[T_i] / W_{tot}^{(i)}$.

In summary, we derived agile analytical tools for the evaluation of network KPIs. In the next section, we address
user's motivations for clustering, based on user's rewards.

\section{Cluster Formation: A Game Theory Approach}
\label{s:game_theory}

This section provides a simple model for the cluster formation process, and sheds light on the impact of clustering when users experience non-stationary channel qualities. The cluster formation in our proposed architecture is modeled using {\it coalitional game theory} \cite{saad2009MagSig}. Here, we treat cluster formation as a game in which users decide to join or to leave a cluster depending on the achievable reward.
We analyze different alternatives to share the clustering gain, i.e., the {\it revenue}, among participating users. The revenue can be expressed in terms of throughput, power, energy efficiency, and so on. We choose energy efficiency so that we can maximize the system capacity with respect to power consumption which is a key issue in today's cellular networks.

\subsection{Definition of the Game}

In the following, $U \!\!= \!\! \left\{u_1,\dots ,u_N \right\}$ denotes the set of users 
in the network and $S\!\!=\!\!\left\{S_1, \dots ,S_l\right\}$ is a partition of $U$, i.e.,  
$\bigcup_{i=1}^l S_n \!\!=\!\!U$ and $S_n \cap S_j \!\!= \!\! \emptyset ~\text{if}~ n \neq j $. 
The utility function $\nu(.)$ defines the value of cluster $S_n$ as: 
%{\color{red} change}
\begin{align}
\label{e:utilFunc}
\!\!\!\!\!\! \nu \! \left(S_n\right) \! = \!\!
 \begin{cases} 
 	\!\sum\limits_{u_i\in S_n} \eta_{u_i} & \!\!\!\text{if} \, d_{S_n} \!\! \leq \!\! d_{\text{m}} \; \& \;
	\eta^{(S_n)}_{u_i} \!\!\geq \!\eta_{u_i}, \! \forall i \!\in\! S_n; \!\!\!\!\!\!\!\!\!\!\!\!\!\!\!\!\!\!\!
	\\
%	& \!\!\!\forall i \in S_n \, ;\\  E[T_{u_i}]
	0 & \!\!\!\text{otherwise}; 
  \end{cases}
 % \vspace{-1cm}
\end{align}
where $d_{S_n}$ and $d_{\text{m}}$ are the distances between the two farthest users in cluster $S_n$, and the maximum 
allowable distance among cluster members, respectively; $\eta_{u_i}^{(S_n)}$ and $\eta_{u_i}$ are the energy efficiencies of user 
$i$ when it joins cluster $S_n$ and when it is not clustered, respectively. In particular, $d_{\text{m}}$ 
accounts for the WiFi transmission range, and can be set  to guarantee that any user inside a cluster can directly 
reach the rest of the cluster members.  The constraint on the energy efficiency guarantees that users form a cluster only if energy efficiency increases.

\subsection{Cluster Formation Algorithm}
%\vspace{-1mm}
The problem of finding optimal coalitions is NP-complete because it requires evaluating all possible partitions of the set of users $U$ in the network.
Obviously, the existing base stations with limited computational resources are not able to handle an NP-complete problem involving a few tens of users. 
Hence, we adapt the simple {\it merge and split} algorithm to solve the coalition formation problem with low complexity~\cite{saad2009MagSig,song2014Commag}. Although merge and split is a trivial 
method for dynamic cluster formation, it was shown to be a good alternative when computational overhead is of concern~\cite{song2014Commag}. The merge and split rules are defined as follows: $(i)$ merge any set $\left\{S_{a_1},..,S_{a_k}\right\}$ into a unique coalition (i.e., cluster), if $\sum_{i = 1}^k  \nu (S_{a_i})  < \nu \left(  \cup_{i=1}^k {S_{a_i}} \right)$;
$(ii)$ if the previous inequality does not hold for a coalition that can be described as $\cup_{i=1}^k S_{a_i}$, then 
split it into its components. %(i.e., split the big cluster into smaller clusters).
Refer to~\cite{saad2009MagSig} for the proof of convergence and $D_{hp}$-{\textit stablity} of this approach.

\subsection{Payoff Allocation}
The {\it payoff} of a cluster member is defined as the amount of throughput which it receives from the total cluster throughput. 
Formally, let $G \in S$ be a cluster of size $|G|$, and $\bar{x}=\{x_1, \dots, x_{|G|}\}$ the payoff vector of members of $G$. 
A payoff vector is called {\it cost efficient} if $\sum_{i\in G} x_i = \nu \left(G\right)$~\cite{saad2008ICC}. Of course, we are only interested in cost efficient payoff vectors.

Here, we chose to compare three mechanisms proposed in the literature, namely  \ac{ES}, \ac{WS}~\cite{saad2008ICC}, and Shapley~\cite{saad2009MagSig}. These mechanisms allow us to illustrate how payoff allocation can impact clustering decisions made by the users.

\textbf{\acf{ES}.} Here, the clustering gain is equally divided among members. The cost efficient payoff distribution used under equal share is formally expressed as follows:
\begin{align}
\label{e:equal_share}
x_i= \frac {\nu (G) - \sum_{j \in G} \nu(\{ j\})} {|G|} + \nu(\{i\}), \quad i \in G.
\end{align}  

\textbf{\acf{WS}.} Here, the payoff distribution is computed based on positive weights $\omega_i$: 
\begin{eqnarray}
\label{e:ws}
x_i\!=\! \frac{\omega_i}{\sum\limits_{j\in G} \omega_j} \cdot [ \nu (G) \!-\! \sum_{j \in G} \nu(\{ j\}) ] \!+\! \nu(\{i\}),  \, i \in G.
\end{eqnarray}  

\textbf{\ac{SS}.}
This is an alternative payoff distribution method that accounts for marginal contribution of each cluster member. The ``Shapley value'' is known to maintain good fairness while considering the contribution of the users in the cluster~\cite{saad2009MagSig}. The Shapley value of user $i$ in cluster G is computed as follows:
\begin{eqnarray}
\label{e:ss}
x_i\!=\!\!\!\!\! \sum\limits_{S\subseteq G \setminus\{i\}} \!\!\!\!\!\! \frac{|S|! \left(|G| - |S|-1 \right)!} {|G|!} \left[ \nu \left(S \cup \{i\} \right ) - \nu \left(S \right ) \right].
\end{eqnarray}  

As shown in~\cite{asadi2013MSWIM}, the clustering gain is mainly due to the presence of {\it good} users, whereas the 
channel state probability distribution of a cluster does not dramatically improve with the addition of a {\it poor} user 
(see Figure~2 in~\cite{asadi2013MSWIM}). 
Hence, \ac{ES} may not strongly motivate {\it good} users to cluster with {\it poor} users. 
In contrast, by adjusting $\omega_i$ in Eq.~\eqref{e:ws}, we can make sure that users with better channel quality receive enough incentive to cluster.  Specifically, in our numerical simulation, we use values of $\omega_i$ equal to the user's throughput achieved without clustering. Note that \ac{WS} better motivates good users to join clusters but it may not achieve a fair payoff distribution (as \ac{SS}) and tuning $\omega_i$ for a complete fair payoff distribution can be challenging in real implementation. Using Shapley value, we do not need  $\omega_i$ because Shapley value is designed in such a way that it distributes the payoffs based on the marginal contribution of each user. Moreover, Shapley value ensures that all clustered users receive at least what they would have received without clustering. Therefore we do not need to add $\nu\{i\}$ in Eq.~\eqref{e:ss}.

We assume that the overhead due to cluster formation is negligible in comparison to the time users spend in the cluster. Nonetheless, both discovery and WiFi connection setup procedures consume time and energy for a few seconds. Thus, our proposal is not suitable for high speed mobile scenarios. 

So far we investigated our proposal analytically, however, the question remains: {\it Is it possible to implement a network-controlled opportunistic D2D system in real world with LTE-A and WiFi?} We answer this question in the next section and in chapters~\ref{ch:clus_proto} and~\ref{ch:clus_exp}.

\section{Performance Evaluation}
\label{s:evaluation}
In this section, we perform numerical and packet simulations to benchmark our proposed D2D schemes (\ac{CL(WRR)} and \ac{CL(MR)}) against \ac{RR} and \ac{PF}~\cite{84} schedulers
in an FDD LTE-A SISO system, whose capacity is $80.64$ Mbps achieved by using a $20$ MHz band and neglecting LTE overheads (which would reduce the capacity to $\sim75$ Mbps).
The \ac{MCS} values in this manuscript are adopted from~\cite{sesia2011lte}.  
For the sake of tractability, in the analysis we assume that mobile users belong to one of three predefined \ac{SNR} {\it classes}, 
which correspond to {\it poor}, {\it average}, and {\it good} mean \ac{SNR}.
The designated \ac{SNR} for different classes are chosen in a manner that the mean achievable
rates for \textit{poor}, {\it average}, and \textit{good} users are $20\%$, $50\%$, and $80\%$ of the
maximum transmission rate achievable in the system, respectively. With the thresholds and \ac{MCS} values adapted from~\cite{sesia2011lte}, the designated \ac{SNR} values are $7$ dB, $16$ dB, and $23$ dB, respectively for {\it poor}, {\it average}, and \textit{good} users. Note also that using non-homogeneous channel qualities allows us to evaluate the long-term system fairness under different (opportunistic) scheduling mechanisms.  
The numerical simulations are based on the results obtained using Mathematica software from the model presented in Sections~\ref{s:system_model} and \ref{s:game_theory}. Each experiment is repeated $2000$ times.
%The numerical simulations are based on the analytical results obtained in Sections~\ref{s:system_model}  implemented using Mathematica software. 
The packet simulations are obtained from our home-grown simulator that reproduces PHY, MAC (i.e., resource allocation and scheduling) and IP operation.
The duration of packet simulations is $60$ s which is repeated with $25$ different seeds.  The values of power related parameters are %reported in Table~\ref{tb:pow_par} and are 
derived from~\cite{huang2012Mobisys, serrano2014TON}. %The average packet size is $L_p\!=\!1500~\text{B}$ and average WiFi net rate $R_{wifi}=48$ Mbps. 
%We leveraged the SDR platform described in Section~\ref{ss:SDR} to prototype the first D2D-assisted clustering scheme to validate the performance of our proposal in a real setup.
The results reported here include average, $25^{th}$ and $75^{th}$ percentiles of the achieved performance figures.
The payoff allocation method is Equal Share Eq.~\eqref{e:equal_share} unless otherwise specified.

%\begin{table}[h!]
%\center
%\vspace{-2mm}
%\caption{Parameters used in the power model}
%\vspace{-2mm}
%\footnotesize
%\label{tb:pow_par}
%\begin{tabular}{|@{ }c@{ }|@{ }c@{ }|@{ }c@{ }|c@{ }|@{ }c@{ }|@{ }c@{ }|@{ }c@{ }|@{ }c@{ }|@{ }c@{ }|}
%%\begin{tabular}{|c|c|c|c|c|c|c|c|c|}
%\hline
%\multicolumn{3}{|c|}{\textbf{LTE}}&\multicolumn{6}{|c|}{\textbf{WiFi}}\\
%\hline
%\hline
%$\beta_{lte}$ & 	$\beta_{lte} ^ {idle}$	 & $\alpha_{tx}$ & $\beta_{wifi}$ & $ \beta_{wifi} ^ {idle}$ & $\zeta_{tx}$ & $\zeta_{rx}$ & $\kappa_{tx}$ &	$\kappa_{rx}$ \\
%\hline
%%$1.29$ & 0.59 & $51.97$ & $0.14$ & 0.08 & $0.46$ & $0.44$ & $0.11$ 	& $0.09$ 	\\
%%\scriptsize [W] & \scriptsize [W] &\scriptsize [nW/bps]& \scriptsize [W] & \scriptsize [W] &\scriptsize [W] &\scriptsize [W] &\scriptsize [mJ] &\scriptsize [mJ] \\
%$1.29$ \scriptsize [W] & 0.59 \scriptsize [W] & $51.97$ \scriptsize [nW/bps] & $0.14$ \scriptsize [W] & 0.08 \scriptsize [W] & $0.46$ \scriptsize [W] & $0.44$ \scriptsize [W] & $0.11$ \scriptsize [mJ] & $0.09$ \scriptsize [mJ] \\
%%\scriptsize [W] & \scriptsize [W] &\scriptsize [nW/bps]& \scriptsize [W] & \scriptsize [W] &\scriptsize [W] &\scriptsize [W] &\scriptsize [mJ] &\scriptsize [mJ] \\
%
%\hline
%\end{tabular}
%\vspace{-10mm}
%\end{table}

%\def\subfigcapskip{-8pt}

\subsection{Performance of Static Clusters}
\label{ss:single_cell}

\begin{figure} [h!]
\centering
		\includegraphics[scale=0.50, angle=0]{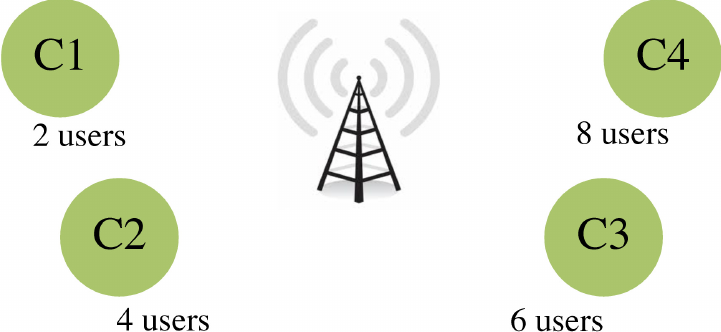} %[width=70mm,angle=0] 
		\caption{Evaluation topology for static clusters.}
		\label{fig:single_cl}
\end{figure}

This subsection provides the evaluation of throughput, fairness, and energy efficiency achievable using D2D clustering schemes. For the sake of clarity, we consider a static scenario, formed by users with heterogeneous average \ac{SNR} (see Figure~\ref{fig:single_cl}). 
In this scenario, clusters C$1$, C$2$, C$3$, and C$4$ have $2$, $4$, $6$, and $8$ users, respectively. In each experiment, the \ac{SNR} class of each user is chosen as {\it poor}, {\it average}, or {\it good} with the same probability. Although the number of users might be higher in a reality, this scenario is intended as a toy example that sheds light on potentials of the proposed schemes. 

\begin{figure*} [t!]
		\centering	
		\subfigure[Throughput and energy efficiency]		
		{
			\includegraphics[width=0.45\columnwidth]{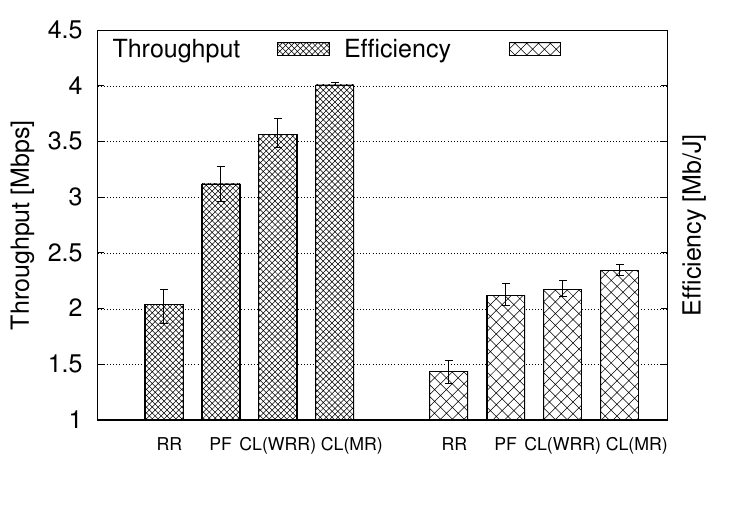} %[width=70mm,angle=0] 
			\label{fig:user_tput_eff}
		}
		\subfigure[Throughput]
		{
			\includegraphics[width=0.45\columnwidth]{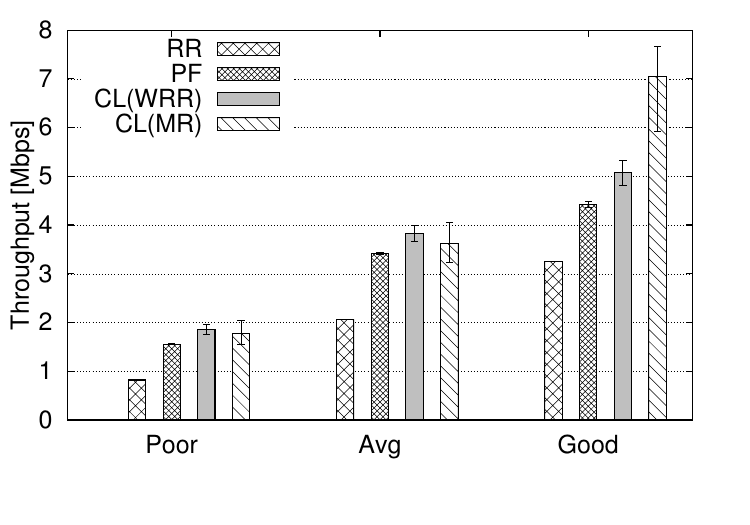} %[width=70mm,angle=0] 
			\label{fig:class_tput}
		}
		\subfigure[Energy efficiency]		
		{
			\includegraphics[width=0.45\columnwidth]{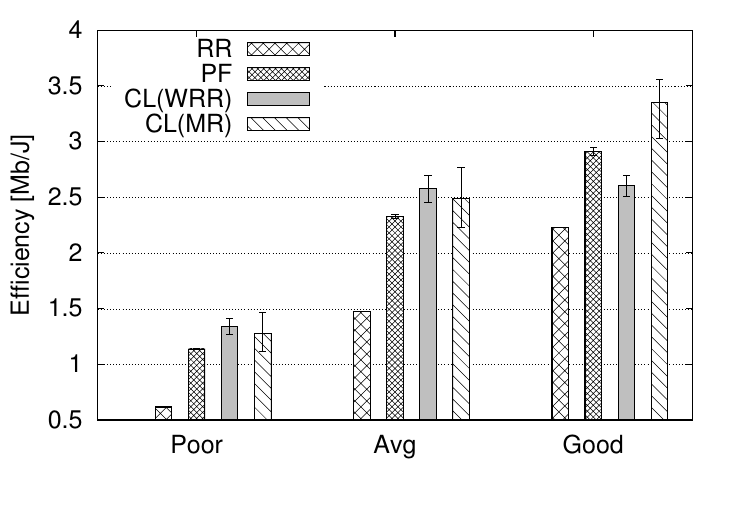} %[width=70mm,angle=0] 
			\label{fig:user_eff}
		}
		\caption{Average per-user and per-class throughput and energy efficiency performance. }
		\label{fig:per-user_static}
\end{figure*}

Figure~\ref{fig:per-user_static} illustrates  the average user performance under different schedulers.
Figure~\ref{fig:user_tput_eff} shows that users receive the lowest throughput under \ac{RR} because they are scheduled irrespective of their channel quality. Instead, \ac{PF} has remarkably better performance in terms of throughput, due to its opportunistic nature. Nevertheless, both \ac{RR} and \ac{PF} are significantly outperformed by D2D-clustering schemes in terms of throughput and energy efficiency. Interestingly, D2D-clustering schemes result in better energy efficiency than PF, although the users should maintain the WiFi interface active, in addition to the cellular interface. This stems from the higher throughput gain achieved by D2D-clusters and the insignificance of WiFi power consumption in comparison with LTE. Since in D2D cluster users with better channel quality are more active than those with poor channel quality, we illustrate the per-\ac{SNR} class user throughput and  user energy efficiency in Figures~\ref{fig:class_tput} and~\ref{fig:user_eff}. 
In terms of throughput, all classes of users enjoy higher throughput than \ac{RR} and PF. D2D-clustering schemes also outperform \ac{RR} and \ac{PF} in terms of energy efficiency with the exception of \ac{CL(WRR)} in which the {\it good} users can obtain higher energy efficiency under \ac{PF} scheduler. Recall than in this scenario the clusters are fixed and users do not decide on the cluster formation. Therefore, the {\it good} users may be forced to form a cluster with low throughput gain which leads to lower energy efficiency. This observation highlights the importance of cluster formation strategies that were discussed in Section~\ref{s:game_theory}. Between D2D clustering schemes, \ac{CL(MR)} has higher throughput and energy efficiency performance due to the adoption of a more aggressive opportunistic cluster selection scheme. 

\begin{figure*} [t!]
		\centering
		\subfigure[Per-cluster throughput]
		{
			\includegraphics[width=0.45\columnwidth]{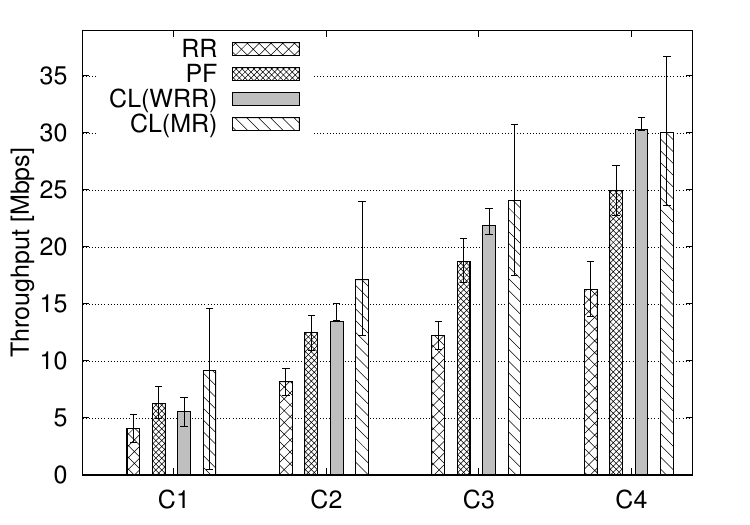} %[width=70mm,angle=0] 
			\label{fig:cluster_tput}
		}	
		\subfigure[Aggregate cell throughput]
		{
			\includegraphics[width=0.45\columnwidth]{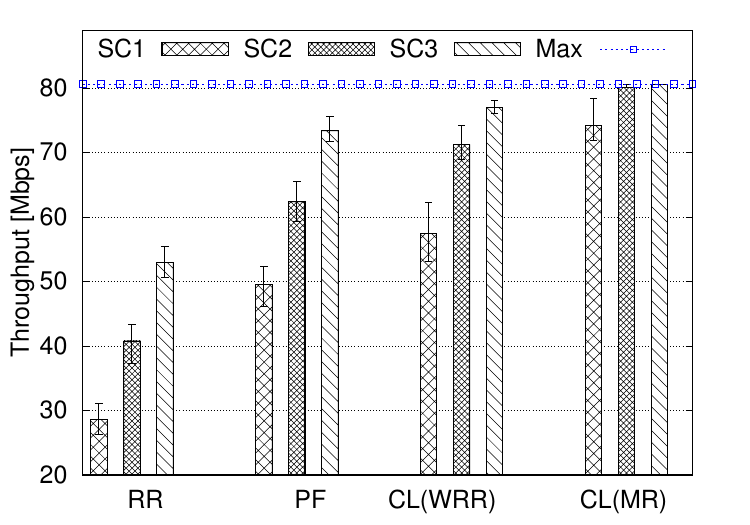} %[width=70mm,angle=0] 
			\label{fig:bs_tput}
			%\vspace{-5mm}
		}
		\subfigure[Jain's fairness indexes]
		{
			\includegraphics[width=0.45\columnwidth]{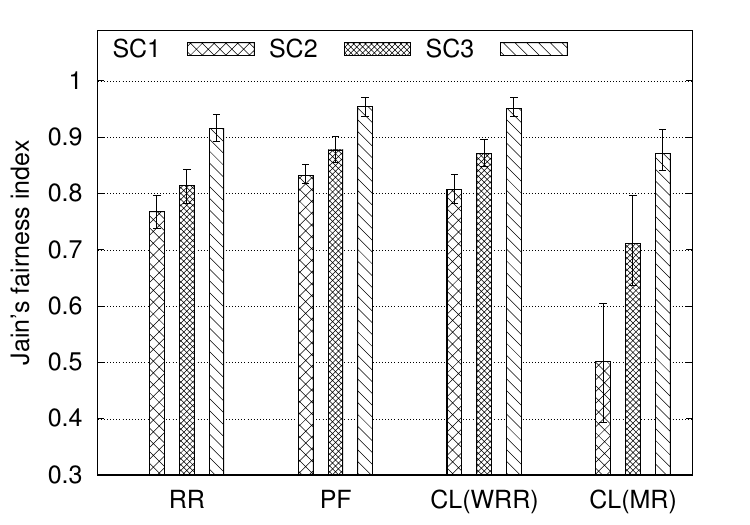} %[width=70mm,angle=0] 
			\label{fig:3-fairness}
			%\vspace{-5mm} 
		}
		\caption{Per-cluster and aggregate performance.}
		\label{fig:aggregate_static}
\end{figure*}

Figure~\ref{fig:cluster_tput} shows the impact of cluster sizes on the throughput of each cluster. For comparison, we also report the aggregate throughput achieved by cluster members if they were scheduled according to \ac{RR} or PF. Therefore, results with \ac{RR} and \ac{PF} scale linearly with the cluster size. Similarly, \ac{CL(WRR)} shows linearity, while the high variability of results for \ac{CL(MR)} does not allow us to confirm or reject the hypothesis that \ac{CL(MR)} scales linearly. This behavior is due to the fact that \ac{CL(MR)}, differently from \ac{CL(WRR)}, does not guarantee a minimum airtime to any cluster, so that clusters not including {\it good} users will receive little throughput.

Figure~\ref{fig:bs_tput} sheds light on the aggregate throughput performance. The figure reports results for three sub-scenarios with varying \ac{SNR} class distribution. SC1 with the  $60$\% {\it poor}, $30$\% {\it average} and $10$\% {\it good} users represents a cell with more low quality channel users. In SC2 there is equal distribution of different \ac{SNR} classes (i.e., $33.\overline{3}$\%). Finally, SC3 represents a cell with more high channel quality users where there are $10$\% {\it poor}, $30$\% {\it average} and $60$\% {\it good} users. 
%The \ac{SNR} class distribution  as stated in Table~\ref{tb:scenarios}. 
The figure also reports the upper bound for the downlink throughput. The aggregate throughput of \ac{RR} and \ac{PF} is outperformed by \ac{CL(WRR)} and \ac{CL(MR)}. \ac{CL(MR)} practically hits the upper bound, while the worst case for \ac{CL(WRR)}, i.e., when the number of {\it poor} users is predominant, outperforms \ac{RR} and \ac{PF} under their best performance.

%\begin{table}[h!]
%\center
%\vspace{-3mm}
%\caption{User \ac{SNR} class distribution in different sub-scenarios}
%\label{tb:scenarios}
%\footnotesize
%\begin{tabular}{|l|c|c|c|}
%\hline
%Scenario sub-case 	& \% of poor users			& \% of avg users			& \% of good users			\\
%%	 			& users				& users				& users 				\\
%\hline
%\hline
%SC I 		& $60\%$ 				& $30\%$ 				& $10\%$  			\\
%\hline
%SC II 	& $33.\overline{3}\%$ 	& $33.\overline{3}\%$ 	& $33.\overline{3}\%$  	\\
%\hline
%SC III	& $10\%$ 				& $30\%$ 				& $60\%$  			\\
%\hline
%\end{tabular}
%\vspace{-3mm}
%\end{table}

So far, \ac{CL(MR)} outperforms all other schedulers. However, considering fairness, \ac{CL(MR)} is always the most unfair, especially when more {\it poor}~~users are~~present, while \ac{CL(WRR)} performs like \ac{PF} in terms of fairness, see Figure~\ref{fig:3-fairness}.

\subsection{Packet Simulation with Static Clusters}
\label{ss:single_pkt}
In the previous subsection, the network performance was studied in a saturated network (i.e., fully backlogged assumption). In order to better analyze the impact of clustering, we evaluate the same scenario (see Figure~\ref{fig:single_cl}) in a non-saturated network (i.e., $50$ Mbps) using our home-grown LTE simulator developed with the Mathematica software tools. Here, in addition to throughput and fairness we focus on the delay in the LTE cell and the load offered to the WiFi network, which provides us with better insight on the practicality of our scheme. The average \ac{SNR} of users is selected randomly with a uniform distribution between $7$ dB to $23$ dB. The instantaneous channel quality of the users follows a Rayleigh distribution. Users have heterogenous Poisson packet arrivals with the total load of $50$ Mbps that allows us to validate the benefits of our D2D-assisted scheme when the network load is below saturation. %point of legacy and D2D-clustering schemes. 
\begin{figure*} [t!]
		\centering	
		\subfigure[Delay \ac{CDF} of delivered packets.]
		{
			\includegraphics[scale=0.44, angle=0]{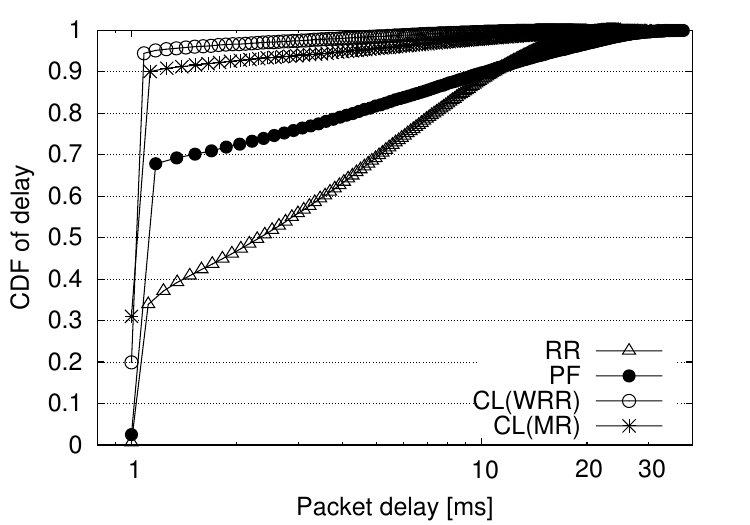} %[width=70mm,angle=0] 
			\label{fig:CDF}
		}
		\subfigure[Packet delivery ratio of the schedulers.]
		{
			\includegraphics[scale=0.44, angle=0]{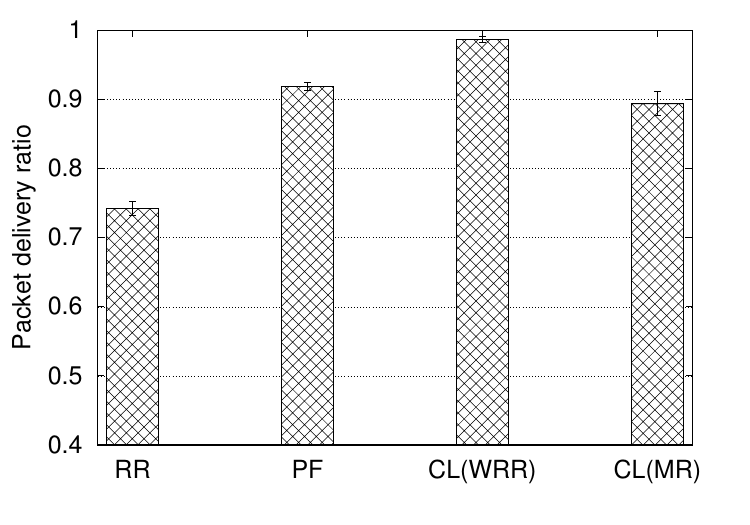} %[width=70mm,angle=0] 
			\label{fig:pktDelivery}
		}		
		\subfigure[WiFi Direct load per user.]
		{
			\includegraphics[scale=0.5, angle=0]{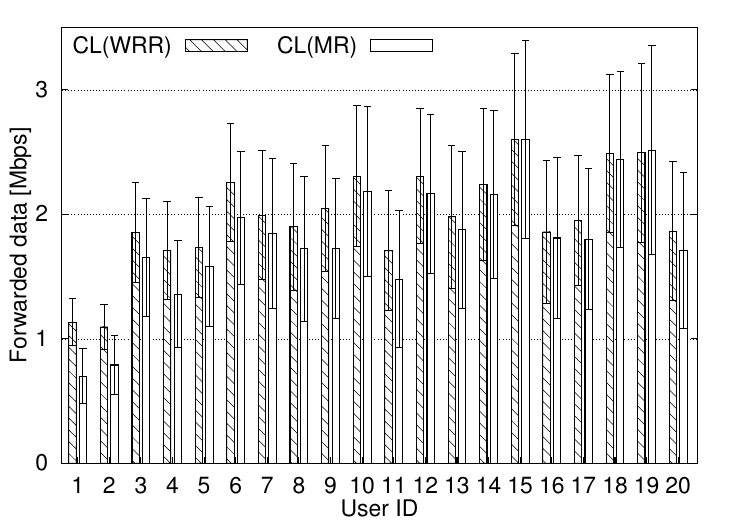} %[width=70mm,angle=0] 
			\label{fig:wifi_traffic}
		}				
		\caption{Delay CDF, packet delivery ratio, and per-user WiFi Direct loads.} %obtained from packet simulations
		\label{fig:pktSim}
		
\end{figure*}

Figure~\ref{fig:CDF} shows the delay \ac{CDF} of the delivered packets. Here, we only account for the packet delivery time from the eNB to the UE, whereas the time to receive the ACK is not counted. 
The figure shows that our proposed schemes maintain a $1$ ms delay with $90\%$ probability while \ac{RR} and \ac{PF} require $10$ ms to reach this threshold. The delay performance of a scheme is mainly affected by the achievable throughput and its prioritizing policy. If the achievable throughput is low, the packet waiting time increases which results in higher delays. On the other hand, a scheme that highly prioritizes a certain class of users (e.g., best instantaneous channel quality) can potentially increase the queue size of the other classes of users. The latter is the reason why \ac{CL(MR)} yeilds higher delay than \ac{CL(WRR)}.
D2D-clustering schemes can guarantee delays lower than $10$ ms with $97\%$ probability or higher, leaving at least $40$ ms of delay budget for WiFi transmissions. Note that the WiFi delay budget is enough to support real time applications such as video conferencing. 

In Figure~\ref{fig:pktDelivery}, we can observe that \ac{CL(WRR)} outperforms other schemes in terms of successful packet delivery ratio. The outstanding results of \ac{CL(WRR)} are because of the throughput gain from D2D-clustering and fair resource allocations which avoids starvation of low priority users. On the other hand, \ac{CL(MR)} and \ac{PF} have comparable performance, although \ac{CL(MR)} can potentially achieve higher throughput than PF. \ac{CL(MR)} cannot outperform \ac{PF} because of its greedy behavior in prioritizing high channel quality users.

Figure~\ref{fig:wifi_traffic} illustrates the load offered to the WiFi network under \ac{CL(WRR)} and \ac{CL(MR)}. This figure confirms that WiFi Direct is not a bottleneck in our proposed architecture. The figure also shows that the maximum load offered to C1 (users 1 and 2), C2 (users 3 to 6), C3 (users 7 to 12), and C4 (users 13 to 20) are less than $4$ Mbps, $12$ Mbps, $20$ Mbps, and $31$ Mbps, respectively, i.e., no more than 4 Mbps per user, on average. The load variation for different users depends on the channel quality. For instance, users $12$ and $15$ relay more traffic because they have higher average \ac{SNR} w.r.t. the other users. In all cases, the traffic to be handled by each cluster is well below typical WiFi capacities.

To summarize the results reported for static cluster evaluation scenarios, we have observed that the clustering proposal not only increases the throughput and the energy efficiency, but it can also increase the fairness level. In particular, \ac{CL(WRR)} achieves similar throughput and energy efficiency results as \ac{CL(MR)}, but it is much fairer. Therefore, the advantage of using \ac{CL(WRR)} is sixfold: $(i)$ it offers the possibility to gain a high throughput with respect to legacy \ac{RR} and \ac{PF} schedulers; $(ii)$ it allows each cluster to exploit the clustering gain proportionally to its size; $(iii)$ it provides nearly perfect fairness among users; $(iv)$ average energy efficiency is increased with respect to \ac{RR} and PF; $(v)$ it has the best delay performance compared with other schemes; $(vi)$ it has much higher packet delivery ratio (almost $100\%$). 
Since numerical and packet simulations showed that \ac{CL(MR)} may lead to poor fairness and packet delivery ratio, we will focus on the \ac{CL(WRR)} in the rest of the evaluation. 

\begin{figure*} [t!]
		\centering
		\subfigure[Aggregate throughput]
		{
			\includegraphics[width=0.45\columnwidth]{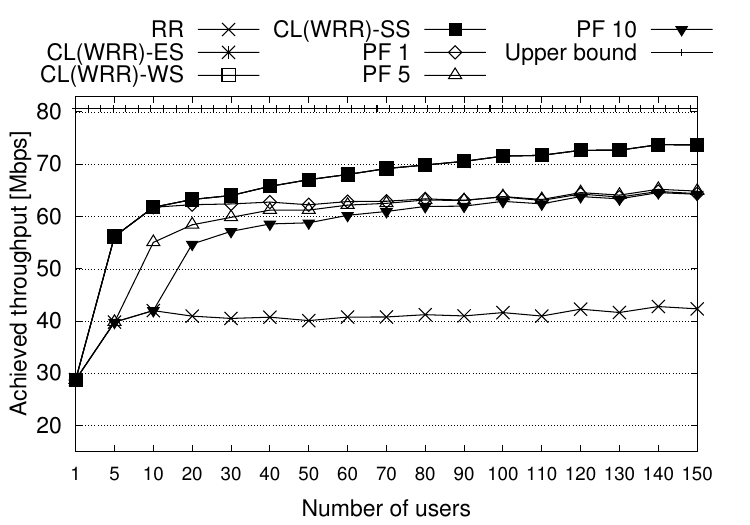} %[width=70mm,angle=0] 
			\label{fig:ms_total_tput}
		}
		\subfigure[Average per-user efficiency]
		{
			\includegraphics[width=0.45\columnwidth]{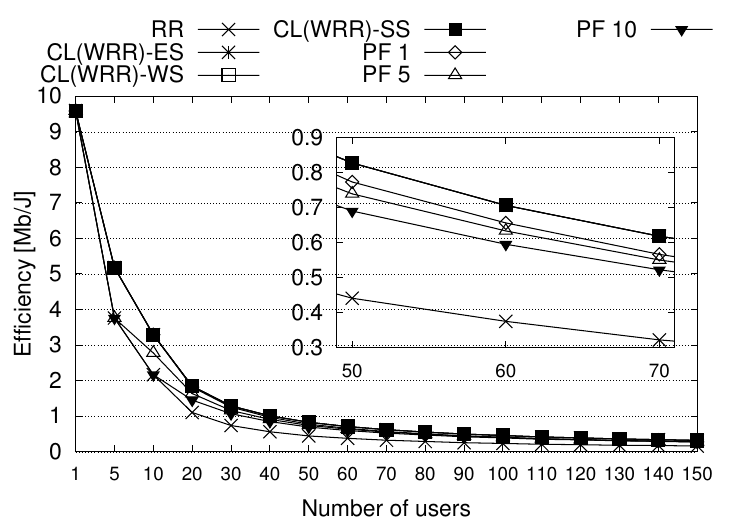} %[width=70mm,angle=0] 
			\label{fig:ms_total_cost}
		}
		\subfigure[Fairness]
		{
			\includegraphics[width=0.45\columnwidth]{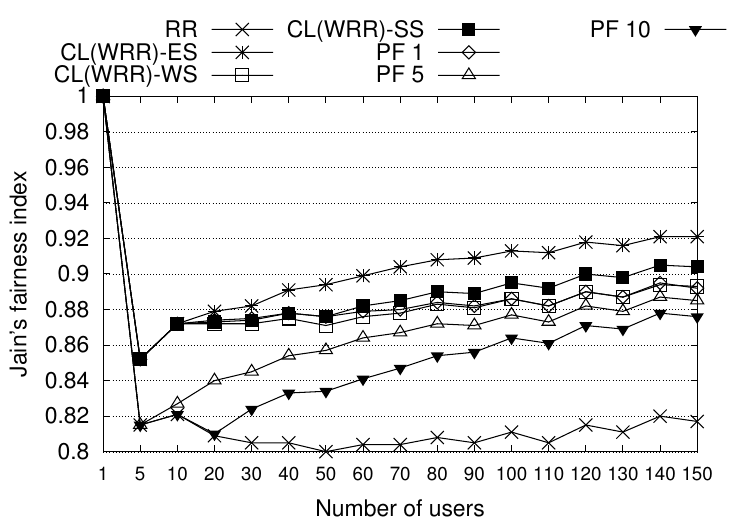} %[width=70mm,angle=0] 
			\label{fig:ms_fair}
		}
		\caption{Throughput, efficiency and fairness under different scheduling mechanisms. }
		\label{fig:ms_total}
\end{figure*}

\subsection{Performance of Dynamic Clusters}

In order to evaluate our proposal in a more realistic setup, we simulate a network with variable number of users (from $1$ to $100$) with varying \ac{SNR}, randomly placed in a circular-shaped cell with $500$ m diameter. The \ac{SNR} class of a user is selected at random with a probability distribution that changes according to the distance from the base station. In particular, the cell area is divided into three areas. In the area close to the base station, the probability of finding {\it good} users is higher than the area far from the base station. The users move with an average pedestrian speed between $0$ and 
$5$ km/h.  The maximum diameter of a cluster $d_{\text{m}}$ is $100$ m. 

Figure~\ref{fig:ms_total} illustrates the performance metrics for different user population sizes. In the figure, we report results achieved with RR, PF, \ac{CL(WRR)} with equal share, namely \ac{CL(WRR)}-ES, \ac{CL(WRR)} with weighted share, namely \ac{CL(WRR)}-WS, and \ac{CL(WRR)} with Shapley share, namely, \ac{CL(WRR)}-SS. Additionally, we report results for \ac{PF} when $n \ge 1$ users are scheduled per frame (PF~$n$ in the figure). We report this comparison since user-based schedulers allocate multiple users per frame, and it is indeed common to schedule tens of users per scheduling interval, even when opportunistic schedulers are adopted. However, \ac{RR} and \ac{CL(WRR)} are not affected by the number of users scheduled per frame, due to the assumption that user's channels are independent and stationary (and so are the channels of cluster heads). 
  
In Figure~\ref{fig:ms_total_tput}, we can observe that the clustering gain rises with the number of users in the system, and as soon as about $30$ users are present, \ac{CL(WRR)} achieves the highest aggregate network throughput, which approaches the upper bound with a reasonable cell population size of $100$ users. Since \ac{CL(WRR)} variants only redistribute the intra-cluster resources, they do not differ over the aggregate network throughput. The throughput of \ac{PF} reduces significantly as the number of scheduled users per frame increases. However, all \ac{PF} curves converge, for high number of users, to a value well below the throughput of \ac{CL(WRR)}. In Figure~\ref{fig:ms_total_cost}, we can observe that the energy efficiency of \ac{CL(WRR)} is the best. Overall, the energy efficiency decreases with the number of users, due to the fact that each additional user incurs a minimum cost due to activating the network interfaces, while the cell capacity is upper bounded. However, e.g., with $70$ users, the efficiency of \ac{CL(WRR)} is higher than \ac{RR} and PF~$5$ by $\sim\!\! 101\%$ and $\sim\!\! 13\%$, respectively. Recall that in Subsection~\ref{ss:single_cell} we observed that {\it good} users may obtain lower energy efficiency than PF. Here, the cluster formation is only allowed if all cluster members can achieve higher energy efficiency than what they would achieve under PF. This reduces the throughput gain of D2D schemes. As regards fairness, Figure~\ref{fig:ms_fair} shows that \ac{CL(WRR)}-ES provides the highest fairness level followed by \ac{CL(WRR)}-SS, while \ac{CL(WRR)}-WS achieves results comparable to the best results achieved by PF. The \ac{ES} method exhibits better fairness because of equal resource distribution. The \ac{SS} method outperforms \ac{WS} because \ac{SS} distributes the resource based on the contribution of each user to the total revenue. The fairness improvement due to clustering with respect to \ac{RR} and PF~$5$ or PF~$10$, which are realistic figures for \ac{PF} performance, is remarkable.  

We also investigate the impact of payoff distribution methods using our LTE packet simulator, the results indicates that payoff distribution methods behave very similarly in non-saturated scenarios because users receive the requested resources. As the load approaches the saturation level, the \ac{WS} results in higher throughput and delay variations compared to \ac{ES} and \ac{SS}. Considering the insignificant impact of payoff distribution method in non-saturated networks, we can use simple payoff distribution methods such as \ac{ES} instead of \ac{SS}  that adds on to the practicality of D2D-clustering.

\section{Summary}

In this chapter, we have analyzed network-controlled opportunistic D2D clustering from a theoretical point of view. We first provide a model for throughput and power consumption of D2D-enabled clusters in LTE-A networks. Next, we have used coalitional game theory technique to devise revenue distribution schemes that encourage mobile users to form clusters. The numerical simulations illustrated that using simple schedulers and game theory techniques, our proposed architecture significantly outperforms legacy schedulers in terms of throughput, delay, energy efficiency, and fairness.

%\acresetall
\chapter{\titForthCh}
\label{ch:clus_proto}

\section{Introduction}
3GPP is actively studying the feasibility and the architecture of \ac{ProSe} (i.e., \ac{D2D} communications) for both {\it inband} and {\it outband} \ac{D2D} techniques. The latest draft of the standard includes the provisional D2D network elements and their expected functionalities. However, there is no concrete signaling and protocol architecture as of today. Despite the unavailability of the standardization for \ac{D2D} communications, possible design could be mapped through the ongoing feasibility studies and technical reports that deal with \ac{D2D} (see in particular~\cite{3GPP23.703, 3GPP23.303, 3GPP36.843}).  
%The outstanding outcome of the theoretic studies on \ac{D2D} drove mobile operators and the 3GPP standardization body to consider the D2D paradigm as a system component for the next generation cellular standards. Indeed, 
In this chapter, we propose a network-controlled protocol and position it  with respect to the existing architecture of LTE-A and WiFi Direct. 
Specifically, this chapter shows how to adapt LTE and WiFi Direct to support our proposed D2D clustering scheme with minimal modification. 
We show how clusters form in WiFi Direct, register to LTE, obtain LTE connectivity and how the corresponding protocol stack for such a system looks like. In addition, other important procedures such as feedbacks, scheduling, security, etc., are elaborated. 
 %These techniques are distinguished based on the communication spectrum and technology employed. Inband \ac{D2D} uses the cellular interface and takes place in the cellular spectrum, while outband \ac{D2D} uses a secondary interface (e.g., the WiFi interface) and  takes place in the unlicensed spectrum.
%A comprehensive state-of-the-art survey for \ac{D2D} communications in~\cite{asadi2014survey} shows that the articles on inband \ac{D2D} outnumber those of outband \ac{D2D}. Indeed, a quick look at the state-of-the-art shows that inband \ac{D2D} is a well-explored topic. Although its standardization is progressing slowly due to the significance of the required modifications for accommodating \ac{D2D} users over the costly cellular spectrum. Hence, the community is moving its attention towards outband \ac{D2D} to leverage the potentials of \ac{D2D} communications as soon as possible~\cite{andreev2014ComMag}.

%\vspace{-4mm}
\section{A D2D Protocol for WiFi Direct in LTE Cells}
%\section{Implementation of D2D clustering using WiFi Direct in LTE cells}
\label{s:protocol}
This section elaborates on the details of our proposed protocol for D2D communications using WiFi Direct in LTE network. 
Here, we refer to clusters as groups in the cluster formation procedure in order to have the coherent terminology with WiFi Direct specifications. 

%As mentioned, there is no protocol for D2D communications in cellular network. 
%
%
%
%
%\com[vm]{Maybe we can restrict the description to downlink only, like the rest of the paper}

%The section concludes with the presentation of our SDR implementation.

\begin{figure*} [t!]
\begin{center}
\includegraphics[scale=0.55]{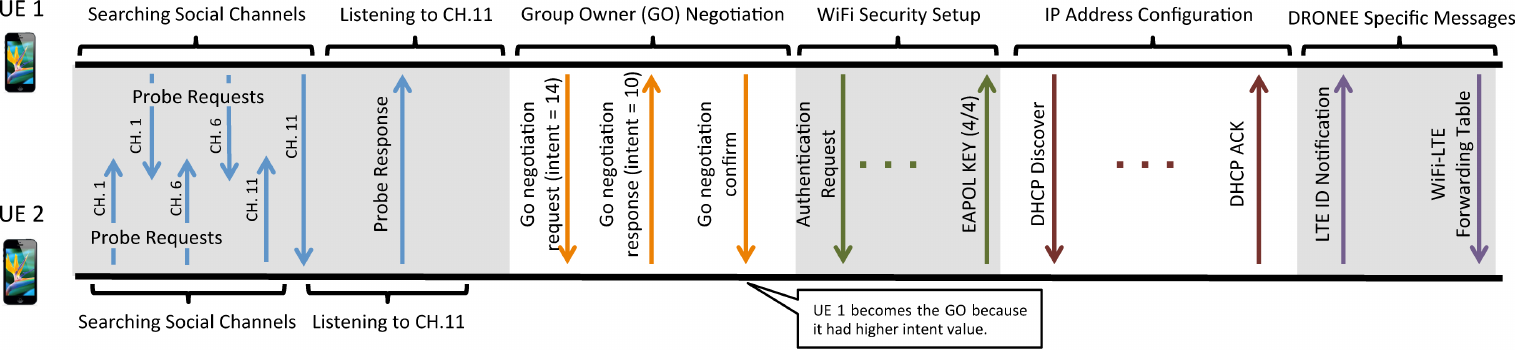}
\caption{WiFi Direct group formation procedure.}
\label{fig:p2p_formation}
\end{center}
\end{figure*}

\begin{figure*} [!t]
\begin{center}
\includegraphics[scale=0.55]{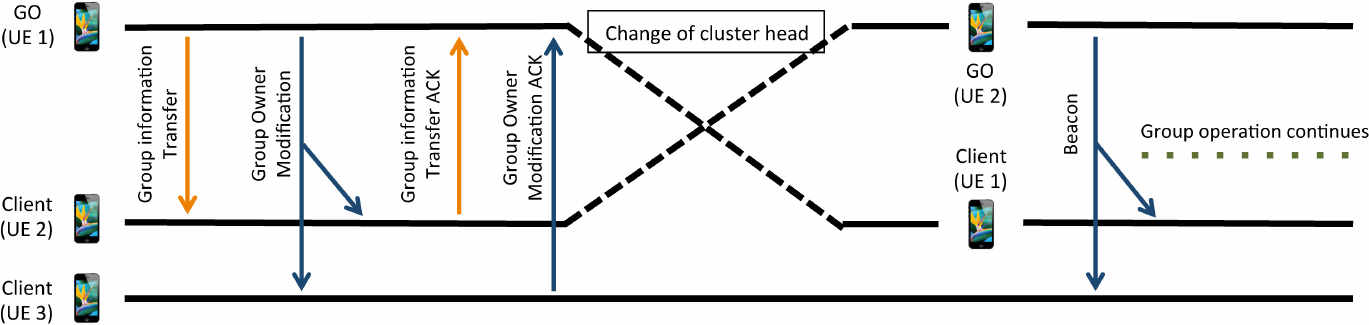}
\caption{Group ownership transfer in WiFi Direct  between UE 1 and UE 2.}
\label{fig:go_transfer}
\end{center}
\end{figure*}

%In this section we show how to implement our proposed scheme using LTE and WiFi Direct. We show that most of the required operations are implementable over the existing wireless protocols with no changes, while a few clustering operations require minor changes to the management of resources in the LTE protocol stack. 
%Specifically, in what follows, we show how clusters form using WiFi Direct, how cluster access the LTE network and obtain connectivity (establish {\it bearers}). We also discuss how to handle the mobility of cellular users belonging to clusters and show the protocol stack to be used in the data plan for the relay of traffic through the cluster head. Finally, we discuss how important LTE procedures, e.g., link adaptation, scheduling, security and policing, have to be adapted to support cluster-based D2D communications. \red{GO=CLUSTER HEAD}

\subsection{Cluster Formation (WiFi Direct)}
\label{s:cluster_formation}
In our proposal, the first step is to form a cluster among \acp{UE} which are willing to use D2D communications. 
The cluster formation procedure is mostly coherent with that defined in WiFi Direct specifications~\cite{WifiDirect2013}. 
The major changes to the existing specifications are: $(i)$ the users/cluster heads also announce their preferred payoff distribution method in the {\it Probe Requests}; $(ii)$ the group ownership is transferable; and $(iii)$ the cluster head receives the LTE ID from its client and shares this information in the form of a forwarding table that contains the LTE and WiFi Direct IDs of all members.
Figure~\ref{fig:p2p_formation} illustrates the required steps for cluster formation between two \ac{UE}s. We briefly explain each step in the following. %with more emphasis on proposed modifications. 

{\bf Search and discovery.}
The \ac{UE} either actively searches to join another \ac{UE}/group by sending {\it Probe Request} over the {\it social channels} (i.e, WiFi channels 1, 6 and 11) ~\cite{WifiDirect2013} or it listens to the social channels for the {\it Probe Requests} sent by other \ac{UE}s/groups. We propose to include a two-bit field to the {\it Probe Request} in order to express information regarding the desired payoff method of the \ac{UE}/group.
Upon reception of a desired {\it Probe Request}, the UE/\ac{GO} responds with a {\it Probe Response} message which initiates the 3-way  group ownership negotiation. 

{\bf Group ownership negotiation.}
In this phase, the negotiating parties exchange their {\it Intent} value and the UE with the highest {\it Intent} becomes the \ac{GO}. In WiFi Direct specification, {\it Intent} value is a number from $0$~to~$15$ which shows the UE's willingness to be \ac{GO}. In our proposal, the {\it Intent} value is the average LTE \ac{CQI} of the \ac{UE}. The {\it Intent} value of an existing \ac{GO}, is the average of  the {\it Intent} values of its group members. Hence, both parties are able to compute an ex-ante revenue prior to group formation. The negotiating parties can leverage the computed revenue to decide whether to terminate or complete the connection setup. 
%If a negotiating party finds out that the group formation leads to loss for itself (or its clients in case of a \ac{GO}), it will terminate the group formation process.  Upon successful Group ownership negotiation, the \ac{GO} initiates the security setup and IP address allocation.

{\bf Security setup and IP address allocation:}
Next, the \ac{GO} initiates the WiFi security setup using {\it \ac{WPS}}.  After the security setup is complete, \ac{GO} assigns IP address to clients following the DHCP protocol.  

{\bf D2D specific messages.}
Our proposed D2D operation requires two additional steps: $(i)$ each group client sends an {\it LTE ID Notification} message to the \ac{GO}, which contains its LTE identity (e.g., \ac{S-TMSI}); and $(ii)$ the  \ac{GO} broadcasts the {\it WiFi-LTE ID Association Table} to all group clients.\footnote{In our proposal, two LTE IDs must be shared among cluster members: $(i)$ the \ac{S-TMSI} which is used in the registration phase; and $(ii)$ \ac{C-RNTI} which is allocated to the UE after it is connected to the eNB. } This message includes the LTE and WiFi Direct IDs of all cluster members. This message can also include WiFi Direct group settings such as power saving parameters, useful to quickly switch the \ac{GO} when needed. 
Therefore, any group client can immediately become the \ac{GO} without tearing down the group. 

{\bf LTE-WiFi mapping.}
Next, each group client sends an {\it LTE ID Notification} message to the \ac{GO} that contains its LTE identity. Finally the  \ac{GO} broadcasts the {\it WiFi-LTE ID Association Table} that includes LTE and WiFi Direct IDs of all cluster members.%\footnote{In our proposal, two LTE IDs must be shared among cluster members: $(i)$ the \ac{S-TMSI} which is used in the registration phase; and $(ii)$ \ac{C-RNTI} which is allocated to the UE after it is connected to the eNB.} 
\footnote{In our proposal, the cluster members should share their \ac{S-TMSI} and  \ac{C-RNTI} with other cluster members.} This message can also include other group settings that are useful to quickly switch the \ac{GO} when needed.

{\bf \ac{GO} transfer.}	
%This step is not a part of the cluster formation in WiFi Direct but it is required for cluster head management. 
In WiFi Direct, the group ownership cannot be transferred. However, our proposal requires the  \ac{GO} to change dynamically. A \ac{GO} transfer occurs when the eNB detects that another cluster member has a better cellular channel quality than the current \ac{GO} (for details, see {\it CSI reporting} and {\it Cluster head selection} in Section~\ref{s:procedures}). 
We define two messages to enable \ac{GO} transfer in WiFi Direct, as shown in Figure~\ref{fig:go_transfer}. 
First, the  \ac{GO} sends the {\it Group Information Transfer} message to the provisioned \ac{GO}. This message contains the updated list of members and their power saving parameters.
Second, the  \ac{GO} sends the {\it  \ac{GO} Modification} broadcast message. Each group client should individually acknowledge this message before the \ac{GO} transfer is completed.

\vspace{-4mm}
\subsection{Cluster Registration in LTE}
\label{s:registration}

\begin{figure*} [!t]
\centering
\includegraphics[scale=0.55]{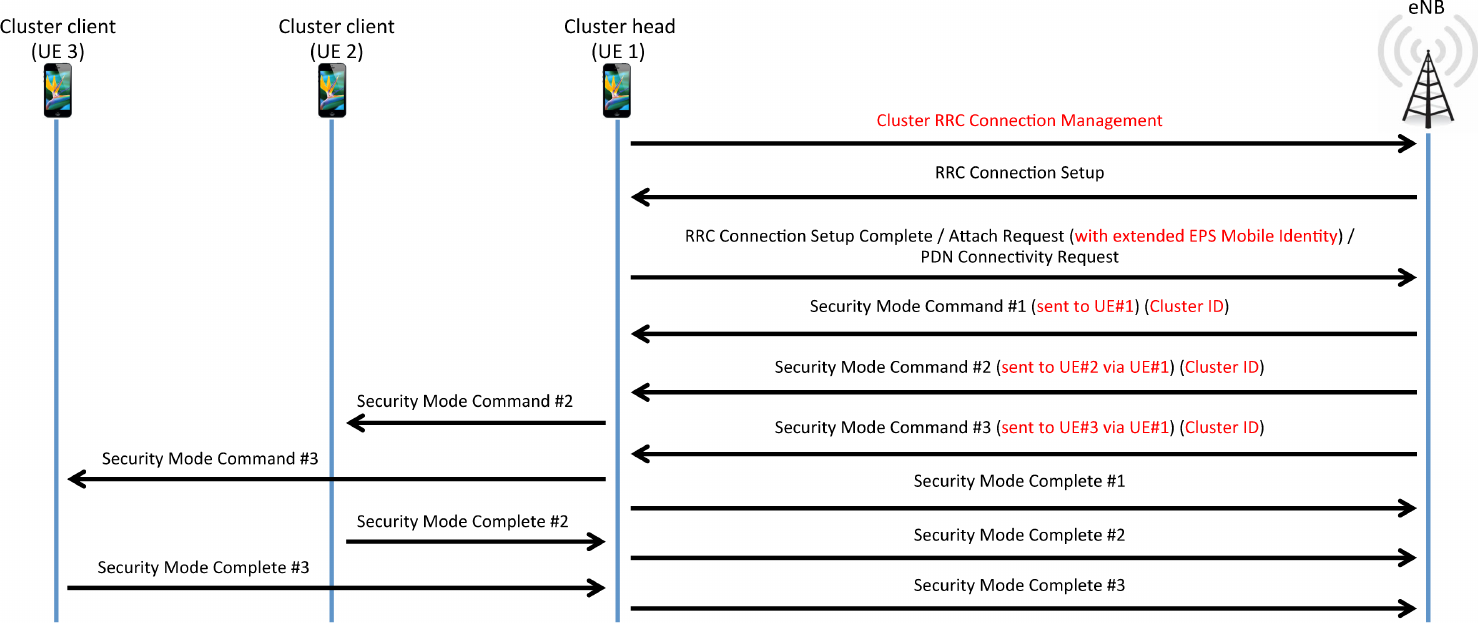}
\caption{Cluster registration procedure in LTE.}
\label{fig:registration}
\end{figure*}

Once a cluster is formed over WiFi Direct, it should register at the LTE network. Cluster registration procedure is shown in Figure~\ref{fig:registration}, which reports the required D2D-enabling modifications in red. This procedure consists of two phases: $( i )$ cluster notification; and $( ii )$ cluster verification. 

{\bf Cluster notification.}
The cluster formation over WiFi Direct is reported to the eNB via {\it Cluster RRC Connection Management} message with {\it Request Cause} set to {\it connection initiation}. This message also contains information such as Identity of the members and their desired payoff allocation method (see Table~\ref{tb:cluster_RRC}). 
The eNB responds to the cluster notification with the {\it RRC Connection Setup} message.
Next, the cluster head sends the {\it RRC Connection Setup Complete} to finish the RRC setup. %Here, the {\it Dedicated NAS Information} field is extended to include the EPS mobile identities of all cluster members. 

\begin{table*}[h!]
\centering
\caption{Contents of  Cluster RRC Connection Management}
\label{tb:cluster_RRC}
\begin{tabular}{| l | l | l |}
\hline
\multicolumn{2}{|c|}{Information Elements} 	 			\\
\hline & \\ [-1 em]\hline
Cluster Identity 			& To be assigned by eNB			 \\
\hline
Cluster Head Identity 	& S-TMSI 						 \\
\hline
Cluster Clients' Identities 	& S-TMSI of Client\#1			 \\
\multirow{1}{10cm}{ (for initiation, all members should be included. Otherwise, only departing/arriving member(s) are listed.)} 			&S-TMSI of Client\#2			\\
					&\hspace{1.5cm} \vdots 			\\
\hline
Request Cause			&\multicolumn{1}{c|}{CHOICE}		\\
\cline{2-2} &\\ [-1 em]\cline{2-2}
					&Connection Initiation			\\
\cline{2-2}
					&Arrival						\\
\cline{2-2}
					&Departure					\\
\cline{2-2}
\hline
\multicolumn{2}{|l|}{Dedicated NAS Information (Attach Request)} \\
\hline
\end{tabular}
\end{table*}

{\bf Cluster verification.}		
Once the RRC connection is established, the eNB sends a {\it Security Mode Command} message to each cluster member via the cluster head. We propose to include the {\it Intent} value (i.e., average CQI) of each cluster member in this message. Since the eNB is aware of real CQI, each member can verify the correctness of the values reported by others. If an anomaly is detected the member can send a negative response and leave the group.  The clients send their response to the cluster head over WiFi and the cluster head forwards the responses to the eNB.  By forcing the security verification to pass through the cluster head, the eNB ensures that all cluster clients are already members of the cluster over WiFi. This step is very important in terms of security because it ensures that any misreported value is detected.

\subsection{Bearer Establishment}

\begin{figure*} [!t]
\vspace{7mm}
\begin{center}
\includegraphics[width=\columnwidth]{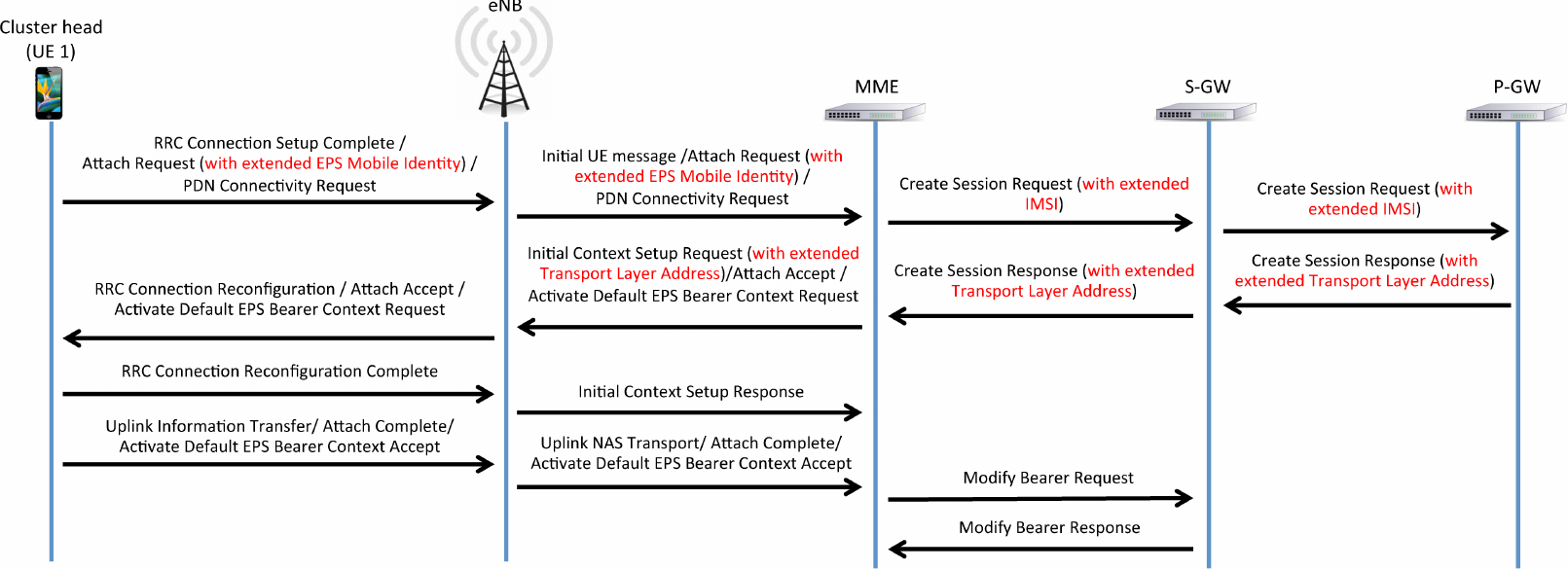}
\caption{Signaling required for  default cluster bearer establishment.}
%\vspace{-5mm}
\label{fig:def_bearer}
\end{center}
\end{figure*}

After cluster registration, the cluster head should initiate the cluster bearer establishment procedure. The difference between cluster bearer and UE bearer is in resource provisioning. The allocated resources for a cluster bearer are equivalent to the aggregate of  resources allocated to all cluster members. LTE standard defines two types of bearers, namely default and dedicated, to support services with different QoS. The default bearer is established once a UE attaches to the network and it remains until the UE leaves the network. 
On the other hand, the dedicated bearer is established for services with specific QoS requirement and it remains active for the life time of the service. For brevity, we suffice to elaborate on the default bearer establishment. The procedure of dedicated bear establishment requires minor changes in the address field in order to accommodate all cluster members.
The procedure for  default bearer establishment is depicted in Figure~\ref{fig:def_bearer}, and it consists of three steps: $(i)$ bearer request,  $(ii)$ bearer request response, and $(iii)$ bearer request confirmation.

{\bf Bearer request.} After cluster registration is completed, the eNB sends the {\it Attach Request} to the \ac{MME}. 
The \ac{MME} determines the \ac{IMSI} of each cluster member from the information provided in {\it EPS Mobile Identity} fields of the {\it Attach Request}. In case the \ac{IMSI} of a member cannot be identified, the \ac{MME} explicitly asks for it. Next, the \ac{MME} sends a {\it Create Session Request} to the \ac{S-GW} which contains information such as \ac{IMSI} of the cluster members and requested \ac{PDN} connectivity. 
The \ac{S-GW} updates its EPS Bearer table  and forwards the {\it Create Session Request} message to the \ac{P-GW}. 
The \ac{P-GW} updates its EPS Bearer Context Table and generates a charging profile for every member which does not have one yet.

{\bf Bearer request response.} The \ac{P-GW} responds to the \ac{S-GW} request with {\it Create Session Response} message. In this message the address (assigned by \ac{P-GW}) and QoS parameters (assigned by \ac{PCRF}\!\!~\footnote{Here, the QoS parameters refer to both per-UE QoS parameter such as \ac{AMBR} and per-bearer QoS parameters such as \ac{QCI}, \ac{ARP},  and \ac{MBR}.}) fields are extended to accommodate all cluster members.  Naturally, QoS parameter of the cluster bearer is equivalent to the aggregate of members' QoS.
Next, the \ac{S-GW} forwards the {\it Create Session Response} message to the \ac{MME} which triggers the {\it Initial Context Setup Request} sent from the \ac{MME} to the eNB.
This message provides the eNB with settings such as the IP address and the QoS parameters of each cluster members. Again, the IP address and QoS parameters fields of the {\it Initial Context Setup Request} message is extended to accommodate all cluster members. Note that this reduces the signaling overhead compared to standard LTE operation because the network does not need to send this information to each UE separately.
Finally, the eNB extracts the {\it Attach Accept} message from the {\it Initial Context Setup Request} and sends it to the cluster head in an {\it RRC Connection Reconfiguration} message.
 
{\bf Bearer request confirmation.} The cluster head updates the cluster clients with information received from the eNB.
It also sends two messages to the eNB in response to {\it RRC Connection Reconfiguration} message. An {\it RRC Connection Reconfiguration Complete} message  which is basically an acknowledgment to the {\it RRC Connection Reconfiguration} and an {\it Uplink Information Transfer} message in order to complete the NAS attach process.
Upon reception of {\it RRC Connection Reconfiguration Complete}, the eNB sends the {\it initial context Setup Response} to the \ac{MME}. This message acknowledges that the \ac{E-RAB} is successfully setup for the default bearer. It also provides an IP address for communication between the eNB and S-GW for downlink data transfer.
After the eNB received the {\it Uplink Information Transfer} message, it sends the {\it Attach Complete} 
message to the \ac{MME}. 
The {\it Attach Complete} and {\it Active Default EPS Bearer Context Accept} messages trigger the \ac{MME} to send the {\it Modify Bearer Request} to \ac{S-GW}. This message mainly serves as an acknowledgement.
Finally, the \ac{S-GW} completes the process by sending {\it Modify Bearer Respond} to the \ac{MME}.

As concerns  IP addressing,  in LTE, each active UE has at least one default bearer and each default bearer has a unique IP address. Therefore, if a cluster member had bearer(s) before cluster formation, the \ac{P-GW} keeps the existing IP address(es) associated to the default bearer(s). Once the cluster bearer is activated, the \ac{P-GW} automatically terminates the old default bearer(s).

\subsection{Mobility}
\label{s:mobility}

In our proposal, UEs may join or leave at any time. In this section, we elaborate the procedures for departure/arrival.  
%However, the cluster membership time should be long enough (a few minutes) in order to compensate for group formation and cluster registration time.

\begin{figure*} [t!]
\begin{center}
\includegraphics[scale=0.65]{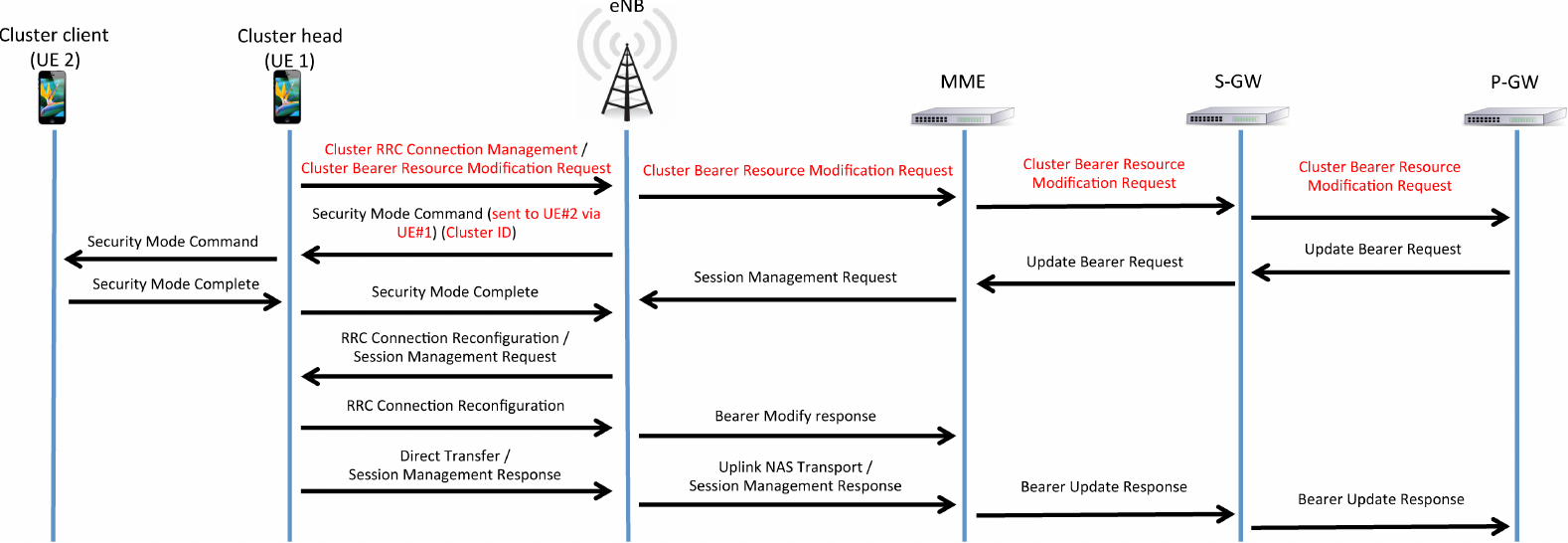}
\caption{Signaling messages required for a new arrival.}
%\vspace{-5mm}
\label{fig:lte_arrival}
\end{center}
\end{figure*}

{\bf Arrival.} Figure~\ref{fig:lte_arrival} depicts the procedure followed by a new arrival. In the event of a new arrival, the cluster head sends a {\it Cluster RRC Connection Management} to the eNB with the {\it Request Cause} set to {\it arrival} (see~Table~\ref{tb:cluster_RRC}). 
In this event, differently from cluster registration (see Section~\ref{s:registration}), the cluster head only sends the \ac{S-TMSI} of the new UE. After the eNB receives the notification of new arrival, it sends a {\it Security Mode Command} to the cluster head to verify the new member. After the verification phase, the eNB sends the {\it Cluster Bearer Resource Modification Request} message, see Table~\ref{tb:bearer_mod}, to the \ac{MME}.
The \ac{MME} replaces the \ac{S-TMSI} identity in  the {\it Cluster Bearer Resource Modification Request} message with \ac{IMSI} and forwards the message to the \ac{S-GW}.
The \ac{S-GW} updates the EPS Bearer Table and sends the  {\it Cluster Bearer Resource Modification Request} message to the \ac{P-GW}. This initiates the standard LTE bearer modification process as described in~\cite{specBearer}. For the sake of brevity, we will not explain the rest of the signaling messages because they follow the standard LTE-defined procedure. 

{\bf Departure.} The signaling procedure for a departure is very similar to that of an arrival. The only procedural difference between arrival and departure is that the eNB does not need to send the {\it Security Mode Command}
in case of departure. 
Moreover, according to WiFi Direct specifications, the members should send a de-authentication message to the cluster head before departure. The de-authentication message triggers the cluster head to initiate the departure procedure.
Nevertheless, if a UE sends an {\it RRC Connection Request} message after it joined a cluster, the eNB assumes that the UE does not belong to a cluster anymore.\footnote{Our proposal does not permit UEs to have user bearer and cluster bearer simultaneously.} In this case, the eNB initiates the departure procedure without receiving the departure notification message from the cluster head. If a UE is reported as a new arrival of a cluster~(i.e., $C_a$) while it is listed as a member of another cluster~(i.e., $C_b$), the eNB initiates the departure procedure for $C_a$ and an arrival procedure for $C_b$.

\begin{table}[h!]
\centering
\caption{Contents of  Cluster Bearer Resource Modification Request}
\label{tb:bearer_mod}
\begin{tabular}{| l | l | l | }
\hline
\multicolumn{2}{|c|}{Information Elements}										\\			
\hline & \\ [-1 em]\hline
Identity of Departing/Arriving Member(s) 			& S-TMSI(s) 				 	\\
\hline
Request Cause								&\multicolumn{1}{c|}{CHOICE}		\\
\cline{2-2} &\\ [-1 em]\cline{2-2}
										&Arrival						 \\
\cline{2-2}
										&Departure					\\
\cline{2-2}
\hline
\end{tabular}
\end{table}

\subsection{Data Plan Operation}
\label{s:stack}

\begin{figure*} [t!]
\begin{center}
\includegraphics[width=\columnwidth]{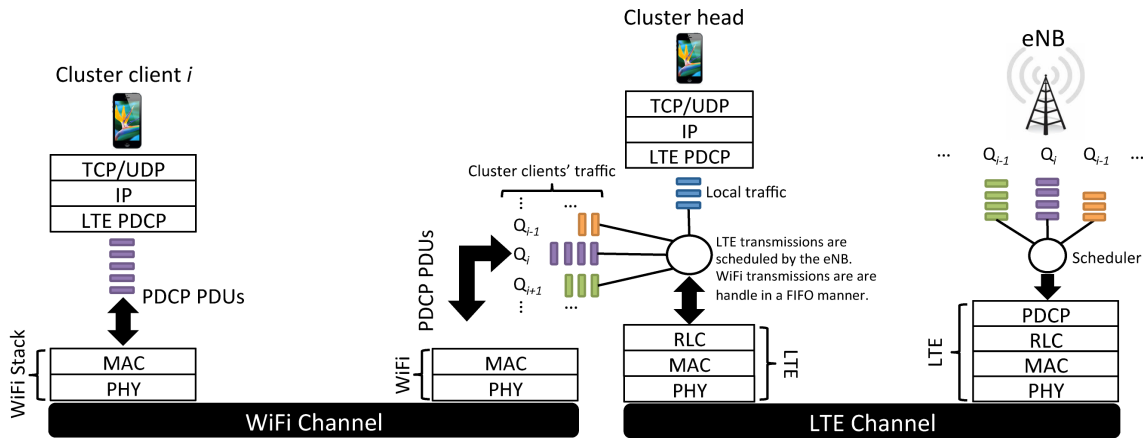}
\caption{Data flow between cluster client $i$ and the eNB.} 
\label{fig:prot_stack}
\end{center}
\end{figure*}

Figure~\ref{fig:prot_stack} illustrates the adaptation of LTE and WiFi Direct data protocol stacks to our proposal. We choose to bridge, at the cluster head, the WiFi Direct MAC and LTE at  \ac{PDCP} layer for three reasons: $(i)$ LTE packets are ciphered and integrity-protected in the \ac{PDCP} layer using keys which are only known to the client and the eNB. Therefore, other UEs cannot decipher the LTE packets traversing the WiFi network; $(ii)$ the cluster head can further process  \ac{PDCP} \ac{PDU}s in RLC layer for concatenation/segmentation according to its LTE physical link quality; 
and $(iii)$ the WiFi Direct MAC provides a robust and secure transmission service, and natively allows to send frames to be relayed at MAC layer.
%Figure~\ref{fig:prot_stack} illustrates the packet flow through the stack in both uplink and downlink directions which are explained below.
Note that the resulting LTE and WiFi Direct data transfer operations are decoupled. Indeed, the cluster head uses the legacy ACK/NACK to secure all the handled LTE traffic, so that ARQ/HARQ operations are managed at the cluster head only, as if the exchanged LTE \acp{PDU} were all belonging to the cluster head. Similarly, the normal ACK and retransmission mechanisms are used by the WiFi Direct interface to transfer LTE \acp{PDU} in a legacy 802.11 payload.

{\bf Uplink.}  
As concerns uplink transmission requests, the clients send their \ac{SR} or \ac{BSR} to the cluster head to be forwarded to the eNB. 
The eNB uses \ac{DCI} to inform UEs regarding their downlink and uplink resource allocation. 
Since the cluster head is the only member which is listening to the LTE channel, it receives the \ac{DCI} and updates the clients with the scheduling decision made by the eNB, using an 802.11 management frame with the same subtype value used by the UEs to encapsulate \ac{SR} and \ac{BSR} messages in the WiFi Direct frame.
As concerns data packets, the scheduled cluster clients encapsulate the LTE \ac{PDCP} PDUs in WiFi frames and send them to the cluster head. 
The cluster head extracts the \ac{PDCP} PDUs and forwards them to the eNB in the designated slot. The cluster head transmits the packets to the eNB with the client's \ac{C-RNTI} address in order to simplify the identification of the real source of the packets for the eNB. Since LTE data packets are ciphered and integrity protected, there are no security concerns and threats in operating LTE relay over WiFi Direct as described in this chapter.

{\bf Downlink.}
The eNB transmits the packets using the client's \ac{C-RNTI} address but it selects the \ac{MCS} according to the cluster head's channel quality. 
Since the cluster head is aware of scheduling plan for its clients, it listens to the downlink channel to receive the packets belonging to all cluster members. 
Next, the cluster head encapsulates the \ac{PDCP} PDUs in regular WiFi data frames that include the source and destination MAC addresses of cluster head and client, and the default MAC address of the eNB.

\subsection{Adaptation of LTE Procedures}
\label{s:procedures}
So far we have defined the required messaging to support our proposed architecture in LTE and WiFi Direct. Here, we elaborate on the adaptation of our proposal to other important operations.

%\vspace{-3mm}
\subsubsection{\ac{CSI} reporting}
In LTE, UEs send \ac{CSI} reports to the eNB for scheduling purposes.  In our proposal, the cluster head sends the \ac{CSI} reports of all cluster members to the eNB. This creates some flexibility which does not exist in the standard LTE operations. For example, the cluster clients can report the \ac{CSI} over all sub-bands to the cluster head over WiFi. Then the cluster head filters these reports and sends the list of top candidates on each sub-band to the eNB. Alternatively,
the cluster head reports the {\it n} highest \ac{CQI} to the eNB. The value of $n$ imposes a trade-off between opportunistic gain and spectral efficiency. 
Note that such a high resolution \ac{CQI} report would not be possible in normal LTE operations.

%\vspace{-3mm}
\subsubsection{Cluster head selection}
%\vspace{-1mm}
The eNB selects the cluster head among the cluster members based on the reported \ac{CSI}s. %reports that are received from the cluster head. 
We propose to add an extra field to the \ac{DCI} so that the eNB can transmit the \ac{C-RNTI} of the new cluster head to the current cluster head, which can trigger the \ac{GO} transfer procedure. %This triggers the current cluster head to initiate the group ownership transfer procedure (see Section~\ref{s:cluster_formation}).
The cluster head selection interval is implementation-specific and it is constrained by the delay of LTE network and group ownership transfer in WiFi Direct. 
This interval introduces a trade-off between signaling overhead and opportunistic gain. On one hand the opportunistic gain is maximized when the cluster head is selected on a per-frame basis (shortest possible interval). On the other hand, per-frame cluster head selection requires higher signaling overhead. 

\subsubsection{Label Switching}
The relay UE experiences high computational overhead due to LTE frame processing. Hence, we propose label switching at LTE \ac{PDCP} layer instead of IP routing which is the current solution in 3GPP. As mentioned earlier, the byproduct of this design choice is the elimination of relay-related security concerns. 

%\vspace{-2mm}
\subsubsection{Scheduling}
\label{ss:scheduling}
The existing LTE scheduler can be adapted to support our proposal with a minor modification. In LTE, the eNB selects the physical layer parameters based on  the \ac{CSI} of the scheduled UE. However, our proposal requires the eNB to select physical layer parameters according to the \ac{CSI} of the cluster head so that the cluster head can decode the packets and forward them to the corresponding clients.
Note that the eNB still uses the \ac{C-RNTI} of the client in the \ac{DCI} so that the cluster head is aware of its transceiving schedule in uplink and downlink. This also eliminates the need for an uplink intra-cluster scheduler in the cluster head.

\subsubsection{Security}
\label{ss:security}
As mentioned earlier, our proposal does not introduce any new security threats to the existing LTE architecture because the LTE packets are ciphered and integrity-protected before forwarding. We also propose to send {\it Security Mode Command}  through the cluster head, so that the cluster head cannot exploit a UE's resources that is not present in the cluster.
The only possible attack is a malicious cluster head that drops packets of its clients. 
The eNB can detect such behavior by tracking communication failures of each cluster head and act accordingly.

\subsubsection{Policy control and charging}
\label{ss:pcc}

Since the cluster head is in charge of the LTE transmissions for all cluster clients, it is important to make sure that the cluster head is not billed for the clients' traffic. 
The policy control and charging of LTE is done via \ac{PCEF} which charges the UEs based on their IP address. Since each cluster member is given a separate IP address, our proposal does not introduce any problem in billing. 
It is also important to ensure that members do not utilize each other resources.
Since the eNB schedules the members individually, utilizing the other cluster members' resources is not a concern. In case a malicious cluster head transmits its own packets on a slot allocated to another member, the eNB discards the cluster head data because it cannot be deciphered.

\section{Summary}

In this chapter, we proposed a practical protocol for supporting \ac{D2D} communications in cellular networks using WiFi Direct and LTE. This protocol, to the best of our knowledge, is the first of its kind. Specifically, we detailed how \ac{D2D} communications can be supported over LTE network with minor modifications to the standard procedures without any infrastructural changes. We first adapted the WiFi Direct group formation procedure to support dynamic cluster formation and LTE specific message exchange. Then, we introduced the necessary signaling messages for cluster registration and bearer setup in LTE. We showed that most of the required pieces are in place and we only need to amend some of existing messages to support \ac{D2D} clustering. Finally, we discussed the impact of \ac{D2D} clusters on LTE operations such as billing and security. Our studies revealed that all native LTE operation work with D2D clusters. Moreover, \ac{D2D} clustering does not impose any security threat which does not already exist in legacy LTE systems.

%\acresetall
\chapter{\titFifthCh}
\label{ch:clus_exp}

\section{Introduction}
Analytical studies on \ac{D2D} communications demonstrate very promising results, but there is no experimental evidence that validates these results to date. Moreover, the 3GPP study is mostly focused on the description of the architecture and procedure for \ac{D2D} communications. As of today, their proposed architecture of ProSe remain unevaluated. The lack of experimental studies is mostly due to tedious process of prototyping complex cellular architecture and expensive equipments that are associated with it. Nevertheless, thanks to the advent of \ac{SDR}, the experimental studies for D2D communications are now made possible.  In this chapter, we elaborate on the development procedure of outband \ac{D2D} communications using a National Instrument's \ac{SDR} platform. Next, we leverage this platform to answer questions such as: How is the cross-platform (LTE to WiFi) communication? What is the delay and the computational overhead? How large is the throughput gain in presence of real channel dynamics?

\section{SDR-based Testbed for Outband D2D Communications}
\label{s:testbed}

In this section, we describe the hardware, the software, and the architecture of our testbed. Our testbed design is utterly beneficial because it is based on an \ac{SDR} platform that uses \ac{FPGA} modules for high-speed \ac{DSP} prototyping. Moreover, it exploits the real-time controllers  which ensures quick protocol prototyping. As today's cellular networks are moving toward functional virtualization and new architecture, \ac{SDR} platforms certainly stand out among the prototyping alternative in today's market. 

%The testbed consists of two main elements, namely, the eNB and the UEs. We choose to use LabView for prototyping outband \ac{D2D} because LabVIEW FPGA allows us to implement high speed DSP operations (e.g.,FFT, iFFT, encoding, and decoding) in FPGA. In addition, operations such as scheduling can be implemented quickly using LabVIEW Real-Time.

\subsection{Software and Hardware}
\begin{figure} [h!]
\centering
		\includegraphics[width=0.5\columnwidth]{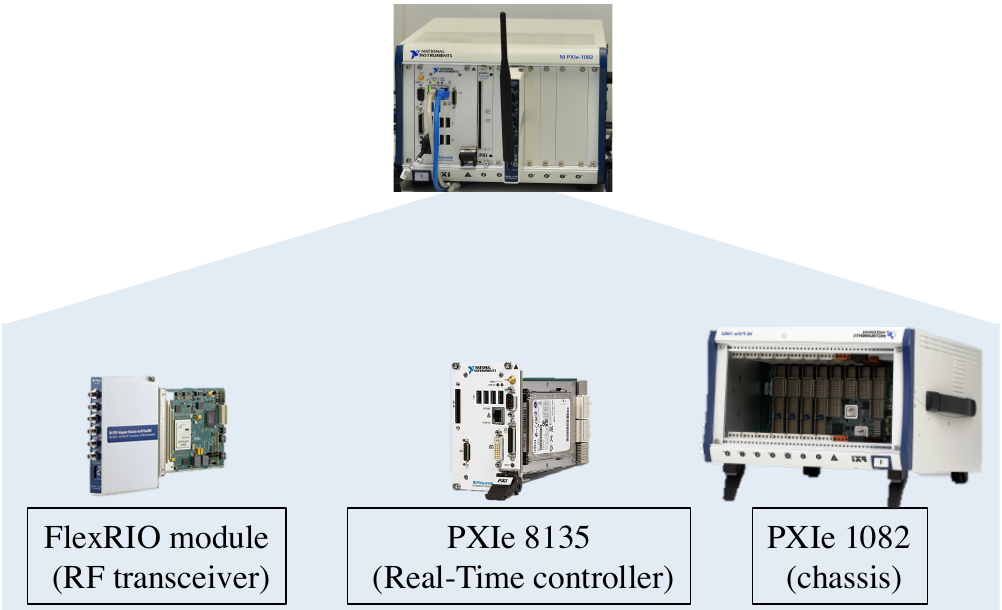} %[width=70mm,angle=0] [scale=0.80, angle=0]
		\caption{Illustration of the hardwares used for the testbed.}
		\label{fig:hardware}
\end{figure}

{\bf Hardware.} The hardware used in this testbed is NI PXI 1082 chassis\footnote{\url{http://sine.ni.com/nips/cds/view/p/lang/en/nid/207346}} that contains: $(i)$ NI PXIe 8135 Real-Time controller\footnote{\url{http://sine.ni.com/nips/cds/view/p/lang/en/nid/210545}} operating on an Intel Core-i7-3610QE CPU. This controller is used to execute LabVIEW Real-Time which runs MAC layer algorithm and physical layer control algorithms. LabVIEW Real-time communicates with the \ac{FPGA} through dedicated direct memory access FIFOs; $(ii)$ The Physical layer is programmed using LabVIEW \ac{FPGA} and runs on NI FlexRIO modules\footnote{\url{http://www.ni.com/flexrio/}} that are equipped with Xilinx Kintex 7 or Virtex 5 \ac{FPGA}s; $(iii)$ NI 5791 \ac{FAM} is used as an RF transceiver operating with a $200$ MHz bandwidth the in frequency range $200$ MHz to $4.4$ GHz. This is used for \ac{DAC}, \ac{ADC}, up-converting  base-band to band-pass, and down-converting band-pass to base-band. 
Figure~\ref{fig:hardware} illustrates all these components. 

{\bf Software.} We found LabVIEW\footnote{\url{http://www.ni.com/labview/}} as a suitable tool to prototype an outband \ac{D2D} platform. On one hand, LabVIEW \ac{FPGA} allows for quick implementation of computationally intensive Physical layer (PHY) operations (e.g., \ac{FFT}, \ac{iFFT}, encoding, and decoding) in Xilinx \ac{FPGA} and meets the stringent nano-second time scale requirement of the digital communication systems. Moreover, it provides the necessary means for high speed communication with CPU and \ac{RF} hardware. On the other hand, LabVIEW Real-Time is suitable for implementing MAC layer operations with micro-second resolution. LabVIEW Real-Time runs on a general purpose CPU and communicates with LabVIEW \ac{FPGA} over high-speed PCIe backplane.

\subsection{Architecture of eNB}
\begin{figure} [h!]
\centering
		\includegraphics[width=0.8\columnwidth]{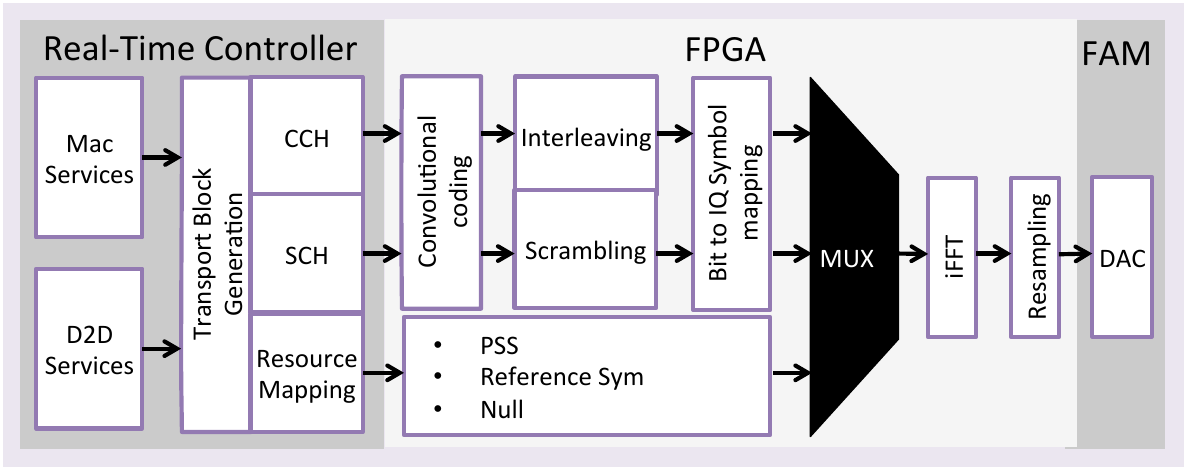} %[width=70mm,angle=0]  [scale=0.80, angle=0]
		\caption{Architecture of the eNB.}
		\label{fig:eNB}
		\vspace{-5mm}
\end{figure}

The eNB consists of a Real-Time controller and a FlexRIO module used for \ac{OTA} LTE transmission. 
Figure~\ref{fig:eNB} shows the important blocks of the eNB. Real-Time controller runs MAC layer services such as scheduling, \ac{D2D} services, and \ac{TB} generation for \ac{CCH} and shared channel \ac{SCH}. \ac{FPGA} executes PHY operations such as convolutional coding, interleaving for \ac{CCH} traffic and scrambling for \ac{SCH} traffic. Finally, the base-band signal is up-converted in the \ac{FAM} module and transmitted to the UE \ac{OTA}. The current testbed only supports \ac{OFDMA} in downlink and the uplink transmissions is performed  over ethernet. However, in future we intend to extend this testbed to support \ac{OFDMA} uplink transmission.

\subsection{Architecture of UE}
As eNB only requires one interface, UE's architecture is thus more complex since it comprises two interfaces: $(i)$~an LTE \ac{OFDMA} downlink receiver for communicating with the eNB; and $(ii)$ a WiFi transceiver to establish the outband \ac{D2D} link with other UEs. 

\begin{figure} [h!]
\centering
		\includegraphics[width=0.7\columnwidth]{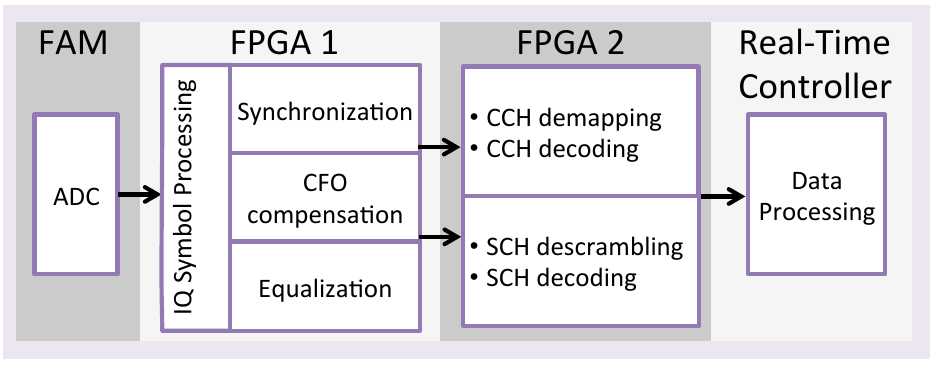} %[width=70mm,angle=0] [scale=0.80, angle=0]
		\caption{Architecture of UE's LTE interface.}
		\label{fig:UE-lte}
\end{figure}

{\bf \ac{OFDMA} receiver.}  Following the design  of the transmitter, the DSP operations are implemented in the \ac{FPGA}. Real-Time controller is only used for processing the received payload and MAC layer \ac{D2D} operations. These operations mainly consist in filtering the relay packets and transmitting them to the WiFi interface for relaying.  Figure~\ref{fig:UE-lte} shows the location of different logical blocks in the system.

\begin{figure} [h!]
\centering
		\includegraphics[width=0.7\columnwidth, angle=0]{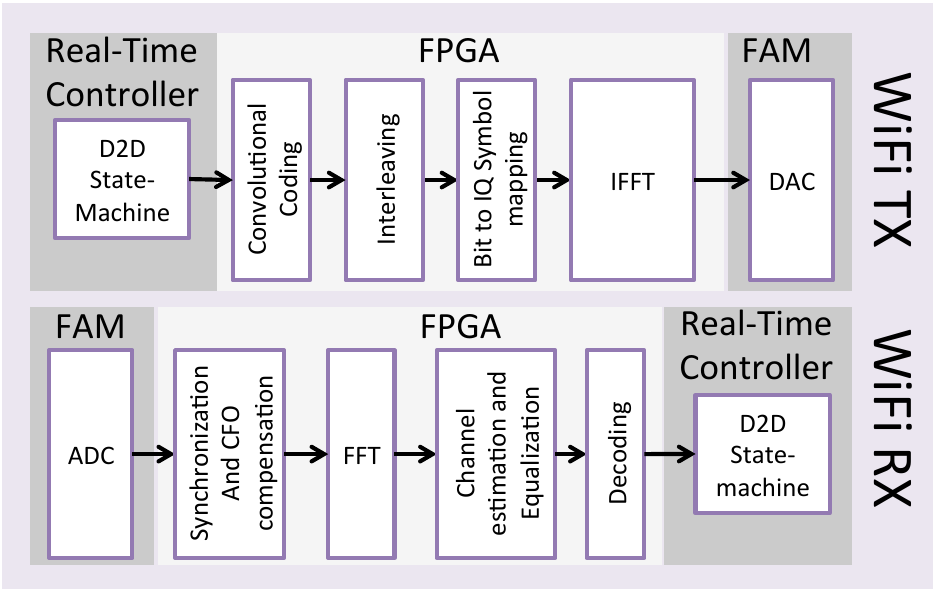} %[width=70mm,angle=0] [scale=0.80, angle=0]
		\caption{Architecture of UE's WiFi interface.}
		\vspace{-5mm}
		\label{fig:UE-wifi}
\end{figure}

{\bf WiFi transceiver.} The majority of the WiFi framework~\cite{wifi2014NI} is implemented in \ac{FPGA}. In addition, the transceiver is implemented within the same \ac{FPGA}. We implemented the \ac{D2D} state-machine and its corresponding logic in the Real-Time controller. The controller is also in charge of feeding data to the \ac{FPGA} transmission processing chain and reading the decoded data from \ac{FPGA} reception processing chain. 
Figure~\ref{fig:UE-wifi} shows the structure of the WiFi framework.

\begin{figure} [t!]
\centering
		\includegraphics[scale=0.3, angle=0]{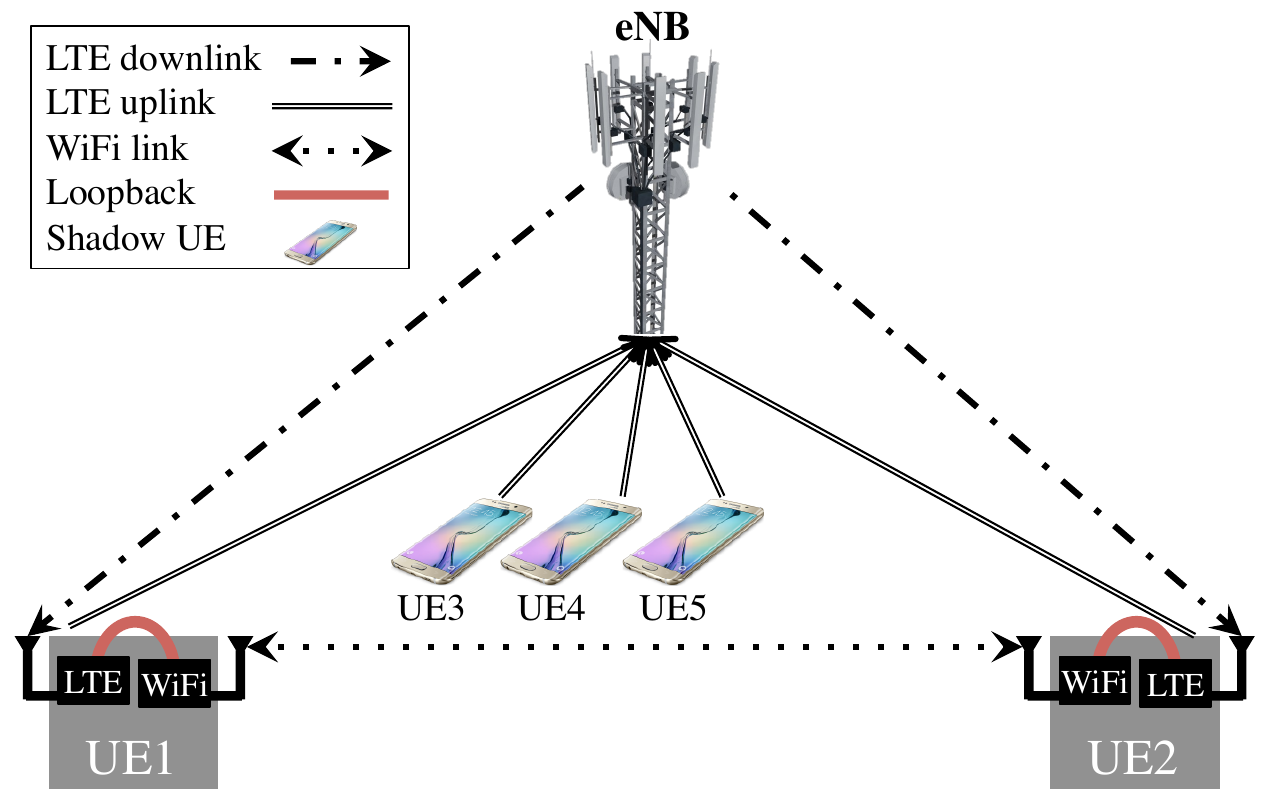} %[width=70mm,angle=0] [scale=0.80, angle=0]
		\vspace{-3mm}
		\caption{Architecture of the testbed.}
		\vspace{-7mm}
		\label{fig:testbed}
\end{figure}

{\bf Communication.} We observe in Figure~\ref{fig:testbed} that UEs receive downlink transmissions over the OFDMA receiver and send the uplink messages over an ethernet link. The WiFi (i.e., D2D) transmission uses an OFDM transceiver. %The transceiver is constantly monitoring the channel and it will transmit only when the medium is free. We use the loopback interface to connect WiFi and LTE interfaces of the UE.

\subsection{Shadow UEs}
\label{ss:shadow}
These non-\ac{SDR} UEs (i.e., UE3, UE4, and UE5 in Figure~\ref{fig:testbed}) are  off-the-shelf android smartphones. We include the shadow UEs in our setup to better capture the performance of outband D2D in a real-world scenario. We developed an android application to obtain real-time cellular channel quality on a millisecond basis.
The application then transmits the channel quality values to an access point, which is connected to the eNB over an ethernet link. Although the shadows do not receive the actual transmission, the eNB schedules them and transmits their data as if they were real UEs. Since the MCS-SNR mapping is constructed to ensure block error rates below $10^{-4}$, we assume that the shadows receive the transmitted blocks with success probability of $0.9999$.

\subsection{Synthetic Fading}

We measure the selected KPIs under different scheduling schemes. Due to the limitation in the number of equipments in our disposal, we should run each experiment  at a separate time instant. In an ideal case, the system can be connected to high-end multi-channel cellular channel emulators to create the same channel variation in each experiment. Since we do not have such a device, we decided to create a repeatable channel variation situation using refractors. 
 
\begin{figure} [h!]
\centering
		\includegraphics[scale=0.5, angle=0]{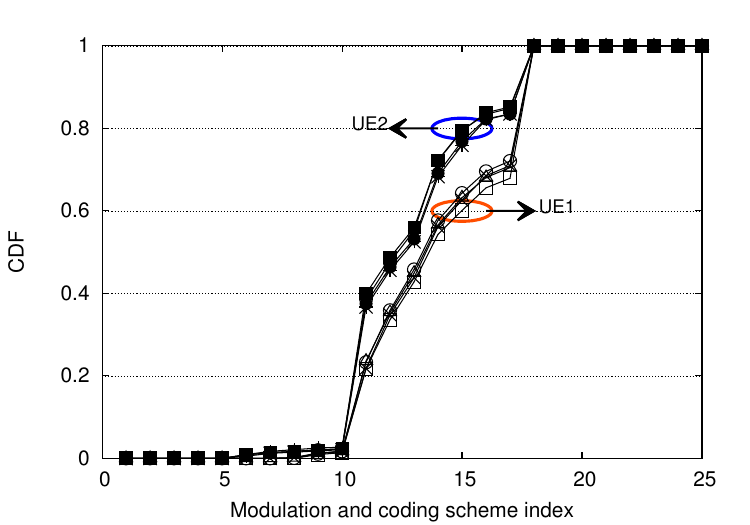} %[width=70mm,angle=0] [scale=0.80, angle=0]
		\vspace{-3mm}
		\caption{CDF of MCS for UE1 and UE2 in each experiment.}
		\vspace{-5mm}
		\label{fig:mcsCdf}
\end{figure}

In order to create repeatable channel variation patterns, we mounted the refractors plates on a step motor that is controlled by an Arduino Uno\footnote{\url{http://www.arduino.cc/en/Main/ArduinoBoardUno}} micro-controller. We generate synthetic channel variation by changing the rotation speed of the step-motor. 
Figure~\ref{fig:mcsCdf} is the proof of concept of this mechanism. We repeated an experiment four times and plotted the CDF of the potential MCS (obtained from channel qualities) for both UEs to ensure stable repetitions of the channel variation. Indeed, the results show that this approach is suitable to re-create the same channel environment for different experiments.  Note that the location/frequency is selected such that unpredictable interference is minimized. 

\section{Experimental Evaluation}
\label{s:eval} 
In this section, we experimentally evaluate the performance of outband \ac{D2D}-relay and our proposed protocol.
We design several experiments to better demonstrate the system behavior in different scenarios.
We first present the performance of a simple outband \ac{D2D}-relay setup. The simple setup is then redesigned to first incorporate channel opportunism and then \ac{QoS}-awareness. We also examine the impact of non-collaborative UEs (i.e., shadow UEs).
The duration of each experiment is $300$ seconds, which is sufficiently long to observe the average system's performance. 
%The detailed description of testbed, its setup and our selected KPIs and measurement methodology are provided in the previous sections. 
%
%The testbed elements and setup is as described in Sections~\ref{s:testbed}~and~\ref{s:setup}. 
In order to provide the reader with a detailed view of the achieved performance, we show minimum, maximum, 25th and 75th percentiles in addition to the average values. Unless otherwise specified, the rotation speed of the refractor is $5$ rpm. Finally, UE2 experiences higher average channel quality than UE1 in all experiments.

\subsection{Algorithm Implementation}
We implement  \ac{DORE}, a channel-opportunistic framework for enhancing network capacity under \ac{QoS} constraints. \ac{DORE} is based on the aforementioned channel opportunistic design in previous chapters. In addition to channel opportunism, we add delay consideration to our implemented algorithm for better \ac{QoS}-awareness. The delay consideration enables our proposed \ac{D2D} scheme to serve a much larger variety of application in particular those with tight delay requirements. Algorithm~\ref{alg:dore} is the pseudocode of \ac{DORE}. In the algorithm, we first calculate the achievable throughput of the cellular link $T_{ lte }^{ ( i ) }$ and the \ac{D2D} link $T_{ d2d }^{ ( ij ) } $. Then Merge and Split algorithm uses the potential throughput information of the links (throughput is zero if the link can not be established) for cluster formation. Next, the algorithm compares the achieved delay $d^{ ( ij ) }$ over the \ac{D2D} link with the application specific threshold $d_{ th }^{ ( i ) }$ to decide whether it should keep the \ac{D2D} link active ($\alpha_{ d2d } = 1$) or it should fall back to cellular link ($\alpha_{ d2d } = 0$).  It should be noted that we do not use the game theory-based dynamic cluster formation algorithm in our experimental evaluation because the testbed consists of only two static UEs. 

\vspace{-2mm}
\begin{algorithm}[h]
\caption{ \texttt{DORE}}
%\scriptsize
\begin{algorithmic}[1]
\Require \\ 
	  	$ T_{ lte }^{ ( i ) }, T_{ d2d }^{ ( ij ) }, d^{ ( ij ) }, d_{ th }^{ ( i ) } \quad \forall i , j \in \{ 0 , \dots , N \} $.
\Ensure $\alpha_{ d2d }$\\
Initialize: $ T_{ lte }^{ ( i ) }=T_{ d2d }^{ ( ij ) }=T_{ gain }^{ ( ij ) }=0$, $d^{ ( ij ) }=\infty$.
\For { k from 1 to N }
	\State{Compute the potential $T_{ lte }^{ ( i ) }$ and $T_{ d2d }^{ ( ij ) } $ based on the received \ac{CSI}.}
\EndFor
\State {{\it Merge \& Split} based on the estimated values for $T_{ lte }^{ ( i ) }$ and $T_{ d2d }^{ ( ij ) } $.}
\If {$d^{ ( ij ) } \leq d_{ th }^{ ( i ) } $}
	\State{Use \ac{D2D} link ($\alpha_{ d2d } = 1$).}
\Else
	\State{Use cellular link ($\alpha_{ d2d } = 0$).}	
\EndIf

\end{algorithmic}
\label{alg:dore}
\end{algorithm}
\vspace{-4mm}

\subsection{Selected KPIs}

%In our evaluation,  the selected KPIs are throughput, delay, CPU load and \ac{D2D} lifetime.
%In order to provide the reader with thorough view of our proposed schemes, we report several KPIs are selected based on their importance towards understanding the characteristic of a practical \ac{D2D} system. The following is the description of each KPI used in our evaluation.

We report several KPIs to examine different aspects of outband \ac{D2D}-relay and \ac{DORE}. The KPIs described below are chosen based on their importance for understanding the characteristics of a practical \ac{D2D} system. 

{\bf Throughput.} Throughput is measured as the number of received bits per second. %Common to the definition of throughput in the same line of work, we also defined throughput  as the number of received bits per second.

{\bf Delay.} We timestamp each packet at the eNB MAC and measure the delay at three points within the path from the eNB to the \ac{D2D} receiver. The delay from the eNB MAC to the relay-UE MAC is referred to as {\it LTE delay}. The delay from the relay's LTE MAC to WiFi MAC is called the {\it cross-platform delay}. {\it WiFi delay} is the time from the relay WiFi MAC to the \ac{D2D} receiver WiFi MAC. The {\it end-to-end delay} is the sum of all these delays. Note that the cross-platform delay is affected by our implementation using multiple FPGA card for WiFi and LTE interfaces. 
%We analyze three kinds of delay: $(i)$ the {\it LTE delay} from the time at which the packet is sent to PHY in the eNB until it arrives at the destination UE, $(ii)$ the {\it cross-platform delay} is the time to forward the LTE frames from the LTE interface to WiFi MAC, and $(iii)$ {\it WiFi delay}s.

{\bf CPU load.} Since the Real-Time controller executes \ac{D2D} related operations, we can calculate the extra CPU load due to \ac{D2D} operations by monitoring the Real-Time module. %\blue{why not? will this then weaken the overhead measurement?}

{\bf \ac{D2D} lifetime.} We examine our proposed design with slow and fast channel variations. In each case, we measure the time during which a UE acts as relay, which we call {\it relay lifetime}. This is an important factor in opportunistic \ac{D2D} because frequent  role switching imposes extra load to the system. 

{\bf \ac{SSIM}}  This is an index of similarity between two images and it is known to be a better estimation of  human eye perception in comparison to other traditional methods such as peak \ac{SNR} or mean squared error. We use this metric for \ac{QoE} measurements in video streaming experiments.

\subsection{Non-opportunistic Outband \ac{D2D} Relay}
We start with the simplest form of outband \ac{D2D}-relay scenario with two UEs. 
Despite the simplicity of this experiment, it provides answers regarding the delay overhead due to multi-hop communication and achievable throughput gain.
%, and the computational overhead at the relay node. 
%In {\it Legacy} scheme, both UE1 and UE2 communicate directly with the eNB. In outband \ac{D2D}, UE1 acts as a relay and UE2 is the receiver, which we call relay and \ac{D2D} receiver, respectively.
%Figure~\ref{fig:simple} illustrates the delay in  and CPU load of each UE with and without \ac{D2D}-relay scheme enabled.

\begin{figure}[!t]
	\centering
		\subfigure[Delay at different parts of the path.]
		{
			\includegraphics[width=0.45\columnwidth]{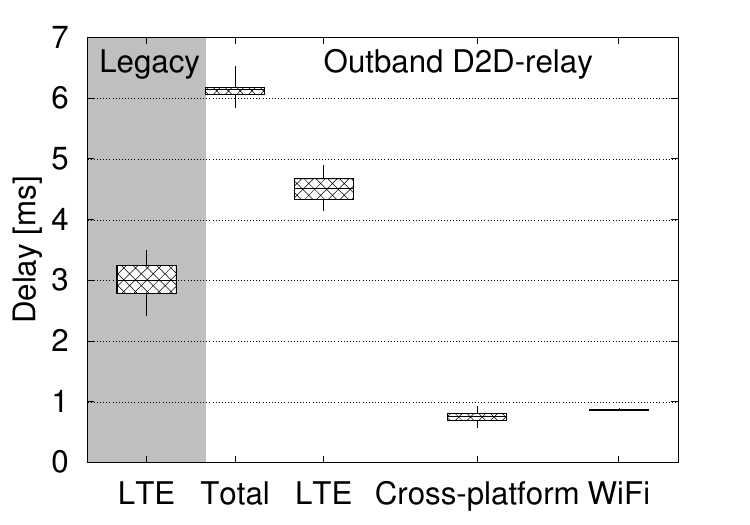} %[width=70mm,angle=0]
			\label{fig:simpDelay}
		}
		\subfigure[Per-UE throughput.]
		{
			\includegraphics[width=0.45\columnwidth]{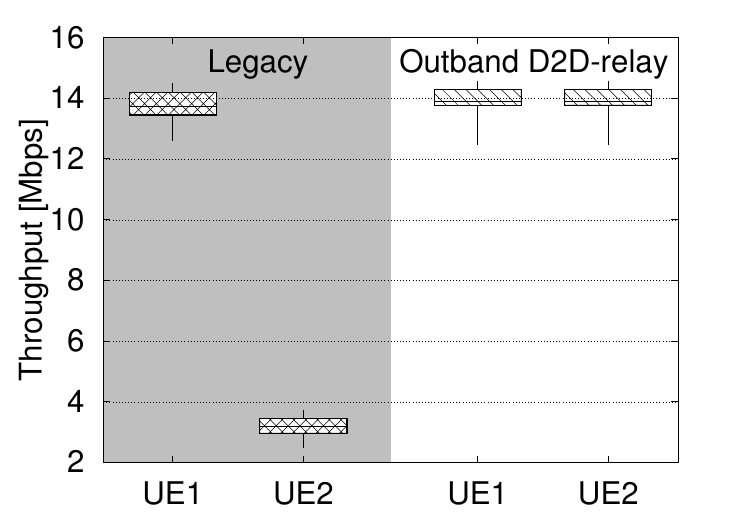} %[width=70mm,angle=0]
			\label{fig:simpTput}
			%\vspace{-5mm}
		}
		\caption{Outband UE-Relay: UE1 relays the traffic from the eNB to UE2.}
		\label{fig:simple}

\end{figure}

Figure~\ref{fig:simpDelay} compares a {\it Legacy} scheme (in which both UEs receive traffic only from the eNB) with an {\it Outband \ac{D2D}-relay}, in which UE1 acts as relay for UE2. We observe that outband \ac{D2D} increases the average end-to-end delay (i.e., Total in the figure) by $3.3$~ms as compared to the Legacy cellular system. 
%The total end-to-end delay for outband \ac{D2D} comes from the LTE transmission delay from eNB to UE1 (i.e., LTE delay), the delay from UE1's LTE MAC to its WiFi MAC (i.e., cross-platform delay), and the WiFi transmission delay from UE1 to UE2 (i.e., WiFi Delay). 
Looking at different delay components of outband \ac{D2D}-relay, we can see that cross-platform delay and WiFi delay are the major contributors to the delay overhead. It is important to note that extra frame processing results in higher LTE delays in outband relay mode. While commonly ignored in the literature, this illustrates that relaying large volumes of traffic comes at a cost. According to the observation from the delay profile, outband relay could be potentially suitable for a large variety of non-mission critical applications. Indeed, outband relay with a total delay of $6.3$~ms meets the 3GPP suggested delay budget of $70$~ms~\cite{3GPP23.203}. The motive for opportunistic \ac{D2D}-relay is vividly depicted in Figure~\ref{fig:simpTput}. The figure shows that UE2 suffers from low channel quality while UE1 experiences a good channel condition. After outband \ac{D2D} activation, UE2's throughput increases significantly because it receives its traffic through a high channel quality relay. 

%\red{Our experiment can be considered optimistic since our UEs, unlike commercial smartphones, do not run multiple applications that compete for processing power and bandwidth. Nevertheless, outband \ac{D2D} is still a valid choice for many applications even if its delay figures increase to $70$ ms, which would be ten times higher than our measured delay.}

%We measured the CPU load of each device with and without outband \ac{D2D}. Our observations show that UE1 (i.e., relay) and UE2 (\ac{D2D} receiver) are subject to $6.3$\% and $4.2$\% CPU load overhead because of outband \ac{D2D} operations. The overhead is negligible at the eNB. Note that running the WiFi code in the idle mode on the Real-Time controller increases the total CPU load by about $5$\%. Hence, if we assume that the UEs WiFi interface is in the idle mode, the overhead due to outband \ac{D2D} is marginal. 

\subsection{DORE with Delay-tolerant Traffic}

%The implementation of opportunistic outband \ac{D2D} is as described in Section~\ref{s:proto}.
Now, we evaluate the performance of opportunistic outband \ac{D2D} using \ac{RR} and \ac{PF} scheduling algorithms. %that are most commonly used as a benchmark: Round Robin (\ac{RR}) and Proportional Fair (\ac{PF}) algorithms. 
We test \ac{DORE} with delay-tolerant traffic
% (i.e., $300$ ms) 
to evaluate the potential throughput gain for such use-cases.
In the figures, we label the legacy schemes as \ac{RR} and \ac{PF}. When used for \ac{DORE} with delay-tolerant traffic, they are labeled as  RR-DT and PF-DT.  %The algorithms are evaluated for their aggregate throughput and CPU load.

\begin{figure}[!t]
	\centering
		\subfigure[Aggregate throughput.]
		{
			\includegraphics[width=0.45\columnwidth]{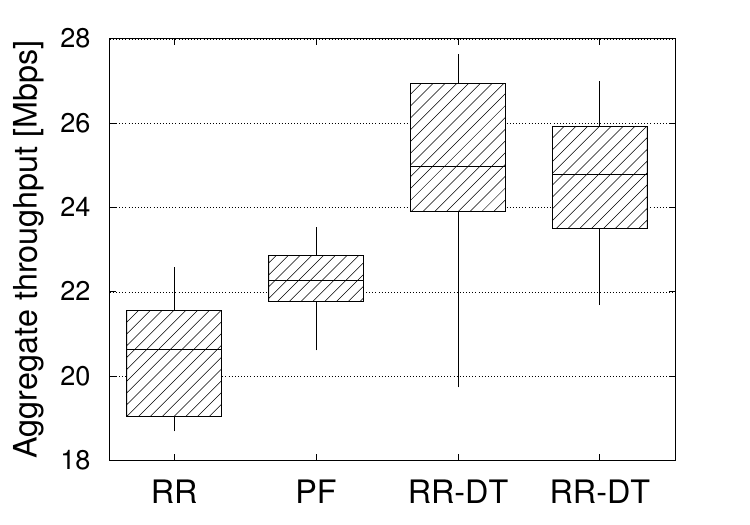} %[width=70mm,angle=0]
			\label{fig:2usrTput}
		}
		\subfigure[Delay]
		{
			\includegraphics[width=0.45\columnwidth]{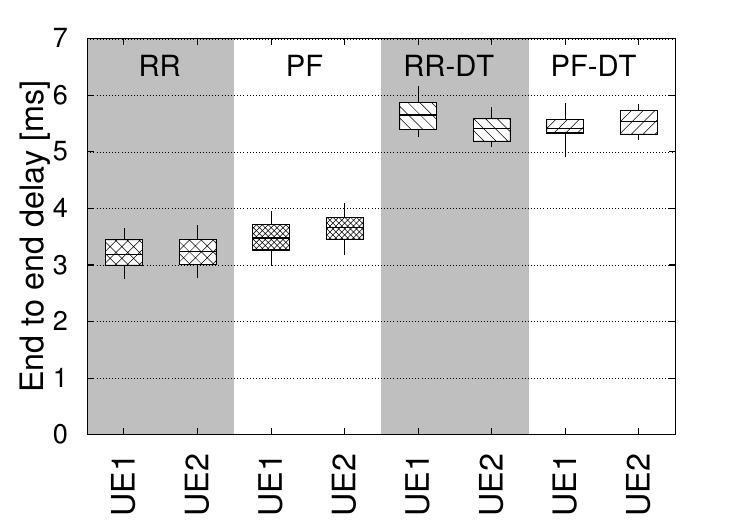} %[width=70mm,angle=0]
			\label{fig:2usrDelay}
		}
		\caption{DORE: the relay UE is chosen according to reported CQI values.}
		\label{fig:twouser-1}
\end{figure}

Figure~\ref{fig:2usrTput} shows the achievable aggregate throughput of RR-DT and PF-DT is $21$\% and $11.2$\% higher than \ac{RR} and \ac{PF}, respectively. As mentioned, opportunistic outband \ac{D2D} leverages the channel diversity between the \ac{D2D} users. Since \ac{PF} harvests part of this opportunism due to its opportunistic nature, the resulting gain reduces by $9.8$\% in comparison to \ac{RR}. Nevertheless, the gain remains relevant for a two-user scenario where there are limited opportunities. We show later in this section that the opportunistic gain increases with the user population. Delay comparison in Figure~\ref{fig:2usrDelay} demonstrates \ac{DORE} causes higher delays. The additional delay stems from WiFi and cross-platform transmission and LTE frame processing. 

%Since UE1 has on average lower channel quality than UE2, it receives most of its traffic through the relay, hence it experience higher delay than UE2.
%Thus, opportunistic outband \ac{D2D} may not results in significant gain with another opportunistic scheduler on top \blue{not clear -- Why?}. 

%With opportunistic outband \ac{D2D}, both UEs act as a relay in a portion of the time, allowing load balancing/sharing between the them. Indeed, our observation shows that UE1 and UE2 have almost $27\%$ CPU occupancy with our proposal. Here, the UEs experience about $6$\% additional CPU load that is mainly due to WiFi operations. In an extreme case where one of the UEs has always the lowest channel quality, we will observe similar results as the previous scenario.

%However, we observed a more balanced load between the UEs because
%Figure~\ref{fig:2usrLoad} depicts the impact of opportunistic outband \ac{D2D} on CPU loads in the eNB and the UEs. The impact of opportunistic outband \ac{D2D} is negligible on eNB. On the other hand, the UEs  experience about $6$\% additional CPU load that is mainly due to WiFi operations. Unlike our observation in the previous scenario (Figure~\ref{fig:simpLoad}), UE1 and UE2 have similar CPU load. In opportunistic outband \ac{D2D}, the relay changes dynamically based on the reported \ac{CQI}. Thus, both UEs act as a relay in a portion of the time, allowing load balancing/sharing between the UEs. In an extreme case where one of the UEs has always the lowest channel quality, we will observe similar results as shown in Figure~\ref{fig:simpLoad}.
 
\subsection{Impact of Fading Speed}

\begin{figure*}[!t]
	\centering
	%\def\subfigcapskip{-3pt}
%		\hspace{-6mm}
		\subfigure[Impact of refractor speeds on \ac{D2D} lifetime.]
		{
			\includegraphics[width=0.45\columnwidth]{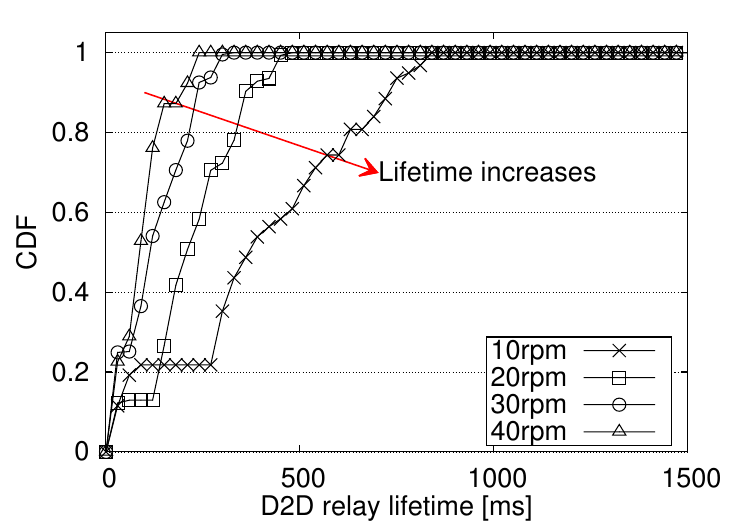} %[width=70mm,angle=0]
			\label{fig:fadingLifecycle}
		}
%		\hspace{-6mm}
		\subfigure[Impact of  threshold on D2D lifetime.]
		{
			\includegraphics[width=0.45\columnwidth]{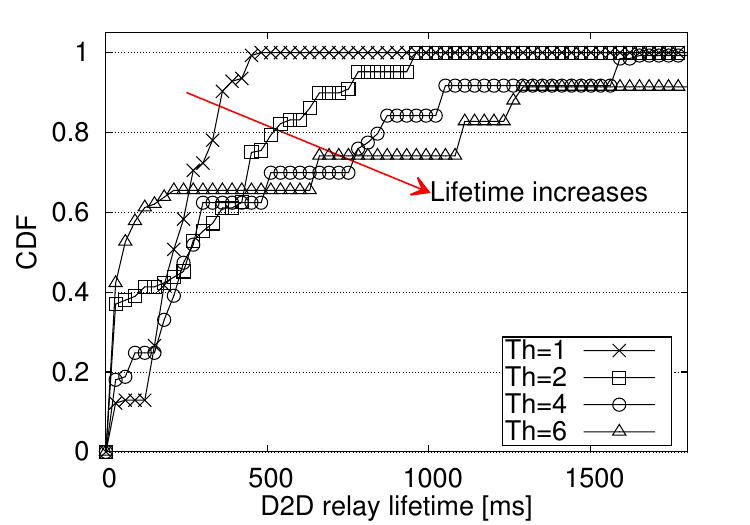} %[width=70mm,angle=0]
			\label{fig:fadingThresh}
		}
%		\hspace{-6mm}
		\subfigure[Impact of switching thresholds.]
		{
			\includegraphics[width=0.45\columnwidth]{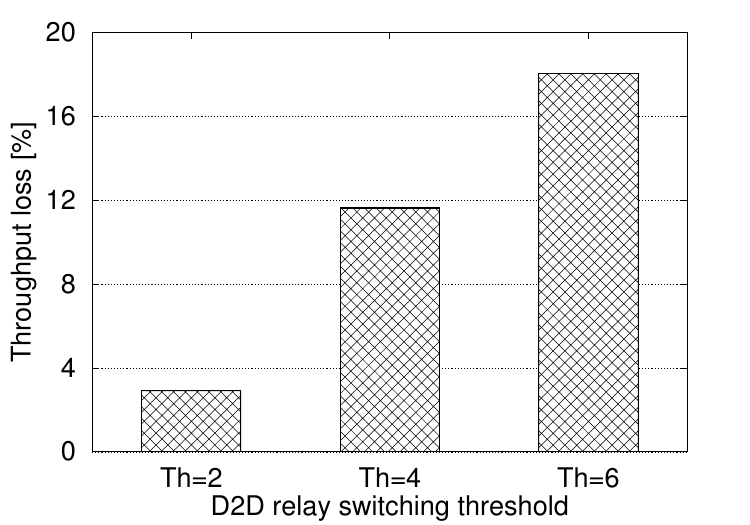} %[width=70mm,angle=0]
			\label{fig:fadingTput}

		}
		\caption{Impact of fading speed on the lifetime of D2D UEs.}
		\label{fig:fading}
\end{figure*}

\begin{figure*}[!t]
	\centering
	%\def\subfigcapskip{-3pt}
%		\hspace{-7mm}
		\subfigure[Per-UE throughput.]
		{
			\includegraphics[width=0.45\columnwidth]{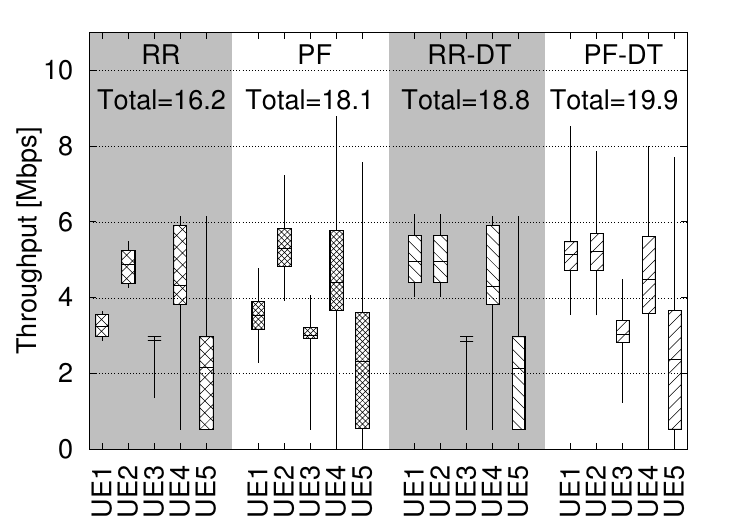} %[width=70mm,angle=0]
			\label{fig:shadowTput}
		}
%		\hspace{-7mm}
		\subfigure[End to end delay for D2D UEs.]
		{
			\includegraphics[width=0.45\columnwidth]{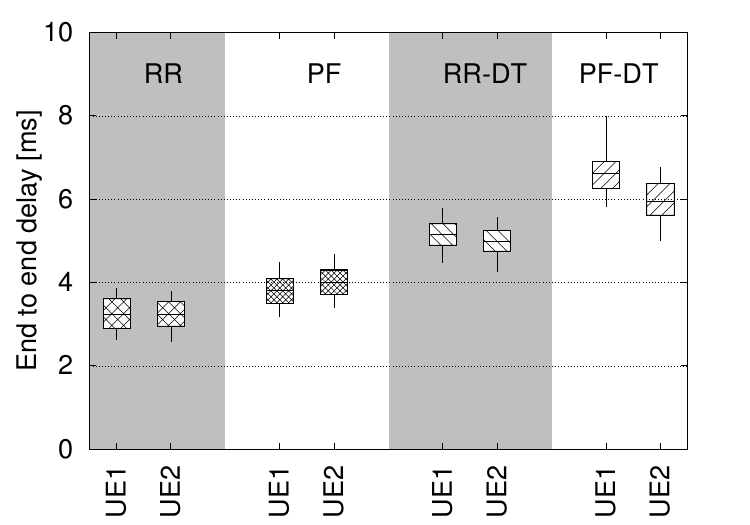} %[width=70mm,angle=0]
			\label{fig:shadowDelay}
		}
		\subfigure[CPU load.]
		{
			\includegraphics[width=0.45\columnwidth]{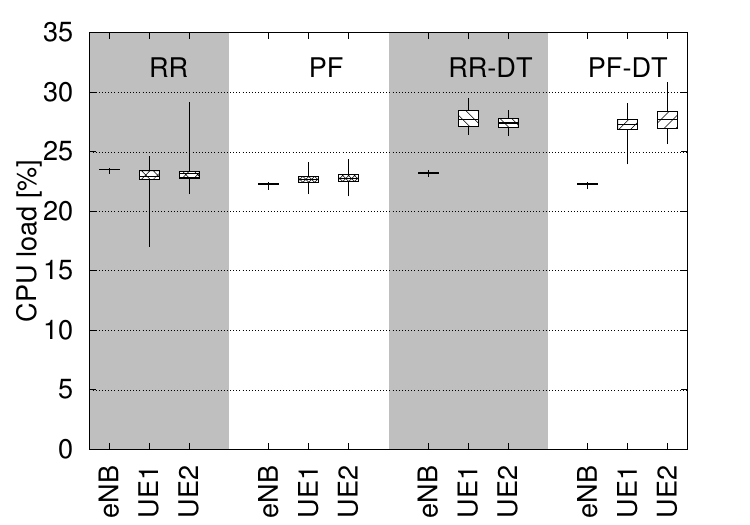} %[width=70mm,angle=0]
			\label{fig:shadowCPU}
		}
		\caption{System KPIs in an experiment with two D2D UEs and three shadow users.}
		\label{fig:shadow-1}
\end{figure*}

This experiment is designed to show the dynamics of \ac{DORE} under different fading scenarios. 
In particular, the change of role in the \ac{D2D} connection (i.e., a UE can be a relay or a \ac{D2D} receiver). We refer to the period in which a \ac{D2D} UE acts as relay as the {\it lifetime}. In this experiment, we shed light on the frequency of these changes and their impact on the system.

Figure~\ref{fig:fadingLifecycle} shows the CDF of the lifetime of UEs when the refractor surface spins at $10$, $20$, $30$, and $40$~rpm. At these rotation speeds, the \ac{MCS} of a UE remains the same for $18.86$~ms,  $15.38$~ms, $13.51$~ms, and $10.82$~ms, on average. We can see that the duration of the lifetimes increases as the fading speed reduces. The results also show that regardless of fading speed, the  lifetime is  shorter than $250$ ms more than $50$\% of the time. This emphasizes on the fact that {\it any implementation of opportunistic outband \ac{D2D} must be capable of handling the relay dynamics on a millisecond timescale}.  

In our implementation of \ac{DORE}, a switch of \ac{D2D} roles occurs as soon as the achievable \ac{MCS} of the \ac{D2D} receiver becomes higher than the one of the relay UE. In other words, the \ac{MCS} difference threshold to switch roles is one \ac{MCS} index. Nevertheless, considering the resulting short lifetimes depicted in Figure~\ref{fig:fadingLifecycle}, we have decided to introduce and test hysteresis in the switching to reduce frequent switching.  Introducing higher switching threshold can avoid role changes due to small \ac{MCS} variations that do not vary much in terms of bit efficiency. Thus, we increase the \ac{MCS} difference that triggers the role switching. 
%This is \blue {natural!!} solution because the nearby \ac{MCS} indices do not vary much in terms of bit efficiency. 
Figure~\ref{fig:fadingThresh} shows that larger thresholds (Th in the figure) increase lifetimes, as expected. However, this increment comes at the cost of reduced throughput. Indeed, Figure~\ref{fig:fadingTput} illustrates that the throughput reduces up to $18$\% when the switching threshold is $6$ \ac{MCS} levels.  Our results indicate that small switching thresholds increase \ac{D2D} lifetime with limited throughput penalty. Therefore, it is not strictly necessary to reconfigure \ac{D2D} links upon any \ac{MCS} change, which reduces the complexity of the implementation.

\subsection{DORE in Presence of Shadows and Delay-tolerant Traffic}

Here, we emulate the presence of additional legacy UEs using the shadow UEs introduced in Section~\ref{ss:shadow}. The shadows do not collaborate in \ac{DORE} but they help us to test \ac{DORE} in presence of non-collaborative UEs. 
%are  added to our previous setup with two UEs (see Figure~\ref{fig:testbed}). 
The shadows send real-time \ac{CQI} reports to the eNB, and the eNB schedules traffic for them, although they cannot decode such traffic.

Per-UE throughput results are presented in Figure~\ref{fig:shadowTput}. We can see that UE1 achieves a $53.2$\% throughput gain with \ac{DORE} (i.e., RR-DT and PF-DT) while UE2 only achieves a mere $1.4$\%  throughput gain.  UE2 achieves lower gain due to its higher average channel quality. 
%Hence, the main contribution to the gain area (see Figure~\ref{fig:2usrGain}) came from UE2. 
We also reported the aggregate throughput of each scheme in Figure~\ref{fig:shadowTput}, marked as {\it Total}. \ac{DORE} results in $10.2$\% and $9$\% throughput gain compared to \ac{RR} and \ac{PF}. The throughput gains are lower than those achieved in the previous scenario ($\sim 20$\%). This is because in a scenario with $5$ UEs, the relay UE receives only a fraction of the total available bandwidth (i.e.,  $2/5$ of the resources can be relayed if RR-DT is used). As a result, the opportunistic scheme can only optimize that portion of the cellular resources. %This highlights the fact that {\it significant network-wide throughput gain can be achieved in presence of more \ac{D2D}-enabled UEs}. 

Figure~\ref{fig:shadowDelay} depicts the end-to-end delay. The delay behavior of the UEs is very similar to the delay behaviors observed in Figure~\ref{fig:simpDelay}. Both UEs experience additional delay under RR-DT and PF-DT w.r.t. \ac{RR} and \ac{PF} because of the aforementioned cross-platform and WiFi delays.  UE1 has a higher delay than UE2 because it has lower channel quality than UE2 and it acts as the \ac{D2D} receiver most of the time. However, the increased delay is quite limited and can be substantially reduced by using dedicated and integrated hardware rather than different FPGA cards for LTE and WiFi interfaces.

Figure~\ref{fig:shadowCPU} compares the CPU load of the eNB and the UEs. The overhead on the eNB is negligible.
% because \ac{DORE} is computed for two UEs. \red{@VM: how to justfiy what happens with more?}. 
The two \ac{D2D}-enabled UEs experience $4.42\%$ and $4.45\%$ higher CPU load due to outband \ac{D2D} operations in WiFi and LTE interfaces. Note that running the WiFi code in the idle mode on the Real-Time controller increases the total CPU load by about $3$\%. Hence, 
%if we assume that the UEs WiFi interface is in idle mode, 
the overhead due to outband \ac{D2D} is marginal. 
%We also show the achieved fairness under different scheduling schemes in Figure~\ref{fig:shadowFair}. The fairness drops under RR-\ac{D2D} and PF-\ac{D2D}. The fairness reduction is due to higher variance among throughput of different UEs. However, this should not be interpreted negatively. In legacy cellular systems, fairness reduction is an indication of the fact that a UE received more resource than another. This does not apply to opportunistic outband \ac{D2D} because the \ac{D2D} UEs receive the same amount of RB under RR and RR-\ac{D2D} or PF and PF-\ac{D2D}. Thus, the throughput increment only comes from better utilization of the existing resources and not receiving more resources. 

\subsection{DORE in Presence of Shadows and Delay-sensitive Traffic}
\label{ss:evalDore}
In this experiment, UE1 and UE2 host a real-time gaming application and a VOIP call with $30$ ms and $80$ ms \ac{OTA} delay threshold, respectively. To highlight the impact of  \ac{DORE}'s \ac{QoS}-awareness, we also show the performance figures when the delay thresholds are set to infinity (i.e., \ac{DORE} ignores the delay constraints). In this scenario, we stressed the WiFi channel (i.e., \ac{D2D} link) by introducing extra non-\ac{D2D} traffic to the network so that the WiFi channel operates near to the congestion point. Therefore, small changes in the instantaneous channel quality provoke non-negligible size queues. 

\begin{figure}[!t]
	\centering
		\subfigure[Aggregate throughput.]
		{
			\includegraphics[width=0.45\columnwidth]{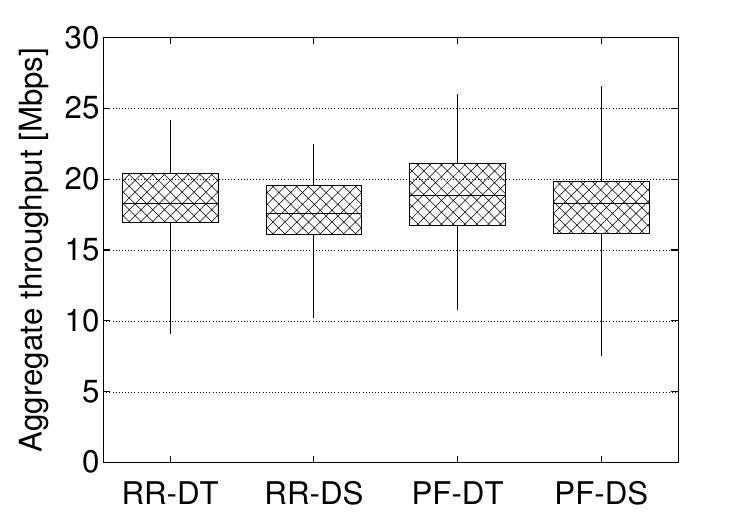} %[width=70mm,angle=0]
			\label{fig:QoSAggTput}
		}		
		\subfigure[End to end delay.]
		{
			\includegraphics[width=0.45\columnwidth]{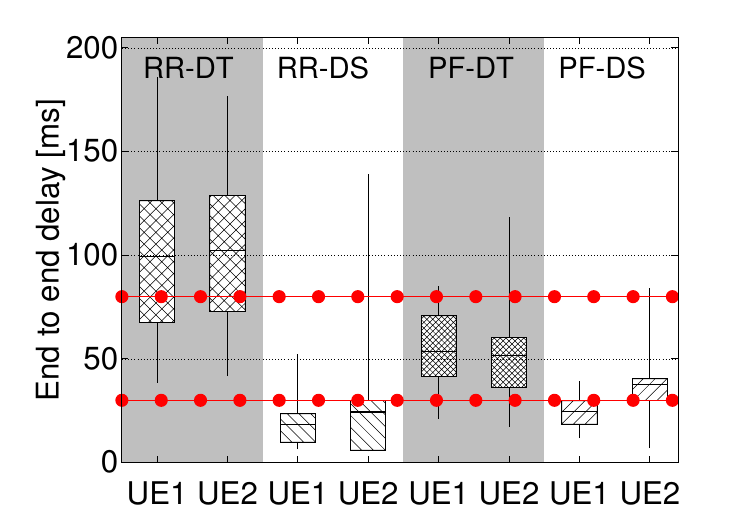} %[width=70mm,angle=0]
			\label{fig:QoSDelay}
			%\vspace{-5mm}
		}		
		\caption{Impact of QoS-awareness of DORE on system performance.}
		\label{fig:QoS}
\end{figure}

Figure~\ref{fig:QoSAggTput} shows the aggregate throughput of \ac{DORE} with \ac{RR} and \ac{PF} but without \ac{QoS} constraints (RR-DT and PF-DT in the figure) and with tight constraints (RR-DS and PF-DS). Both RR-DS and PF-DS achieve slightly lower throughput ($3\%$) w.r.t RR-DT and PF-DT because the \ac{QoS}-awareness of \ac{DORE} prevents opportunistic relay when delay constraints are violated. However, the $3\%$ throughput loss is a small price to pay to maintain the \ac{QoS} requirements of the time-sensitive applications. Indeed, we observe in Figure~\ref{fig:QoSDelay} that \ac{DORE} can successfully cap the average delay below 30 ms and 80 ms. The effectiveness of \ac{DORE} is especially seen when it reduces the packet delay of the voice traffic from $100$ ms to $23$ ms and $30$ ms. Since \ac{DORE} delay control mechanism relies on UE feedbacks, it cannot avoid the delay caused by dramatic channel variations. As a result, the maximum delay under RR-DS and PF-DS can be higher than the delay thresholds.

\subsection{QoE  with DORE}
\begin{figure}[!h]
	\centering
%	%\def\subfigcapskip{-6pt}
		\subfigure[\ac{SSIM} between streamed videos. ]
		{
			\includegraphics[width=0.44\columnwidth]{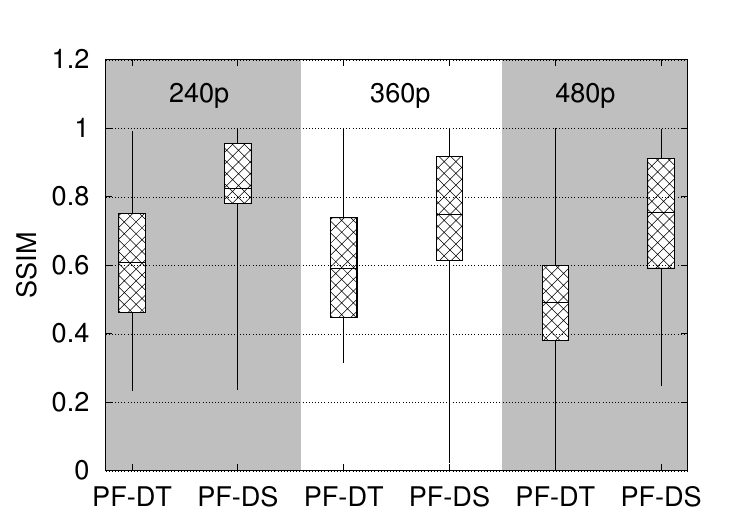} %[width=70mm,angle=0]
			\label{fig:ssim}
		}
%		\hspace{-8mm}
		\subfigure[Snapshots of streamed video at the receiver.]
		{
			\includegraphics[width=0.52\columnwidth]{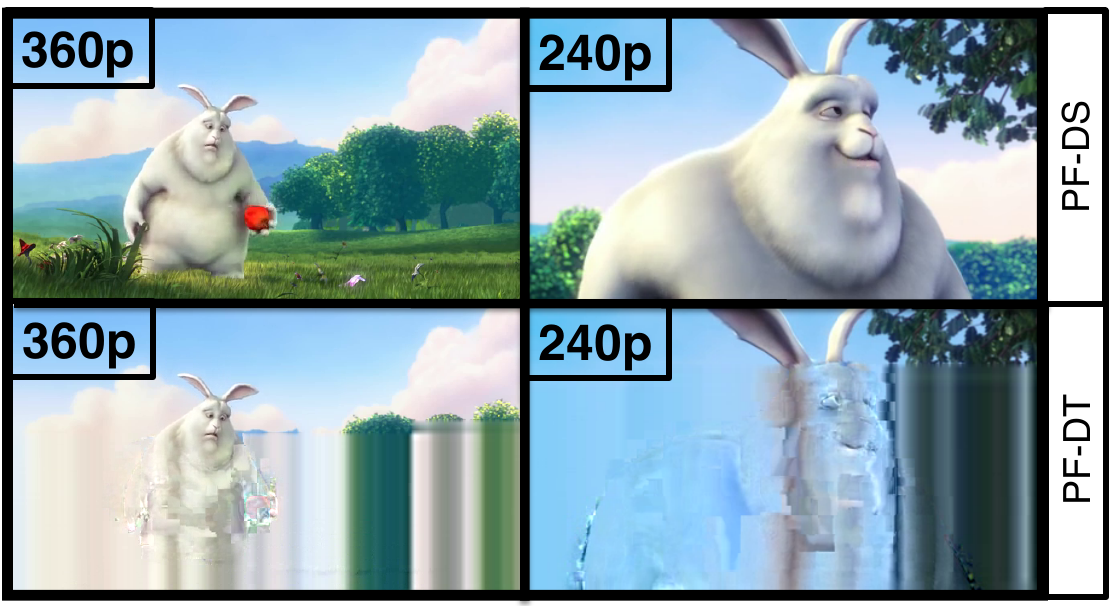} %[width=70mm,angle=0]
			\label{fig:snapshot}
		}		
%		\hspace{-8mm}
%		\vspace{-2mm}
		\caption{QoE performance of DORE.}
		\label{fig:video}
\end{figure}

Good \ac{QoS} does not necessarily corresponds to good \ac{QoE}. Thus, we design a video streaming scenario using VLC\footnote{\url{http://www.videolan.org/vlc}}to measure the \ac{QoE} in terms of \ac{SSIM}. We use {\it AviSynth} to measure \ac{SSIM}. Both \ac{PF} and \ac{RR} demonstrated similar trend hence we only show the result for \ac{PF}, for brevity.  Again, we show in Figure~\ref{fig:ssim} the \ac{SSIM} of the received video with $30$ ms delay constraint (i.e., PF-DS) and with an infinite one (i.e., PF-DT). We repeat the experiment for three different videos with 240p, 360p, and 480p resolutions. The results indicate that the \ac{QoS} awareness of \ac{DORE} results in up to $26\%$ \ac{SSIM} improvement. The \ac{SSIM} values degrade with higher resolution videos because they are more sensitive to channel impairments. We also demonstrate a snapshot of the received video for 240p and 360p resolutions, in Figure~\ref{fig:snapshot}. As expected, tight \ac{QoS} constraints result in better image quality. 

 %the changes $53$, $65$, $74$, and $92$ times per second.  The speed change changes the frequency of channel variation. 

\subsection{Opportunistic Relay within Large Relay Groups}
\begin{figure} [t!]
\centering
		\includegraphics[width=0.45\columnwidth]{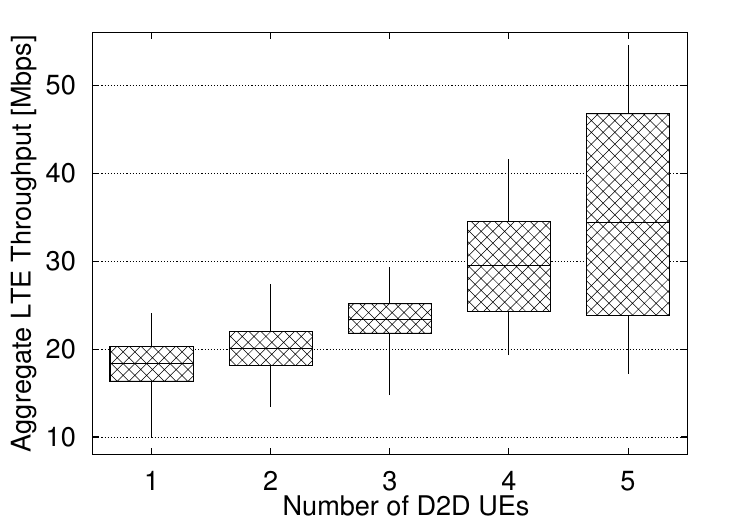} %[width=70mm,angle=0] [scale=0.80, angle=0]
		\caption{Aggregate throughput versus the number of UEs in the same opportunistic outband \ac{D2D} group.}
		\label{fig:cluster}
		
\end{figure}

In the previous experiments, only two UEs were allowed to collaborate in \ac{DORE}. Our observation in Figure~\ref{fig:shadowTput} showed that the impact of opportunistic outband relay with only two users is limited.
Since one-to-many communication is also present in 3GPP \ac{ProSe} services, we can increase the size of the outband \ac{D2D} group in order to achieve higher throughput. This experiment is designed to illustrate the impact of larger \ac{D2D} UEs groups. Here, all UEs report their \ac{CQI}s to the eNB and we only measure the throughput at the LTE because the shadows are commercial smartphone, which are unable to decode the messages of our experimental eNB.
%\red{the setup is similar to \ref{ss:evalDore}, but all also shadows are allowed to join the \ac{D2D} groups and can act as relays. Thus, the throughput of shadows is measure at the eNB because they cannot the decode the communication from the eNB.}
Figure~\ref{fig:cluster} shows the aggregate system throughput. Our results confirm that by enlarging the outband \ac{D2D} group from $2$ to $5$ UEs, the network throughput increases up to $71.8$\%. The result is critical to confirm the potentials of opportunistic \ac{D2D} schemes. Indeed, we are the first to assess the opportunistic gain with multiple UEs relaying traffic among each other with a real implementation of an eNB scheduler and real-time \ac{CQI} acquisition from multiple UEs. The reported results are obtained under \ac{PF} scheduling. The achievable gains are even higher with \ac{RR}, as shown in prior subsections. 
% \blue{SHALL WE MOVE TO DISCUSSIONS---->In addition, this result is very promising and might suggest to increase relay groups as much as possible. However, in virtue of the observations we have made on the extra load due to relay operations, it is plausible to suggest that each relay group should not include more than a handful of users, which is enough to provide a factor $1.5$ or $2$ to network capacity.  }

\section{Discussion}
\label{s:disc} 
This chapter provides in-depth intuitions to understand the practicality of integrating outband \ac{D2D} communications in cellular networks. 
Here, we discuss the feasibility of such integration in order to enlighten some key requirements for developing an experimental setup as well as for designing possible use-cases.

%\red{ the feasiblity in terms of delay\\ feasibility in terms of achitecture \\ dynamic protocol to support fast \ac{D2D} sw \\ larger groups \\ ease of implementation}
{\bf Feasibility.}
%As demonstrated in Section~\ref{s:eval}, outband \ac{D2D} is feasible and it operates within reasonable delay bounds while imposing minimal complexity. 
The \ac{SDR}-based implementation of \ac{DORE} is the proof-of-concept for the feasibility of outband \ac{D2D} schemes with more complex and dynamic scenarios than non-opportunistic and \ac{QoS}-unaware UE to UE communications. 
%\red{can be leveraged for more complex and dynamic scenarios. \red{NOT SURE WHAT ELSE TO SAY NOW}}

{\bf Implementation.}
There are several challenging issues to solve for \ac{SDR} implementation of a \ac{D2D} system. Here, we point out the most critical ones. The relay UE experiences high computational overhead due to LTE frame processing. Hence, we use label switching at LTE \ac{PDCP} layer instead of IP routing which is the current solution in 3GPP. During the course of \ac{DORE} implementation, we realized that \ac{D2D} UEs switch role with high frequency (in order of milliseconds). Thus, we place \ac{DORE} at the eNB instead of \ac{ProSe} Function/Server to meet timing constraints and to avoid the additional overhead on the backhaul links.

%\red{
%First, the relay UE is subjected to extra computational load due to frame processing at LTE interface. This is commonly ignored in the analytical/simulation studies. This fact inspired us to propose label switching instead of IP relaying in outband \ac{D2D}, so that the load on the relay is minimized.
%Second, relaying always raises security concerns. This was another reason to opt for relaying LTE \ac{PDCP} SDUs, which are enciphered. 
%Third, our experiments indicated that \ac{D2D} UEs can switch role with high frequency (on a millisecond timescale). Therefore, any practical implementation must account for high relay dynamics in the architecture. This is the reason for proposing the eNB to take over opportunistic \ac{D2D} scheduling operation after the activation phase. In fact, \ac{ProSe} Function or \ac{ProSe} Server are not suitable for this purpose due the timing constraints and the additional overhead on the backhaul links. 
%}

{\bf Choice of platform.}
To date, there are a few \ac{SDR} platforms with {\it `simultaneous'} LTE and WiFi capability, namely, Open Air Interface, and LabVIEW. We choose LabVIEW for its modular and graphical programming structure that allows for quick real-time and \ac{FPGA} code development without stepping into complex low-level programming languages. The choice of the NI PXI-based platform over USRP is due to the real-time capability of the PXI system that speeds up MAC layer algorithm prototyping and testing. 

{\bf Capacity.}
\ac{DORE} is key for boosting network capacity in one-to-many relay scenarios. This result is very promising and it may suggest to increase the size of the relay groups as much as possible. However, in virtue of our observations on the extra load due to relay operations, it is plausible to suggest that each relay group should not include more than a handful of users, which is enough to enhance the network capacity by $70\%$. 
%factor $1.5$ or $2$ to the network capacity. 

{\bf \ac{QoS}.}
\ac{QoS} provisioning is a concern in outband \ac{D2D} due to the use of unlicensed spectrum. As a result, we designed \ac{DORE} and the surrounding protocol with necessary feedback and handlers to enable \ac{QoS} monitoring in our testbed. The experiments confirmed that \ac{DORE} achieves the \ac{QoS} requirements using a simple monitoring and feedback scheme.

{\bf Use-cases.}
Our experimental evaluation showed that (opportunistic) outband \ac{D2D} schemes have low latency and ameliorate the throughput substantially. Hence, these schemes suit a large variety of applications including voice calls, video streaming, real-time gaming, and content sharing.

\section{Summary}

We prototyped the first \ac{SDR} platform for outband \ac{D2D} communications. We leveraged Xilinx FPGAs and the NI Real-Time OS to develop realistic experiments with LTE-like millisecond \ac{CQI} reporting, scheduling, and high-speed LTE- WiFi interaction. Our experimental evaluation using several \ac{QoS} and \ac{QoE} metrics confirmed the feasibility and potentials of opportunistic outband D2D communications. 
%In particular, we designed \ac{DORE} which is a 3GPP ProSe-compliant and QoS-aware opportunistic outband D2D framework. 
The results revealed that experimental performance figures are lower than the reported values in the prior analytical studies, although still notable (up to 20\% with just two users). Nevertheless, high throughput gains are achievable if the number of participating UEs in opportunistic outband D2D increases (up to 71\% with five users). Finally, we are in the process of providing public access to our D2D SDR implementation so that the research community can benefit from deeper study of such a system.

%\acresetall
% Part 3: Understanding and Reducing energy consumption in Data centers
\part{\titSecondPart}
\chapter{\titSixthCh}
\label{ch:mode}

\section{Introduction}

\begin{figure} [!h]
\centering
\includegraphics[width=.7\columnwidth]{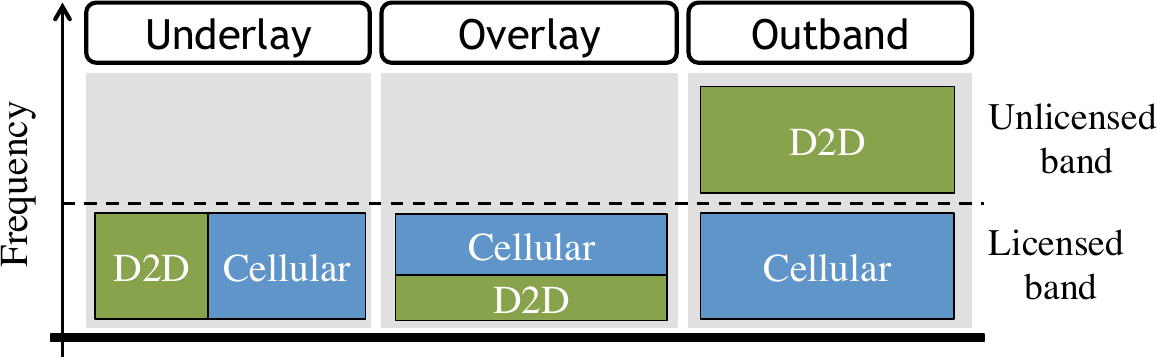}
\caption{Schematic representation of overlay inband, underlay inband, and outband \ac{D2D} for cellular scenarios.}
\label{fig:in-outband2}
\end{figure}

There have been extensive research efforts in both academia and industry to explore \ac{D2D} techniques~\cite{asadi2014survey}. \ac{D2D} communications have been
considered for a large variety of use-cases such as 
%multicasting~\cite{zhou_intracluster_2013}, 
cellular offloading~\cite{Zhang2014TWC}, 
mobile relaying~\cite{asadi2013MSWIM}, 
%load balancing~\cite{liu2014WC}, 
and video streaming~\cite{Kim2013Mobicom}. 
These studies indicate the potential outstanding
gain of \ac{D2D} communications in cellular networks. Indeed, the high performance gain motivated  leading telecommunication companies such as Qualcomm to perform
experimental studies on this paradigm using early stage prototypes~\cite{wu2013TON}. Standardization bodies such as 3GPP have also joined this front by
considering \ac{D2D} communications as a public safety feature in the next release of LTE-A~\cite{lin2014ComMag}. These efforts from academia, industry, and
standardization bodies confirm that the society regards \ac{D2D} communications as a crucial feature for next generation networks. Nevertheless, there is still no
concrete agreement on \ac{D2D} operational details such as which medium access control to adopt, or which spectrum allocation schemes, connection setup, and
resource management protocols are to be implemented. As described in Chapter~\ref{ch:background}, initial proposals for \ac{D2D} communications aimed at re-using the same resources that are used for conventional
cellular communications (i.e., {\it inband underlay} \ac{D2D} mode)~\cite{doppler2009COMMAG}. The significance of the \ac{D2D} gain had led to proposals in which a part
of the cellular resources is dedicated only to \ac{D2D} communications (i.e., {\it inband overlay} \ac{D2D} mode).  Finally, the scarcity and the high price of cellular
spectrum motivated some researchers to explore \ac{D2D} communications over the unlicensed band (i.e., {\it outband} \ac{D2D} mode). The use of resources with these D2D modes is illustrated schematically in Figure~\ref{fig:in-outband2}.
%\cyan{These \ac{D2D} modes exhibit advantages and disadvantages which are inherited from the choice of spectrum and medium access method.}

%\begin{figure} [!t]
%\centering
%\includegraphics[scale=0.45]{./figs/in-outband}
%%\vspace{-3mm}
%\caption{Schematic representation of overlay inband, underlay inband, and outband \ac{D2D} for cellular scenarios.}
%\label{fig:in-outband}
%%\vspace{-5mm}
%\end{figure}

The majority of the existing studies on \ac{D2D} communications select one of the aforementioned modes,
then propose a method for resource allocation/interference management in order to handle the resulting complications, and finally illustrate the achievable performance improvement~\cite{zulhasnine2010WiMob, asadi2013MSWIM, doppler2009COMMAG, zhou2013TVT, Zhang2014TWC}.
However, single mode \ac{D2D} significantly limits the system performance to the interference profile of the network.
Existing multi-mode \ac{D2D} systems only focus on inband \ac{D2D} modes, i.e., fully dependent on cellular spectrum.
Other proposals focus on joint scheduling and mode selection~\cite{Doppler2010WCNC,Phun2013WC}, although they are extremely complex (more complex than scheduling, which is already proven to be NP-hard for cellular systems such as LTE~\cite{lee2009Infocom}) and introduce unnecessarily frequent mode selection decisions.

Interestingly, while some researchers limit \ac{D2D} communications to cellular spectrum, the standards have a more liberal view of \ac{D2D}. In fact, 3GPP defines \ac{D2D} as ``the communication between two users in proximity using a direct link between the devices without traversing the eNB(s)
or the core network''~\cite{lin2014ComMag}.
We also remark that network-assisted outband \ac{D2D} is accounted for in 3GPP \ac{ProSe}~\cite{3GPP23.303}.  Although both inband and outband \ac{D2D} are considered valid options for ProSe services, there is no indication on how to select between the two. Hence, given the fast-track emergence of \ac{D2D} communications in cellular networks, the need for an adaptive \ac{D2D} mode selection scheme is beyond question.

In this chapter, we propose a flexible framework to adaptively select \ac{D2D} mode or operating band and technology. % ADDED BY PETER
In particular, we first discuss the practical
implications of each \ac{D2D} mode based on the latest standard releases of LTE-A and WiFi-Direct. This discussion clarifies that {\it there is no superior \ac{D2D}
mode} and the potential of each mode is highly scenario/use-case dependent.
After discussing practical implementation issues of \ac{D2D}-enabled networks, we provide analytical insights into the mode selection problem in an innovative multi-mode multi-band setup, which accounts for both achieved {\it throughput} and {\it energy costs}.
We call such a novel approach {\it Floating Band \ac{D2D}}, because \ac{D2D} transmissions can occur on either inband or outband modes.
The problem is formulated as a non-linear integer programming problem. Given the NP-hardness of the problem and time-stringent requirements of future cellular networks, e.g., 5G networks, we propose three practical heuristics with near-optimal performance and low complexity.
Finally, we evaluate the performance of the proposed heuristics in a multi-cell scenario using a realistic setup designed based on the ITU-R guidelines for
evaluating IMT-Advanced networks~\cite{ITU2009}. Our results confirm that the coexistence of \ac{D2D} modes ameliorates the performance of the system in terms of the key performance factors such as throughput and utility (up to one order of magnitude), and near complete fairness.

\section{Pros and Cons of D2D Modes}

As mentioned, 3GPP's definition does not restrain \ac{D2D} implementations to a specific technology or spectrum. To date, the available commercialized technologies that suit \ac{D2D} communications are either in the family of 3GPP standards such as  LTE-A  or in the family of IEEE standards such as WiFi. The former are suitable candidates for inband \ac{D2D} and the latter match the requirements of outband \ac{D2D}. Indeed, the feasibility of \ac{D2D} communications with the aforementioned technologies has recently been theoretically proven not only by us in~\cite{asadi2013WD} but also by other independent studies~\cite{raghothaman2013ICNC}.

%\subsection{Definitions}
%\label{ss:def}

We refer to users that communicate with the eNB as {\it cellular users} and to those who communicate with other neighboring users as {\it \ac{D2D} users}. The following describes the list of \ac{D2D} modes available are  \emph{Underlay inband},  \emph{Overlay inband}, and  \emph{Outband}. 
%for \ac{D2D} users~\cite{asadi2014survey}, as illustrated in Figure~\ref{fig:in-outband}:
%\begin{itemize}
%\item \emph{Underlay inband}: \ac{D2D} communications occur over the same licensed spectrum simultaneously used for legacy cellular communications.
%\item \emph{Overlay inband}: A fixed fraction of the licensed spectrum is reserved for \ac{D2D} users.
%\item \emph{Outband}: \ac{D2D} users exploit unlicensed spectrum to communicate with each other.
%\end{itemize}
Note that inband \ac{D2D} users are allowed {\it to share the same resource} (i.e., simultaneously transmit over the same frequency), while
outband \ac{D2D} users adopt a WiFi MAC and contend for channel access.
The differences among available \ac{D2D} modes
%(see~Figure~\ref{fig:in-outband})
pose advantages and disadvantages for each mode, as summarized in Table~\ref{tb:adv-disadv-ch6}.
 %provides a summary of pros and cons of different \ac{D2D} modes.
For completeness, we also include legacy cellular communication in the table.
%\blue{Furthermore, inband \ac{D2D} communication allows several \ac{D2D} pairs to transmit over the same frequency simultaneously.}
Interestingly, none of the available \ac{D2D} modes can simultaneously guarantee features like controlled interference, spectrum efficiency, and QoS. So, when it comes to electing a specific mode for implementing \ac{D2D} in a network, there is no clear winner.

\begin{table}[t!]
%\scriptsize
\centering
\caption{Pros and cons of each \ac{D2D} mode}
\label{tb:adv-disadv-ch6}
%\resizebox{0.8\columnwidth}{!}{
%\begin{tabular}{|@{}p{8cm}@{}|@{} c@{}c@{}c@{}c@{}|}
\begin{tabular}{| l | c c c c|}
\cline{2-5}
%\multicolumn{1}{c|}{}							&\multicolumn{3}{|c|}{Inband} 			&\multicolumn{1}{|c|}{Outband}	 		\\
\cline{2-5}
\multicolumn{1}{c|}{}								&~~Underlay				&~~Overlay				&~~WiFi~~				&Cellular~~	\\	
\hline
\hline
~Interference between \ac{D2D} and cellular users~~ 		&$\checkmark$				&$\times$					&$\times$					&$\times$	\\
\hline
~Interference among \ac{D2D} users		 			&$\checkmark$				&$\checkmark$				&$\times$					&$\times$	\\
\hline
~Needs dedicated resources for \ac{D2D} users 			&$\times$					&$\checkmark$				&$\times$					&$\times$	\\
\hline
~Controlled interference environment 				&$\checkmark$				&$\checkmark$				&$\times$					&$\checkmark$	\\
\hline
~Simultaneous \ac{D2D} and cellular transmission~~		&$\times$					&$\times$					&$\checkmark$				&$\times$		\\
\hline
~Increased spectral efficiency						&$\checkmark$				&$\checkmark$				&$\checkmark$				&$\times$		\\
\hline
~Requires additional wireless interface				&$\times$					&$\times$					&$\checkmark$				&$\times$		\\
\hline
~QoS guarantee								&$\checkmark$				&$\checkmark$				&$\times$					&$\checkmark$	\\
\hline
~Energy cost									&Eq.\eqref{eq:einband}		&Eq.\eqref{eq:einband}		&Eq.\eqref{eq:eoutband}		&Eq.\eqref{eq:eCell}	\\
\hline
\end{tabular}
%}
\end{table}

%\subsection{Which \ac{D2D} Mode is the Best?}
Looking at the pros and cons of the available \ac{D2D} modes, one can observe that none of the available \ac{D2D} modes is ideal. So the question remains: {\it Which \ac{D2D} mode is the best?}
Let us look at a few examples to better address this question. The use of underlay  in micro cell scenarios, where users are in short range, results in intolerable co-channel interference to cellular users. In such scenarios, overlay and outband modes better facilitate \ac{D2D} communications. On the other hand, using overlay in a macro cell with many cellular users can result in underutilization of network resources if the number of \ac{D2D} users is small.
Here, underlay and outband potentially perform better because of the sufficient distance among users.
%Here, the sufficient distance among users allows for underlay  and outband \ac{D2D} which potentially enhances the spectrum efficiency.
Finally, places with high occupancy of unlicensed band are not suitable for outband mode, due to well known congestion problems of contention-based MAC protocols.

One can observe that an eNB may face the above-mentioned scenarios on a daily basis, e.g. as different groups of users (workers/students, residents, shoppers, etc.) become dominant at particular times of the day.  % ADDED BY PETER
Thus, {\it an adaptive scenario-independent \ac{D2D}-enabled system cannot be tied to a specific mode or band}. Indeed, we propose a multi-band mode selection
scheme in order to facilitate such high level of adaptiveness in real implementations.

%\vspace{-4mm}
\section{System Model}
\label{s:system}
In this section, we describe our reference system, our proposed mode selection approach, and its practical implications. %along with their feasible solution.
\subsection{System}
\begin{figure}[t]
\centering
  \includegraphics[scale=0.35]{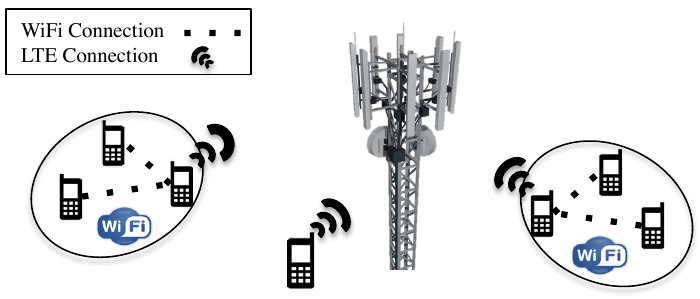}
  \caption{Our system model that consists of a cell with its first-tier neighbors.}
  \label{fig:network}
\end{figure}

We consider a hexagonal multi-cell LTE-A network with a reference cell in the center and its first-tier neighbors as shown in Figure~\ref{fig:network}. The cell
consists of $ N $ users labelled as $ n \in \setN := \{ 1, 2, \dots, N \}$. Downlink and uplink channels are separated and each one has a fixed bandwidth.
Users may communicate with other users in the cell or with those outside the cell. If a user wants to communicate with another user in her proximity, she can
use \ac{D2D} communications.
%\cyan{We call such a pair of users a \ac{D2D} pair.}
Inband \ac{D2D} communications use uplink cellular spectrum~\cite{lin2014ComMag}.  It is assumed that each user communicates with (at most) one user at any given
time. Each connection between users $ n $ and $ m $ is referred to as $ (n, m) $, $\forall n,m \in \setN $. For notational convenience, the eNB is addressed as
user $N+1$. Here, we assume that the outband \ac{D2D} exploits WiFi Direct technology.
%
%There exist four communication modes in the described system:
%\begin{itemize}
%\item   Mode $ 0 $: the communication mode between cellular users and the eNB (cellular);
%\item   Mode $ 1 $: the communication mode in which \ac{D2D} users reuses cellular resource (underlay);
%\item   Mode $ 2 $: the communication mode in which \ac{D2D} users receive dedicated resources (overlay);
%\item   Mode $ 3 $: the communication mode in which \ac{D2D} occurs on unlicensed spectrum, i.e., WiFi (outband).
%\end{itemize}
%
With the above, we use four communication modes operating:
\begin{itemize}
\item   Mode $ 0 $ $\leftrightarrow$ cellular;
\item   Mode $ 1 $ $\leftrightarrow$ inband underlay \ac{D2D};
\item   Mode $ 2 $ $\leftrightarrow$ inband overlay \ac{D2D};
\item   Mode $ 3 $ $\leftrightarrow$ outband \ac{D2D} (WiFi).
\end{itemize}

Our system operates in discrete time units and the eNB is in charge of mode selection and scheduling. The eNB makes the scheduling decisions on a per-frame
basis. Each \textit{frame} consists of $s$ {\it subframes}. In each subframe, only one cellular user is scheduled, while
the number of concurrent \ac{D2D} transmissions is not limited {\it a priori}. Therefore, there is no interference among cellular users (i.e., mode $ 0 $), but
underlay users (i.e., mode $ 1 $) interfere with the cellular and other underlay users (i.e., modes $ 0 $ and $ 1 $). Overlay users only interfere with each
other, while outband \ac{D2D} users simply contend for the WiFi channel. A fixed portion of cellular bandwidth is dedicated to overlay \ac{D2D} users. This portion is
released to cellular and underlay users if there is no user in overlay mode.

\subsection{Mode Selection and Scheduling}

As mentioned, mode selection and scheduling decisions by nature require decision making schemes with a different time-scale resolution. Thus, we propose to decouple the mode selection and scheduling problems. The decoupling is mainly inspired by the fact that \ac{D2D} connections last more than a few frames in a real world scenario and scheduling them on a per-frame basis is unnecessary and possibly inefficient. The inefficiency is due to the high signaling overhead, which is caused by such a high resolution mode selection (see~Subsection~\ref{ss:practicality}). Moreover, the channel quality of \ac{D2D} links is potentially less time variant in comparison to that of the cellular links due to the short-range nature of \ac{D2D} communications. The decoupling also simplifies the integration of \ac{D2D} communications into current cellular systems as it minimizes the changes to the scheduler.
Although mode selection and scheduling are decoupled, they are still highly intertwined. On one hand, the scheduling is affected by the interference, which is unknown before mode selection. On the other hand, mode selection depends on the set of cellular users scheduled along with \ac{D2D} users. Hence, we choose the eNB to perform mode selection, because it is already in charge of scheduling.

We propose a mechanism in which the eNB handles these decisions in two steps: $(i)$~\emph{mode selection} and $(ii)$~\emph{scheduling}.
First, in mode selection, each \ac{D2D} pair is assigned a mode (modes $1$ to $3$), and the assignment is
repeated at regular {\it mode intervals} of length $ T $ seconds.
The eNB selects \ac{D2D} modes with the assumption of a worst-case interference scenario. This approach helps to reduce the system complexity and to avoid disruptive co-channel interference. Second, in the scheduling phase, the eNB schedules users and assigns them a Modulation and Coding Scheme (MCS). %\red{The scheduling is done on a per-frame basis. }
%\red{Any legacy LTE-A scheduler can be used, although we choose Proportional Fair (PF), which provides a good tradeoff between throughput and fairness, and is currently implemented in 3GPP systems~\cite{Margolies2014Infocom}.}
%the existing cellular systems are strongly centralized and do not favor distributed schemes due to security vulnerabilities.
%\red{should we say we can adapt if cellular systems became distributed?}
%In any \ac{D2D}-enabled network, efficient
Mode selection and scheduling both rely on the accuracy of \ac{CSI} data gathered at the eNB, which can be challenging in terms of signaling overhead and scheduling.

%\blue{is tx rate of \ac{D2D}  determined locally?}. Specifically, cellular users are prioritized based on their instantaneous or expected instantaneous channel quality.
%Note that RBs allocated to inband overlay are released to for modes $0$ and $1$ if no user uses mode $2$.
%In each {\it subframe} (lasting $ 1 $ ms), only one cellular user is scheduled, while the number of concurrent \ac{D2D} transmissions is not limited {\it a priori}. Therefore, there is no interference among cellular users (i.e., mode $0$) users, but underlay users (i.e., mode $ 1 $) interfere with the cellular and other underlay users (i.e., modes $ 0 $ and $ 1 $), and finally overlay users only interfere with each other.
%However, one goal of our ongoing work consists in identifying how our solution can be implemented in a distributed way.

\subsection{Practical Implications}
\label{ss:practicality}
%\red{The majority of the literature on \ac{D2D} communications propose centralized mechanisms because any inband \ac{D2D} implementation overlaps with the cellular system which is a strongly centralized system. Therefore, it seems unlikely that operators would allow their resources  to be utilized without supervision. }
In a \ac{D2D}-enabled network, the eNB requires \ac{CSI} between each pair of users (i.e., user-to-user \ac{CSI}) in addition to user-to-eNB \ac{CSI} in order to perform MCS assignment and scheduling. However, the existing cellular technologies do not have the means to obtain user-to-user \ac{CSI}. Hence, we need a mechanism to obtain and send this information to the eNB efficiently because the addition of user-to-user \ac{CSI} imposes high signaling overhead to the system.

{\bf \ac{CSI} measurement}. In LTE, the eNB-to-user \ac{CSI} is estimated by active measurements from the received signal strength. However, there is no signaling message exchange between the users. Therefore, some researchers propose probing techniques to perform \ac{CSI} estimation among users~\cite{Doppler2010WCNC}. This approach imposes even higher signaling overhead to the system. In contrast, we propose an adaptive passive \ac{CSI} estimation between users as explained in what follows. In LTE, each user has a unique ID (i.e., \ac{C-RNTI}~\cite{johnson2010lte}) and this ID is included in the frame header. Thus, the users can detect the ID of the source of interference at each frame. Alternatively, the user can read \ac{C-RNTI}s from the broadcasted scheduling map to identify the interfering user's ID. The latter does not require users to sniff and decode other users' frame headers. The \ac{CSI} is then reported to the eNB. The eNB builds an interference table, whose elements $I_{n,m}\! \ge \!0$ represent the interference caused by user $ n $ to user $m$ $\left( \forall n,m \!  \in \! \setN \! \cup \! \{\! N\!+\!1\! \}\right)$. In case two users do not detect each other for physical/timing reasons, the failure only causes an interruption on a millisecond scale.  Once an interruption occurs, the user will report it to the eNB which will update the interference matrix. As for outband \ac{D2D}, each user reports the last achieved rate over WiFi. In case of an inaccurate report due to long inactivity period, the users can send an updated report before the next mode interval.

{\bf Signaling overhead}. The maximum number of \ac{CSI} reports in LTE-A (with wideband \ac{CSI} reporting~\cite{johnson2010lte}) is equal to $ N $. This number increases to  $N+ 2 |\setN_d| |\setN_c| +  |\setN_d| \left(|\setN_d| - 1 \right)$ in a \ac{D2D}-enabled network, where $\setN_c$ and $\setN_d$ are the sets of cellular and \ac{D2D} users, respectively. For instance, a \ac{D2D}-enabled network with $ 4 $ cellular and $ 6 $ \ac{D2D} users may require up to $ 88 $ \ac{CSI} reports, which is almost $9$ times higher than its equivalent in a legacy system. Fortunately, the \ac{CSI} feedbacks can be considerably reduced using the state-of-the-art feedback reduction techniques~\cite{Li2014TWC}. Moreover, we will see in Section~\ref{s:mode-eval} that the \ac{D2D} signaling overhead is negligible as compared to the resulting gain.
This overhead is further reduced by our proposal because we decouple scheduling from mode selection, hence the \ac{D2D} related \ac{CSI}s are obtained less frequently.

\section{Floating Band \ac{D2D} Framework}
\label{s:problem}

In this section we describe our proposed Floating Band \ac{D2D} framework and formulate the problem of mode selection at the
beginning of each mode interval $ j $, i.e., each $T$ seconds. The utility function in our problem formulation depends on throughput and energy costs, for
which we provide a general model in which specific schedulers can be plugged in. Note that, although the general model can be used with specific schedulers to evaluate the performance of various
%scheduling
strategies, the formulation of the problem does not depend on the scheduler actually implemented,
% in the network, 
and is not affected by resource allocation strategies for either cellular or \ac{D2D} connections. 

\textbf{Throughput and energy costs.}
The transmitted data $ \theta^{ i }_{ n , m } ( j ) $  for a connection $ ( n , m )$ in mode $ i \in \{0, 1, 2 \} $ during mode interval $ j $ is formulated as follows:

\begin{align}
\theta^{ i }_{ n , m } ( j ) & = B^{ i }_{ n , m } ( j ) R^{ i, \text{CSI} }_{ n , m } ( j ), \label{eq:tinband}
%E^{ i }_{ n , m } ( j ) & =  2\left( \beta_{lte}+   \beta^\texttt{WiFi}_{idle}  \right)  +\left( p^{ i, \text{TX} }_{ n } +   p^{ i, \text{RX} }_{ m } \right) t_{B^{ i }_{ n , m }} ( j ) ,  \label{eq:einband} \\
%E^{ 0 }_{ n , m } ( j ) & =    \beta_{lte}   +  p^{ 0, \text{TX} }_{ n } \cdot  t_{B^{ 0 }_{ n , m }} ( j ) , \label{eq:einband}
\end{align}
%\vspace{-1mm}
where $ B^{ i }_{ n , m } ( j ) $ is the number of Resource Blocks (RBs) allocated to connection $ ( n , m ) $ in mode interval $ j $.  $ R^{ i, \text{CSI} }_{ n , m } ( j ) $ is the number of transmitted bits per RB of connection $ (n , m) $ in mode $ i $ during mode interval $ j $, computed based on the channel gain between users $n$ and $m$, and the interference matrix $\bf I$.

The energy consumption of  a cellular user $ E^{ 0 }_{ n , m } ( j ) $ and the energy consumption of a \ac{D2D} pair $ E^{ i }_{ n , m } ( j ) $ in inband mode $ i \in \{1, 2 \}$ are given by:
\begin{align}
%\theta^{ i }_{ n , m } ( j ) & = B^{ i }_{ n , m } ( j ) R^{ i, \text{CSI} }_{ n , m } ( j ), \label{eq:tinband}\\
E^{ 0 }_{ n , m } ( j ) & =    \beta_{lte}   +  p^{ 0, \text{TX} }_{ n } \cdot  t_{B^{ 0 }_{ n , m }} ( j ) \qquad m=N+1, \label{eq:eCell}\\
E^{ i }_{ n , m } ( j ) & =  2\left( \beta_{lte}+   \beta^\texttt{WiFi}_{idle}  \right)  +\left( p^{ i, \text{TX} }_{ n } +   p^{ i, \text{RX} }_{ m } \right) t_{B^{ i }_{ n , m }} ( j ) ,  \label{eq:einband}
\end{align}
where $ \beta_{lte} $ and $\beta^{\texttt{WiFi}}_{idle} $ are the baseline energy consumed in a mode interval by an active cellular interface and an idle WiFi
interface, respectively. The WiFi interface is kept idle in inband modes to speed up WiFi connection setup. Here, $ p^{ i, \text{TX} }_{ n } $ and $ p^{ i,
\text{RX} }_{ m } $ are the energy consumed for transmission and reception in one subframe, respectively. $ t_{B^{ i }_{ n , m } } ( j ) $ is the transmission time associated with $
B^{ i }_{ n , m } ( j ) $. Here, we do not calculate the energy per RB, because it is shown that the transmission/reception power mainly depends on time rather
than
%bandwidth~\cite{Jensen2012VTC,dusza2013infocom}.
bandwidth~\cite{dusza2013infocom}.
%We solve the problem hierarchically at the beginning of each mode interval $ j $, i.e., each $T$ seconds. Let $ \setL ( j ) $ be the set of all existing arcs
%during mode interval $ j $, i.e., such that $ Z_{ n , m } ( j ) = 1 $.
%For an active arc $ ( n , m ) $ under an LTE mode $ i \in \{0, 1, 2 \} $ in mode interval $ j $ we define the energy consumption $ E^{ i }_{ n , m } ( j ) $
%and the transferred data $ \theta^{ i }_{ n , m } ( j ) $ (both per mode interval $T$) as follows:

%where
%%$\beta^{\texttt{LTE}}_{(.)}$ is the baseline energy consumed by a user in one mode interval,
%we do not consider the baseline energy consumed by a user in LTE in one mode interval, since it cannot be changed unless the node is switched off,
%$ p^{ i, \text{TX} }_{ n } $ and $ p^{ i, \text{RX} }_{ m } $ are the energy consumed by user $ m $ per transmitted and received RB, respectively, $ B^{ i }_{ n , m } ( j ) $ is the number of RBs allocated to arc $ ( n , m ) $, and $ R^{ i, \text{CSI} }_{ n , m } ( j ) $ is the
%number of transmitted bits per RB of arc $ (n , m) $ under mode $ i $ during mode interval $ j $.
%%Note that we have omitted the energy consumed for maintaining the connection
%%with eNB and the energy consumption of eNB itself, since these are incurred independently of being active or passive.

The expression of transmitted data $ \theta^{ 3 }_{ n , m } ( j ) $ and the energy consumption $ E^{ 3 }_{ n , m } ( j ) $ for connection $ ( n , m ) $ under outband mode (i.e., mode $ 3 $) in mode interval $ j $ is as follows:
%\vspace{-1mm}
\begin{align}
\theta^{ 3 }_{ n , m } ( j ) &= T \cdot R^{ 3, \text{CSI} }_{ n , m } ( j ), \label{eq:toutband}\\
E^{ 3 }_{ n , m } ( j ) &= 2 (\beta_{lte} + \beta^{ \texttt{WiFi} }_{active} ) + \left( p^{ 3, \text{TX} }_{ n } + p^{ 3, \text{RX} }_{ m } \right) \theta^{ 3 }_{ n , m } ( j ), \label{eq:eoutband}
\end{align}
where $R^{ 3, \text{CSI} }_{ n , m }$ is the WiFi rate and $ \beta^{ \texttt{WiFi} }_{active} $ is the baseline WiFi energy consumed by a user in a mode interval. $ p^{ 3, \text{TX} }_{ n } $ and $ p^{ 3, \text{RX} }_{ m } $ are the energy consumed by user $ m $ per transmitted/received bit. Note that the energy consumption as defined here can incorporate both the consumption due to transmission/reception and packet processing (see~\cite{asadi2013MSWIM}). The $\beta_{lte}$ is due to the dependence of outband users to the eNB signaling.

We define a utility function for connection $ ( n , m ) $ under mode $ i $ in mode interval $ j $ as follows:
\begin{equation}
\label{eq:u}
U_{ n , m }^i ( j ) = \theta^{ i }_{ n , m } ( j ) - \alpha  E^{ i }_{ n , m } ( j ),
\end{equation}
where $ \alpha $ is the relative cost of energy. The utility accounts for both throughput and energy consumption. The value of $ \alpha $ determines whether the system is biased towards higher throughput or lower energy consumption. %also We can change the system bias towards higher throughput or lower energy consumption by tuning the value of $ \alpha $ .
In our model, the impact of schedulers is summarized in $ B^{ i }_{ n , m }$ and $ t_{B^{ i }_{ n , m }} $. Those parameters have to be computed in each mode interval and for each possible mode selection decision. 
%\subsection{Optimal solution}
%\label{ss:optimize}

\textbf{Problem Formulation.}
Let $ \setL ( j ) $ be the set of all existing connections during mode interval, $ j $, $\{Y^i_{n,m}(j)\}$ be the set of binary decision variables, and
$\gamma_{n}$ be the tolerable interference threshold that allows for a non-zero reception rate by user $ n $. We formulate the problem of mode selection for
mode interval $ j $ as a binary programming problem (the dependency on $ j $ is omitted for readability):
%We use a set of binary decision variables $\{Y^i_{n,m}(j)\}$,to formulate the problem of mode selection for mode interval $j$, preceding the RB allocation procedure in the above described system (note that at mode selection time it is not yet possible to predict the exact interference caused by/to \ac{D2D} users, so we account for the worst-case interference).
%The problem is formulated as follows (we omit ):
%\vspace{-2mm}
%\vspace{-3mm}
\begin{align}
\!\!\!\!\! \begin{cases}
\max	  \!\!	& U_{sum}:= \sum_{i=0}^{3} \sum_{(n, m)\in\setL}  U_{ n, m }^{ i } Y_{ n, m }^{ i }  \quad \!\!\!\! \forall (n, m) \in \setL  \!\!\!\! \label{eq:NLBIP}\\
\text{s.t.} 	&\!\!\!\!\!\!\! \sum_{i=0}^{3} \sum_{n|(n, m)\in \setL} Y_{ n , m }^{ i } \le 1  \quad \forall m \in \setN \\
		&\!\!\!\!\!\!\! \sum_{i=0}^{3} \sum_{m|(n, m)\in \setL} Y_{ n , m }^{ i } \le 1 \quad \forall n \in \setN \\
		&\!\!\!\!\!\!\! \sum_{\substack {(n, m)\in \setL }} Y_{ n , m }^{ 1 } I_{ n,x} \le \gamma_x \quad \forall x\in  \setN_{c} \cup \{N+1\} \\
		&\!\!\!\!\!\!\! \sum_{i \in \{0, 1\}} \sum_{(x, y) \in \setL \setminus \{ (n, m) \} }  Y_{ x , y }^{ i } Y_{n, m}^ {1} I_{ x, m} \le \gamma_m \\
		&\!\!\!\!\!\!\! \sum_{(x, y) \in \setL \setminus \{ (n, m) \} } Y_{ x , y }^{ 2 } Y_{n, m}^ {2} I_{ x, m} \le \gamma_m \quad \!\!\! \forall (n, m)\! \in \! \setL 	\!\!\!\!	
\end{cases}
\end{align}
Problem~\eqref{eq:NLBIP} maximizes the sum of utilities $U_{sum}$
over all possible combinations of users and modes. Our assumption on single instantaneous connectivity is enforced with the first and second constraints (the
eNB, which is labeled as $N+1$, is an exception). The third constraint ensures that the co-channel interference from underlay users to cellular users and to
the eNB is kept below the threshold. The fourth constraint limits the interference from cellular and inband underlay users to other inband underlay users. The
interference of overlay transmissions is limited
%below  the tolerable threshold
by the fifth constraint.

%Alternatively, Problem~\eqref{eq:NLBIP} can be transformed into a linear programming problem, by introducing new variables in order to replace all the quadratic terms \red{AA: Can we say this in a way that it looks like a contribution?}. This allows for solving the problem relatively efficiently by standard approaches, such as Branch \& Bound~\cite{lawler1966branch}. Nevertheless, we deem such an approach impractical as the system requires a solution in milliseconds.

{\bf Complexity.}  Problem~\eqref{eq:NLBIP} is NP-hard and non-linear since it can be reduced to the {\it longest path problem} (e.g., for a weighted directed and possibly disconnected graph), which is NP-hard~\cite{LongPath}. This reduction is obtained when we consider Problem~\eqref{eq:NLBIP} for a single mode $i = 3$ (outband), in which the objective is to activate \ac{D2D} pairs so as to achieve the maximum utility possible with the two restrictions on at most one incoming and at most one outgoing transmission for every user.
%Indeed, it
Problem~\eqref{eq:NLBIP} requires the computation of  $\{ U_{ n, m }^{ i } \}$, which is based on \ac{SINR}
and its optimal solution
%to Problem~\eqref{eq:NLBIP}
can be achieved by brute force: exploring the consequences of assigning
modes $1$, $2$, or $3$ to any of the $ \frac { |\setN_{ d }| } { 2 }  $  \ac{D2D} pairs. Hence, the resulting complexity is $O(N \cdot 3^\frac{|\setN_{d}| }{2} )$, which grows exponentially with the number of \ac{D2D} pairs. The optimal solution to the above maximization problem is computationally expensive and practically unfeasible in dense networks. However, the non-linear constraints can be linearized so that the problem can be solved relatively efficiently by standard approaches, such as Branch \& Bound~\cite{lawler1966branch}.
%The linearized problem is omitted for brevity, but can be found in~\cite{extended}.
Nevertheless, we deem such an approach impractical as the system requires a solution in milliseconds. Thereby, we propose efficient heuristics in what follows.
%\red{A proof of NP-hardness and linearization details are omitted for brevity, and can be found in~\cite{extended}.}

\section{Heuristics}
\label{s:heuristics}

The exact solution to Problem~\eqref{eq:NLBIP} is computationally expensive and does not allow for a fast and scalable mode selection.
Given the similarity of the problem to the longest path problem and the knapsack problem, we propose three practical heuristics.
%which reduces the complexity of the problem from exponential to linear. %THE FINAL COMPLEXITY IS NOT LINEAR
These heuristics explore the achievable utilities of the users in an iterative manner. Note that these utilities are computed assuming that the system is fully
utilized (i.e., users' queues are fully backlogged) so that they do not require the knowledge of the actual user's offered load.

\begin{algorithm}[h!]
\caption{ \texttt{Social}}
\footnotesize
\begin{algorithmic}[1]
\Require \\
%		$R_{n,m}^{CSI,i}$: transmission rates between all users in each mode.\\ %$\quad \forall n \in \setN, \forall m \in (\setN \cap N+1), i \in \{0,1,2,3\} $ \\
		$\setN_{d,\text{TX}}$: set of \ac{D2D} transmitters (randomized order).\\
		$I_{n,x}$:  interference between each pair of users. % $\quad \forall x\in \setS_{N+1} \cup \{N+1\};$
\Ensure $Y_{(n,m)}^i,  \forall n \in \setN_{d,\text{TX}}$\\
initialize: ${\bf Y}\!=\! {\bf Y}_{old} \!=\!\emptyset$; $Y_{(c,N+1)}^{0} \! = \!1, \forall c \! \in \! \setN_c$; $Y_{(n,m)}^{3} \!=\!1, \forall n \! \in \! \setN_{d,\text{TX}}$; $\text{max}= U_{sum}$
%, $Y_{(n,m)}^{i} =0$
\While{${\bf Y} \neq {\bf Y}_{old}$}
\State{${\bf Y}_{old}={\bf Y}$}
\For{$ n   \in \setN_{d,\text{TX}}$ }
%	\State{$ mode = 3$}
	\For{$ j   \in \{1,2,3\}$ }
%		\State{$Y_{(i,m)}^{j}=1$; $Y_{(i,m)}^{k}=0, k\in\{1,2,3\}\setminus\{j\}$}
		\State{Calculate: $ U_{sum} | n \text{ is in mode } j$}
%		\If{$ U_{total} ^ j  > U_{max} ^ j $ }
		\If{$ U_{sum}  > \text{max}$ }
			\State{$ \text{max} =  U_{sum} $ }
			\State{$Y_{(n,m)}^{j}=1$; $Y_{(n,m)}^{k}=0, k\in\{1,2,3\}\setminus\{j\}$}
%			\State{mode = j }
		\EndIf	
	\EndFor
%	\State{$Y_{(i,m)}^{mode}=1$}
\EndFor
\EndWhile
\end{algorithmic}
\label{social}
\end{algorithm}
%
%\vspace{-3mm}
\subsection{Heuristic~1. \texttt{Social}}
%\vspace{-1mm}
The eNB iterates over the set of \ac{D2D} transmitters $\setN_{d,TX}$, and it selects the mode that maximizes the aggregate utility (lines $7$-$13$ in
Algorithm~\ref{social}). Note that the mode for user $i$ is selected based on the modes selected for the precedent users. Initially, all \ac{D2D} pairs are assigned
to mode $3$ (outband), to minimize the impact on cellular users. For better fairness~\cite{wu2013TON}, the order of users in $\setN_{d,TX}$ is randomized at
any mode interval. The mode selection repeats until the algorithm converges to a decision. We name this heuristic as \texttt{Social} because it decides based
on social welfare. Since the utility of \texttt{Social} cannot decrease with mode selection decisions, the heuristic always converges.
%Algorithm~\ref{social} illustrates the pseudocode of this heuristic.

%\vspace{-3mm}
\subsection{Heuristic~2. \texttt{Greedy}}
%\vspace{-1mm}
The \texttt{Greedy} heuristic is similar to \texttt{Social}. Unlike \texttt{Social}, \texttt{Greedy} selects the mode which maximizes the user's individual utility (line $10$ in Algorithm~\ref{greedy}).
The drawback of \texttt{Greedy} is that it might not converge. However, we can index each decision since the algorithm is running in the eNB.
Once a duplicate index (stored in $\setD$) is found, the algorithm stops the iteration.
%Algorithm~\ref{greedy} illustrates the pseudocode of the heuristic.
\begin{algorithm}[h]
\caption{\texttt{Greedy}}
\footnotesize%\scriptsize
\begin{algorithmic}[1]
\Require \\
%		$R_{ n , m } ^ { CSI , i }$: transmission rates between all users in each mode.\\ %$\quad \forall n \in \setN, \forall m \in (\setN \cap N+1), i \in \{0,1,2,3\} $ \\
		$\setN_{d,\text{TX}}$: set of \ac{D2D} transmitters (randomized order).\\
		$I_{ n , x }$:  interference between each pair of users.% $\quad \forall x\in \setS_{N+1} \cup \{N+1\};$
%		$\setD$: set of indices for all decision.
%\Ensure $Y_{ ( n , m ) } ^ i $\\
\Ensure $Y_{(n,m)}^i,  \forall n \in \setN_{d,\text{TX}}$\\
%initialize: $Y_{ ( n , m ) }^{ i }\!\!  =\! 0 $, $Y_{(n,m)}^{3} \!=\!1$, $ \text{max}_i  \!=\! U_{(i,m)} $, $exit \!=\! False, \setD=\emptyset$
initialize: ${\bf Y} = \emptyset$; $Y_{(c,N+1)}^{0} = 1, \forall c \! \in \! \setN_c$; $Y_{(n,m)}^{3} =1, \forall n \! \in \! \setN_{d,\text{TX}}$; $ \text{max}_i =U_{(i,m)} $;  $exit \!=\! False, \setD=\emptyset$

\While{$exit = False$}
\For{$ i   \in \setN_{d,\text{TX}}$ }
%	\State{$ mode = 3 $}
	\For{$ j   \in \{1,2,3\}$ }
		\State{Calculate: $ U_{(i,m)}^j | i \text{ is in mode } j$}
		\If{$U_{(i,m)}^j > \text{max}_i $ }
		\State{$\text{max}_i =  U_{(i,m)}^j$ }
		\State{$Y_{(i,m)}^{j}=1$; $Y_{(i,m)}^{k}=0, k\in\{1,2,3\}\setminus\{j\}$}
%		\State{mode=j}
		\EndIf	
	\EndFor
%	\State{$Y_{(i,m)}^{mode}=1$}	
\EndFor
\State{$dec$ = Index of current $\bf Y$}
\If{$dec \in \setD$}
\State{exit = True}
\EndIf
\State{Add $dec$ to $\setD$}
\EndWhile
\end{algorithmic}
\label{greedy}
\end{algorithm}
%

%\vspace{-3mm}
\subsection{Heuristic~3. \texttt{Ranked}}
%\vspace{-1mm}
Both \texttt{Social} and \texttt{Greedy}  operate on a list of \ac{D2D} transmitters with a randomized order. In contrast, \texttt{Ranked} heuristic
sorts this list based on the achievable utility of each user without considering the impact of other users (PHASE 1).
%Next, the \ac{D2D} pair list is sorted based on their utility in a descending order.
In PHASE 2, the pre-ordered list $\setN_{d,\text{TX}}^{(\text{ranked})}$  is evaluated using \texttt{Greedy}, which makes the heuristic {\it greedier} than \texttt{Greedy}.  This helps to evaluate the ability of our approach to withstand unfair conditions. Algorithm~\ref{ranked} illustrates the pseudocode of the heuristic.
\begin{algorithm}[h]
\caption{\texttt{Ranked}}
\begin{algorithmic}[1]
\footnotesize%\scriptsize
\Require \\
%		$R_{ n , m } ^ { CSI , i }$: transmission rates between all users in each mode.\\ %$\quad \forall n \in \setN, \forall m \in (\setN \cap N+1), i \in \{0,1,2,3\} $ \\
		$\setN_{d,\text{TX}}$: set of \ac{D2D} transmitters (randomized order).\\
		$I_{ n , x }$:  interference between each pair of users. % $\quad \forall x\in \setS_{N+1} \cup \{N+1\};$
\Ensure $Y_{(n,m)}^i$\\
%initialize: $Y_{(n,m)}^{i} =0$, $Y_{(n,m)}^{3} =1$
initialize: ${\bf Y} = \emptyset$; $Y_{(c,N+1)}^{0}   =  1, \forall c  \in   \setN_c$; $Y_{(n,m)}^{3}  = 1, \forall n  \in   \setN_{d,\text{TX}}$\vspace{1mm}
\Algphase{PHASE 1: Sorting \ac{D2D} pairs based on their utility}
\For{$ i   \in \setN_{d,\text{TX}}$ }
	\For{$ j   \in \{1,2,3\}$ }
		\State{Calculate $U_{(i,m)}^j$}
	\EndFor
%	\State {mode= $\max \{ U_{(i,m)}^j \},  \forall  j \in \{1,2,3\} $}
	\State {$\text{mode}_i=\arg\max \{ U_{(i,m)}^j; j \in \{ 1,2,3 \} \}$ }
%	\State{$Y_{(i,m)}^{mode_i}=1$}	
\EndFor
\State{sort the $\setN_{d,\text{TX}}$ based on utilities $U_{(i,m)}^{\text{mode}_i}$ \& store in $\setN_{d,\text{TX}}^{(\text{ranked})}$.  }
\vspace{1mm}
\Algphase{PHASE 2: Executing \texttt{Greedy} heuristic}
\State{Do \texttt{Greedy} with $\setN_{d,\text{TX}} =  \setN_{d,\text{TX}}^{(\text{ranked})}$.}
\end{algorithmic}
\label{ranked}
\end{algorithm}

%With proposed heuristics, the eNB performs mode selection according to patterns which result in near optimal results without the complexity of the optimal solutions.
% The complexity of  our proposed heuristic is $O (3n)$ which even allows for online implementation of the algorithm.

%\vspace{-1mm}

%\vspace{-3mm}
\subsection{Complexity Analysis}
\label{ss:complexity_analysis}
%\vspace{-1mm}
Our proposed heuristics compute $N-|\setN_{d,\text{TX}}|$ utilities $\{ U_{ n, m }^{ i } \}$ for each mode and for
every \ac{D2D} transmitter in a sequential manner, i.e., $ 3 \left ( N | \setN_{d,\text{TX}}| - | \setN_{d,\text{TX}}|^2 \right )$ utilities
per round of evaluation.
%Note that the evaluation sequence is randomized in both \texttt{Social} and \texttt{Greedy}.
In each mode interval, the evaluation cycle is repeated $ r_i $ times, $ r_i \geq 1 $, until the algorithm converges to a decision. Therefore, the complexity of \texttt{Social} and \texttt{Greedy} is $ O \left(3 r_i N | \setN_{d,\text{TX}}| \right) $, $i\in\{1,2\}$.
%The complexity of \texttt{Greedy} is the same as \texttt{Social} because in both algorithms, all utilities have to be recomputed for each mode.
\texttt{Ranked} has an additional sorting procedure before the mode selection in which the utility of each \ac{D2D} pair is computed in isolation. Thus, the algorithm only needs to compute $ 3 |  \setN_{d,\text{TX}}| $ utilities in PHASE 1, which can be neglected with respect to the number of utilities to be computed in PHASE 2. Hence, the complexity of  \texttt{Ranked}  is $ O \left(3 r_3 N | \setN_{d,\text{TX}}| \right) $.
Therefore, the three proposed heuristics have the same complexity, except for a constant factor $r_i$ that we will quantify experimentally later.

\section{Evaluation}
\label{s:mode-eval}

In this section, we use numerical simulations to evaluate the performance of our proposed heuristics. The evaluation scenario consists of a hexagonal
multi-cell network with a reference cell in the middle and its first-tier neighbors (see Figure~\ref{fig:network}). The results reported in this chapter pertain to
the reference cell, and the neighboring cells model the impact of inter-cell interference. Error bars in the results are the $95\%$ confidence intervals.
Although our approach can be tested with any scheduler, here we refer to the Proportional Fair (PF) scheme for scheduling cellular users, since it represents the state of the art for schedulers used in real implementations~\cite{lee2009Infocom, Margolies2014Infocom}. In addition to our heuristics, we evaluate three benchmark schemes, namely, \texttt{Forced-LTE},
\texttt{Forced-WiFi}, and \texttt{Optimal}. In  \texttt{Forced-LTE}, \ac{D2D} users are forced to use legacy cellular communications (i.e., mode $ 0 $). In
\texttt{Forced-WiFi}, \ac{D2D} users are forced to communicate over WiFi (i.e., mode $ 3 $). \texttt{Optimal} results are based on the exact solution to
Problem~\eqref{eq:NLBIP} obtained by brute force. The benchmarks allow to  compare our proposals with the legacy cellular system, to measure the gain due to extra
WiFi bandwidth, and to see how far the heuristics are from the optimum.

%We use brute-force method to illustrate the optimal solution to maximization problem introduced in \autoref{ss:optimize}. Although the brute-force approach is not scalable and practical, it is used to show the upper bound. Next, the performance of the proposed heuristic is benchmarked against the optimal solution.

\subsection{Simulation Setup}

User placement follows the uniform distribution. The number of \ac{D2D} users is on average $30\%$ of the cell population. The simulation parameters
are chosen according to the evaluation guidelines of ITU-R~\cite{ITU2009} which are reported in Table~\ref{tb:simPar}. In the simulation, we show both the
\emph{packet simulation results} (i.e., performance under finite offered load and in the presence of probabilistic arrival processes) and the \emph{achievable
performance} (i.e., performance at capacity-level utilization, under infinite offered load conditions). In the later, the transmission queues of the users are always fully backlogged. Unless otherwise specified, the default values for
$\alpha$ and overlay resource portion are those reported in Table~\ref{tb:simPar}, with an aggregate \ac{D2D} and cellular load of $ 30 $ Mbps and $ 90 $ Mbps,
respectively. Since the \ac{D2D} capacity is higher than the cellular one, due to proximity of \ac{D2D} users and availability of outband resources, we deemed fair to
impose higher load to \ac{D2D} users. Note that the default value of $\alpha$ is selected based on a rough estimate of the current relative price of bit per Joule
(b/J) in the market.

Besides the values of Table~\ref{tb:simPar}, we investigate the impact of user density $ N $, overlay resource portion,
relative cost of energy $\alpha$, and \ac{D2D} load on the system performance. Moreover, we shed light on the convergence time of our heuristics and their flexibility in different environments.
%The duration of each simulation is $ 30 $ s and the mode selection is done at the beginning of each mode interval.

%The \ac{D2D} communications occurs over uplink resources.
%Therefore, the maximum bandwidth for a \ac{D2D} connection is $20$ MHz. The simulation parameters can be found in \autoref{tb:simPar}.

\begin{table}[h]
\setlength\extrarowheight{2pt}
\caption{The parameters used in the evaluation}
\centering
%\scriptsize
\label{tb:simPar}
%\begin{tabular}{|@{}c@{}| p{2.8in}@{} |}
%\resizebox{0.6\columnwidth}{!}{
\begin{tabular}{|l|c|}
\hline
\textbf{ Parameter }			 		&\textbf{ Value } 		\\
\hline
\hline
\multicolumn{2}{|c|}{Cellular}\\
\hline
Cellular uplink bandwidth			& 20 MHz	 \\
\hline
Cell radius					& 250 m	\\
\hline
eNB, cellular user TX power					& 44 dBm, 24 dBm	 \\
\hline
%Cellular user TX power			& 24 dBm	 \\
%\hline
Thermal noise power 			& -174 dBm/Hz	 \\
\hline
Mode interval length $T$			& 2 s	 \\
\hline
Fading, shadowing, pathloss  		& Reyleigh, 6 dB, UMa~\cite{ITU2009}	 \\
%\hline
%Fading	 					& Reyleigh	 \\
%\hline
%Pathloss 						& Urban Marco NLOS\\
\hline
Buffer size 					& 500 packets	\\
\hline
$\beta_{lte}$ 					&1288.04 mW\\
\hline
\multicolumn{2}{|c|}{WiFi}\\
\hline
WiFi bandwidth 				& 22 MHz	 \\
\hline
WiFi effective range				& 150 m	\\
\hline
WiFi TX power					& 20 dBm	 \\
\hline
$\beta^{ \texttt{WiFi} }_{active} $,~$\beta^{ \texttt{WiFi} }_{idle} $ 	&132.86 mW, 77.2 mW\\
\hline
%$\beta^{ \texttt{WiFi} }_{idle} $ 		&77.2 mW\\
%\hline
\multicolumn{2}{|c|}{D2D}\\
\hline
Underlay max bandwidth			& 20 MHz	 \\
\hline
Overlay resource portion				& 30\% 	 \\
\hline
D2D maximum distance			& 20 m	 \\
\hline
D2D inband TX power	 		& 10 dBm	 \\
\hline
Relative cost of energy $\alpha$		& 1 bit/Joule		\\
\hline
\end{tabular}
%}
\end{table}

\subsection{Simulation Results}

\begin{figure*} [t]
		\centering
		\subfigure[Aggregate cell throughput.]
		{
			\includegraphics[width=0.45\columnwidth]{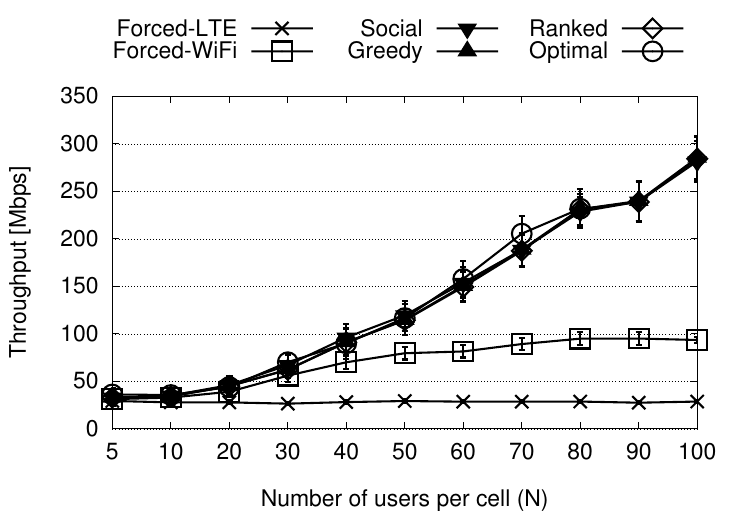} %[width=70mm,angle=0]
			\label{fig:totalTput}
		}
		\subfigure[Aggregate cell power.]
		{
			\includegraphics[width=0.45\columnwidth]{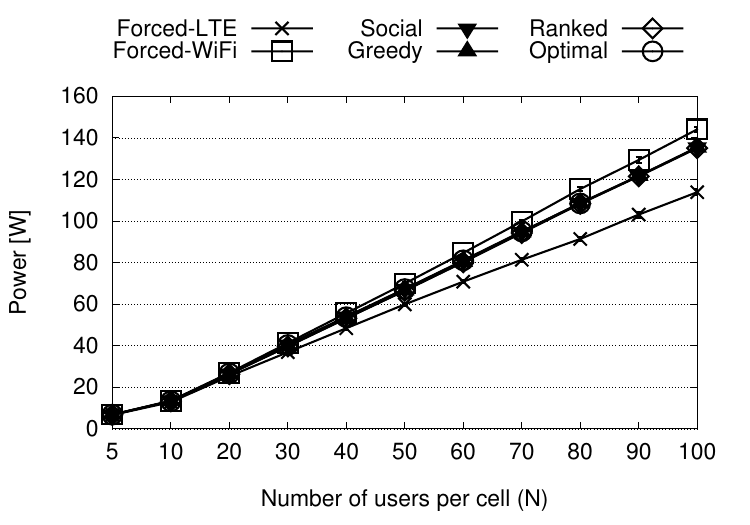} %[width=70mm,angle=0]
			\label{fig:totalEnergy}
			%\vspace{-5mm}
		}
		\subfigure[Aggregate cell Utility.]
		{
			\includegraphics[width=0.45\columnwidth]{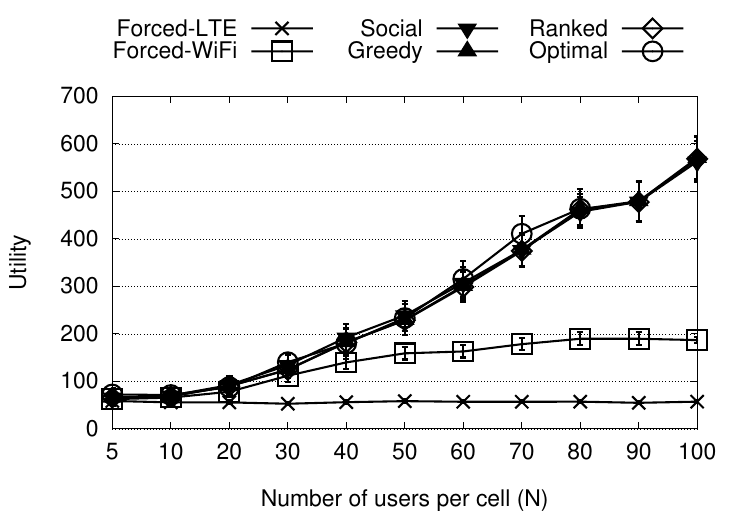} %[width=70mm,angle=0]
			\label{fig:totalUtility}
			%\vspace{-5mm}
		}
		\caption{The impact of user population on the system performance with fully backlogged queues ({\it achievable performance}).}
		\label{fig:aggregate_static-mode}
\end{figure*}

\begin{figure*} [t]
		\centering
		\subfigure[Fairness among cellular users.]
		{
			\includegraphics[width=0.45\columnwidth]{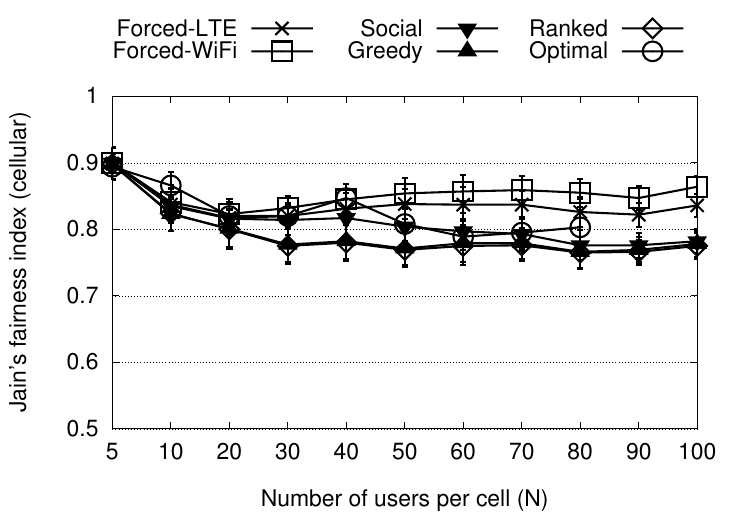} %[width=70mm,angle=0]
			\label{fig:cellFair}
		}
		\subfigure[Fairness among D2D users.]
		{
			\includegraphics[width=0.45\columnwidth]{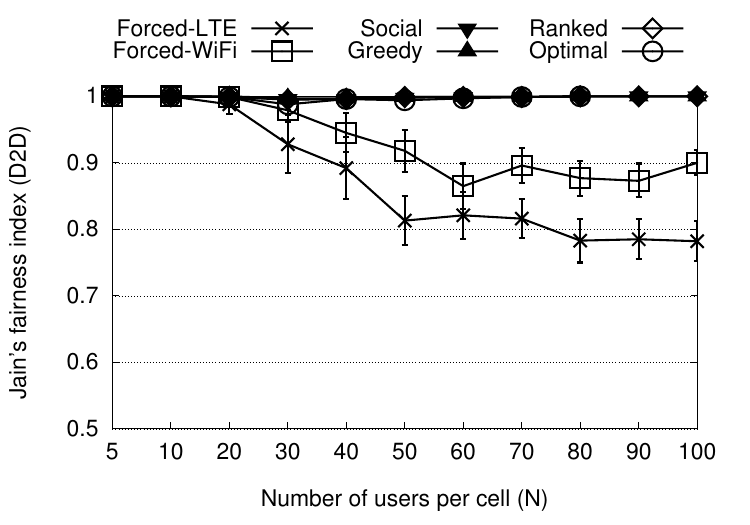} %[width=70mm,angle=0]
			\label{fig:d2dFair}
			%\vspace{-5mm}
		}
		\subfigure[Aggregate cell Utility.]
		{
			\includegraphics[width=0.45\columnwidth]{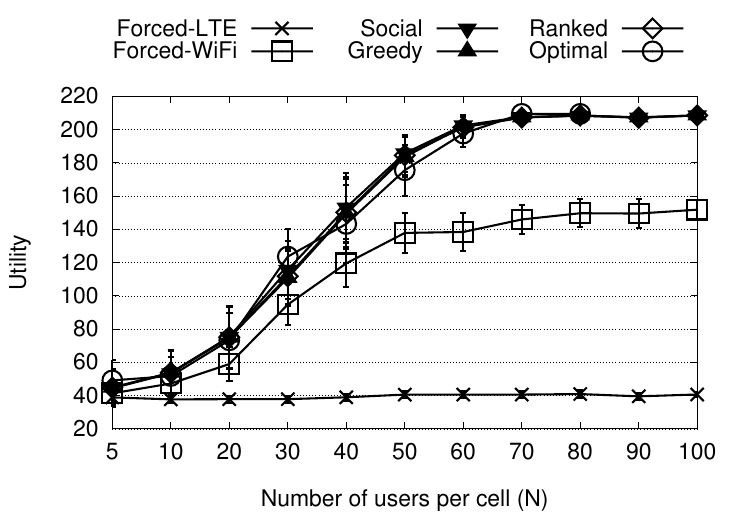} %[width=70mm,angle=0]
			\label{fig:totalPktUtility}
			%\vspace{-5mm}
		}
		\caption{The impact of user population on the system performance evaluated through packet simulation.}
		\label{fig:pktSimUsrDensity}
\end{figure*}

{\bf Impact of the number of users $ N $.} Figure~\ref{fig:aggregate_static-mode} illustrates the impact of $ N $ on achievable system performance.
%Here, the overlay portion is fixed to $30\%$ and $\alpha=2$.
We can observe the achievable throughput in Figure~\ref{fig:totalTput}. The aggregate throughput has a negligible change with $ N $ under \texttt{Forced-LTE}
%because the density of users does not change the distribution of channel qualities in the cell,
because the distribution of channel qualities in the cell remains the same for different density of users,
and therefore the average aggregated throughput. The throughput of the rest of schemes increases with $ N $ because there are probabilistically more \ac{D2D} pairs in a denser cell, hence \ac{D2D} throughput is higher. In \texttt{Forced-WiFi}, the throughput grows slowly due to the contention-based nature of WiFi, in which the MAC overhead increases with the number of contending users.
Since some of the outband \ac{D2D} pairs do not interfere with each other (i.e., they are more distant than $ 150 $ m), the aggregate throughput of \texttt{Forced-WiFi} in our experiments reaches up to $ 98 $ Mbps. More importantly, not only the simple proposed heuristics greatly
outperform \texttt{Forced-LTE} and \texttt{Forced-WiFi}, but they also perform very close to \texttt{Optimal}
%(for which we only have results up to $70$ users, due to the excessive time needed to compute results with such an ideal scheme).
(due to the computational complexity of such an ideal scheme, we only have the results up to $80$ users).

In terms of energy cost, the aggregate cell power increases with $ N $, as shown in Figure~\ref{fig:totalEnergy}, mainly due to the baseline energy consumption of wireless interfaces. \texttt{Forced-WiFi} has higher energy consumption because outband users have to maintain two active wireless interfaces instead of one.

Figure~\ref{fig:totalUtility} shows that the trend for system utility is similar to that of throughput because the throughput is the dominant factor with the current value of $ \alpha $. Our results show that, with a reasonable population, say $ 100 $ users per cell, the aggregate throughput gain over \texttt{Forced-LTE} is tenfold. This gain comes from both the frequency re-use of inband modes and additional spectrum provided by the outband mode, as shown in the figure. The significant contribution of both outband and inband modes to this gain highlights the importance of Floating Band \ac{D2D}. Moreover, this gain can easily compensate for the infrequent \ac{D2D} \ac{CSI} feedbacks sent to the eNB (user-to-user \ac{CSI}). Note that in LTE-A systems with millisecond feedback reporting, the \ac{CSI} contributes to less than $20$\% of the total bandwidth.
% \texttt{Forced-WiFi} and \texttt{Optimal} shows that both inband and outband \ac{D2D} significantly contribute to this $ 600\% $ gain.

In Figure~\ref{fig:pktSimUsrDensity}, we can observe the accuracy of our mode selection and its performance using packet simulation. Figure~\ref{fig:cellFair}
shows that cellular users have comparable throughput performance under all schemes due to PF scheduling. If the data rate of a cellular user degrades due to
co-channel interference, the PF compensates for it by allocating more resources to that user. In Figure~\ref{fig:d2dFair}, it is observed that the fairness among
\ac{D2D} users drops under \texttt{Forced-LTE} and \texttt{Forced-WiFi}. Under \texttt{Forced-LTE}, \ac{D2D} users are scheduled as cellular users, hence they achieve
similar fairness performance as cellular users (but not equal because their fairness is computed over a different set and their load is different). The
fairness reduction under \texttt{Forced-WiFi} is due to topologically uneven distribution of contending outband users.  In Figure~\ref{fig:totalPktUtility}, we
can observe that utilities of all \ac{D2D}-enabled schemes grow until $ N $ reaches $50$. The reason for this behavior is that the network operates under saturation
up to this point. In fact, one can observe in Figure~\ref{fig:totalTput} that the achievable throughput with $ 50 $ users or less is below $ 120 $ Mbps which is
equal to the total offered load (i.e., $\! 30\! + \!90 $) in the scenario of Figure~\ref{fig:totalPktUtility}.  For $N\!>\!50$, the utility in the packet
simulation is limited by the adopted load.

\begin{figure*} [t]
		\centering
		\subfigure[Aggregate achievable utility.]
		{
			\includegraphics[width=0.45\columnwidth]{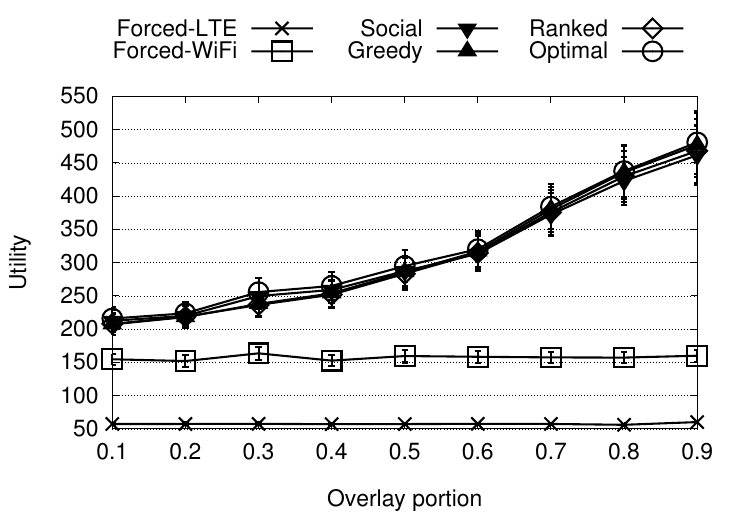} %[width=70mm,angle=0]
			\label{fig:overlayUtil}
		}
		\subfigure[Average delivery ratio of cellular users.]
		{
			\includegraphics[width=0.45\columnwidth]{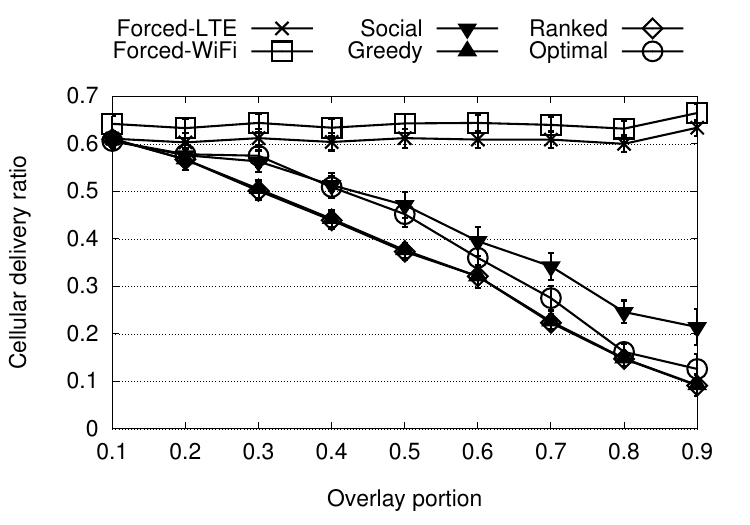} %[width=70mm,angle=0]
			\label{fig:overlayCellDr}
			%\vspace{-5mm}
		}
		\subfigure[Average delivery ratio of \ac{D2D} users.]
		{
			\includegraphics[width=0.45\columnwidth]{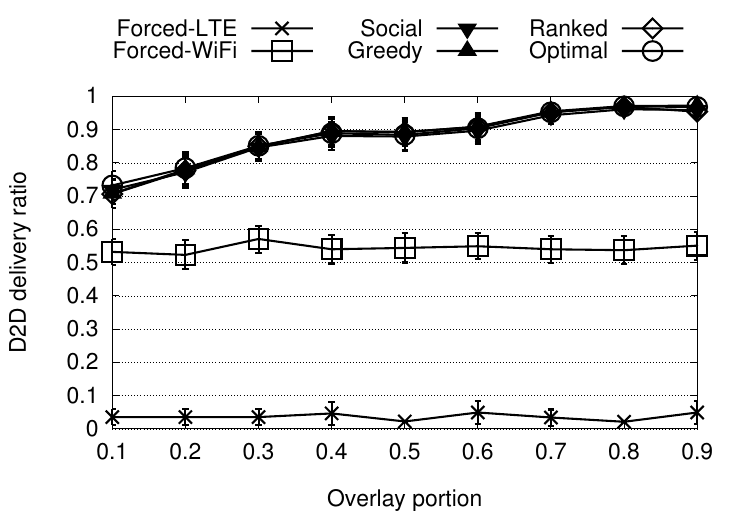} %[width=70mm,angle=0]
			\label{fig:overlayD2dDr}
			%\vspace{-5mm}
		}
		\caption{The impact of overlay portion on system performance $\left(  N = 50  \right)$. }
		\label{fig:overlay}
\end{figure*}

{\bf Impact of overlay resource portion.}
%\cyan{The system behavior versus the portion of dedicated overlay bandwidth is shown in Figure~\ref{fig:overlay}.}
Here, the number of users per cell is fixed to $ 50 $. Figure~\ref{fig:overlayUtil} shows that the utilities of multi-band schemes
increase with the overlay bandwidth. This increment is due to  throughput  improvement under mode 2.
This implies that \ac{D2D} users tend to receive more interference from cellular users than from other \ac{D2D} users, hence, the spectral efficiency is higher in overlay than underlay.
%This figure also indicates that the \ac{D2D} users have higher spectral efficiency in overlay rather than underlay (i.e., \ac{D2D} users receive interference of cellular users than that of other \ac{D2D} users). Hence, \ac{D2D} users achieve higher utilities when they share the medium only with other \ac{D2D} users.
As mentioned, the overlay portion is given to modes $0$ and $ 1 $ if there are no overlay users. As a result, the utilities of \texttt{Forced-LTE} and \texttt{Forced-WiFi} remain unchanged here.
%Figure~\ref{fig:overlayCellDr} helps to better illustrate the impact of dedicated overlay bandwidth on cellular user.

Although the aggregate utilities are improved, we should also investigate the impact of overlay bandwidth on cellular users.
Figure~\ref{fig:overlayCellDr} illustrates that the delivery ratio of cellular users degrades as the overlay bandwidth grows because there is less bandwidth at their disposal.
Figure~\ref{fig:overlayCellDr} also sheds light on the differences among multi-band schemes.
Cellular users experience higher packet delivery ratio with \texttt{Social}. Indeed, \texttt{Social} is the only scheme that aims to maximize the aggregate utility, which includes the utility of cellular users.
Finally, Figure~\ref{fig:overlayD2dDr} shows how the delivery ratio of \ac{D2D} users approaches $ 1 $ with higher overlay bandwidths, as expected.

\begin{figure*} [t!]
		\centering
		\subfigure[D2D delivery ratio.]
		{
			\includegraphics[width=0.45\columnwidth]{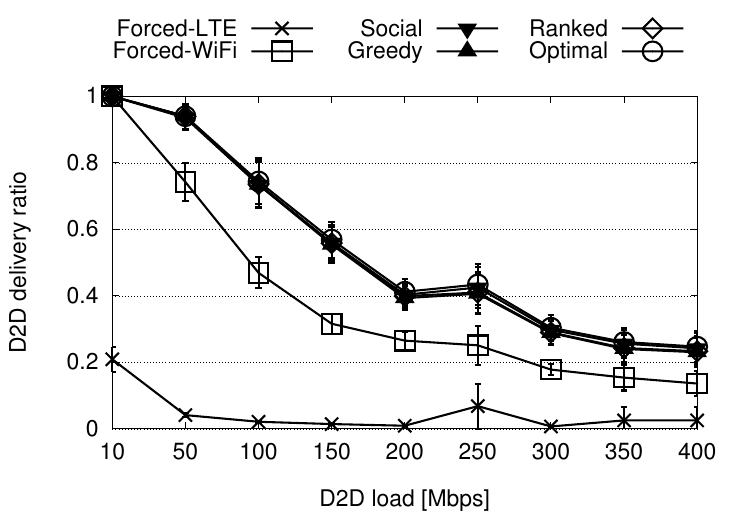} %[width=70mm,angle=0]
			\label{fig:loadD2dDr}
			%\vspace{-5mm}
		}
		\subfigure[Cellular delivery ratio.]
		{
			\includegraphics[width=0.45\columnwidth]{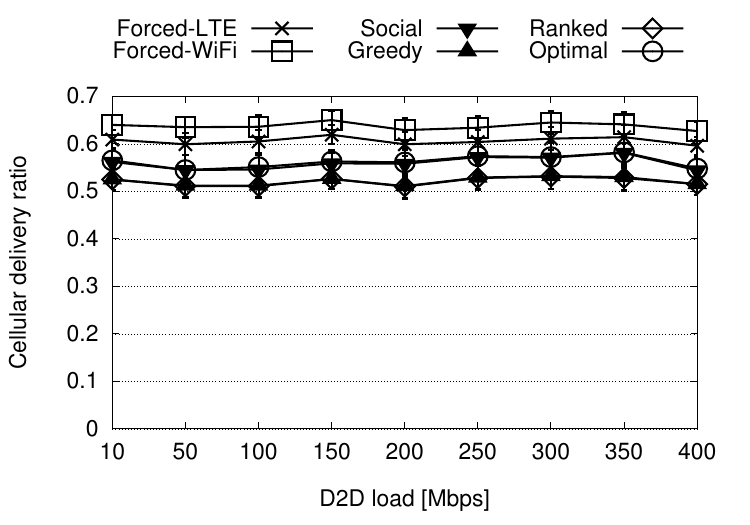} %[width=70mm,angle=0]
			\label{fig:loadCellDr}
		}
		\subfigure[Aggregate utility.]
		{
			\includegraphics[width=0.45\columnwidth]{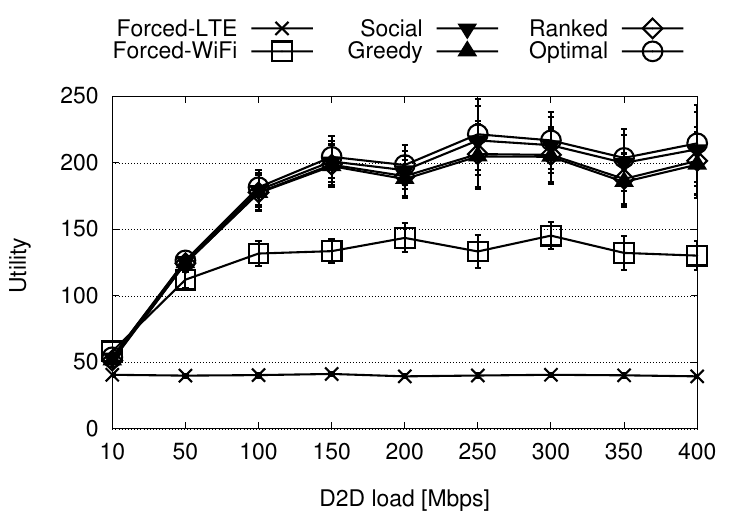} %[width=70mm,angle=0]
			\label{fig:loadUtil}
			%\vspace{-5mm}
		}
		\caption{The impact of D2D load on system performance evaluated through packet simulation ($ N = 50 $). }
		\label{fig:load}
\end{figure*}

\begin{figure*}
	\centering
		\subfigure[Aggregate achievable throughput.]
		{
			\includegraphics[width=0.45\columnwidth]{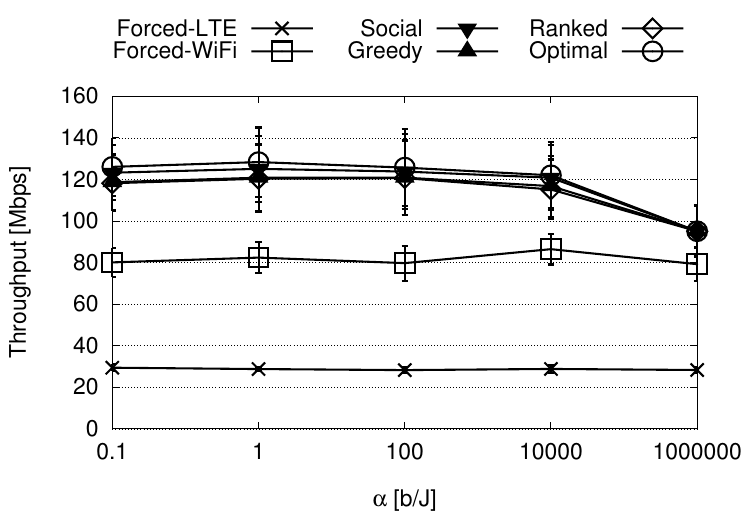} %[width=70mm,angle=0]
			\label{fig:alphaTput}
		}
		\subfigure[Aggregate achievable utility.]
		{
			\includegraphics[width=0.45\columnwidth]{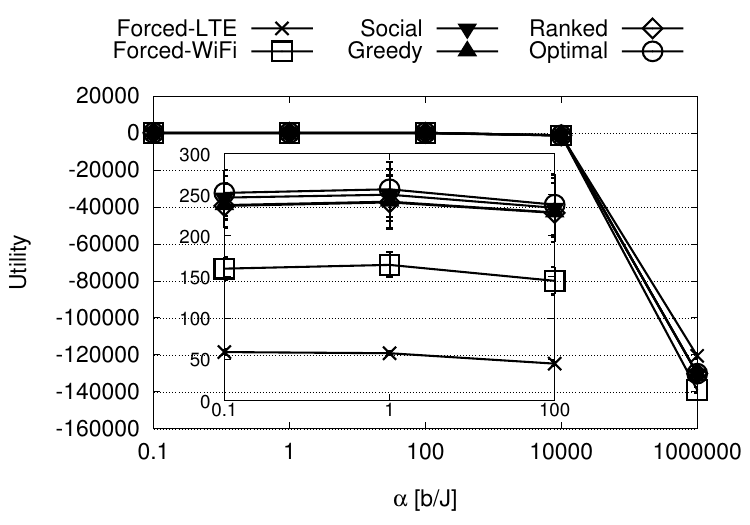} %[width=70mm,angle=0]
			\label{fig:alphaUtil}
			%\vspace{-5mm}
		}
		\caption{The impact of $\alpha$ on system throughput and utility $\left(  N = 50  \right)$.}
		\label{fig:alpha}
\end{figure*}

{\bf Impact of \ac{D2D} load.} The impact of \ac{D2D} load is shown in Figure~\ref{fig:load} for $ N = 50 $. The packet delivery ratio for \ac{D2D} users drops as the load increases, as shown in Figure~\ref{fig:loadD2dDr}. This is the expected behavior of systems in saturation. However, as we see in Figure~\ref{fig:loadCellDr}, our schemes are designed in such a way that saturation of \ac{D2D} users does not impact the cellular users. This shows that our proposal can be a candidate for distributed \ac{D2D} mode selection implementations in which {\it cellular users are protected from mode selection decisions of \ac{D2D} users}.
It is observed in Figure~\ref{fig:loadUtil} that system utility approaches its achievable limit ($220$) when the \ac{D2D} load is almost $250$ Mbps (see Fig~\ref{fig:totalUtility}, $N \!=\! 50$). In Fig~\ref{fig:totalTput}, we observe that the achievable capacity for $N \!= \!50 $ is almost $\!120\!$ Mbps. Indeed,  by multiplying the packet delivery ratios (see Figs.~\ref{fig:loadD2dDr} and \ref{fig:loadCellDr}) with the aggregate network load ($250\!$ Mbps for \ac{D2D} users and $\! 30 \!$ Mbps for cellular users), we observe that the achieved throughput is almost $120$ Mbps (i.e., $250 \cdot 0.4 + 30 \cdot 0.6 \!=\! 118$). With similar calculations, one finds that  \texttt{Forced-WiFi} saturates almost at $ 100 $ Mbps.

{\bf Impact of energy cost $\alpha$.} Recall that with the current relative energy cost $ \alpha $ our system is biased towards throughput. Hence, we
investigate the impact of $ \alpha$ in Figure~\ref{fig:alpha}, for $N=50$. We start with Figure~\ref{fig:alphaTput}, in which a $ 20\% $ throughput reduction is
observed at $\alpha = 10^6$ b/J. This shows the system's bias shifts towards energy minimization as $ \alpha $ increases. In Figure~\ref{fig:alphaUtil},
the utility reduces as $ \alpha $ grows, although the behavior of the curves is not linear at all. In particular, for very large values of $ \alpha $, our
system prefers \texttt{Forced-LTE} (i.e., mode $ 0 $) because it only powers one interface. Note that \ac{D2D} users need to keep the second interface in idle mode to be able to quickly switch among different modes. 
%Moreover, achieves higher utility than \texttt{Optimal}. This happens because the the energy consumption is the lowest when everyone is in LTE mode.
Since we disallow multi-band schemes to assign mode $ 0 $ to D2D users, \texttt{Forced-LTE} might
achieve utilities higher than that of \texttt{Optimal} when $ \alpha $ is very large (e.g., with $\alpha = 10^6$ b/J, which is
too unrealistic as of today and for the near future due to the high cost of electricity).

{\bf Convergence.} In Table~\ref{tb:conv}, we report the convergence of our proposed heuristics in terms of time and the number of iterations. The heuristics
are tested on Mathematica\texttrademark~on a machine with a $3.6$ GHz processor and $8$ GB memory.  \texttt{Greedy} and \texttt{Social} have a very similar
convergence time. \texttt{Greedy} is slightly slower than \texttt{Social} due to decision indexing. Interestingly, notwithstanding its ranking operations,
\texttt{Ranked} has a better performance. This happens because \texttt{Ranked} converges to a decision with less iterations (which is what we have indicated as
factor $r_i$ in Section~\ref{ss:complexity_analysis}). 
\begin{table}[h!]
\setlength\extrarowheight{2pt}
\caption{Convergence of the heuristics ($ N = 100$)}
\centering
%\scriptsize
\label{tb:conv}
%\begin{tabular}{|@{}c@{}| p{2.8in}@{} |}
%\resizebox{0.7\columnwidth}{!}{
\begin{tabular}{|c|c|c|c|}
\cline{2-4}
\multicolumn{1}{c|}{}				&\texttt{ Social } 	&\texttt{ Greedy } 	&\texttt{ Ranked } \\
\hline
\hline
 Average convergence time [s]		& 1.61 		 	& 1.62 		 	& 1.43  \\
\hline
Average number of iterations $r_i$ 		&  2.69  			& 2.80  			&  1.46    \\
\hline
\end{tabular}
%}

\end{table}

{\bf Flexibility.}
As mentioned, Floating Band \ac{D2D} is key to flexible \ac{D2D} architectures. We emphasize this fact by evaluating our proposal in various cellular environments, according to ITU-R guidelines~\cite{ITU2009}. Table~\ref{tb:flex} shows that, moving from micro-cell to rural macro-cell, the system relies more on the cellular spectrum as density reduces. As a consequence, the number of underlay connections increases. For denser environments, we observe that a significant part of connections is served using outband \ac{D2D}.

\begin{table}[h!]
\setlength\extrarowheight{2pt}
%\vspace{-1mm}
\caption{Percentage of each mode in different environments ($ N = 100$)}
\centering
%\scriptsize
\label{tb:flex}
%\begin{tabular}{|@{}c@{}| p{2.8in}@{} |}
%\resizebox{0.7\columnwidth}{!}{
\begin{tabular}{|c|c|c|c|c|}
\cline{2-5}
\multicolumn{1}{c|}{}			&Urban 				&Urban 				&Suburban 				&Rural \\
\multicolumn{1}{c|}{}			&micro-cell 			&macro-cell 			&macro-cell  				&macro-cell\\
\hline
\hline
Inband underlay 			& 4\% 		 		& 8\% 		 		& 29\% 				  	& 31\%\\
\hline
Inband overlay 			&  63\%   				& 66\% 				&  66\%  					& 67\%\\
\hline
Outband					&  33\%   				& 26\%  				&  5\%   					& 2 \%\\
\hline
\end{tabular}
%}
\end{table}

\section{Summary}
We have shown that the performance of D2D modes is highly scenario-dependent. Thus, the most convenient mode in one scenario, say in a macro-cell, could be a poor choice in another, say in a micro-cell. 
To cope with this issue, we proposed the {\it Floating Band D2D} framework along with practical heuristics suitable for quick and adaptive mode selection in such a complex setup. Unlike existing schemes, we allow D2D users to communicate over inband or outband modes, depending on network load and channel conditions. Our results demonstrate the impressive potentials of multi-band mode selection.
Remarkably, our simple heuristics result in fair operation and achieve near optimal performance by dramatically ameliorating network utility, which accounts
for both throughput and energy consumption.

%\acresetall
\chapter{\titSeventhCh}
\label{ch:tie}

\section{Introduction}
It has been shown that the cellular throughput can be dramatically improved by using opportunistic schedulers such as
MaxRate~\cite{knopp95ICC} and Proportional Fair~\cite{bender2000ComMag}.
The opportunistic schedulers proposed for cellular networks face a trade off between throughput and fairness when it comes to prioritizing the users based on their channel qualities~\cite{18,86}.
Hence, with the existing cellular architectures, opportunistic schedulers cannot achieve maximum throughput and fairness at the same time, unless all mobile users experience the very same channel quality~\cite{AyestaEtal2010time-varying}.

In contrast, in this chapter, we show how to evolve the cellular architecture by leveraging \ac{D2D} communications to achieve
maximum throughput and maximum fairness. In particular, we explore the possible gain from having \ac{D2D} connections within {\it clusters} of mobile users, as presented in Part I of this dissertation. 

%as shown in
%Figure~\ref{fig:single_cl_oth}, where each cluster is treated by the base station as a regular mobile user in a cell.
%We propose to change the normal cellular operation (e.g,. in LTE-A) as follows: at each scheduling frame, the scheduled mobile user is responsible for
%the traffic of its {\it entire} cluster, i.e., it acts as {\it cluster head}.
%The cellular traffic managed by the cluster head is then immediately exchanged within the cluster
%via \ac{D2D} communications on a secondary wireless interface (e.g., an 802.11-based protocol like WiFi-Direct or WiGig).
%Note that the cluster head is, in principle, {\it opportunistically} different at any frame, thereby achieving maximum throughput.

However, fairness is achieved in the following way.
The schedulers select some of the connections for transmission, and there are situations in which two or more users are in {\it tie},
i.e., they can be scheduled with the same transmission rate. These ties are usually ignored or broken randomly~\cite{neely2008TON}.
In contrast, we show that a smart {\it tie-breaking} strategy allows to compensate for the channel quality differences experienced by the
data connections which are active in the cellular network.

So far, nobody has investigated the possibility of enhancing fairness of opportunistic schedulers by
utilizing tie-breaking methods. Thus, we are the first to explore tie-breaking mechanisms for improving the  fairness achieved with a MaxRate
scheduler, though other opportunistic schedulers could also be enhanced with our approach.
We select the MaxRate scheduler since it maximizes system throughput, and we show that improved fairness levels can be achieved
without paying any throughput cost, i.e., {\it we study how to break ties in order to improve connection fairness while
maintaining the maximum cell throughput.}

\section{System model}
\label{s:systemS}
%\vspace{-1mm}

In this section we present our \ac{D2D}-based opportunistic scheduling system that can be leveraged to
improve both throughput and fairness in the system.

\subsection{Connections} %Basic
%\vspace{-1mm}

We consider a cellular network with a set $\setN$ of $ N $ persistent connections of users to a base station, using dedicated wireless
channels. In the following, we focus on downlink communications, though the model is applicable also to synchronized uplink
communications. The base station operates in a synchronous time-slotted way, and its task is to schedule connections for
transmission in every {\it frame} $ t = 0, 1, \dots $. We also assume that the base station has a queue for storing packets to
be delivered to each connection, and queues are never empty (fully backlogged assumption), so that we can evaluate the behavior
of the system in {\it saturation}, i.e., under the worst scheduling operational conditions.

Connection channels are \emph{heterogeneous}, i.e., {\it not} satisfying the i.i.d. assumption common to many works on cellular networks. The communication
channel for connection $ n $ is characterized by stationary Rayleigh fading.
Therefore, the \ac{SNR} for connection $ n $ can be described as a random process $ C_{ n } ( t ) $ with mean $ \gamma_{ n } $ and \ac{CDF} given by:
\begin{align}
\vspace{-1mm}
\label{e:CDF_user}
%	f_n(z) = \frac{1}{\gamma_n} e^{-\frac{z}{\gamma_n}} u(z),
%	\quad
	F_n(z) = 1 - e^{-\frac{z}{\gamma_n}}, \quad z \ge 0, \; \gamma_n \in \Gamma, \; n \in \setN;
\end{align}
where, for the sake of tractability, we have introduced $\Gamma$ as the discrete set of average values for the \ac{SNR} of connections.

We assume that the information available at the base station corresponds to the steady-state distribution of \ac{SNR}. Note that,
for practical systems, in which measured channel conditions form a discrete set, existing
patents~\cite{ChaponniereEtal2002patent,Bonald2004patent,AyestaJacko2013patent} propose to keep track of historical
observations of the \ac{SNR} of every connection in order to provide an estimate of the steady-state distribution of \ac{SNR}.
We assume that the selection of \ac{MCS} is perfect (i.e., transmissions are affected by negligible error rate), so that we ignore
retransmissions. Eventually, we consider a system with no power control, which is typical for realistic downlink
transmission schemes in LTE.

\subsection{Scheduling of Clusters of Users}
\vspace{-1mm}

%\begin{figure*}[!t]
%\centering
%	\includegraphics[scale=.7, angle=0]{./figs/fig01}
%	\caption{Channel quality heterogeneity affects the probability that perfect fairness can be achieved by means of MaxRate scheduling with non-random tie-breaking. Large heterogeneity practically prevents fairness.}
%	\label{fig:achievability01}
%\end{figure*}

In the previous discussion we had implicitly associated every connection with a single user. However, mobile users may form
clusters using cooperative \ac{D2D} communications, in which only one of the users, namely the {\it cluster head}, connects at a
given frame to the base station and relays traffic for the other users. %A cluster is formally defined as follows:

%\begin{defn}{\bf (Cluster)}
%A cluster is a group of mobile users that can communicate with each other using an acceptable data rate, typically more
%advantageously (in some sense that may depend on each user) than with the cellular base station. Only one cluster member,
%namely the cluster head, is allowed to receive data from the base station within each cellular transmission interval.
%The downlink traffic received at the cluster head can belong to any of the cluster members.
%\end{defn}

%{\red We emphasize that the composition of a cluster can vary in time, as some users may turn on/off their devices or migrate to
%another cluster or another cell.}

The scheduling algorithm is MaxRate, i.e., the cluster that contains the user with the highest \ac{MCS} is scheduled, so we propose
to operate clusters in an opportunistic way: $(i)$ the cluster head can change on a per-frame basis, as it is
opportunistically selected as the cluster member with the highest current \ac{MCS} rate; $(ii)$ an entire cluster is scheduled as an
individual user whose \ac{MCS} is the highest among members; and $(iii)$ the cluster head relays the downlink packets to the final
destination (intra-cluster communications) on a secondary wireless interface, using \ac{D2D} communications. Therefore, in this
chapter, a {\it connection} $n$ can be a user or a cluster (also indicated as CL$_n$) composed by $m_n$ mobiles, and its instantaneous \ac{SNR} is
the {\it highest} \ac{SNR} among the mobile users composing the cluster. In particular, the probability $p_{\text{CL}_n,k}$ that a scheduled connection
$ n $ (i.e., cluster CL$_n$) receives data encoded according to the $ k $-th \ac{MCS} can be computed based on the \ac{SNR} and the \ac{MCS} thresholds used in LTE (see Appendix~B).%~\ref{s:appendix_\ac{MCS}}).

%can be written similarly to
%Eq.\eqref{e:channel_state_prob} :
%%\begin{align}
%%\label{e:channel_state_prob_cluster}
%%p_{\text{CL}_n,k} & = \int_{c_k}^{c_{k+1}} dF_{\text{CL}_n}(z),
%%\end{align}
%where $F_{\text{CL}_n}(z)$ is the \ac{CDF} of the maximum of $m_n$ random variables representing the \ac{SNR} values of each of the
%$m_n$ mobiles forming cluster CL$_n$:
%\begin{equation}
%\label{e:\ac{CDF}_cluster}
%F_{\text{CL}_n}(z)   =  \prod_{j \in {\text{CL}_n} } F_j(z) =   \prod_{j \in {\text{CL}_n} }  \left (  1 - e ^ { - \frac { z } { \gamma _ j} } \right ), \; z \ge 0.
%\end{equation}

For simplicity of notation, we omit the ``CL'' index in the formulas in the reminder of this chapter, so that expressions like
$p_{n,k}$ and $F_n(z)$ can be equivalently used for connection $n$ and cluster CL$_n$. Similarly, we will use the notation
``connection $n$'' to address either the mobile user $n$ or the cluster CL$_n$, according to the context.

As detailed in Subsections~\ref{ss:cluster2}, \ref{ss:clusterN}, and~\ref{ss:clustering_gain}, scheduling clusters instead of users not only brings advantages in
terms of system throughput, but also in terms of fairness. However, we will mainly focus on inter-cluster fairness, since the
actual per-user fairness depends on the way resources are shared within a cluster. Accordingly, when we refer to per-user
throughput and fairness, we assume that cluster resources are divided equally between cluster members.
Note that user throughput unfairness due to heterogeneous channel qualities within the same cluster is smoothed by the adopted
cooperative \ac{D2D} communications mechanisms. 
%However, the particular mechanism to manage intra-cluster fairness is left out of the scope of this manuscript.
%Here, we rather focus on studying possible gains in throughput and in fairness among clusters.

\section{Maximal Fairness with MaxRate Scheduling}
\label{s:model}

Maximum throughput is achieved in our setting by using MaxRate, which in each frame transmits data to a connection with
the highest instantaneous \ac{SNR}, i.e., the process of selected connection $ A ( t ) $ must satisfy for every $ t $ that $ A ( t )
= n $ implies $ R_{ n } ( t ) \ge R_{ m } ( t ) $ for all connections $ m = 1,\dots, N $. Therefore, by definition, MaxRate
is throughput-optimal, and so is our proposal.

The objective of this section is to study when it is possible, and how, to achieve the {\it perfect} fairness given
that the scheduler be MaxRate. We focus on fairness in the sense of equalizing the expected time-average throughput across
connections, independently of their average channel quality. The only degree of freedom that the system offers to play with
fairness consists in the occurrence of {\it ties} in the scheduling mechanism, which is frequent in systems using only few
discrete \ac{MCS} values. This degree of freedom can be exploited by
designing a {\it tie-breaking rule} to employ if at least two connections compete for scheduled with the same highest
instantaneous transmission rate. Formally, we use the following definitions:

\begin{defn}{\bf (Best set $\cal M$ and best MSC $\kappa$)}
${\cal M}(t, \kappa(t))$ is the set of connections that can be scheduled with the $\kappa$-th \ac{MCS} in frame $t$, and $\kappa$ is the best \ac{MCS} that can be
used in the system in frame $t$, according to the \ac{SNR} of the connections. We will use $\cal M$ as short for ${\cal M}(t, \kappa(t))$.
\label{d:M}
\end{defn}

\begin{defn}{\bf (Tie)}
A tie occurs when, in a given frame $t$, two or more connections can be scheduled by the MaxRate mechanism with the same
\ac{MCS} $\kappa$, which is the best possible \ac{MCS} in the system at that scheduling epoch, that is: $|{\cal M}(t, \kappa(t))| > 1$.
\end{defn}

\begin{defn}{\bf (Tie-breaking)}
A tie-breaking mechanism is a procedure to select exactly one connection $i \in {\cal M}(t, \kappa(t))$ to be scheduled when a tie occurs at time $t$.
\end{defn}

%Note that, in systems with discrete \ac{MCS} values, ties can occur with high probability, and legacy schedulers in such systems
%often use randomization as a predefined tie-breaking rule \cite{neely2008TON}.

In what follows, we first examine the multiple-connection case and show that it is intractable to solve exactly. Then we study the two-connection case in detail, for which we provide complete answers. Using the outcome of the analysis for two-connections case, we develop the fundamental intuition for tie-breaking. In Section~\ref{s:mrmf_multi}, inspired by this intuition, we design low complexity heuristics for the multi-connection case. Subsequently, we show how clustering can be beneficial in achieving perfect fairness via tie-breaking, without paying in terms of throughput. 

In many of the arguments below we will rely on the fact that the expected long-term fairness (throughput distribution over an indefinitely long interval) is
equivalent to the expected one-slot fairness (average per-slot throughput distribution), due to the stationarity channel assumption we made earlier.

\subsection{Analysis of the Multiple-Connections Case}
\label{s:model_multi}

Let us denote by $ \displaystyle Q_{ n, k } := \sum_{ l = 1 }^{ k - 1 } p_{ n, l } $ the probability that connection $ n \in \setN $ has an \ac{MCS} strictly worse
than $ k $. Note that $ Q_{ n, 1 } = 0 $. Let $ h_{ n } \in \{ 0 , 1 \} $ denote whether the current \ac{MCS} of connection $ n $ is higher than or equal to the
current \ac{MCS}' of all the other connections ($ h_{ n } = 1 $) or not ($ h_{ n } = 0 $). Then, vector $ \vech := \left( h_{ n } \right)_{ n \in \setN } $
identifies with elements $ 1 $ precisely those connections that are allowed to transmit under MaxRate (only one of them is transmitting at any moment). Note
that, by definition, $ \vech = \veczero $ cannot happen.

We can now define
\begin{align}
\label{e:Rh}
R^{ \vech } &:= \sum_{ k = 1 }^{ K } \left( r_{ k } \prod_{ n : h_{ n } = 1 } p_{ n, k } \prod_{ n : h_{ n } = 0 } Q_{ n, k } \right),
\end{align}
which represents the expected (both one-slot and time-average)  cumulative throughput of the system in situation $ \vech $ (i.e., when exactly the connections specified by $\vech $ are in a tie situation). $R^{\vech}$ is therefore the ``tie throughput'' to be shared between connections for which $h_n=1$.

The number of $ 1 $'s of $ \vech $ is denoted by $ L_{ 0 } ( \vech ) $ (so-called \emph{zero ``norm''}). If $ L_{ 0 } ( \vech ) = 1 $, then there is a single
connection with highest \ac{MCS}, so it will be scheduled. If $ L_{ 0 } ( \vech ) \ge 2 $, then there are several connections in a tie, and the scheduler must
decide who to serve. Without loss of generality this can be done randomly. Then we need to define $ L_{ 0 } ( \vech ) $ parameters $ 0 \le \alpha_{ n }^{
\vech } \le 1 $ for every connection $ n $ such that $ h_{ n } = 1 $, denoting the probability of serving connection $ n $ in situation $ \vech $.

Let us denote by $ \setH_{ 1 } := \{ \vech : L_{ 0 } ( \vech ) \ge 1 \} $ the set of all vectors $ \vech \neq \veczero $. Then the aggregate throughput of the
system under MaxRate scheduler is equal to $ \sum_{ \vech \in \setH_{ 1 } } R^{ \vech } $, and the perfectly fair share is thus $ R^{ * } := \sum_{ \vech \in
\setH_{ 1 } } R^{ \vech } / N $. We further denote by $ \setH_{ 2 } := \{ \vech : L_{ 0 } ( \vech ) \ge 2 \} $ the set of all vectors representing ties
(of at least two connections). Then achieving perfect fairness means that the following equalities hold:
\begin{align}
R^{ \vece_{ n } } + \sum_{\vech \in \setH_{ 2 } : h_{ n } = 1 } \alpha_{ n }^{ \vech } R^{ \vech } &= R^{ * } &&\text{ for all } n \in \setN \\
\sum_{ n \in \setN } \alpha_{ n }^{ \vech } &= 1 &&\text{ for all } \vech \in \setH_{ 2 }
\end{align}
where $ \vece_{ n } $ is the unit vector with $ 1 $ at $ n $-th position, and $ 0 $'s otherwise, representing the situation when connection $ n $ is the
unique connection achieving the highest transmission rate.

It is easy to calculate that there are $ 2^{ N } - 1 $ vectors belonging to $ \setH_{ 1 } $ and $ 2^{ N } - N - 1 $ vectors belonging to $ \setH_{ 2 } $.
Therefore, there are $ 2^{ N } - 1 $ constraints (out of which one is redundant), while having $ N 2^{ N - 1 } - N $ unknowns, which is significantly more
(except for $ N = 2 $).

We can reformulate the above problem formally as a linear programming (LP) problem: %primal
\begin{align}
\max 0 &&& \\
\sum_{\vech \in \setH_{ 2 } : h_{ n } = 1 } \alpha_{ n }^{ \vech } R^{ \vech } &= R^{ * } - R^{ \vece_{ n } } &&\text{ for all } n \in \setN \label{restriccion1}\\
\sum_{ n \in \setN : h_{ n } = 1 } \alpha_{ n }^{ \vech } &= 1 &&\text{ for all } \vech \in \setH_{ 2 } \label{restriccion2}\\
\alpha_{ n }^{ \vech } &\ge 0 &&\text{ for all } \vech \in \setH_{ 2 } \text{ and } n \in \setN \text{ such that } h_{ n } = 1 \label{restriccion3}
\end{align}

The constant objective ($ \max 0 $) indicates that we are in fact interested in finding whether there is a feasible solution satisfying all the constraints.
Because of the non-negativity of $ \alpha_{ n }^{ \vech } $ (cf. \eqref{restriccion3}), feasibility of \eqref{restriccion1} necessarily requires having $ 0 \le
R^{ * } - R^{ \vece_{ n } } $ for all $ n $. Moreover, it is easy to see that every $ \vech \in \setH_{ 2 } $ gives one necessary condition (by adding up
\eqref{restriccion1} for all $ n $ such that $ h_{ n } = 1 $, and simplifying using \eqref{restriccion2} for all vectors $ \vecg \le \vech $ and using
\eqref{restriccion3} for the remaining unknowns). For instance, a 2-connection tie $ \vech = \vece_{ n , m }$ gives
\begin{align}
R^{ \vece_{ n , m } } \le ( R^{ * } - R^{ \vece_{ n } } ) + ( R^{ * } -  R^{ \vece_{ n } } ).
\end{align}
%This means that connections $ n $ and $ m $ together cannot get more than fair throughput on their own and from their $ 2 $-connection tie.
This means that the tie throughput associated to ties of connection $ n $ and $ m $ cannot be higher than what $n $ and $m$ need to reach fairness.

In order to find a feasible solution if there is any (assuming $ R^{ * } - R^{ \vece_{ n } } \ge 0 $ for all $ n $), we
consider the following relaxed LP-associated problem in which we allow for partial utilization of tie throughput:

\begin{align}
\max \sum_{\vech \in \setH_{ 2 } } \sum_{ n \in \setN : h_{ n } = 1 } \alpha_{ n }^{ \vech }  &&& \\
\sum_{\vech \in \setH_{ 2 } : h_{ n } = 1 } \alpha_{ n }^{ \vech } R^{ \vech } &\le R^{ * } - R^{ \vece_{ n } } &&\text{ for all } n \in \setN \\
\sum_{ n \in \setN : h_{ n } = 1 } \alpha_{ n }^{ \vech } &\le 1 &&\text{ for all } \vech \in \setH_{ 2 } \\
\alpha_{ n }^{ \vech } &\ge 0 &&\text{ for all } \vech \in \setH_{ 2 } \text{ and } n \in \setN \text{ such that } h_{ n } = 1
\end{align}
Obviously, a feasible initial solution to this problem is $ \alpha_{ n }^{ \vech } = 0 $ for all $ \vech \in \setH_{ 2 } $ and $ n \in \setN $ such that $ h_{
n } = 1 $, and can be solved using standard LP algorithms (simplex method, interior-point, etc.). Note that the optimal objective value is equal to $ 2^{ N } -
N - 1 $ (i.e., the number of vectors belonging to $ \setH_{ 2 } $) if the original LP problem is feasible.

\subsection{Analysis of the Two-Connections Case}
\label{ss:2users}

If there are many connections, the size of the problem becomes too large to be solved at milliseconds scale in a real base station. Therefore, we analyze the case of two
connections with the aim to get more insight into the problem in order to design heuristics. The two-connections case can be analyzed and solved without the need of using numerical methods. To make the notation more intuitive, we denote the following quantities:
\begin{align}
\label{e:R1}
R^{ ( 1 ) } &:= \sum_{ k = 2 }^{ K } r_{ k } p_{ 1, k } Q_{ 2, k }, \\
\label{e:R2}
R^{ ( 2 ) } &:= \sum_{ k = 2 }^{ K } r_{ k } p_{ 2, k } Q_{ 1, k }, \\
\label{e:RX}
R^{ ( X ) } &:= \sum_{ k = 1 }^{ K } r_{ k } p_{ 1, k } p_{ 2, k },
\end{align}
which represent the expected (both one-slot and time-average) transmission rates in the following three cases: Eq.~\eqref{e:R1}
expresses the rate of connection~$ 1 $ when it has an \ac{MCS} strictly better than connection~$ 2 $; Eq.~\eqref{e:R2} is for
connection~$ 2 $ having an \ac{MCS} strictly better than connection $ 1 $; and Eq.~\eqref{e:RX} is for the case of tie.
Note that the aggregate throughput of the system under MaxRate scheduler is equal to $ R^{ ( 1 ) } + R^{ ( 2 ) } + R^{ ( X ) }
$.

In the following proposition we give a sufficient and necessary condition for a scheduler that achieves both maximal throughput
and fairness.

\begin{proposition}
\label{prop:Existence}
The MaxRate scheduler can achieve both one-slot and time-average fairness if and only if
\begin{align}
\left| R^{ ( 1 ) } - R^{ ( 2 ) } \right| \le R^{ ( X ) }. \label{e:cond2}
\end{align}
\end{proposition}

The proof is given in Appendix~B. % \ref{s:appendix}. 
Here, it is worth to discuss when such a condition might hold. Indeed, there are
some intuitive sufficient conditions stated next, which are independent of the transmission rates $ r_{ k } $.

\begin{proposition}
\label{prop:suffcond}
The MaxRate scheduler can achieve both one-slot and time-average fairness if any of the following conditions hold:
\begin{enumerate}

\item $ p_{ 1, k } = p_{ 2, k } $ for all $ k \ge 2 $ (i.e., the channels of the two connections are statistically equal);

\item $ | p_{ 1, k } Q_{ 2, k } - p_{ 2, k } Q_{ 1, k } | \le p_{ 1, k } p_{ 2, k } $ for all $ k \ge 2 $;

\item $ p_{ 1, k } \ge p_{ 2, k } $ for all $ k \ge 2 $ and $ p_{ 2, K } \ge 1/2 $;

\end{enumerate}
\end{proposition}

The proof is presented in Appendix~B. %~\ref{a:p2}.  
Moreover, there may be weaker conditions which make it likely
that fairness be achievable. For instance, if one of the following conditions holds, perfect fairness is achievable:
\begin{enumerate}

\item $ p_{ 1, k } p_{ 2, k } $ is large enough for all $ k $ large enough;

\item $ | p_{ 1, k } Q_{ 2, k } - p_{ 2, k } Q_{ 1, k } | $ small enough for all $ k $ large enough;

\item probabilities $ p_{ n, k } $ for all $ k $ large enough are approximately equal for the two connections;

\item the expression $ p_{ 1, k } Q_{ 2, k } - p_{ 2, k } Q_{ 1, k } $ often changes sign as $ k $ grows.

\end{enumerate}
Finally, taking into account that transmission rates $ r_{ k } $ grow somewhat exponentially with $ k $ (see
Table~\ref{tb:MCS}), it is much more important that the two connections be statistically similar in the upper \ac{MCS} range rather
than in the lower \ac{MCS} range.

%CAN WE SAY HOW LIKELY IT IS THAT A RANDOMLY GENERATED TWO-USER SYSTEM WILL GET ALPHA BETWEEN 0 AND 1?

Let us define now the \emph{MaxRate-MaxFair scheduler} for two connections, by introducing a bias in the tie-breaking rule of
the MaxRate scheduler as follows:
\begin{defn}\label{d:mrmf}{\bf (MaxRate-MaxFair scheduler)}
In case a tie occurs under MaxRate scheduling, serve connection $ 1 $ with probability $ \alpha^{ ( X ) } $ and serve connection $ 2 $ with probability $ 1
- \alpha^{ ( X ) } $, where
\begin{align}
\alpha^{ ( X ) } := \frac{ 1 }{ 2 } + \frac{ R^{ ( 2 ) } - R^{ ( 1 ) } }{ 2 R^{ ( X ) } }.
\label{e:alphaX}
\end{align}
Moreover, if $ \alpha^{ ( X ) } \notin [ 0, 1 ] $, then it is not a proper probability value, thus we cut such values off:
\begin{equation}
\label{e:alphaX_norm}
	\begin{cases}
		\text{if } \alpha^{ ( X ) } < 0  \text{ then } \alpha^{ ( X ) } := 0;\\
		\text{if } \alpha^{ ( X ) } > 1  \text{ then } \alpha^{ ( X ) } := 1.	
	\end{cases}
\end{equation}
\end{defn}

The following proposition establishes when $ \alpha^{ ( X ) } $ is a proper probability value, so that no cut-off is needed.
The proof is immediate, therefore we omit it.

\begin{proposition}
Condition \eqref{e:cond2} is equivalent to $ \alpha^{ ( X ) } \in [ 0, 1 ] $ as defined in \eqref{e:alphaX}.
\end{proposition}

Using MaxRate-MaxFair, connection $1$ receives $R^{(1)} \!+\!  \alpha^{ ( X ) } R^{ ( X ) }$, while the throughput of connection $2$
is  $R^{(2)} \!+\! (1\!-\! \alpha^{ ( X ) }) R^{ ( X ) }$.
Such a throughput distribution is the fairest possible, and the aggregate is maximum, as stated in the following Proposition, which
is the main result of this section and validates the MaxRate-MaxFair name of the above scheduler.

\begin{proposition}\label{prop:Scheduler2users}
If \eqref{e:cond2} holds, then the MaxRate-MaxFair scheduler achieves maximum throughput, and both one-slot and time-average
fairness. If \eqref{e:cond2} does not hold, then the MaxRate-MaxFair scheduler achieves maximum throughput, and the difference
between individual throughputs is the minimum achievable with tie-breaking schemes.
\end{proposition}

The proof is presented in Appendix~B. %~\ref{a:p4}. 
According to Proposition~\ref{prop:Scheduler2users}, when
condition~\eqref{e:cond2} does not hold, the scheduler still achieves maximum throughput, but will not be perfectly fair
anymore. Nevertheless, the difference between individual throughputs will be the minimum possible, and may significantly
outperform randomized tie-breaking. Note that, from the proof of Proposition~\ref{prop:Scheduler2users}, it follows that, using
the Jain's fairness index as metric~\cite{jain1984}, the MaxRate-MaxFair scheduler achieves the smallest possible distance from the perfectly
fair throughput distribution. In fact, the Jain's fairness index is maximized when differences are minimized. The result is
formalized in the following corollary.

\vspace{-1mm}

\begin{corollary}
MaxRate-MaxFair scheduler achieves the highest possible Jain's fairness index achievable
by means of any tie-breaking mechanism.
\end{corollary}

\begin{figure} [t!]
\begin{center}
%\vspace{-0.6cm}
	\includegraphics[scale=0.3, angle=0]{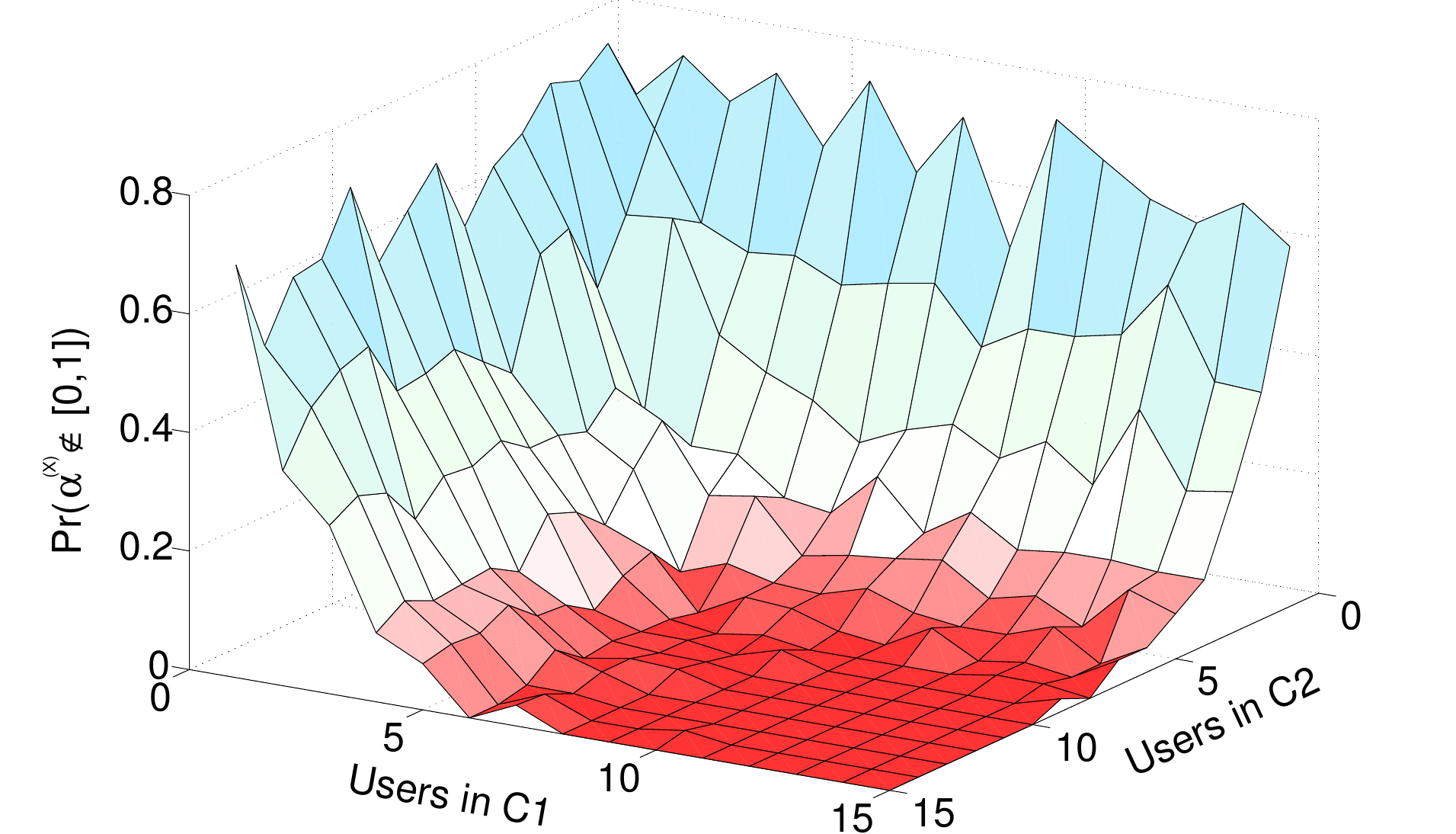} %[width=70mm,angle=0]
%\vspace{-2mm}
	\caption{The probability that perfect fairness cannot be achieved, i.e., $\alpha^{(X)} \notin [ 0, 1 ]$, vanishes as the cluster sizes grow.}
%	\vspace{-0.9cm}
	\label{fig:alpha-tie}
\end{center}
\end{figure}

%\subsection{Impact of Cluster Composition on the Two-Connection Case}
\subsection{Impact of Cluster Composition}
\label{ss:cluster2}

Under MaxRate scheduling, both user-based or cluster-based scheduling is throughput-optimal. However, the
advantage of clustering  is twofold: $(i)$ opportunistic cluster head selection yields higher channel qualities used to transmit in the system, which contributes to equalize transmission rates among connections (note that efficiency curves increase logarithmically according to Shannon's results); and $(ii)$ clustering offers the possibility to re-distribute the throughput among cluster members, thus yielding potentially higher fairness levels. We exemplify this effects by considering the specific case of MaxRate-MaxFair with two clusters. Each cluster can be regarded as a single connection, and the exact analysis of Subsection~\ref{ss:2users} applies.
%consists in the possibility to re-distribute the cluster throughput
%not only among clusters, but also
%among cluster members, thus yielding potentially higher fairness levels.
%We exemplify this effect by considering the specific case of {\it MaxRate-MaxFair}
%with two clusters of possibly different sizes. Each cluster can be regarded as a single connection, and the exact analysis of
%Subsection~\ref{ss:2users} applies. Specifically, thanks to clustering with opportunistic cluster head selection, \ac{SNR}
%statistics of scheduled connections of both clusters are improved, which decreases the probability of $\alpha ^{ (X) } $
%falling out of range $[0, 1]$.
% as cluster sizes increase.

Figure~\ref{fig:alpha-tie}, which is the result of $20,000$ random instances for two clusters of random size and composition, is in
line with our intuition on the effect of cluster size on  $ \alpha ^{ (X) } $. The figure shows the probability that $ \alpha
^{ (X) } $ be outside the acceptable range $[0,1]$, when the perfect fairness can not be achieved. The figure reveals that
perfect fairness can be achieved almost surely when both clusters consist of more than $5$ users. The probability of fairness
non-achievability radically increases as the cluster size drops below $ 5 $ users, since the average cluster qualities can be
very unbalanced and yield large $|R^{(1)}-R^{(2)}|$. In contrast, when clusters are large enough (i.e., with $5-10$
members), the fact that each cluster head exhibits the highest \ac{SNR} in its cluster makes the probability to use the best \ac{MCS}
practically $1$, and thus  $|R^{(1)}-R^{(2)}|$ approximates $0$, while increasing the probability of ties. Therefore,
condition~\eqref{e:cond2} is met with high probability.
%due to the lack of clustering gain (see, Subsection~\ref{ss:clustering_gain}).

\subsection{How Much Throughput Lies in Ties?}
\label{ss:tieThrouput}
\begin{figure} [t!]
\begin{center}
%\vspace{-0.2cm}
\subfigure[The probability that $n$ users are in a tie.]
	 {
	\includegraphics[scale=0.55, angle=0]{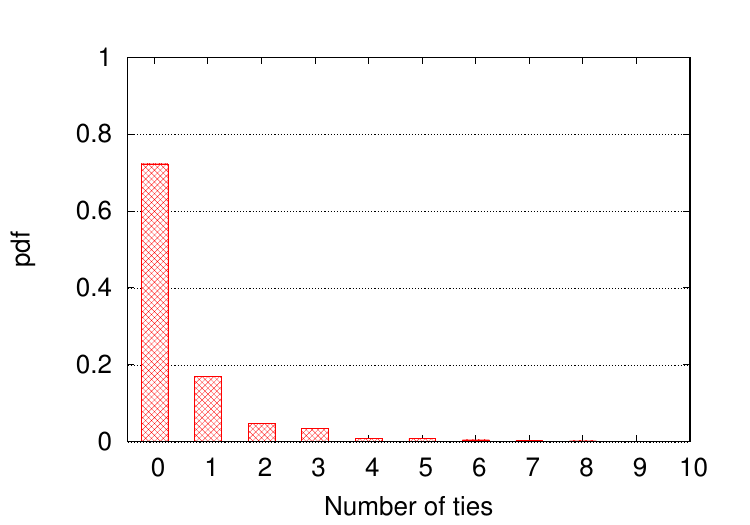} %[width=70mm,angle=0]
	\label{fig:ties}
}
%\hspace{2cm}
\subfigure[The probability of ties in different MCS indices.] 
	{
	\includegraphics[scale=0.55, angle=0]{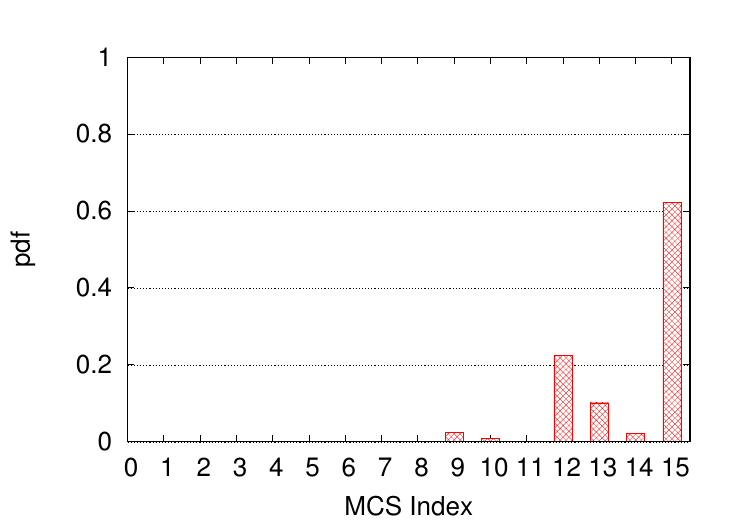} %[width=70mm,angle=0]
	\label{fig:mcsTies}
}
%\vspace{-0.2cm}
\caption{Results based on traces obtained within one working day (without clustering). }
%\vspace{-0.8cm}
	\label{fig:traces}
\end{center}
\end{figure}

In previous sections, we observed that providing fairness by leveraging scheduling ties can be challenging and complex. Now the question remains: is there enough gain in smart tie-breaking to justify the complexity? In order to answer this question, we setup a measurement campaign with $15$ mobile users that are subscribed to the same service provider and spread over an area of $500$ m$^2$. We use an android application (i.e., G-MoN\footnote{http://www.wardriving-forum.de/wiki/G-MoN}) to record the channel qualities of the smartphones in short intervals (one second) during a working day at the office. Note that given the good coverage in our premises, the users generally experience high channel quality in this experiment. Figure~\ref{fig:traces} illustrates the results of our measurements. We can observe the pdf of having ties of different size in Figure~\ref{fig:ties} (e.g., $0$ when there is no tie and $1$ for a tie between two users). The results show that with a $0.28$ probability, there is a tie between at least two users in the network. This is a motivating observation that highlights the significance of ties in cellular networks. Next, Figure~\ref{fig:mcsTies} shows the pdf of the ties versus the \ac{MCS} in which a tie occurs. Again, the measurements reveal that the majority of the ties occur in high \ac{MCS} levels where there is higher throughput in ties.

\begin{figure} [t!]
\begin{center}
%\vspace{-0.2cm}
\subfigure[The probability that $n$ users are in a tie.]
	 {
	\includegraphics[scale=0.55, angle=0]{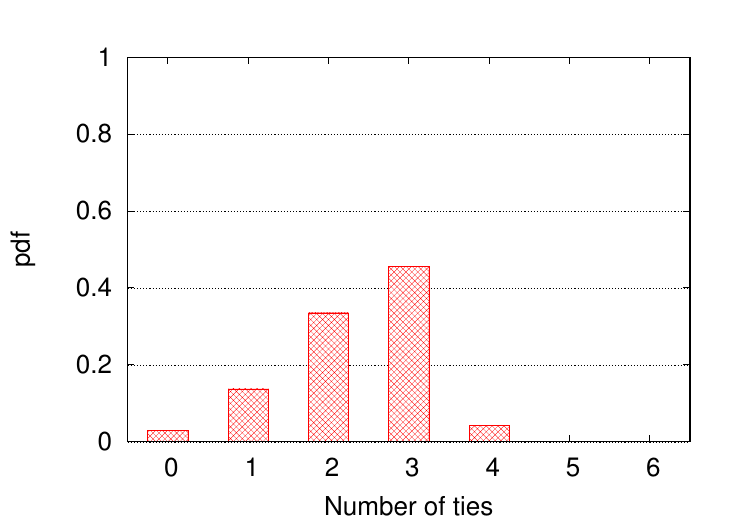} %[width=70mm,angle=0]
	\label{fig:tiesCL}
}
%\hspace{2cm}
\subfigure[The probability of ties in different MCS indices.] 
	{
	\includegraphics[scale=0.55, angle=0]{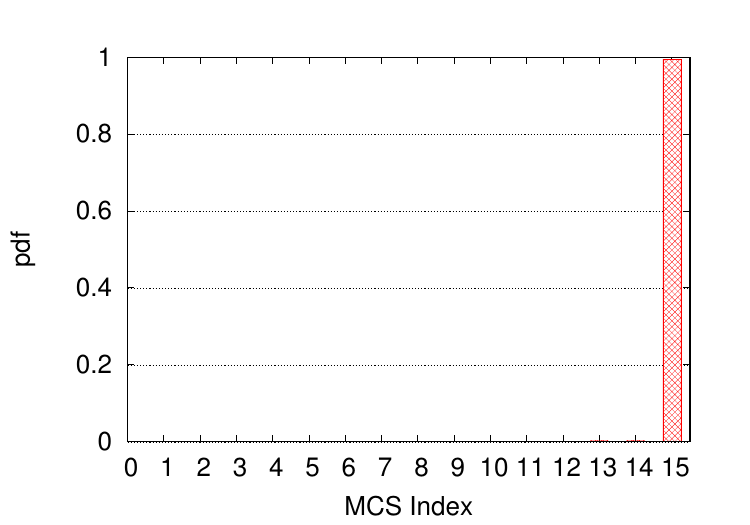} %[width=70mm,angle=0]
	\label{fig:mcsTiesCL}
}
%\vspace{-0.2cm}
\caption{Results based on traces obtained within one working day (with five clusters). }
%\vspace{-0.8cm}
	\label{fig:tracesCL}
\end{center}
\end{figure}

In order to observe the impact of clustering, we divided the users into five clusters (i.e., $3$ users in each cluster) based on their physical proximity. In this experiment, we fed the same measurements into a script which applies the aforementioned clustering concept on the traces. In particular, the script looks at the traces of each cluster at each time interval and assigns the user with the highest \ac{MCS} as the cluster head. Figure~\ref{fig:tracesCL} is produced using the results obtained from the clustering script that demonstrates impact of clustering. It can be seen in Figure~\ref{fig:tiesCL} that clustering significantly increases the probability of having ties. Moreover, the \ac{MCS} in which a tie occurs is shifted to the right which maximizes the throughput in ties (see Figure~\ref{fig:mcsTiesCL}).

%\vspace{-1mm}
\section{Heuristics to Achieve Maximal Fairness with MaxRate Scheduling of Multiple Connections}
\label{s:mrmf_multi}
%\subsection{Multiple-user Case}
%\vspace{-1mm}

%SHOULD THIS SUBSECTION BE ON userS OR USERS? MAYBE IF IT IS ON USERS, THEN GROUPS APPEARING IN
%THE HEURISTICS CAN BE SOLD AS  THE NATURAL WAY OF userING. THEN MOVE BEFORE EVALUATION SECTION

Having developed fundamental intuition based on exact results for the case of two connections, we now set out to design MaxRate schedulers for
generic number of connections N in a set $\cal N$, achieving better fairness than with randomized tie-breaking. Note that, as discussed in Section~\ref{s:model_multi}, extending the analytical approach from the
two-connection case would require $ N 2^{ N\! -\! 1 }\! - N $ tie-breaking parameters for all possible ties of $2, 3,\!\dots\!, N$ connections, which grows too
fast to be implementable ($ 2, 6, 28, 75,\!\dots $\!). Instead, we focus on scalable solutions that rely on at most $ N $ tie-breaking parameters.

\subsection{WRR Tie-breaking}

We design a scheduler in which tie-breaking is resolved as if Weighted Round Robin (WRR) was implemented.
%, which is analogous to the generalized processor sharing (GPS).
Thus, we assume that for each connection $ n $ there is a non-negative parameter $ \alpha_{ n } $ used as follows.

\begin{defn}
{\bf (WRR Tie-breaking)}
If, at a given scheduling epoch $t$, $ {\cal M } $ is the set of connections that are currently tied in
the highest \ac{MCS} (see Definition~\ref{d:M}),
%such that no other connections are in upper MCS's,
then the probability (or, the average fraction of time) that connection $ n \in {\cal M } $ is served in such situations is as follows:
\begin{align}
\label{e:wrr}
\frac{ \alpha_{ n } }{ \displaystyle \sum_{ m \in {\cal M } } \alpha_{ m } }, \quad \alpha_m \ge 0 \quad \forall m \in {\cal M}.
\end{align}
The expected throughput of connection $ n $  under MaxRate with WRR tie-breaking is then:
\begin{align}
\sum_{ k = 1 }^{ K } r_{ k } \sum_{ {\cal M } \ni n } \left[ \frac{ \alpha_{ n } }{ \displaystyle \sum_{ m \in {\cal M } } \alpha_{ m } } \prod_{ m \in {\cal M } } p_{ m, k } \prod_{ m \notin {\cal M } } Q_{ m, k } \right], \label{e:tput}
\end{align}
where $ {\cal M } \ni n $ denotes any set of connections that includes $ n $.
\label{d:wtb}
\end{defn}
Using \eqref{e:tput} it is, however, intractable to obtain values of $ \alpha_{ n } $ which would equalize the expected
individual throughput of all connections, since it leads to a system of non-linear equations. Hence, in the following we
propose heuristics to obtain $ \alpha_{ n } $.

%\textbf{Heuristic 1: All-For-One (AFO).}
\subsubsection{Heuristic 1: Best Leaf First (BeLF)}

This method is based on the results of the two-connection case, and on the use of binary trees. We have shown in
Section~\ref{ss:2users} the exact way to compute $\alpha$ for two connections in a tie. Now, two connections can represent two
mobile users, as well as two clusters. More in general, the approach of Section~\ref{ss:2users} is valid for any two {\it
groups} of users for which an \ac{SNR} \ac{CDF} is available to compute the probabilities to use the various \ac{MCS} values.
%
%For instance, if we were interested in the fairness between two groups of users, without further considerations about the fairness between each group member, we could have defined those two groups as two users whose channel quality corresponded to the maximum of the channel qualities within the group members.
Therefore, we can use the results presented in Section~\ref{ss:2users} to compute the optimal tie-breaking probability for any
two disjoint and not empty user groups covering the entire set $\cal N$, say subsets ${\cal N}_{1,0} \neq \emptyset$ and ${\cal
N}_{1,1} \neq \emptyset$, ${\cal N}_{1,0} \cup {N}_{1,1} = {\cal N}$, ${\cal N}_{1,0} {\cap N}_{1,1} = \emptyset$. Let us call
$\beta_{1,0}$ and $\beta_{1,1} = 1-\beta_{1,0}$ the tie-breaking probabilities of ${\cal N}_{1,0}$ and ${\cal N}_{1,1}$,
respectively. These priorities are computed as per Eqs.~\eqref{e:alphaX} and \eqref{e:alphaX_norm} given in
Definition~\ref{d:mrmf}. Any of the subsets grouping at least two users can be further split into two subsets, e.g., if $|{\cal
N}_{1,0}| \ge 2$, $\exists {\cal N}_{2,0} \neq \emptyset, {\cal N}_{2,1} \neq \emptyset$, for which ${\cal N}_{2,0} \cup
{N}_{2,1} = {\cal N}_{1,0}$, ${\cal N}_{2,0} {\cap N}_{2,1} = \emptyset$. To these smaller subsets, we can associate again two
tie-breaking probabilities $\beta_{2,0}$ and $\beta_{2,1} = 1 - \beta_{2,0}$, computed as per Definition~\ref{d:mrmf}. Each
subset with at least two users can be recursively split into two subsets, and each subset receives a tie-breaking probability
$\beta_{i,j}$, where $i$ is the level of recursion, and $j$ is a sequential index within a recursion level. This binary
splitting procedure can be represented with a binary tree, as shown in Figure~\ref{fig:GT}. The tree root $n_{0,0}$ represents
the entire network ${\cal N}$, and its tie-breaking probability is formally set to $1$. Leaves represent connections (either
individual mobile users or clusters), and each node $n_{i,j}$ at level $i$ in the tree represents the group of connections that
appear as leaves in the branches originating at that node. We finally associate a WRR priority $\alpha_n$ to each of the $N$
leaves of the tree: since each $\beta_{i,j}$ represents a conditional tie-breaking probability (given that there is a tie
between two groups),  the WRR priority of a connection is computed from the corresponding leaf as the product of the
$\beta_{i,j}$ values on the path from the root to the leaf. With the above procedure, the sum of WRR priorities $\alpha_n$ is
exactly $1$, so that they can be directly interpreted as tie-breaking probabilities for a multi-connection case.

The proposed heuristic might not work well when two leaves in the same branch are associated to users with very different
average channel qualities, e.g., in very heterogeneous network conditions.
%In fact, the proposed binary tree scheme will
%compute priorities, for nodes that are not leaves, based on groups of connections.
This effect is due to the adoption of an expression similar to Eq.~\eqref{e:CDF_cluster} for the computation of the \ac{SNR} \ac{CDF} associated to a node in the tree, i.e., the \ac{SNR} of a node in the tree corresponds to the highest \ac{SNR} among the users (leaves) connected to that node.
In particular, at a given node of the binary tree, the presence of a branch without users with statistically good channel, namely {\it good} users, is kept ``hidden'' by the
presence of another branch departing from the same node in which {\it good} users are present. Thus, qualitatively speaking,
the grouping mechanism described here fails in distinguishing groups that differ in the number of poorly performing members.

\subsubsection{Heuristic 2: Worst Leaf First (WoLF)}

\begin{figure} [t!]
\begin{center}
%\vspace{-0.2cm}
\subfigure[Computation of $\alpha_n$ with BeLF.]
	 {
	\includegraphics[scale=0.53, angle=0]{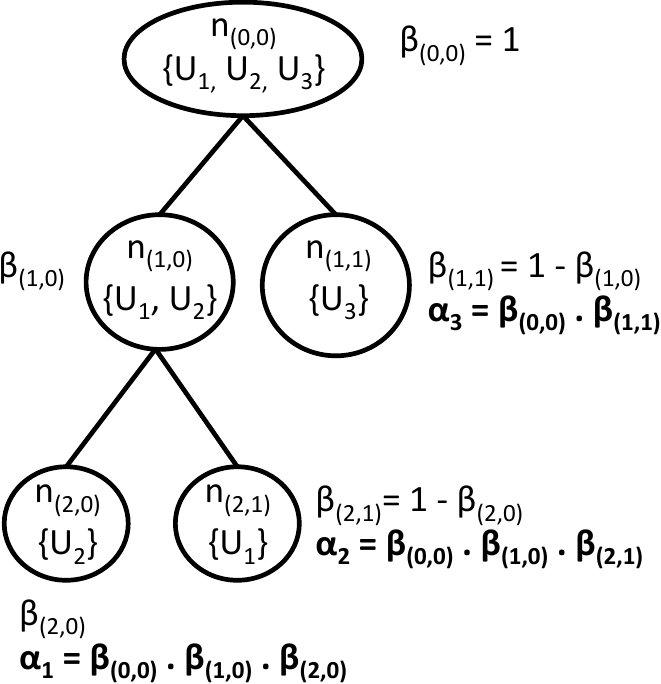} %[width=70mm,angle=0]
	\label{fig:GT}
}
\hspace{2cm}
\subfigure[Computation of $\alpha_n$ with WoLF.] {
	\includegraphics[scale=0.53, angle=0]{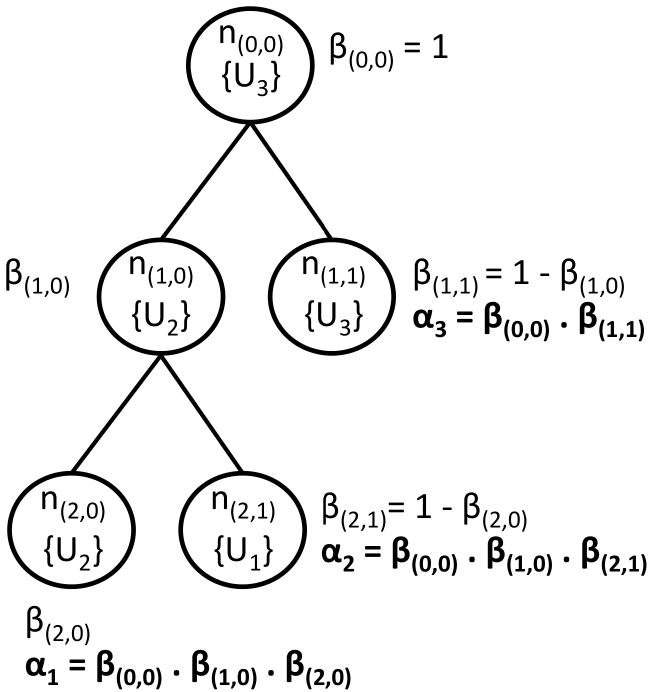} %[width=70mm,angle=0]
	\label{fig:MT}
}
%\vspace{-0.2cm}
\caption{Example of BeLF and WoLF with $3$ users (numbers in brackets represent users, $U_1$ being the best user and $U_3$ being the worst). }
%\vspace{-0.8cm}
	\label{fig:binarytree}
\end{center}
\end{figure}

The previous discussion suggests an alternative way of building priorities by means of a binary tree.  In particular, one could
proceed as for the BeLF heuristic, but consider each node in the tree as a group represented by  the {\it worst} channel
quality among all users associated to leaves on the branches originating at that node (see Figure~\ref{fig:MT}).
%{\bfcyan [SHOULD WE GIVE A FORMULA TO COMPUTE THE \ac{SNR} \ac{CDF} FOR THIS CASE?]}
Therefore, the presence of a user with statistically poor
channel on a leaf provokes a shift in the distribution of priorities towards the branch that contains that user.

\subsubsection{Heuristic 3: Fair Individual Share (FISh)}

To compute $\alpha_n$ for each connection $n$ in the system as the $ \alpha^{ ( X ) } $ of a
two-connection system (see Eq.~\eqref{e:alphaX}) in which connection $ n $ competes with the rest of connections.
%, i.e., the group representing all users but $n$.
In this case, we assume that the target of connection $n$ is to achieve a portion $ 1 / N $ of the cell throughput.

Following the ideas from the two-connection case, we first define $ Q_{ -n, k } $ as the \ac{CDF} for the best \ac{MCS} of all the
connections except connection $ n $, $ \displaystyle Q_{ -n, k } := \prod_{ m = 1, m \neq n }^{ N } Q_{ m, k } $, and $ p_{ -n
, k } $ as the probability that at least one of the connections (except connection $ n $) is in MCS $ k $ and no other
connection is in a better \ac{MCS}, formally, $ p_{ -n , k } = Q_{ -n, k + 1 } - Q_{ -n, k } $. Then, \eqref{e:R1} to \eqref{e:RX}
can be rewritten as follows, for $ n = 1 \dots N $:
\begin{align}
R^{ ( n ) } &:= \sum_{ k = 2 }^{ K } r_{ k } p_{ n, k } Q_{ -n, k },\\
R^{ ( -n ) } &:= \sum_{ k = 2 }^{ K } r_{ k } p_{ -n, k } Q_{ n, k },\\
R_n^{ ( X ) } &:= \sum_{ k = 1 }^{ K } r_{ k } p_{ n, k } p_{ -n, k }.
\end{align}
The total cellular throughput of MaxRate is
$ R_{Tot} = R^{ ( n ) } + R^{ ( -n ) } + R_n^{ ( X) } $,
which is the same for all values of $ n $, as it can be easily verified.

\begin{proposition}
\label{prop:ExistenceMulti}
The MaxRate scheduler for a system with $N \ge 2$ connections can achieve both one-slot and time-average fairness if and only if
\begin{align}
\left| R^{ ( n ) } - R^{ ( -n ) } \right| \le R^{ ( X ) }, \quad \forall n=1 \dots N. \label{e:cond_multi}
\end{align}
\end{proposition}
The proof of the Proposition~\ref{prop:ExistenceMulti} derives from the proof of Proposition~\ref{prop:Existence}.

The value of the tie-breaking probability $ \alpha_n $ is computed from the following equation:
\begin{align}
R^{ ( n ) } + \alpha_n R_n^{ (X) } = \frac{1}{N}  R_{Tot}.
\end{align}

The resulting value of $ \alpha_n $ is then:
\begin{align}
\label{e:alpha}
\alpha_n = \frac{ 1 }{ N } + \frac{ \sum\limits_{ k = 1 }^{ K } r_k \left[ p_{ -n, k } Q_{ n, k } - \left( N - 1 \right) p_{ n, k } Q_{ -n, k } \right] } { N \sum\limits_{ k = 1 }^{ K } p_{ n, k } p_{ -n, k } r_{ k } }.
\end{align}

However, when the expected transmission rate of connection $ n $ is strictly higher than the average fair
individual share (i.e., $ R^{ ( n ) } > \frac{1}{N} R_{Tot} $), then $\alpha_n$ is negative, which is not acceptable for the
WRR scheduling mechanism. This corresponds to a situation in which the amount of resources in ties are not enough to equalize
the connection throughputs without loss of throughput maximality.
Therefore, we propose the following transformation which
preserves the order of $\alpha_n$, i.e., preserves the {\it priority list} among connections:
\begin{align}
\label{eq:FISh}
\alpha_{ n } := \alpha_{ n } - \min_{ m \in {\cal N } } \alpha_{ m }, \quad \forall n \in {\cal N}.
\end{align}

%\begin{proposition}
%\label{prop:maxfair_multi_fish}
%The MaxRate scheduler for a system with $N \ge 2$ users, adopting FISh with the WRR priorities $\alpha_n$ given by
%Eqs.~\eqref{e:alpha} and \eqref{eq:FISh} achieves higher fairness than a random tie-breaking strategy.
%\end{proposition}

Note that, since we use $ \alpha_n $ values as described in \eqref{e:wrr}, the proposed transformation is equivalent to
normalizing the values of $ \alpha_n $ in the interval $[0,1]$,  as if using $ \alpha_{ n } := \frac { \displaystyle \alpha_{ n
} - \min_{ m \in {\cal N } } \alpha_{ m } } { \displaystyle \max_{ m \in {\cal N } } \alpha_{ m } - \min_{ m \in {\cal N } }
\alpha_{ m } } $. We remark that, in practice,  we do not need to enforce any transformation if some $ \alpha_{ n } > 1 $,
since WRR normalizes such values.

\subsubsection{Heuristic 4: Priority Keying (PIKe)}

Forcing the values $\alpha_{ n }$ in the interval $[0,1]$ might result in one or more connections not benefitting from
tie-breaking at all (i.e., connections with $\alpha_{ n } = 0$). However, considering that $\alpha_{ n }$ represents the excess
throughput received by connection $n$, setting $\alpha_{ n } = 0$ should be allowed only for connections receiving more than
the fair share, i.e., connections for which $ R^{ ( n ) } > \frac{1}{N} R_{Tot} $. Therefore, we propose a modified version of
FISh, namely PIKe, in which priorities $\alpha_{ n }$ are shifted only if negative values are present:
\begin{align}
\label{eq:PIKe}
\text{if } \min_{ m \in {\cal N } } \alpha_{ m } < 0, \quad \text{ then }  \quad \forall n \in {\cal N}, \quad \alpha_{ n } := \alpha_{ n } - \min_{ m \in {\cal N } }
 \alpha_{ m }.
\end{align}

As for the case of FISh, the proposed transformation is equivalent to normalizing the values of $ \alpha_n $ in the interval
$[0,1]$. However, the transformation is operated only when there exist negative values of $\alpha_n$.

%{\red
%\begin{proposition}
%\label{prop:maxfair_multi_pike} The MaxRate scheduler for a system with $N \ge 2$ connections, adopting PIKe with the WRR
%priorities $\alpha_n$ given by Eqs.~\eqref{e:alpha} and \eqref{eq:PIKe} achieves a fairness level at least as good as the one
%achieved with a random tie-breaking strategy.
%\end{proposition}
%\begin{proof}
%The value $ \alpha_n $ is proportional to the difference $ \frac{1}{N}  R_{Tot} - R^{ ( n ) } $, which represents the excess
%(possibly negative) throughput achieved by node $n$ with respect to the fair share point. Therefore, using $\alpha_n$,
%connections can be sorted in descending order, from the highest value of $\alpha_n$, corresponding to the connection with the
%worst average channel quality, to the lowest, corresponding to the connection with the best average channel quality. Since the
%throughput of a connection $n$ increases with $\alpha_n$, and our scheme reduces or maintains constant the gap between the
%received throughout and the fair individual share point of every connection, thereby our scheme ensures that fairness level in
%the system cannot decrease with respect to a random tie-breaking strategy. {\bf [NOT SURE WHETHER THIS IS CORRECT]}
%\end{proof}
%}

In Section~\ref{s:validation}, we will quantify the level of fairness achieved in the system with the four proposed heuristics.
We will also show that, on average, the PIKe heuristic performs better than the others.
Although we will use static clustering scenarios to illustrate the advantages of our proposal, we remark that our methodology and findings
apply to clusters whose composition varies in time, as some users may turn on/off their devices or migrate to another cluster or another cell.
In particular, we note that in presence of {\it flows} with limited duration, probabilities $p_{n,k}$ described in Session~\ref{s:systemS} can be
adapted to represent the steady-state  probabilities of flow $n$ given the arrival and flow-size distributions.

%\begin{figure*}[!t]
%\centering
%	\includegraphics[scale=.7, angle=0]{./figs/fig01}
%	\caption{UNCLEAR TEXT AND X-AXIS Channel quality heterogeneity affects the probability that perfect fairness can be achieved by means of MaxRate scheduling with non-random tie-breaking. Large heterogeneity practically prevents fairness.}
%	\label{fig:achievability01}
%	\includegraphics[scale=.7, angle=0]{./figs/fig02}
%	\caption{UNCLEAR TEXT AND X-AXIS Clustering increases the probability that perfect fairness can be achieved by means of MaxRate scheduling with non-random tie-breaking (clusters with fixed number of users and user's mean \ac{SNR} uniformly distributed in the range $[7, 23] \, \textit{dB}$).}
%	\label{fig:achievability02}
%\end{figure*}

\begin{figure*}[!t]
	\centering
	\begin{minipage}[t]{0.49\textwidth}
	\begin{minipage}[t]{0.95\textwidth}	
		\includegraphics[scale=.55, angle=0]{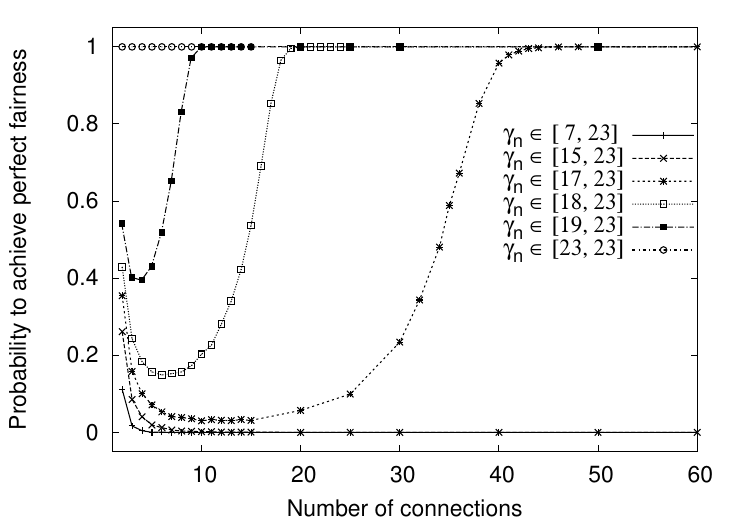}
		\vspace{-4mm}
		\caption{Probability to achieve perfect fairness with MaxRate without clustering under different levels of connection quality heterogeneity
				(under different ranges for $\gamma_n$).}
		\label{fig:achievability01}
	\end{minipage}
	\end{minipage}
	\begin{minipage}[t]{0.49\textwidth}
	\begin{minipage}[t]{0.95\textwidth}	
		\includegraphics[scale=.55, angle=0]{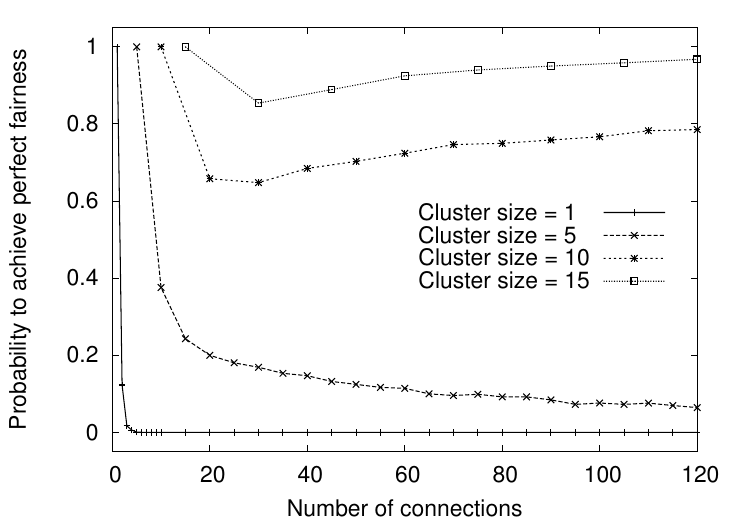}
		\vspace{-4mm}
		\caption{Probability to achieve perfect fairness with MaxRate with different cluster sizes, under large connection quality heterogeneity
			($\gamma_n \in [7, 23] \, \textit{dB}$).}
		\label{fig:achievability02}
	\end{minipage}
	\end{minipage}
%	\vspace{-5mm}
\end{figure*}

\subsection{Impact of Clustering on WRR Tie-breaking with Multiple Connections}
\label{ss:clusterN}
%\vspace{-1mm}

Perfect fairness could be impossible to achieve under MaxRate scheduling due to the heterogeneity of user's channel qualities
(see Propositions~\ref{prop:Existence} to \ref{prop:Scheduler2users}).
However, as mentioned earlier, clustering {\it
enough} users, i.e., as few as $5-10$ mobile users, in practice, causes a very high probability to use the highest \ac{MCS} value only.
Therefore, clustering reduces the heterogeneity of the channel quality as observed by a scheduled connection (i.e., a cluster
head).

To appreciate the impact of channel heterogeneity and clustering on a system with multiple connections, we depict in
Figures~\ref{fig:achievability01} and \ref{fig:achievability02} the probability to achieve perfect fairness as a function of
connection quality distribution and cluster size, for a variable number of connections in the system.
Figure~\ref{fig:achievability01} shows simulation results using the MaxRate scheduler when the mean \ac{SNR} $\gamma_n$ of user $n$ is
picked from a uniform distribution.

Different intervals for $\gamma_n$ are considered in the figure, to show the impact of
different degrees of channel heterogeneity.
For each interval of $\gamma_n$, we tested $10,000$ random instances of $N \in \{1\dots60\}$ connections, and checked whether
any tie-breaking strategy could lead to perfect fairness or not ({\it brute force search of the optimal tie-breaking}). Observing Figure~\ref{fig:achievability01}, we can deduce
that, depending on the interval in which the mean \ac{SNR} can range, having a few tens of users in the cell can enable perfect fairness
via tie-breaking. However, under typical heterogeneous conditions, in which the range for the mean \ac{SNR} $\gamma_n$ is several {\it dB} units,
the probability to achieve perfect fairness is almost zero even when the number of users per cell is very high.
Therefore, reducing heterogeneity in channel qualities is key to achieve fairness.

The impact of clustering on such quality heterogeneity is shown in Figure~\ref{fig:achievability02} for large
levels of heterogeneity ($\gamma_n \in [7, 23] \, \textit{dB}$). The Figure
illustrates that clustering heterogenous users results in increased probability to
achieve perfect fairness under MaxRate scheduling.

%Specifically, the Figure reports simulation results for perfect fairness
%achievability when the mean \ac{SNR} ranges in a large interval, $\gamma_n \in [7, 23]\,\textit{dB}$.
Notably, not using clustering makes the probability to achieve fairness practically negligible
(see Figure~\ref{fig:achievability02} when Cluster size is $1$). In contrast, using small clusters (as few as $5$ users)
dramatically increases the probability to achieve perfect fairness from $\sim\! 0\%$ to $10\%$ or more when the number
of users in the cell ranges from $20$ to $80$. Larger cluster sizes (e.g., $10$
or $15$) further boost perfect fairness achievability to $70\%$ or $90\%$ with a reasonable number of users in the system.
Therefore, the potential impact of clustering on the fairness performance is paramount.

\section{Evaluation}
\label{s:validation}

In this section, we validate via numerical evaluation of our \ac{D2D}-based cluster scheduling proposals. We use the Jain's fairness index to compare the
effectiveness of our \ac{D2D}-based schemes as compared to ET and PF scheduling with unclustered mobile users. As for the throughput, we
normalize throughput results in terms of cell capacity. Since our work does not investigate intra-cluster mechanisms, 
assume that the total cluster throughput can be equally shared among cluster members, using cooperative \ac{D2D} communications. PF performance figures
are obtained by simulating a scheduling process in which the average user throughput is computed with an autoregressive filter
with exponential time constant equal to $1000$ frames. However results computed with time constant in the range $50$ to $5000$
do not significantly differ.

For the sake of tractability, in what follows we assume that mobile users belong to one of three predefined \ac{SNR} {\it classes},
which correspond to {\it poor}, {\it average}, and {\it good} average \ac{SNR}, i.e., set $\Gamma$ contains three elements only. The designated  average \ac{SNR} for different classes are chosen
in a manner that the mean achievable rates for \textit{poor}, {\it average}, and \textit{good} users are $20\%$, $50\%$, and
$80\%$ of the maximum transmission rate achievable in the system, respectively. Therefore, with the \ac{MCS} values reported in
Table~\ref{tb:MCS} and the assumed Rayleigh fading model, the average \ac{SNR} values to be used in Eq.~\eqref{e:channel_state_prob} for
\textit{poor, average} and \textit{good} users are $\gamma_n = 7\,\textit{dB},\, 16\,\textit{dB},\, 23\,\textit{dB}$,
respectively.

\subsection{Impact of Clustering on System Performance}
\label{ss:clustering_gain}
\vspace{-1mm}

The potential clustering gain versus the conventional cellular architecture is discussed here. First, we illustrate the incentive for
clustering with simple numerical calculations that show the average clustering gain. We assume that MaxRate schedules clusters and breaks
ties at random (i.e., any cluster/connection has the same probability to be selected when a tie occurs). We refer to this particular version of MaxRate as MR. Second, we evaluate throughput and fairness performance.
We evaluate the impact of clustering against two baseline schedulers. The first scheduler is \emph{Equal Time} (ET), a simple
and largely deployed round robin scheduler which guarantees the same fraction of airtime to each user. The second scheduler
that we consider here is PF. In the figures presented in this subsection,  results are averaged over $2000$ random instances, and
user qualities are uniformly distributed among {\it poor}, {\it average}, and {\it good}.

\begin{figure} [t!]
\begin{center}
	\includegraphics[scale=0.65, angle=0]{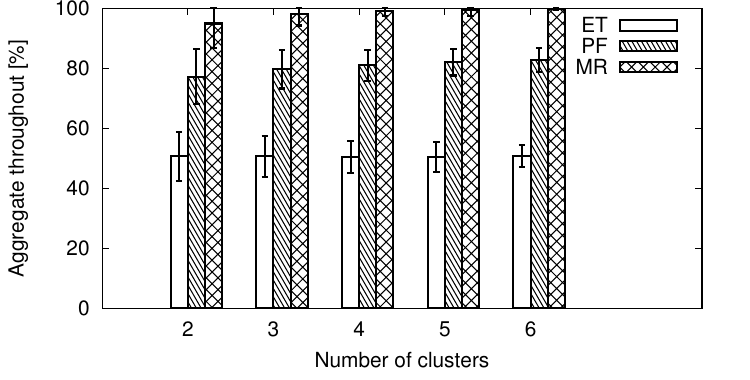}
	\caption{Throughput comparison (average and standard deviation) of clusters of $1$ to $10$ users, with uniform quality distribution.}
	\label{fig:T}
\end{center}
\end{figure}

\begin{figure} [t!]
\centering
\subfigure[Throughput per user.] {
	\hspace{-0.7cm}
	\includegraphics[scale=0.28, angle=-90]{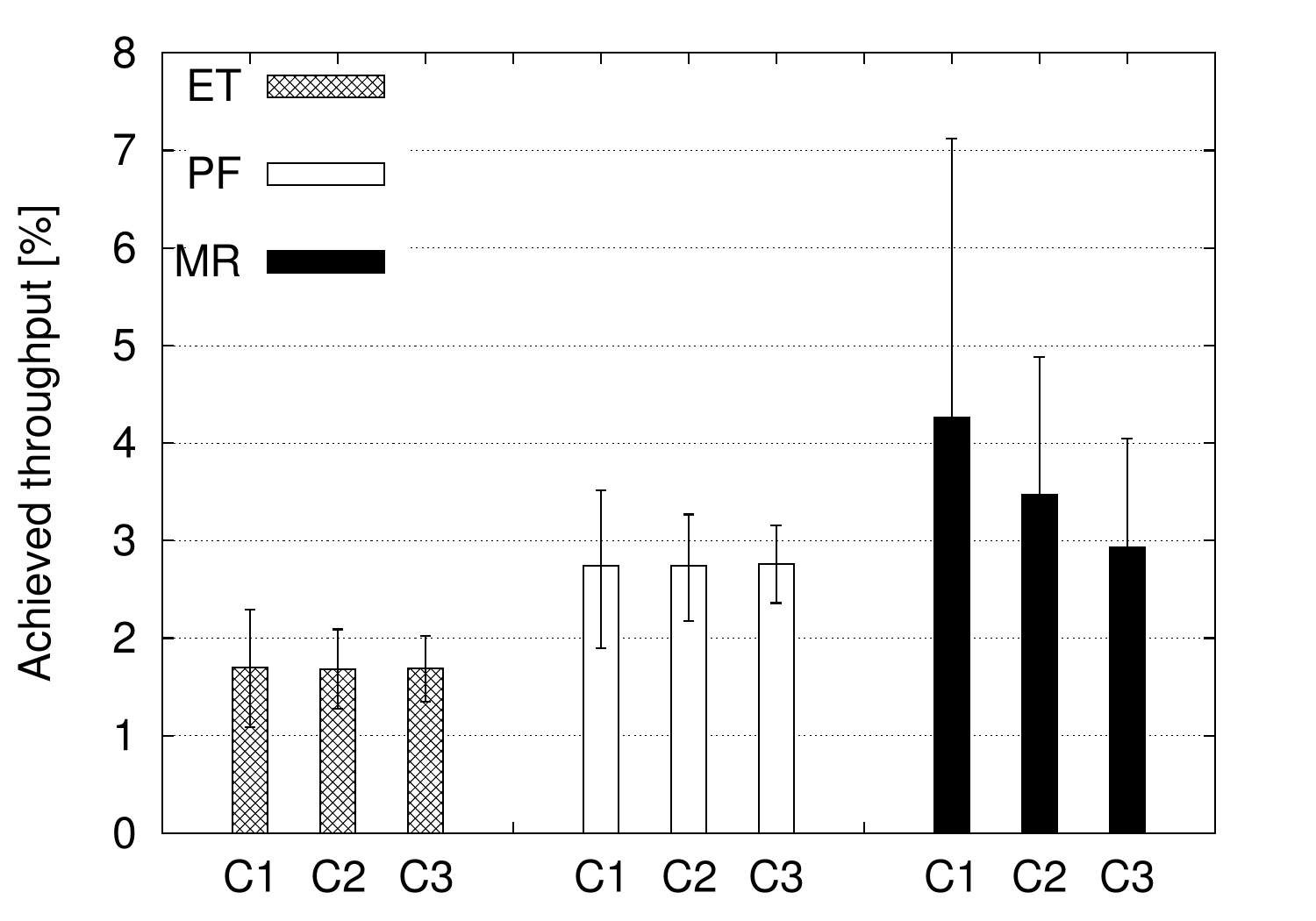} %[width=70mm,angle=0]
	\label{fig:1bs_user_tput}
}
\subfigure[Throughput per cluster.] {
	\hspace{-0.5cm}
	\includegraphics[scale=0.28, angle=-90]{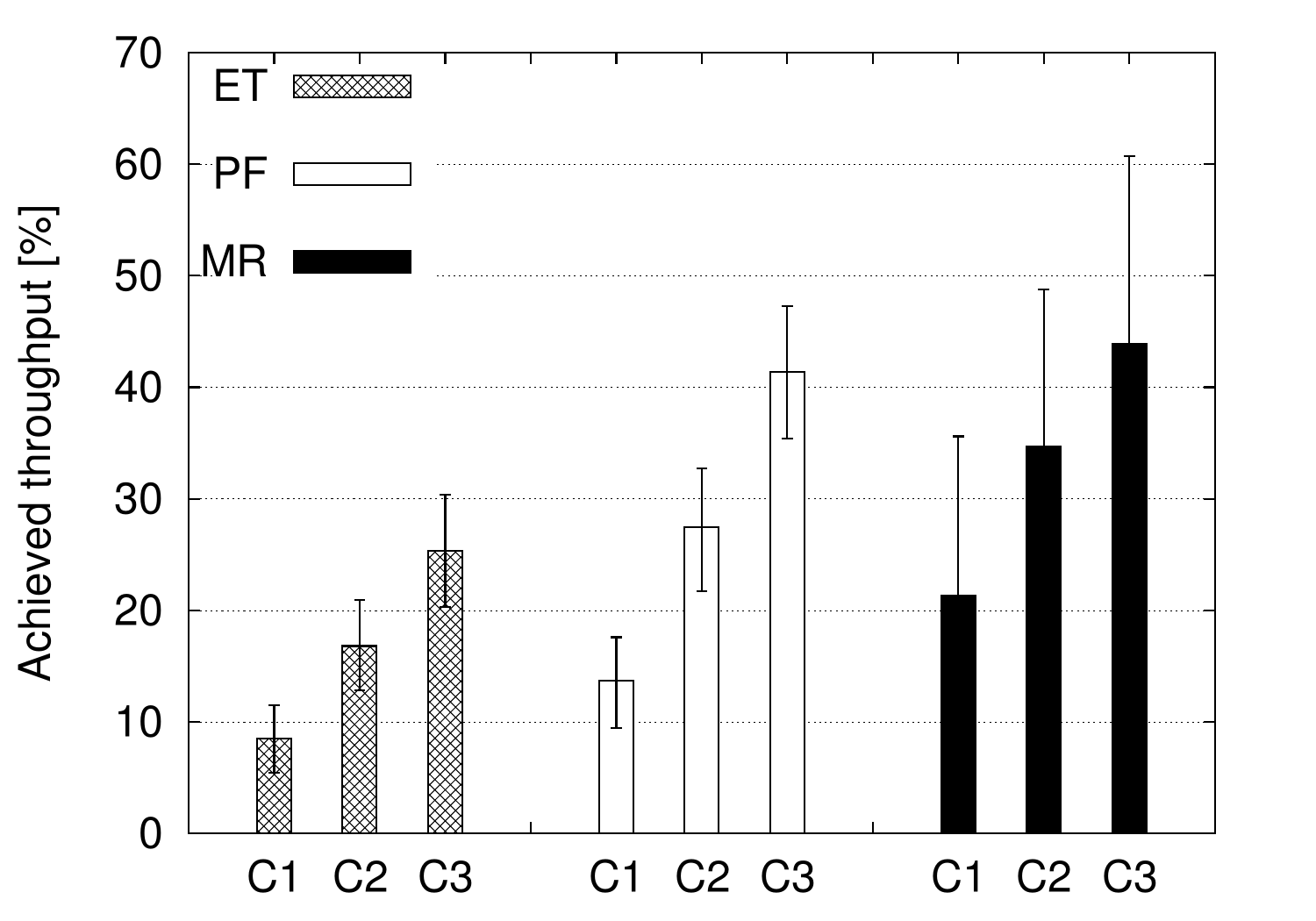} %[width=70mm,angle=0]
	\label{fig:1bs_cluster_tput}
	\hspace{-0.8cm}
}		
\vspace{-1mm}
\caption{Throughput under different schedulers (average plus $5th$ and $95th$ percentiles),
assuming resources are divided equally among cluster members, and ties are broken at random. }
\label{fig:1bs}
\vspace{-2mm}
\end{figure}
\begin{figure}[t!]
\centering
\vspace{-2mm}
\subfigure[User fairness index under different schedulers.]		
{
	\hspace{-0.4cm}
	\includegraphics[scale=0.30, angle=-90]{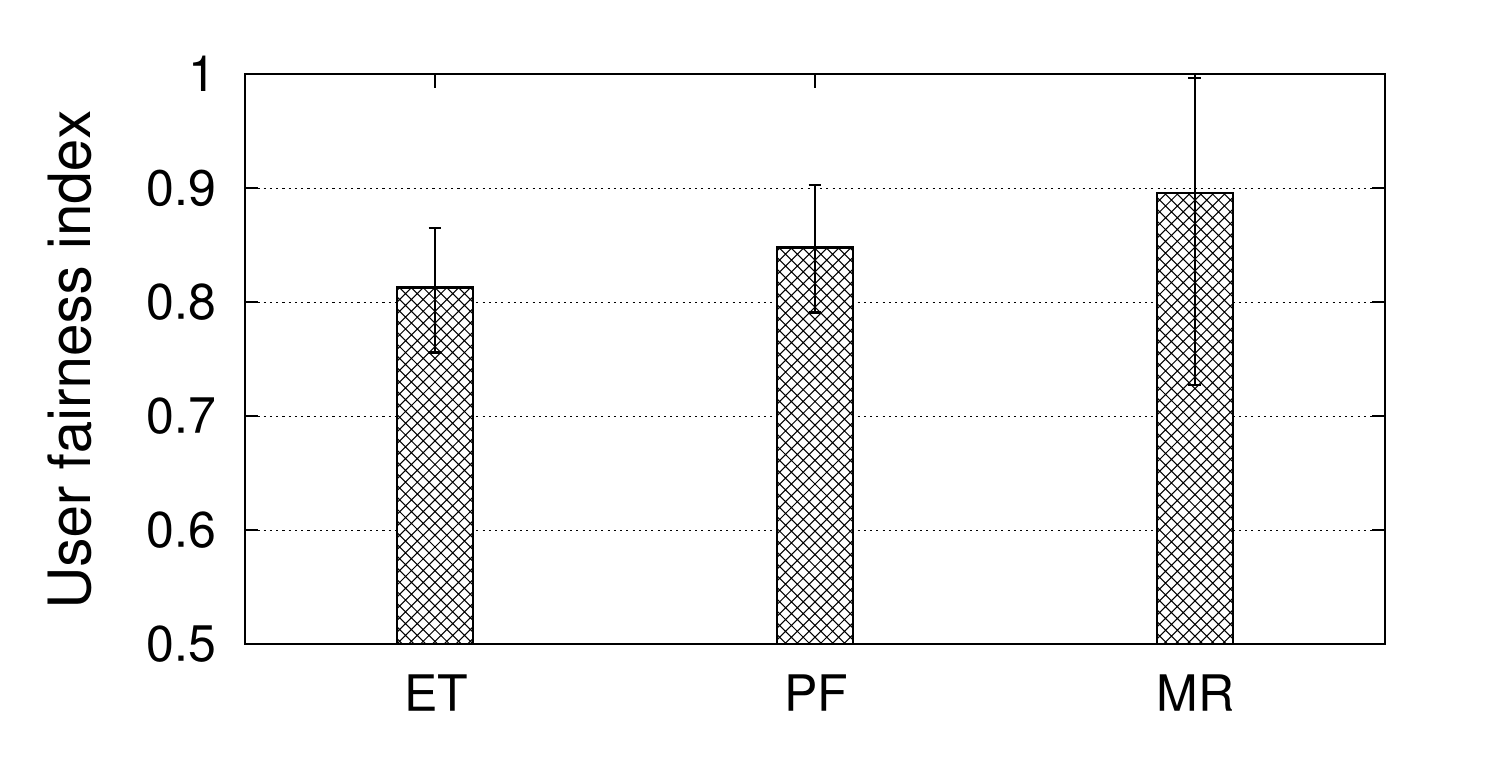} %[width=70mm,angle=0]
	\label{fig:user_fairness}
}
\subfigure[Cluster fairness index under different schedulers.]
{
	\hspace{-0.5cm}
 	\includegraphics[scale=0.30, angle=-90]{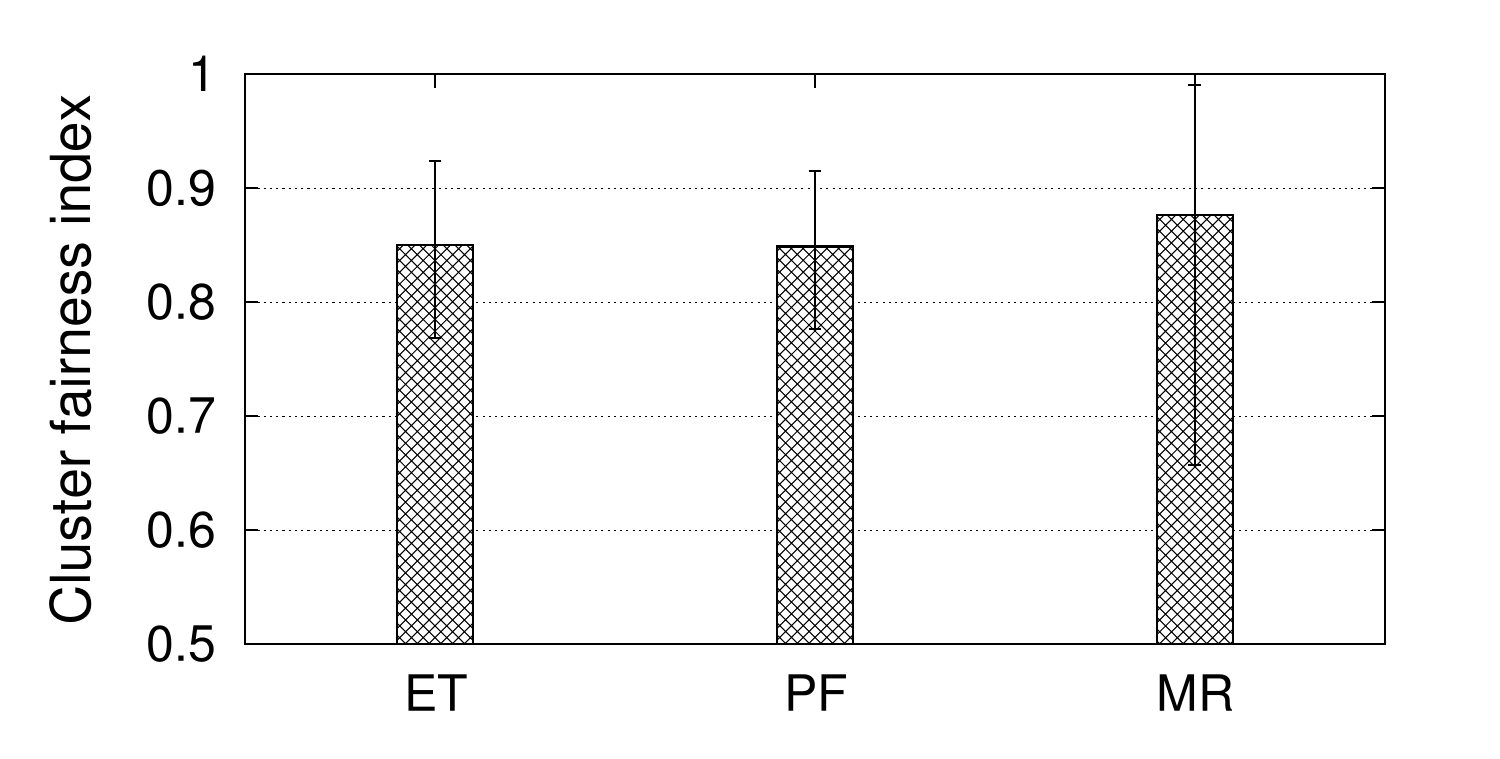} %[width=70mm,angle=0]
	\label{fig:cluster_fairness}
	\hspace{-0.8cm}
}	
\vspace{-1mm}
\caption{Fairness achieved under different schedulers (average plus $5th$ and $95th$ percentiles), with randomized tie-breaking.}	
\vspace{-4mm}
\end{figure}

For reference, Figure~\ref{fig:T} shows the difference in throughput achieved by ET, PF and MR schedulers as a function of the number of clusters in the network. Cluster sizes are
chosen at random, ranging uniformly from $1$ to $10$ members. Of course, the throughput of ET and PF only depends on the number of mobile
users and their channel qualities, but we keep using the number of clusters as reference. Interestingly, MR can double the throughput of ET and outperform PF by more than $20\%$. Most
importantly, MR can nearly achieve $100\%$ of the achievable throughput.

\begin{figure}[t!]
\centering
        \includegraphics[scale=0.4, angle=0]{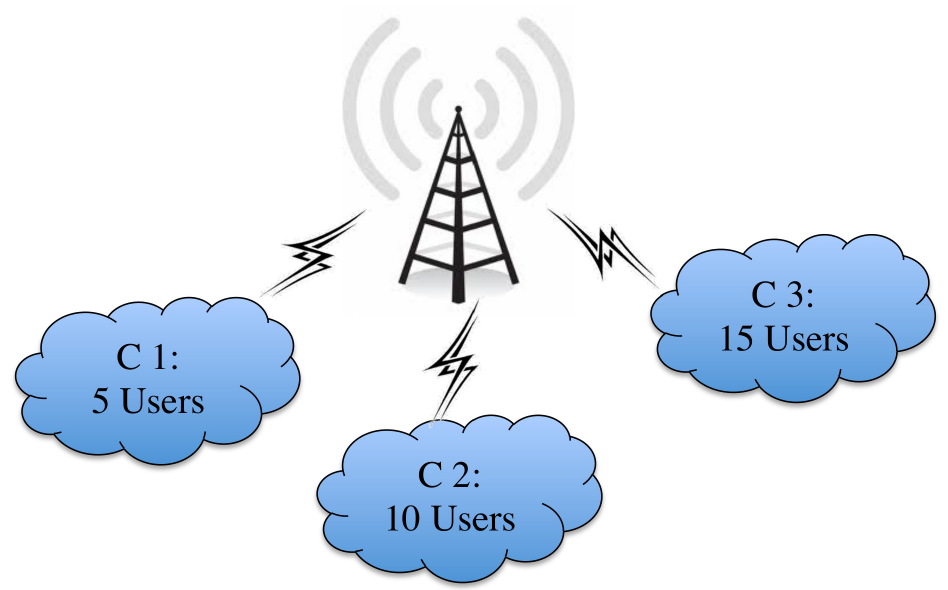} %[width=70mm,angle=0]
                %\vspace{-0.5em}
                \caption{Example scenario: three clusters in a base station, with five, ten and fifteen mobile users, respectively.}
                %\vspace{-4mm}
                \label{fig:single_cl_oth}
\end{figure}

We now zoom into the performance experienced in the different clusters. Specifically, we consider a fixed topology, which
is the simple one depicted in Figure~\ref{fig:single_cl_oth}. Clusters C$1$, C$2$ and C$3$ share the same base station and have $5$,
$10$ and $15$ mobile users, respectively. 
Figure~\ref{fig:1bs_user_tput} shows the average throughput achieved by mobile users belonging to different clusters.
%, when the MR is adopted to schedule the three clusters. In the Figure, throughput achieved by a cluster is distributed equally by the mobile
%users forming the cluster, so it represents the average per-user throughput. 
The aggregate throughput per cluster is shown in
Figure~\ref{fig:1bs_cluster_tput}. For reference, Figure~\ref{fig:1bs} also reports the throughput achieved with ET and PF
schedulers without clustering (per-cluster throughput is then computed as the sum of throughputs achieved by each member
separately). 
The high variability exhibited by MR is a drawback due to its greedy behavior (i.e., its random
tie-breaking strategy),  and to the occurrence of unbalanced clusters in our simulations (e.g., clusters with only {\it good} users will achieve extremely high throughput as compared to clusters with only {\it poor} users).
Furthermore, clustering helps in terms of {\it user} fairness, as shown in Figure~\ref{fig:user_fairness}, 
where the Jain's fairness index~\cite{jain1984} among users is graphically depicted. As it can be seen in the figure, ET and PF are both 
outperformed by MR both in terms of per-user and per-cluster fairness. In practice, not only MR exhibits the highest throughput by far, but it also reduces unfairness. %, in which throughput is sensibly higher and unfairness is halved. 
%It is interesting to observe that our clustering proposal not only increases the throughput and fairness between users, 
%but it also increases, on average, the fairness index among {\it clusters}, see Figure~\ref{fig:cluster_fairness}. 
However, the variability shown by a pure MaxRate approach with random tie-breaking (MR) is
high, which means there are great potentials for improvements.
We next evaluate tie-breaking alternatives to further improve fairness.

\subsection{Mapping Clusters to Leaves in BeLF and WoLF}

Before proceeding with the full evaluation of the heuristics proposed in Section~\ref{s:mrmf_multi}, let us recall that in our
binary tree-based schemes, BeLF and WoLF, the leaves represent the entities to be scheduled. In principle, the order in which
leaves in the binary tree are associated to connections (either users or clusters) is not necessarily related to topological considerations. In fact,
nodes represent fictitious groups, not necessarily clusters. The only aim of defining fictitious groups consists in allowing a
simple computation of WRR tie-breaking priorities for the real entities to be scheduled.
%Therefore, BeLF and WoLF heuristics  can be used to break ties in cluster scheduling.
However, the way in which connections are grouped in the binary tree affects the achieved fairness level.

\begin{figure} [t!]
\begin{center}
%\vspace{-0.6cm}
\subfigure[3 clusters.]
{
	\includegraphics[scale=0.7]{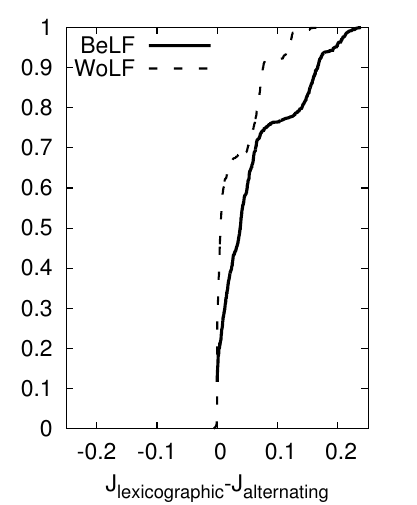}
	\label{fig:JDiffTree-3cl-1to10}
%	\vspace{-0.3cm}
}
\subfigure[6 clusters.]
{
	\includegraphics[scale=0.7]{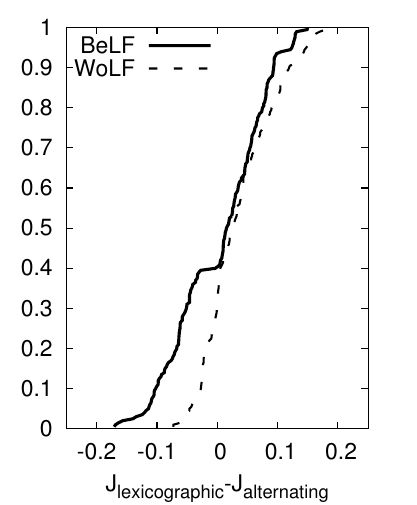}
	\label{fig:JDiffTree-6cl-1to10}
%		\vspace{-0.3cm}
}
%\vspace{-0.2cm}
\caption{CDF of the difference in the Jains' fairness indexes computed with the lexicographic mapping and with the alternating mapping (random clusters with size: $1$ to $10$ users, each of which an be {\it poor}, {\it average} or {\it good} with the same probability).}
\label{fig:JDiffTree-1to10}
%\vspace{-1.2cm}
\end{center}
\end{figure}

Let us refer to connections as clusters, and associate each cluster with a {\it goodness} metric, which consist in counting the number of {\it good} users first,
then the number of {\it average} users and eventually the number of {\it poor} users.
%One point that we left open in the description of our binary tree-based schemes (BeLF and WoLF) is the way clusters
%are mapped onto leaves of the binary tree.
Using this metric in our simulations, we have noticed that the highest fairness is achieved under one of two particular
mappings. The first is a lexicographic mapping, i.e., connections are sorted from the best to the worst, and mapped in  this order
onto the leaves of the tree, from left to right. The second mapping consists in sorting connections according to an alternating
order, i.e., the best  connection  first, then the worst, then the second best, followed by the second worst, and so on.
Figures~\ref{fig:JDiffTree-3cl-1to10} and~\ref{fig:JDiffTree-6cl-1to10} depict two examples of \ac{CDF} of the difference in
Jain's indexes achieved by lexicographic and alternating mappings, respectively $J_{lexicographic}$ and $J_{alternating}$.
Figure~\ref{fig:JDiffTree-3cl-1to10} shows that with $3$ connections both BeLF and WoLF yield better results with the lexicographic
mapping. In contrast, Figure~\ref{fig:JDiffTree-6cl-1to10} shows that when the number of  connections increases to $6$, it is
not clear which mapping is better, even though the difference is quite limited with very high probability (the \ac{CDF} grows quite
sharply around $0$).

However, since we are interested in the potential performance of \ac{D2D}-based clustering systems, in the following, in order to
compare BeLF and WoLF to the other proposed schemes, we will show results computed with the best mapping of clusters to tree
leaves, which we found by testing all possible permutations. 
%Of course, this constitutes a serious obstacle towards the adoption of BeLF and WoLF in real systems, although these schemes remain of interest due to their simplicity.

%\vspace{-2mm}
\subsection{Comparison of the Heuristics for MaxRate with WRR Tie-breaking}

We now compare our proposed MaxRate variants based on the four heuristics introduced to compute the weights for the WRR
tie-breaking. Specifically, we compare BeLF, WoLF, FISh, and PIKe in terms of per-cluster fairness. As a reference, we report
the fairness indexes achieved by a plain MaxRate scheme where ties between connections are broken randomly (MR in the figures). We
also compare results achieved by ET and PF. 
Note that, according to the common understanding, PF should have much higher
fairness than MaxRate-based approaches~\cite{kwan2009SPL}, while we show that the opposite is true under cooperative \ac{D2D}
communications approaches,  in which connections represent clusters. 
For ET and PF, we numerically simulate the scheduling of single users, then we sum up the
throughput of users according to which cluster they belong to.
% in the computation of MaxRate variants.

\begin{figure*} [t!]
\centering
%	\vspace{-0.8cm}
	\hspace{-3mm}	
	\includegraphics[scale=0.78, angle=0]{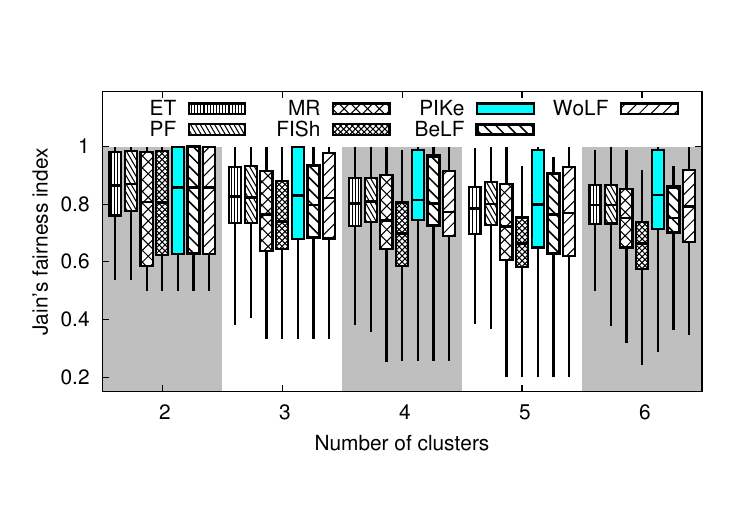}
%	\vspace{-1.5cm}
	\caption{Comparison of fairness achieved with ET, PF, MR, and MaxRate with WRR tie-breaking (cluster size: $1$ to $10$ users).}
%	\vspace{-0.7cm}
	\label{fig:JBW-1-10}
\end{figure*}
\begin{figure*} [t!]
\centering
%	\vspace{-0.7cm}
		\hspace{-3mm}	
	\includegraphics[scale=0.78, angle=0]{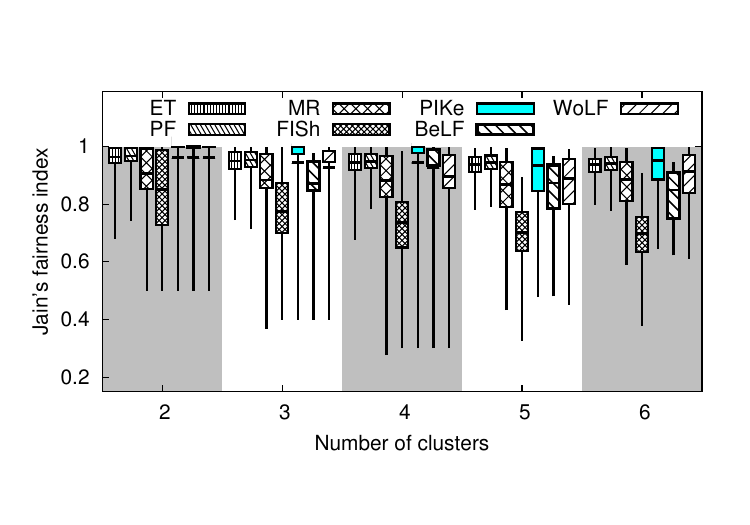}
%	\vspace{-1.5cm}
	\caption{Comparison of fairness achieved with ET, PF, MR, and MaxRate with WRR tie-breaking (cluster size: $5$ to $10$ users).}
%	\vspace{-0.7cm}
	\label{fig:JBW-5-10}
\end{figure*}

Figure~\ref{fig:JBW-1-10} reports fairness indexes for the different scheduling schemes for systems with $2$ to $6$ clusters,
each formed by $1$ to $10$ members. Fairness indexes are shown in terms of box and whiskers plots, reporting minimum and
maximum values recorded over the set of simulations performed, the $25th$ and $75th$ percentiles of their values, and the
average (the solid dashes in the boxes reported in the figure). Interestingly, the level of fairness achieved by our proposed
schemes is very high, and PIKe achieves as much fairness as PF. Recalling that the throughput achieved by PF is much lower than
the one achieved by PIKe, this result is very encouraging.

Even more interestingly, simulations accounting for clusters of at least $5$ users reveal that PIKe can outperform ET and PF in
terms of fairness, as depicted in Figure~\ref{fig:JBW-5-10}, where the boxes delimited by the $25th$ and $75th$ percentiles are
very close to $1$ for the PIKe scheme.

Figures~\ref{fig:JBW-1-10} and \ref{fig:JBW-5-10} show that PIKe, BeLF, and WoLF clearly outperform the MR cluster scheduler,
which justifies the work carried out in this manuscript. Comparing MR and PIKe, it can be seen that PIKe reduces the distance from
perfect fairness (i.e.,~$\!1$) by $50\%$.
%one can notice how results achieved with the latter
%scheme reduce by $50\%$ the distance from perfect fairness (i.e.,~$\!1$) in comparison to MR results.
Note also that FISh achieves significantly poorer performance than the benchmarking policies, especially as the number of
clusters increases.

\begin{figure} [t!]
\begin{center}
	%\vspace{-0.8cm}
	\includegraphics[scale=0.7, angle=0]{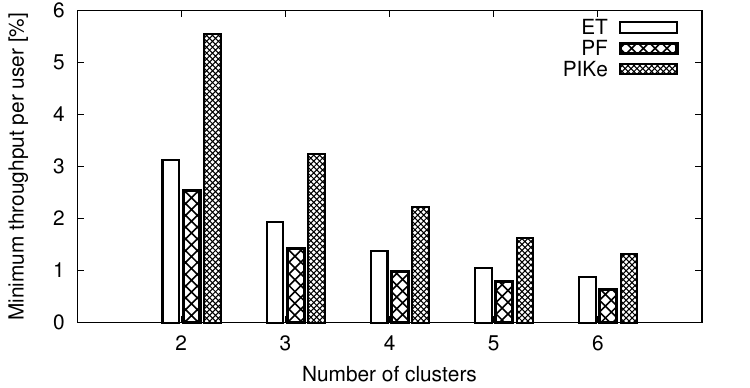}
%	\vspace{-0.5cm}
	\caption{Minimum throughput attained by a cluster member (average over $2000$ simulations---Cluster size: $5$ to $10$ users).}
%	\vspace{-1.1cm}
	\label{fig:minimalia}
\end{center}
\end{figure}

To conclude, we remark that our proposed schemes, and in particular PIKe, are beneficial for both throughput and
fairness, which means that they would allow {\it better worst-case performance} in comparison to ET and PF. Indeed,
Figure~\ref{fig:minimalia} shows that the minimum throughput received by a cluster member in the system, using PIKe, is much
higher that the one achieved with ET (by a factor $\sim 1.5$ or more), and PF (by a factor $\sim 2$).

\section{Summary}
In this chapter, we have shown how to attain maximal throughput in cellular networks without paying any fairness penalty. 
Although scheduling ties are usually ignored and uniformly random tie-breaking is an accepted practice, we have shown that this practice is rather inefficient. Our simulations indicate that there is a great potential for fairness and throughput enhancements in customized tie-breaking as soon as tie probabilities become relevant, e.g., by using opportunistic clustering strategies.
%To enhance performance, we have proposed to leverage scheduling strategies in which \ac{D2D} communications play a fundamental role. 
We have rigorously formulated a tie-breaking mechanism that achieves maximal fairness {\it and} maximal throughput. The complexity of the optimal approach is not tractable for more than two connections, so  we have proposed PIKe, a heuristic inspired by the optimal solution for the case of $2$ connections. Our results confirm that PIKe achieves almost perfect fairness and maximal throughput and largely benefits of \ac{D2D}-enabled opportunistic clustering schemes.

%\acresetall
\chapter{\titEighthCh}
\label{ch:conclusion}

This dissertation is dedicated to study and to explore different aspects of \ac{D2D} communications in cellular networks. In Chapters~\ref{ch:intro} and~\ref{ch:background}, we provided the reader an introduction to \ac{D2D} communications and opportunistic scheduling, and a thorough survey of the-state-of-the-art in the field. In our survey, we shed light on the open research problems in \ac{D2D} communications, some of which are addressed in this dissertation. In Chapter~\ref{ch:clus_theo}, we proposed an opportunistic clustering scheme leveraging outband \ac{D2D} communications and we provided a theoretical analysis of throughput and energy consumption of our proposal. In Chapter~\ref{ch:clus_proto}, we devised a protocol for our proposed architecture in Chapter~\ref{ch:clus_theo}. Chapter~\ref{ch:clus_exp} elaborated our testbed setup and experimental evaluation of our proposal. In Chapter~\ref{ch:mode}, we demonstrated the true potential of \ac{D2D} communications when both inband and outband modes are present. In Chapter~\ref{ch:tie}, we exploited \ac{D2D} communications and smart tie-breaking techniques to achieve maximal fairness using MaxRate scheduling algorithm.  

%First, we analyze network-controlled opportunistic \ac{D2D} clustering from a theoretical and practical point of view. 
Our theoretical results illustrate that using simple schedulers and game theory approaches, our proposed architecture significantly outperforms legacy schedulers in terms of throughput, delay, energy efficiency, and fairness. In particular, our opportunistic scheme achieves up to $100$\% throughput gain and $53$\% energy efficiency gain with respect to legacy schemes. Our results also confirmed that our D2D-clustering scheme results in almost perfect user fairness. We also analyzed the practicality of implementing opportunistic \ac{D2D} communications in cellular networks using WiFi Direct and LTE-A. Our proposed protocol proved that not only \ac{D2D}-assisted cellular communications are practical, but also that they can be achieved with minimal modifications to the current architecture of LTE-A and WiFi Direct. 
We prototyped the first \ac{SDR} platform for outband \ac{D2D} communications. We leveraged Xilinx \acp{FPGA} and the NI Real-Time OS to develop realistic experiments with LTE-like millisecond \ac{CQI} reporting, scheduling, and high-speed LTE-WiFi interaction. Our experimental evaluation using several \ac{QoS} and QoE metrics confirms the feasibility and potentials of opportunistic outband \ac{D2D} communications. In particular, we designed \ac{DORE} which is a 3GPP ProSe-compliant and QoS-aware opportunistic outband \ac{D2D} framework. The results reveal that experimental performance figures are lower than the reported values in the prior analytical studies, although still notable (up to $20$\% with just two users). Nevertheless, high throughput gains are achievable if the number of participating UEs in opportunistic outband \ac{D2D} increases (up to $71$\% with five users). 

We did not confine our research to a single \ac{D2D} mode and we investigated the problem of \ac{D2D} mode selection in a multimode \ac{D2D} setup to better evaluate the true potential of \ac{D2D} communications. We showed that the performance of \ac{D2D} modes is highly scenario-dependent. To cope with this issue, we proposed the {\it Floating Band \ac{D2D}} framework along with practical heuristics suitable for quick and adaptive mode selection in such a complex setup. Unlike existing schemes, we allow \ac{D2D} users to communicate over inband or outband modes, depending on network load and channel conditions. Our results demonstrate the impressive potentials of multi-band mode selection. Remarkably, our simple heuristics result in fair operation and achieve near optimal performance by dramatically ameliorating network utility, which accounts for both throughput and energy consumption. In particular, we achieved up to $231$\% throughput gain in comparison to legacy cellular networks. 

Finally, we showed that scheduling ties are significant but forgotten resources that can be exploited to reduce unfairness among mobile users. To this aim, we proposed novel solutions for leveraging the capabilities of \ac{D2D} communications to achieve higher fairness without affecting the throughput optimality of MaxRate scheduler. We first analyzed the tie-breaking problem mathematically. Next, we designed smart tie-breaking mechanisms to exploit these ties to improve the fairness of MaxRate scheduler. Our evaluation results indicate that using \ac{D2D} communications and smart tie-breaking, we ameliorate the network performance in terms of throughput and fairness.

The research performed in this dissertation resulted in eight conference/workshop papers~\cite{asadi2013WD, asadi2013MSWIM, asadi2015wowmom, asadi20155garch, asadi2015WInnComm, asadi2014Mama,vitale2015Greencom}, four journal articles~\cite{asadi2014survey, asadi2013survey, asadi2014ComCom, asadi2014Per}, one magazine article~\cite{asadi2015Commag},  four posters~\cite{asadi2013energy, asadi2013IMDEA, asadi2014IMDEA, asadi2015IMDEA} and one demo~\cite{asadi2015EUCNC}.

%Therefore, large clusters are not desirable as the subject the cluster-head to high computational load and subject other members to higher delay.
%\blue{The achieved performance is evaluated under ideal conditions; however, the real world implementation may or may not perform as good. We are now developing a flexible LTE testbed in the framework of CROWD project in order to evaluate the real performance of our proposal and its corresponding  implementation effort. }

%\label{part:energy}
%\include{./05-rateAdapt/rateAdaptHead}  % Rate Adaptation for Green Networking
%\clearemptydoublepage
%\include{./06-servers/serversHead}      % Server characterization
%\clearemptydoublepage
%\include{./07-vma/vmaHead}			    % VMA Assignment
%\clearemptydoublepage
%\include{./concFutW/conclusions}        % Conclusions
%\clearemptydoublepage

% Appendixes
\chapter*{Appendices}
\addcontentsline{toc}{chapter}{Appendices}
\appendix

\section*{Appendix A}
\label{s:appA}
\addcontentsline{toc}{section}{Appendix A}

\subsection*{Proof of Proposition~\ref{p:Ph-W}}
\begin{proof}
In CL$($WRR$)$, each cluster $CL_n$ is scheduled with probability $w_n$. When cluster $CL_n$ is selected by the scheduler, 
its user with the highest SNR is actually scheduled. For each possible MCS $k$, user $i  \in CL_n$ is the one experiencing 
the highest SNR in its cluster with probability $P_h^{(i|k)} = Pr(C_i > Y_i | MCS_i  = k)$, $Y_i  =  \max_{j \in {\textit CL_n}\setminus\{i\}}\{C_j\}$. 
Since channels are independent, using the total probability formula yields $P_h^{(i|k)} = \int_0^{\infty} \left[ 1 - F_i(z|MCS_i  =  k)\right] dF_{Y_i}(z)$.
Given that $\pi_k^{({\textit CL_n})}$ represents the probability that cluster $CL_n$ can be scheduled with the $k$-th MCS, 
the result follows by applying again the total probability formula for the discrete set of MCS values. 
\end{proof}

\subsection*{Proof of Proposition~\ref{p:Tc-M}}
\begin{proof}
The proof is similar to the proof of Proposition~\ref{p:Ph-W}. In CL$($MR$)$, the scheduled cluster $CL_n$ receives  all resources $S_{tot}$, given that the cluster contains the user with the highest SNR in the system. 
For each possible MCS $k$, cluster $CL_n$ is the one containing the user experiencing 
the best SNR with probability $P_h^{({\textit CL_n}|k)} = Pr(X_n > Y_n | MCS_{\textit CL_n}  =  k)$, with $X_n  =  \max_{j \in {\textit CL_n}}\{C_j\}$, 
and $Y_n  =  \max_{j \not \in {\textit CL_n}}\{C_j\}$. 
Since channels are independent, using the total probability formula yields $P_h^{({\textit CL_n}|k)} = \int_0^{\infty} \left[ 1 - F_{X_n}(z|MCS_{\textit CL_n}  =  k)\right] dF_{Y_n}(z)$.
Given that $\pi_k^{({\textit CL_n})}$ represents the probability that cluster $CL_n$ is scheduled with the $k$-th MCS, with which the transmitted bits per symbol are $b_k$, the result follows by applying
the total probability formula for the discrete set of MCSs. 
\end{proof}

\subsection*{Proof of Proposition~\ref{p:Ph-M}}
\begin{proof}
In CL$($MR$)$ a user is scheduled when it has the highest SNR. 
Therefore, for each possible MCS $k$, user $i$ is scheduled with probability $P_h^{(i|k)} = Pr(C_i > Y_i | MCS_{i} = k)$, with $Y_i  =  \max_{j \not = i}\{C_j\}$. 
Since channels are independent, using the total probability formula yields $P_h^{(i|k)} = \int_0^{\infty} \left[ 1 - F_i (z|MCS_{i}  =  k)\right] dF_{Y_i}(z)$.
Given that $\pi_k^{(i)}$ represents the probability that user $i$ can be scheduled with the $k$-th MCS, the result follows by applying
the total probability formula for the discrete set of MCS values. 
\end{proof}

\subsection*{Proof of Proposition~\ref{p:RWIFI}}
\begin{proof}
Due to our stationary traffic and channel quality assumptions, the traffic distribution over a scheduling interval is the same as the long term 
distribution of throughputs within the cluster $CL_n$.
Let us denote by $\delta_i$ the ratio between the user's throughout $E[T_i]$ and the total cluster throughput $E[T_{{\textit CL_n}}]$. 
Therefore, the traffic sent over WiFi, $R_{tx}^{(i, wifi)}$, is a fraction $1-\delta_i$ of the traffic received over the LTE interface
by user $i$, which yields Eq.~\eqref{eq:RWIFI-rx}
Similarly, the WiFi transmission data rate $R_{rx}^{(i, wifi)}$ corresponds to a fraction $\delta_i$ of all the traffic delivered by 
LTE, when user $i$ is not the cluster head, which yields Eq.~\eqref{eq:RWIFI-tx}.
\end{proof}

\subsection*{Proof of Proposition~\ref{p:Pa}}
\begin{proof}
$P_a^{(i)}$ is the sum of two terms: the probability that user $i$ is the cluster head and
sends traffic to other cluster members, and the probability that user $i$ 
is not cluster head and receives its packets from the cluster head.
Since such probabilities can be interpreted as the average fraction of time spent in either in reception or transmission over the WiFi interface, 
we have $P_a ^ { ( i ) } = 	( 1 - \delta _ i ) \frac { R _ { rx } ^ { ( i, \, lte ) } } { R _ {wifi} }	 + \delta _ i  \frac { E [ T _ { C _ n } ] - R _ { rx } ^ { ( i, \, lte ) } }  { R _ {wifi} }$,
which leads to the result.
\end{proof}

\newpage

\section*{Appendix B}
\label{s:appB}
\addcontentsline{toc}{section}{Appendix B}

\subsection*{Transmission Rates and Cluster MCS Selection}
\label{s:appendix_mcs}

The instantaneous achievable rate of connection $ n $ at slot $ t $,  $ R_n(t) = r_k $, depends on the adopted \ac{MCS} $ k = 1, 2, \dots, K $. We assume that the actual MCS for connection $ n $ at slot $ t $ is selected as a function of the instantaneous SNR $ C_{ n } ( t ) $, i.e.:
\begin{align}
& R(t) = r_k \iff MCS_{ n } ( t ) = k  \iff C_{ n } ( t ) \in \left[c_{ k }; c_{ k + 1 } \right[, &
\\
& 0 = c_{ 1 } < c_{ 2 } < \dots < c_{ K } < c_{ K + 1 } = \infty, &  \nonumber
\\
& 0 = r_{ 1 } < r_{ 2 } < \dots < r_{ K }. & \nonumber
\end{align}

Therefore, the probability $ p_{ n , k } $ that a scheduled connection $ n $ receives data encoded according to the $ k $-th MCS is:
\begin{align}
\label{e:channel_state_prob}
p_{ n , k } & = \int_{c_k}^{c_{k+1}} dF_n(z)
	 = e^{-\frac{c_k}{\gamma_n}} -  e^{-\frac{c_{k+1}}{\gamma_n}}.
\end{align}

%We assume that the selection of Modulation and Coding Scheme (MCS) is perfect (i.e., transmissions are affected by negligible error rate), so that we ignore
%retransmission mechanisms. Eventually, we consider a system with no power control, which is typical for realistic downlink
%transmission schemes.

Table~\ref{tb:MCS} shows a list of possible modulation and coding schemes for LTE-like networks~\cite{sesia2011lte}, their
coding rate, and the SNR threshold (in {\it dB}) that has to be reached to achieve a negligible error rate. The table
also contains the net transmission rate, in bits per symbol, achieved with each MCS. The  \ac{IM} in
Table~\ref{tb:MCS} is a value that represents the noise due to non-ideal receiver. In our simulation, MCS thresholds $c_k$ include both SNR and IM.

%For the sake of tractability, in what follows we assume that mobile users belong to one of three predefined SNR {\it classes},
%which correspond to {\it poor}, {\it average}, and {\it good} average SNR. The designated  average SNR for different classes are chosen
%in a manner that the mean achievable rates for \textit{poor}, {\it average}, and \textit{good} users are $20\%$, $50\%$, and
%$80\%$ of the maximum transmission rate achievable in the system, respectively. Therefore, with the MCS values reported in
%Table~\ref{tb:MCS} and the assumed Rayleigh fading model, the SNR to be used in Eq.~\eqref{e:channel_state_prob} for
%\textit{poor, average} and \textit{good} users is $\gamma_n = 7\,\textit{dB},\, 16\,\textit{dB},\, 23\,\textit{dB}$,
%respectively.

\newcolumntype{H}{>{\setbox0=\hbox\bgroup}c<{\egroup}@{}}%this is used to delete the SINR column from the table

\begin{table}[h!]
\centering
\caption{Modulation and coding schemes and their thresholds}
\label{tb:MCS}
\begin{tabular}{|c|c|c|c|c|H c|}
\hline
\textbf{Modulation} & \textbf{Coding} & \textbf{SNR} & \textbf{IM} & \textbf{SNR+IM} & \textbf{SNR} &\textbf{Bits per} \\
 & \textbf{Rate} &({\textit dB})&({\textit dB})&({\textit dB}) & & \textbf{symbol} \\
\hline
\multirow{8}{*}{QPSK}
& 1/8 & -5.1 &\multirow{8}{*}{2.5} & -2.6 & 0.54 &0.25\\
& 1/5 & -2.9 &                               & -0.4 & 0.91 &0.4\\
& 1/4 & -1.7 &                               & 0.8 & 1.2 &0.5\\
& 1/3 & -1 &                                  & 1.5 & 1.41 &0.67\\
& 1/2 & 2 &                                   & 4.5 & 2.81 &1 \\
& 2/3 & 4.3 &                                & 6.8 & 4.78 &1.3\\
& 3/4 & 5.5 &                                & 8.0 & 6.3 &1.5\\
& 4/5 & 6.2 &                                & 8.7 & 7.41 &1.6\\
\hline
\multirow{4}{*}{16QAM}
& 1/2 & 7.9 & \multirow{4}{*}{3}& 10.9& 12.30 & 2 \\
& 2/3 & 11.3 &                          &14.3 & 26.91 & 2.66\\
& 3/4 & 12.2 &                          &15.2 & 33.13 & 3\\
& 4/5 & 12.8 &                          &15.8 & 38.01 & 3.2\\
\hline
\multirow{3}{*}{64QAM}
& 2/3 & 15.3 &\multirow{3}{*}{4} & 19.3 &  85.11  & 4\\
& 3/4 & 17.5 & 			          & 21.5 & 141.25 &4.5\\
& 4/5 & 18.6 &                            & 22.6 & 181.97 &4.8\\
\hline
\end{tabular}
\end{table}

%\subsection{Per-cluster MCS Selection}
%\vspace{-1mm}

%\begin{figure*}[!t]
%\centering
%	\includegraphics[scale=.7, angle=0]{./figs/fig01.eps}
%	\caption{Channel quality heterogeneity affects the probability that perfect fairness can be achieved by means of MaxRate scheduling with non-random tie-breaking. Large heterogeneity practically prevents fairness.}
%	\label{fig:achievability01}
%\end{figure*}

%In the previous discussion we had implicitly associated every connection with a single user. However, mobile users may form
%clusters using cooperative D2D communications, in which only one of the users, namely the {\it cluster head}, connects at a
%given frame to the base station and relays traffic for the other users. A cluster is formally defined as follows:
%
%\begin{defn}{\bf (Cluster)}
%A cluster is a group of mobile users that can communicate with each other using an acceptable data rate, typically more
%advantageously (in some sense that may depend on each user) than with the cellular base station. Only one cluster member,
%namely the cluster head, is allowed to receive data from the base station within each cellular transmission frame.
%The downlink traffic received at the cluster head can belong to any of the cluster members.
%\end{defn}

In our scheme, we adopt the MaxRate scheduling algorithm to maximize the utilization of cellular resources. The cluster that contains the user with the highest MCS is scheduled, so we propose
to operate clusters in opportunistic way: $(i)$ the cluster head can change on a per-frame basis, as it is
opportunistically selected as the cluster member with the highest current MCS rate; $(ii)$ an entire cluster is scheduled as an
individual user whose MCS is the highest among members; $(iii)$ the cluster head relays the downlink packets to the final
destination (intra-cluster communications) on a secondary wireless interface, using D2D communications. Therefore, in this
work, a {\it connection} $n$ is a cluster (also indicated as CL$_n$) composed by $m_n$ mobiles, and its instantaneous SNR is
the {\it highest} SNR among the mobile users composing the cluster. In particular, the probability $p_{\text{CL}_n,k}$ that a scheduled connection
$ n $ (i.e., cluster CL$_n$) receives data encoded according to the $ k $-th MCS can be computed based on the SNR CDF and the MCS thresholds used in LTE, similarly to Eq.\eqref{e:channel_state_prob}:
\begin{align}
\label{e:channel_state_prob_cluster}
p_{\text{CL}_n,k} & = \int_{c_k}^{c_{k+1}} dF_{\text{CL}_n}(z),
\end{align}
where $F_{\text{CL}_n}(z)$ is the CDF of the maximum of $m_n$ random variables representing the SNR values of each of the
$m_n$ mobiles forming cluster CL$_n$:
\begin{equation}
\label{e:CDF_cluster}
F_{\text{CL}_n}(z)   =  \prod_{j \in {\text{CL}_n} } F_j(z) =   \prod_{j \in {\text{CL}_n} }  \left (  1 - e ^ { - \frac { z } { \gamma _ j} } \right ), \; z \ge 0, \; n \in \mathcal{C}.
\end{equation}

%For simplicity of notation, we omit the ``CL'' index in the formulas in the reminder of this paper, so that expressions like
%$p_{n,k}$ and $F_n(z)$ can be equivalently used for connection $n$ and cluster CL$_n$. Similarly, we will use the notation
%``connection $n$'' to address either the mobile user $n$ or the cluster CL$_n$, according to the context. 
%This use of notation
%stems from the fact that, in our system, a cluster is scheduled as if it were a user, and the only thing formally
%differentiating users from clusters is the structure of the SNR CDF (Eq.~\eqref{e:CDF_user} holds for a single user, while
%Eq.~\eqref{e:CDF_cluster} holds for an entire cluster). 

%We note that in presence of flows, the probabilities $p_{n,k}$ can be
%adapted to represent the steady-state probability given the arrival and flow-size distributions.

\subsection*{Proof of Proposition~\ref{prop:Existence}}
\label{s:appendix}
%\vspace{-1mm}

\begin{proof}
Observe first that $ R^{ ( n ) } $ is the minimum achievable transmission rate for user $ n = \{1, 2\} $ under the MaxRate
scheduler, since this user must be served if it has an MCS strictly better than the other user. Moreover,
$ R^{ ( n ) } + R^{ ( X ) } $ is the maximum achievable transmission rate for user $ n = \{1, 2\} $ under the MaxRate
scheduler, since this user must be served if it has an MCS strictly better than the other user and, in addition, it can be served at most in all the ties.
Then, fairness cannot be achieved if either $ R^{ ( 1 ) } > R^{ ( 2 ) } + R^{ ( X ) } $ or $ R^{ ( 2 ) } > R^{ ( 1 ) } + R^{ (
X ) } $, which is equivalent to $ \left| R^{ ( 1 ) } - R^{ ( 2 ) } \right| > R^{ ( X ) } $.
%In contrast, fairness can be achieved if \eqref{e:cond2} holds, as shown in \autoref{prop:Scheduler2users}. Therefore,
In contrast, fairness can be achieved if \eqref{e:cond2} holds, as shown in~\eqref{prop:Scheduler2users}. Therefore,
the equivalence holds.
\end{proof}

\subsection*{Proof of Proposition~\ref{prop:suffcond}}
\label{a:p2}

\begin{proof}
Item 1) is a special case of item 2), which holds because
\begin{align}
\left| \sum_{ k = 2 }^{ K } r_{ k } \left( p_{ 1, k } Q_{ 2, k } - p_{ 2, k } Q_{ 1, k } \right) \right| \le \sum_{ k = 2 }^{ K } r_{ k } \left| p_{ 1, k } Q_{ 2, k } - p_{ 2, k } Q_{ 1, k } \right|.
\end{align}
Item 3) can be proved as follows: $ p_{ 2, K } \ge 1/2 $ is equivalent to $ p_{ 2, K } \ge Q_{ 2, K } $, therefore by
non-negativity of probabilities it is true that
\begin{align}
- \sum_{ k = 2 }^{ K } p_{ 2, k - 1 } Q_{ 1, k } - \sum_{ k = 2 }^{ K - 1 } p_{ 2, k } Q_{ 1, k } \le p_{ 2, K } - Q_{ 2, K }.
\end{align}
By adding $ Q_{ 2, K } - Q_{ 1, K } p_{ 2, K } $ we have
\begin{align}
Q_{ 2, K } - \sum_{ k = 2 }^{ K } p_{ 2, k - 1 } Q_{ 1, k } - \sum_{ k = 2 }^{ K } p_{ 2, k } Q_{ 1, k } \le ( 1 - Q_{ 1, K } ) p_{ 2, K }.
\end{align}
By expanding $ Q_{ 1, K } $ and $ Q_{ 2, K } $ we further obtain
\begin{align}
\sum_{ k = 2 }^{ K } p_{ 2, k - 1 } ( 1 - Q_{ 1, k } ) - \sum_{ k = 2 }^{ K } p_{ 2, k } Q_{ 1, k } \le p_{ 1, K } p_{ 2, K }.
\end{align}
Realizing that the first sum equals $ \displaystyle \sum_{ k = 2 }^{ K } p_{ 1, k } Q_{ 2, k } $, we have
\begin{align}
r_{ K } \sum_{ k = 2 }^{ K } ( p_{ 1, k } Q_{ 2, k } - p_{ 2, k } Q_{ 1, k } ) \le r_{ K } p_{ 1, K } p_{ 2, K }.
\end{align}
For each $ k \ge 2 $, term $ p_{ 1, k } Q_{ 2, k } - p_{ 2, k } Q_{ 1, k } \ge 0 $ since $ p_{ 1, k } \ge p_{ 2, k } $,
therefore
\begin{align}
\left| \sum_{ k = 2 }^{ K } r_{ k } ( p_{ 1, k } Q_{ 2, k } - p_{ 2, k } Q_{ 1, k } ) \right| \le \sum_{ k = 1 }^{ K } r_{ k } p_{ 1, k } p_{ 2, k }.
\end{align}
\end{proof}

\subsection*{Proof of Proposition~\ref{prop:Scheduler2users}}
\label{a:p4}

\begin{proof}
Consider the MaxRate scheduler with randomized tie-breaking with bias $ \alpha \in [ 0, 1 ] $ for user $ 1 $. Then, the
expected time-average and one-slot individual throughputs are $ R^{ ( 1 ) } + \alpha R^{ ( X ) } $ and $ R^{ ( 2 ) } + ( 1
- \alpha ) R^{ ( X ) } $, respectively. It is straightforward to verify that, if \eqref{e:cond2} holds, plugging $ \alpha^{ ( X
) } $ for $ \alpha $ the throughput of each user is equal to $ \left( R^{ ( 1 ) } + R^{ ( 2 ) } + R^{ ( X ) } \right) / 2 $.

If \eqref{e:cond2} does not hold, then suppose that $ R^{ ( 1 ) } > R^{ ( 2 ) } + R^{ ( X ) } $ (case $ R^{ ( 2 ) } > R^{ ( 1 )
} + R^{ ( X ) } $ is analogous). The difference in the individual throughputs is $ R^{ ( 1 ) } + \alpha R^{ ( X ) } - ( R^{ ( 2
) } + ( 1 - \alpha ) R^{ ( X ) } ) = R^{ ( 1 ) } - R^{ ( 2 ) } - R^{ ( X ) } + 2 \alpha R^{ ( X ) } $, which is minimized if $
\alpha = 0 $. Indeed $ \alpha^{ ( X ) } = 0 $, because of the cut-off of a negative value given by \eqref{e:alphaX}.
\end{proof}

%\include{5.appendix02}

%\newpage

% bibliograf�a le�da y usada en la tesis

\bibliographystyle{IEEEtran}%
\bibliography{./biblio/references}
\addcontentsline{toc}{chapter}{References}
\clearemptydoublepage

\end{document}